%% file: Paper_IOP.tex
\documentclass[12pt]{iopart}
\usepackage{iopams,graphicx}
\usepackage{color}
\usepackage{graphicx}
\usepackage[pdftex,colorlinks,citecolor=blue,bookmarks]{hyperref}
\usepackage{bm}
\usepackage{url}
\usepackage{longtable}

\begin{document}

\review[Mean field and beyond with the Gogny force]%
{Mean field and beyond description of nuclear structure with the Gogny force: A review}

\author{L.M. Robledo$^{1,2}$, T.R. Rodr\'{\i}guez$^{2}$, and R. R. Rodr\'{\i}guez-Guzm\'an$^{3}$}

\address{$^{1}$ Center for Computational Simulation,
Universidad Polit\'ecnica de Madrid,
Campus de Montegancedo, Boadilla del Monte, 28660-Madrid, Spain}
\address{$^{2}$ Departamento de F\'{\i}sica Te\'{o}rica y 
Centro de Investigaci\'on Avanzada en F\'{\i}sica Fundamental, 
Universidad Aut\'onoma de Madrid, E-28049 Madrid, Spain}
\address{$^{3}$ Physics Department, Kuwait University, Kuwait 13060}
\vspace{10pt}

\ead{luis.robledo@uam.es,tomas.rodriguez@uam.es,raynerrobertorodriguez@gmail.com}

\begin{indented}
\item[] May  2018
\end{indented}

% -------------------------------------------------------------------------------------------
%                                                                        Abstract
% -------------------------------------------------------------------------------------------
\begin{abstract}
Nowadays, the Gogny force is a referent in the theoretical description 
of nuclear structure phenomena. Its phenomenological character 
manifests in a simple analytical form that allows for implementations 
of techniques both at the mean field and beyond all over the nuclide 
chart. Over the years, multiple applications of the standard many-body 
techniques in an assorted set of nuclear structure applications have 
produced results which are in a rather good agreement with experimental 
data. The agreement allows for a simple interpretation of those 
intriguing phenomena in simple terms and gives confidence on the 
predictability of the interaction. The present status on the implementation
of different many body techniques with the Gogny force is reviewed with a
special emphasis on symmetry restoration and large amplitude collective
motion.
%\begin{description}
%\item [{Usage}] Secondary publications and information retrieval purposes.{\small \par}
%\item [{PACS~numbers}] May be entered using the environment \textsf{PACS~numbers}.{\small \par}
%\item [{Structure}] You may use the \texttt{Description} environment to
%structure your abstract.{\small \par}
%\end{description}
\end{abstract}


\maketitle

% ------------------------------------------------------------------------------------------------------
%                                                                                I n t r o d u c t i o n 
% ------------------------------------------------------------------------------------------------------

\include{Introduction}

% ------------------------------------------------------------------------------------------------------
%                                                                                        The Gogny force
% ------------------------------------------------------------------------------------------------------

\include{GognyForce}

% ------------------------------------------------------------------------------------------------------
%                                                                                    M e a n   F i e l d   
% ------------------------------------------------------------------------------------------------------

\include{MeanField}

% ------------------------------------------------------------------------------------------------------
%                                                             B e y o n d   t h e   M e a n   F i e l d  
% ------------------------------------------------------------------------------------------------------

\include{BMF1}

% ------------------------------------------------------------------------------------------------------
%                                                                                  Symmetry restoration 
% ------------------------------------------------------------------------------------------------------

\include{SR}

% -------------------------------------------------------------------------------------------------------
%                                                                       Large amplitude collective motion
% -------------------------------------------------------------------------------------------------------

\include{GCM}

% -------------------------------------------------------------------------------------------------------
%                                                                       Summary
% -------------------------------------------------------------------------------------------------------
\include{Summary}

% -------------------------------------------------------------------------------------------------------
%                                                                                        Acknowledgements
% -------------------------------------------------------------------------------------------------------

\ack{We would like to express our gratitude to all our collaborators 
along the years who helped us to improve our understanding of the 
physics of nuclear structure and its applications with the Gogny force. 
Special attention deserves J. F. Berger who was extremely helpful at the 
beginning of our involvement with the Gogny force. With his help, 
encouragement and friendship one of us, LMR, was able to enter into the 
intricate world of the Gogny force. We owe a lot to Luis Egido who had 
the patience to be the Ph. D. advisor of the three of us and the long 
standing collaboration. We also appreciate very much the collaboration 
with M. Baldo, G. F. Bertsch, J. Dukelsky, J. Engel, K. Langanke, G. 
Mart\'inez-Pinedo, A. Poves, P. Ring, P. Sarriguren, K. Schmid, P. Schuck 
and X. Vi\~nas. The exchange of ideas with the people at 
Bruy\`eres-le-Ch\^atel including the old and young generations is also 
greatly appreciated. The exchange of ideas and discussions with our 
colleagues from the Skyrme and Relativistic community is also very much
appreciated. In this respect we are indebted to M. Bender, J. 
Dobaczewski, T. Duguet, P.-H. Heenen and D. Vretenar. We are also very 
grateful to  other people that along the years have collaborated with 
us in different facets of our research: M. Anguiano, A. Arzhanov, B. 
Bally, R. Bernard, M. Borrajo, M.A. Fern\'andez, F. de la Iglesia, E. 
Garrote, S. A. Giuliani, N. L\'opez-Vaquero, V. Martin, J. Men\'endez, 
K. Nomura, S. P\'erez, A. Valor, A. Villafranca, N. Schunk, Y. Sun, M. 
Warda and A. Zdeb.

The work of LMR was partly supported by  Spanish MINECO grant Nos 
FPA2015-65929 and FIS2015-63770. TRR acknowledges the support from the 
Ministerio de Econom\'ia y Competitividad (Spain) under contracts 
FIS2014-53434-P and Programa Ram\'on y Cajal 2012 number 11420. 

}

% -------------------------------------------------------------------------------------------------------
%                                                                                              Appendices
% -------------------------------------------------------------------------------------------------------

\appendix{}

% -------------------------------------------------------------------------------------------------------
%                                                                           M a t r i x   E l e m e n t s
% -------------------------------------------------------------------------------------------------------

\include{MatElem}

% -------------------------------------------------------------------------------------------------------
%                                                                                          Computer codes
% -------------------------------------------------------------------------------------------------------

\include{Codes}

% -------------------------------------------------------------------------------------------------------
%                                                                                Generalized Wick Theorem
% -------------------------------------------------------------------------------------------------------

\include{GWT}

% -------------------------------------------------------------------------------------------------------
%                                                                                           Abbreviations
% -------------------------------------------------------------------------------------------------------

\include{Abbreviations}

% -------------------------------------------------------------------------------------------------------
%                                          B I B L I O G R A P H Y      
% -------------------------------------------------------------------------------------------------------

\bibliographystyle{unsrt}
\bibliography{out2}

\end{document}

%% file: Introduction.tex
% ------------------------------------------------------------------------------------------------------
%                                                                           I n t r o d u c t i o n 
% ------------------------------------------------------------------------------------------------------

\section{Introduction}

The Gogny force, named after the renowned French physicist Daniel Gogny 
\cite{Berger2017}, has been used to describe many different facets of 
nuclear structure since its inception back in the early seventies of 
the past century. It has been mainly used in a mean field framework 
including pairing correlations where the Hartree Fock Bogoliubov (HFB) 
mean field is the basic entity used to obtain the quantities appearing 
in the theory. In this category we include not only the calculation of 
potential energy surfaces (PES) using constraints on the relevant 
collective degrees of freedom so useful to describe the shape of the 
ground state or the path to fission, but also extensions like the collective 
Schr\"odinger equation (CSE), the Bohr Hamiltonian approach focused on
quadrupole degrees for freedom and their coupling to rotations, the QRPA 
or the Interacting Boson Model (IBM) with parameters determined by HFB 
PES. As it stands, the mean field only provides wave functions in the 
intrinsic frame of reference where the mean field wave function is 
allowed to break symmetries of the Hamiltonian. In order to compute 
physical observables like mean values or transition probabilities it is 
very important to use wave functions in the laboratory frame 
which have the quantum numbers of the symmetries preserved by the nuclear
interaction. Therefore, in addition to the mean field, mechanisms to restore 
the spontaneously broken symmetries have to be applied to perform the
passage from the intrinsic to the laboratory frame. These 
mechanisms require the evaluation of overlaps between different HFB 
wave function that are subsequently used in integrals over the symmetry 
groups. This is referred to as "symmetry restoration" and its 
implementation with the Gogny force has been an active field of 
research since its first implementation in the early nineties. These 
ideas can also be used to deal with large amplitude collective motion 
in the spirit of the Generator Coordinate Method (GCM) often in 
combination with symmetry restoration. The purpose of this review is to 
describe the status of the implementation of all these techniques with 
the Gogny force putting special emphasis in those aspects related to 
symmetry restoration and large amplitude collective motion. There are 
already two long papers in the literature that overview some aspects of 
the applications of the Gogny force in nuclear structure but in our opinion they 
only offer partial views of the whole picture. In Ref \cite{Peru2014} 
the authors review applications of the mean field, the 5D Collective 
Hamiltonian and the QRPA with the Gogny force. However, the coverage of 
the mean field is rather limited and fails to account for important 
applications like the study of high spin physics, finite temperature 
HFB or the applications to odd mass nuclei and multi-quasiparticle 
excitations as in the physics of high-K isomers. On the other hand, the 
review of \cite{Egido2016a} focuses only on aspects of symmetry 
restoration and large amplitude motion, not paying much attention to 
other applications beyond the mean field. Also the formal difficulties 
encountered in the implementation of symmetry restoration techniques 
with density dependent forces are scarcely discussed and also relevant 
technical details are overlooked. In our review we have tried to give 
a complete, albeit not very deep, description of all different 
techniques used in nuclear structure paying some attention to some 
relevant technical details like evaluation of matrix elements, Pfaffian 
techniques to evaluate overlaps or computer code implementations. 
Although not belonging to the family of Gogny forces we have decided to 
include the description of translational invariance restoration with 
the Brink-Boecker interaction as an illustration of this important, and 
often overlooked, aspect of symmetry restoration. The present review and
the one of \cite{Peru2014} can be considered as complementary as we do not
cover in too much details the subjects treated by Peru and Martini. 

The theoretical description of nuclear reactions with the Gogny force 
is not described in this review. In the last years and with the help of 
increasing computing power, more and more of the nuclear structure 
microscopic input required for reactions can be obtained from sound 
theoretical models with the Gogny force 
\cite{HilaireS.2016,BlanchonG.2017,Dupuis2017}. All these models are 
described in this review but their connection with reaction theory is 
scarcely discussed in the text.

The review focuses almost exclusively on applications with the Gogny 
force. The many body techniques used to describe nuclear structure have 
also been used with other interactions/functionals like the 
non-relativistic family of Skyrme energy density functionals or the 
relativistic models with great success. Although those calculations are 
similar and in many cases complementary to the ones presented here we 
are not going to discuss them and we refer the interested reader to the 
vast literature already available in the form of reviews.

The review is divided in six sections including the Introduction, the 
second section is devoted to the description of the different 
parametrizations of the Gogny force and several recent 
improvements/departures from it. In Section 3 the mean field method, 
adapted to deal with density dependent interactions is discussed and 
several examples of application with the Gogny force are presented. In 
Section 4 two methods beyond the mean field but not requiring 
Hamiltonian overlaps are described: namely the QRPA and the IBM mapping 
procedure. In Section 5, the issue of symmetry restoration is discussed 
in general and later applications to the most common types of symmetry 
restoration (parity, particle number projection, angular momentum 
projection, linear momentum projection) are presented along with 
several applications with the Gogny force. The difficulties encountered 
in the application of the symmetry restoration techniques to the case 
of phenomenological density dependent interactions is also addressed. 
Finally, in Section 6 the standard method to deal with large amplitude 
collective motion is discussed. Among the applications, we discuss the 
application of the method with symmetry restored wave functions as well 
as the mixing of multi particle-hole excitations intimately connected 
with the Configuration Interaction method. Also approximate methods 
based on the Gaussian overlap approximation are discussed. We conclude 
with a summary and perspectives section. Several appendixes with a 
more technical information are also included.

%% file: GognyForce.tex
% ------------------------------------------------------------------------------------------------------
%                                                                           The Gogny force
% ------------------------------------------------------------------------------------------------------

\section{The Gogny force: its origins, motivation and present implementations}~\label{Sec:Gogny_force}

In this section a historical overview of the origins and motivation of 
the Gogny force is presented with special emphasis in the fitting 
protocols used in each of the different main parametrizations considered 
(D1, D1', D1S, D1N, D1M). A few paragraphs will also be devoted to the newly 
proposed D2 Gogny force with a finite range density dependent 
interaction. Finally, we also discuss less known parametrizations including
some specific terms and some forces inspired by the Gogny interaction.

\subsection{The Gogny force: guiding principles}

The Gogny force was conceived in a period of time when the Skyrme 
interaction had started to become fashionable mostly because of its 
ability to describe nuclear properties at the simple Hartree Fock (HF) 
mean field level \cite{PhysRevC.5.626,PhysRevC.7.296}. At that time, the fact that in Skyrme forces  
different interactions had to be used in the pairing and particle-hole 
(ph)  channel was considered as a drawback. Also the necessity to 
consider a window around the Fermi level where the pairing interaction 
was active, was often considered as an annoying characteristic. In order to 
have a pairing force derived from the same central potential than the 
particle-hole (ph) channel, a finite range interaction, with its natural 
ultraviolet cutoff, had to be implemented.  This is the main reason why 
the Gogny force was created: it had a finite range central potential 
that could also be used to obtain the pairing interaction. The central 
potential was inspired by early attempts by Brink and Boecker 
\cite{Brink19671}  to derive a finite range central potential with a Gaussian form for 
nuclear structure calculations. Going finite range was a technical challenge for the 
computers available at that time. However, combining together the 
simplicity of Gaussian shape for the central potential and a nice 
property of the harmonic oscillator wave functions 
\cite{TALMAN1970273}, to be discussed below, gave the opportunity to
get a reasonable implementation of  the HF  or the Hartree Fock 
Bogoliubov (HFB) mean fields  on those days computers \cite{Gog75}. 

The Gogny force consists of four terms
\begin{equation}\label{eq:force-4t}
v(1,2)=v_c(1,2)+v_\mathrm{LS}(1,2)+v_\mathrm{DD}(1,2)+v_\mathrm{Coul}(1,2)
\end{equation}
A central term $v_c$ of finite range which is a linear combination of two Gaussians and contains
the typical spin and isospin channels with the Wigner (W), Barlett (B), Heisenberg (H) and 
Majorana (M) terms
\begin{equation} \label{eq:central}
v_c (1,2)  = \sum_{i=1,2}  e^{-\frac{|\vec{r}_1-\vec{r}_2|^2}{\mu_i^2}} \left( W_i + B_i
P_\sigma -H_i P_\tau -M_i P_{\sigma} P_{\tau} \right).
\end{equation}
A two body spin orbit for zero range is taken directly from the Skyrme functional
\begin{equation}
v_\mathrm{LS}(1,2)=i W_\mathrm{LS} (\vec{\nabla}_{12} \delta (\vec{r}_1-\vec{r}_2) \wedge \vec{%
\nabla}_{12}) (\vec{\sigma}_1+\vec{\sigma}_2),
\end{equation}
a pure phenomenological  density dependent term, strongly repulsive, introduced to make the force 
fulfill the saturation property of the nuclear interaction
\begin{equation}
v_\mathrm{DD}(1,2) = t_3(1+P_\sigma x_0) \delta(\vec{r}_1-\vec{r}_2) \rho^{\alpha}((\vec{r}%
_1+\vec{r}_2)/2).
\end{equation}
This ``state dependent" part of the interaction has to be handled properly in the application
of the variational principle which is behind the HF or HFB procedures and gives rise
to a so-called rearrangement potential to be discussed below.
Finally, the standard Coulomb potential $v_\mathrm{Coul}(1,2)$ between protons is added to
the interaction. Usually, the Coulomb
potential is taken only into account in the direct channel of the HF or HFB procedures. The
exchange term, which is rather involved due to the infinite range of the interaction, is considered
in the local Slater approximation \cite{PhysRev.81.385,TITINSCHNAIDER1974397}  that comes 
in the form of an additional term to be added to the energy
\begin{equation}
E_{\rm Slater} = -\frac{3}{4} e^2 \left(\frac{3}{\pi}\right)^{1/3}
\int d^3 \; \vec{r} \left[ \rho^P (\vec{r} ) \right]^{4/3}
\label{COULEXE}
\end{equation}
and depending on the proton's density alone. This term also gives a "rearrangement" 
contribution to the HF or HFB potentials when treated appropriately in 
the application of the variational principle.

The traditional center of mass correction to the mean field energy, including both the one body and two
body components, is fully considered in all Gogny parametrizations and included
in the variational procedure. Both the contributions to the HF and pairing
(anti-pairing) fields is taken into account.

The Gogny interaction depends on 15 adjustable parameters that are obtained
after performing a fit to experimental data and nuclear matter properties.
Different parametrizations have been obtained throughout  the years
depending on the set of data and the quality of the approaches used to
solve the nuclear many body problem. For a recent discussion of the
fitting protocol see Ref. \cite{Pillet2017}.

In the recent literature it is common to catalog the Gogny force as an
Energy density functional (EDF) due to its density dependent term. In this
review we will use indistinctly the term force, interaction and EDF to refer to 
the Gogny force.

As mentioned before, one of the main assets of the Gogny force is the 
use of the same central interaction both for the Hartree Fock 
(particle-hole) and the pairing (particle-particle)part of the HFB 
procedure. This property, however, is questioned by several authors 
using several arguments (see, for instance Ref 
\cite{PhysRevC.69.054317}). The first argument has to do with the fact 
that the bare nucleon-nucleon interaction can be used in the pairing 
channel whereas for the p-h part a regularization of the repulsive core 
in the spirit of the Brueckner method is required. As a consequence, 
both the effective p-p and p-h channels  of the interaction to be used 
at the mean field can be considered as unrelated and the consistency 
between the two channels is not required. The second argument is 
related to the modern view of the nuclear interactions as tools to 
generate energy density functionals (EDF) in the spirit of the EDF in 
condensed matter physics. In the nuclear physics case, those 
functionals must also include a pairing part that can be taken, in the 
spirit of the EDF, as completely uncorrelated from the rest of the EDF 
as long as it is able to grasp all the relevant correlations. These two 
arguments are mostly invoked by practitioners of the relativistic mean 
field and also in the non-relativistic case when zero range forces 
(Skyrme like) are used. It this way the use of a different interaction 
in the pairing channel is justified and therefore the use of the same 
interaction in the Gogny force can be considered more as a limitation 
than as an advantage. Usually a phenomenological density dependent zero 
range force is used in the pairing channels, although in some cases, 
see below, the central part of the Gogny force is used for the p-p 
channel. On the other hand, the Gogny force is often considered as a 
benchmark concerning pairing properties in finite nuclei (see below). 
Also the S=0, T=1 gap in nuclear matter behaves very much the same as 
the gap of realistic interactions as a function of $k$ 
\cite{KUCHAREK1989249,PhysRevC.60.064312}. Therefore, we can conclude 
that the pairing channel of Gogny is competitive with other pairing 
interactions. Unfortunately, the same analysis can not be carried out 
for the p-h channel of the central force, but at least in nuclear 
matter (see below) it provides (depending on the parametrization) more 
than reasonable equations of state in nuclear matter. In addition, the 
freedom to consider a different interaction in the pairing channel 
comes at a price: as it will be discussed latter, beyond mean field 
approaches require the evaluation of Hamiltonian overlaps that contain 
three contributions, direct, exchange and pairing when a Hamiltonian 
operator is considered. Under some circumstances those contributions 
turn out to be divergent. Due to the magic of the symmetrization 
principle and the associated Pauli exclusion principle the divergences 
cancel out to render the overlap a finite quantity. When one or two of 
the contributions are omitted spurious non physical results are 
obtained for the overlap. Regularization procedures have been proposed 
(see below) to handle those common situations but they are of limited 
applicability. This is the main argument to use an interaction, like 
the Gogny force, that provides at the same time the p-h and p-p 
channels. 

The finite range of the central part of the Gogny force is another of 
its differentiating aspects. This is in opposition to the Skyrme like 
EDFs which contain zero range contact interactions only. However, 
gradients of the density are often introduced in those EDFs to simulate 
the effect of a finite range. So far, it is not clear whether those 
gradient terms or the finite range of the Gogny force are absolutely 
necessary to reproduce the rich and vast nuclear phenomenology. On the 
other hand, the simplifications implied by contact interactions in the 
numerical implementation of the HF or HFB methods with those forces is 
nowadays irrelevant due to the advances in computational resources.

% ------------------------------------------------------------------------------------------------------
%                                                                           D1 and D1''
% ------------------------------------------------------------------------------------------------------

\subsection{D1 and D1$^\prime$}

The first parametrization of the Gogny force was denoted D1 
\cite{Gogny75,Decharge1975} and the fitting protocol used included nuclear 
matter properties like the binding energy per nucleon, the Fermi 
momentum $k_F$ at saturation or the symmetry energy, the binding 
energies of a couple of  spherical nuclei ($^{16}$O and $^{90}$Zr), the 
energy splitting $\epsilon_{p_{3/2}}-\epsilon_{p_{1/2}}$ of single 
particle levels in $^{16}$O and a couple of relevant pairing matrix 
element. The first calculations showed a good reproduction of basic 
nuclear properties both in the ph as well as the pairing channels. 
Pairing properties were analyzed in Ref \cite{decharge1980} in the tin 
isotopic chain along with several bulk properties of spherical nuclei 
like binding energies and radii. In this reference,  the good 
performance of D1 regarding pairing correlations was clearly stablished. 
A minor readjustment 
of the spin-orbit strength $W_{LS}$ introduced to improve the 
description of binding energies of spherical nuclei led to the 
D1$^{\prime}$ parametrization \cite{decharge1980}. Some calculations of 
quadrupole deformed nuclei also seemed to indicate a good reproduction 
of experimental data \cite{Decharge1975}. The D1 parametrization was 
also used in the RPA calculations of Ref \cite{Blaizot1977}. However, 
when D1 was applied to fission barrier height calculations it became 
clear that its surface properties were not appropriate, leading to a 
too high fission barrier in the prototypical calculation of $^{240}$Pu 
potential energy surface. A refitting of D1 was in order as to reduce the 
surface coefficient in nuclear mater. The new fit, including a fission 
barrier height target, led to the D1S  parametrization 
discussed below. Since the advent of D1S, the D1 and D1$^{\prime}$ 
parametrizations were abandoned and just used to study the sensitivity 
of the results to a change in the interaction.

% ------------------------------------------------------------------------------------------------------
%                                                                           D1S
% ------------------------------------------------------------------------------------------------------

\subsection{D1S}

When the parametrization D1 (and D1') were
applied to the calculation of excited states in the framework of the 
RPA they produced too high excitation energies in the few examples 
studied. This drawback was originally associated to too strong pairing 
correlations, leading to too low collective inertias \footnote{The 
excitation energy of a collective excitation can be estimated using the 
results of the simple harmonic oscillator potential. Assuming the 
nuclear potential energy surface around the ground state as being 
characterized by a curvature $K$ and a collective inertia $M$, the 
excitation energy is given by $E_{ex}\sim \sqrt{K/M}$}. In addition, 
preliminary fission studies in $^{240}$Pu using a two centered harmonic 
oscillator basis, led to a too high second fission barrier height that 
was attributed to a too large value of the nuclear matter surface 
energy coefficient $a_{s}$. Both difficulties motivated a new 
parametrization of the force in order to reduce the amount of pairing 
correlations and the value of $a_{s}$. In this way D1S (S stands for 
surface) was proposed \cite{berger1984}. Since then, this 
parametrization has been used in a very large set of calculations aimed 
to study many different nuclear structure phenomena. Just to mention a 
bunch of relevant calculations we can mention the fission studies of 
Refs \cite{berger1984,War02,Delaroche2006} , cluster emission in Ref 
\cite{War11}, $(\beta, \gamma)$ potential energy surfaces \cite{Gir83} 
and the subsequent Bohr Hamiltonian calculations \cite{Libert1999,Delaroche2010}, 
survey of octupole properties in the GCM framework 
\cite{Robledo2011}, studies of high spin physics 
\cite{EgidoRobledo93} or finite temperature \cite{Egi00}. We can also 
mention sophisticated QRPA calculations \cite{Peru2014} or state of the 
art symmetry restoration plus the GCM to describe $^{44}$S 
\cite{Egido2016}. The general consensus nowadays is that D1S 
performs rather well in the description of experimental data in most of 
the analyzed phenomena. As a consequence, this parametrization is 
considered to have a strong ``predictive power" around the stability 
valley and it has become a ``de facto" standard in Gogny 
like calculations. In the uncharted region of very neutron rich nuclei, 
however, there is no guarantee about its performance mostly due to its 
poor behavior in describing the neutron matter equation of state. 

% ------------------------------------------------------------------------------------------------------
%                                                                           D1N
% ------------------------------------------------------------------------------------------------------

\subsection{D1N}

One of the deficiencies of the D1S parametrization was the drifting in 
binding energies along isotopic chains that was thought to be a 
consequence of the not so satisfactory neutron matter equation of state 
of D1S, as compared to more realistic calculations like the one of Ref \cite{FRIEDMAN1981502}. With this in mind, 
a new parametrization of the Gogny interaction was proposed in Ref 
\cite{chappert2008}. It was denoted D1N (N for neutron) and it 
reproduces quite well the realistic neutron matter equation of state of 
Friedman and Pandaripande (FP) \cite{FRIEDMAN1981502}. As a consequence 
of this new constraint in the fitting protocol, the drifting in binding 
energies is severely reduced while  other properties of D1S like 
pairing gaps in the tin isotopes, fission barriers in $^{240}$Pu, 
moments of inertia in rare earth nuclei or $2^+$ excitation energies 
all over the periodic table, are preserved. This parametrization, 
however, has not been used much in the literature, with some exceptions 
\cite{Rodriguez-Guzman2010b,Robledo2011,Rodriguez-Guzman2014,Robledo2015}.  

% ------------------------------------------------------------------------------------------------------
%                                                                           D1M
% ------------------------------------------------------------------------------------------------------

\subsection{D1M}

Astrophysical applications require the knowledge of the properties of 
nuclear systems which are so neutron-rich that  no experimental access 
to them can be expected in the foreseeable future. Therefore, an 
accurate modeling of the properties of those exotic nuclear systems, 
like their masses, is mandatory in order to improve  astrophysical 
predictions \cite{ARNOULD200797}. On the microscopic side, the 
Hartree-Fock-Bogoliubov (HFB) approximation based on Skyrme 
interactions (see, for example, Ref.~\cite{PhysRevLett.102.152503} and 
references therein) has already been able to reproduce 2149 
experimental masses \cite{AUDI2003337} with a root mean square 
(rms) deviation at the level of the best droplet like models 
\cite{CHAMEL200872}.

Though the parametrization D1S  \cite{berger1984} of the Gogny 
interaction reproduces a wealth of low-energy nuclear data, it is not 
suited for an accurate estimate of the nuclear masses. The same holds 
for the parameter set D1N \cite{chappert2008}. In particular, the 
parameter set D1S is well known to exhibit a pronounced under-binding 
in heavier isotopes \cite{chappert2008}. Those Gogny-like interactions 
cannot account for nuclear masses with an rms better than 2 MeV. 
Therefore, a new parametrization of the Gogny interaction, i.e., D1M 
was introduced in Ref.~\cite{goriely2009}. In addition to nuclear 
masses, other constraints were used in its fitting protocol to provide 
reliable nuclear matter and neutron matter properties but also radii, 
giant resonances as well as fission properties. A unique aspect of the 
fitting protocol of the Gogny D1M interaction is that for the first 
time, correlations beyond the mean field level, i.e., zero point 
rotational and vibrational corrections, have been taken into account 
in the binding energy via a five dimensional collective Hamiltonian (5DCH) \cite{Libert1999}.

The fitting strategy and the parameters corresponding to the Gogny 
force D1M can be found in Ref.~\cite{goriely2009}. Here, we will just 
comment on some key aspects of the fitting protocol. Both axial and 
triaxial codes were employed in the calculations to obtain the 
parametrization D1M \cite{goriely2009}. In particular, a triaxial code 
was employed to estimate the zero-point vibrational and rotational 
energy corrections to the mean field binding energies and the charge 
radii, obtained in the framework of axially symmetric calculations, via 
a 5DCH model. However, the 5DCH model leads to a wrong (negative) 
zero-point energy correction in the case of closed shell nuclei. 
Therefore, for those systems it is simply set to zero \cite{Hilaire.05}. In addition, 
an infinite-basis correction is introduced to account for the finite 
size of the  single-particle basis 
\cite{Hilaire.05}. No phenomenological Wigner terms were 
considered to obtain the D1M parameter set.

%%%%%%%%%%%%%%%%%%%%%%%%%%%%%%%%%%%%%%%
\begin{figure}
\begin{center}
\includegraphics[width=0.45\textwidth]{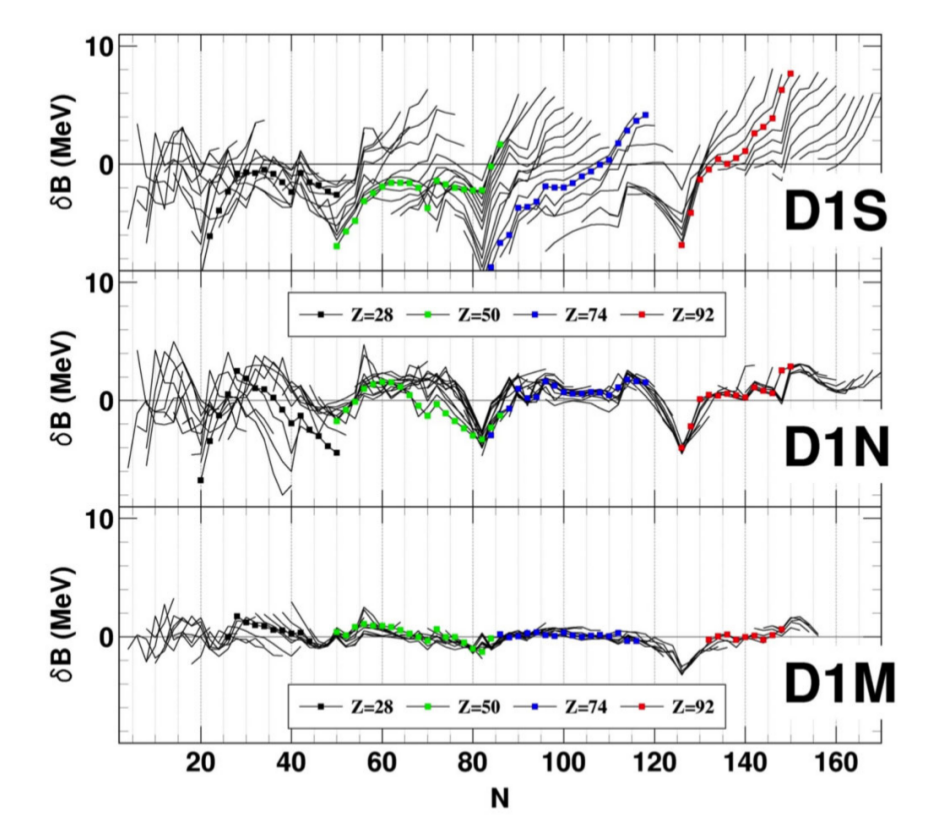}%
\includegraphics[width=0.55\textwidth]{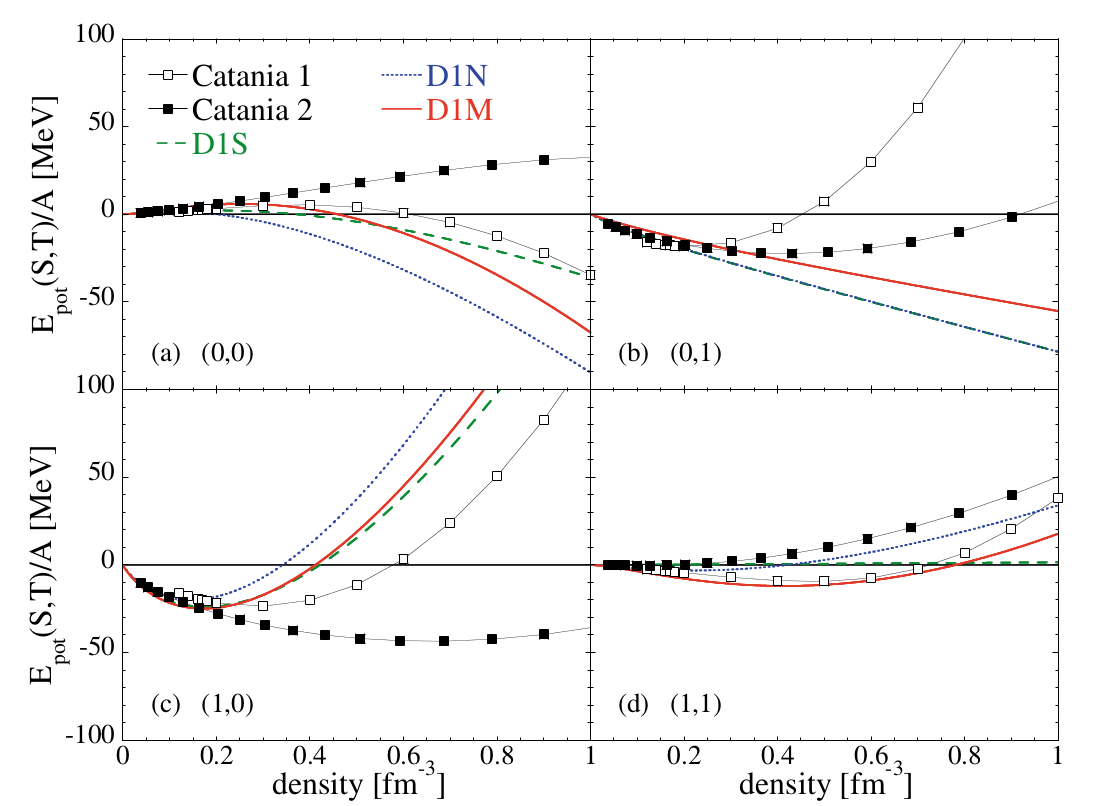}
\caption{ On the left hand side, the binding energy differences $\delta B$ in
MeV is represented as a function of neutron number for the theoretical
predictions of D1S, D1N and D1M. On the right hand side, the binding
energy per particle $E/A$ in nuclear matter for the four different $(S,T)$
channels is shown for D1S, D1N and D1M as well as microscopic results 
(Catania 1 and 2). Figures taken from Ref.~\cite{Goriely2016a}. }
\label{Fig:D1SD1ND1M}
\end{center}
\end{figure}
%%%%%%%%%%%%%%%%%%%%%%%%%%%%%%%%%%%%%%

As a result of the adopted fitting protocol  the 
parametrization D1M exhibits an impressive rms deviation, with respect 
to the measured 2149 nuclear masses \cite{AUDI2003337}, of 
0.798 MeV. This accuracy can be compared with the best nuclear mass 
models.  The largest deviations occur around magic numbers and, in 
particular, for nuclei with neutron number  N $\approx$ 126. 
Furthermore, the neutron matter equation of state (EOS) corresponding 
to D1M agrees well with the one of D1N but also with the one obtained 
in realistic calculations \cite{FRIEDMAN1981502}. Regarding the potential 
energy per particle for symmetric nuclear matter, the comparison with 
realistic  Brueckner-Hartree-Fock (BHF) calculations 
\cite{goriely2009,PhysRevC.77.034316} reveals a fair agreement in each of the 
four two-body spin-isospin (S,T) channels. In particular, the set D1M 
accounts for the repulsive nature of the (S=0,T=0) channel as well as 
for the isovector splitting of the effective mass in the case of 
neutron-rich matter. A complete mass table has been built 
with the Gogny interaction D1M, for nuclei located in between the proton and 
the neutron drip-lines \cite{Goriely2016a}.

In the left panels of Fig \ref{Fig:D1SD1ND1M} the binding energy difference between the
theoretical predictions and the experimental data $\delta B$ is shown as a function
of neutron number for the three most popular parametrizations of the Gogny 
force, namely D1S, D1N and D1M \cite{Goriely2016a}. As discussed before, we clearly observe for D1S the
drift in $\delta B$ for heavy systems that makes this interaction unsuitable
for binding energy predictions. The parametrization D1N mostly corrects
the drift of D1S for large values of N but some deviations still remain.
Finally, D1M gets a very good agreement with experimental data with values
of $\delta B$ much smaller than the ones of the other two interactions.
On the right hand side panels, the binding energy per nucleon $E/A$ in
nuclear matter is plotted as a function of the density for the four different
spin, isospin (ST) channels \cite{Goriely2016a}. The results are compared with the ones obtained
with sophisticated many body techniques and realistic interactions. The
agreement is rather good up to twice saturation density but from there 
one there are large deviations with some unphysical behaviors at large
densities like in the (0,1) channel. The relevance of such disagreement
for finite nuclei densities remains to be assessed.

The Gogny D1M force has  been tested with respect to kinetic moments of 
inertia for Eu and Pu nuclei, giant monopole, dipole and quadrupole 
resonances as well as with respect to the  519 experimentally known 
2$_{1}^{+}$ excitation energies of even-even nuclei \cite{goriely2009}. 
For global calculations of 2$_{1}^{+}$ excitation energies in the 
framework of the symmetry-projected Generator Coordinate Method (GCM) 
with the Gogny D1S and D1M interactions, the reader is also referred to 
Ref.~\cite{Rodriguez2015}. Previous studies for even-even 
\cite{Rodriguez-Guzman2012,Rob12l,Rodriguez-Guzman2010b,Rob09a} 
but also for odd-mass 
\cite{Rod16y,Rod10,Rodriguez-Guzman2010a,Rodriguez-Guzman2010,Rodriguez-Guzman2011} 
nuclei have shown that the parametrization D1M essentially keeps the 
same predictive power of the well tested D1S set to describe a wealth 
of low-energy nuclear structure data while improving the description of 
the nuclear masses. In particular, several calculations 
\cite{Rodriguez-Guzman2014,Rod14z,Rod16z,Rod16y} 
suggest that the D1M parametrization represents a reasonable starting 
point to describe fission properties in heavy and super-heavy nuclear 
systems. However, much work is still needed to further support this 
conclusion. Other applications of the Gogny D1M force can be found, for 
example, in Refs.~\cite{Robledo2012,Martini2014a}.

%%%%%%%%%%%%%%%%%%%%%%%%%%%%%%%%%%%%%%%
\begin{figure}
\begin{center}
\includegraphics[width=0.95\textwidth]{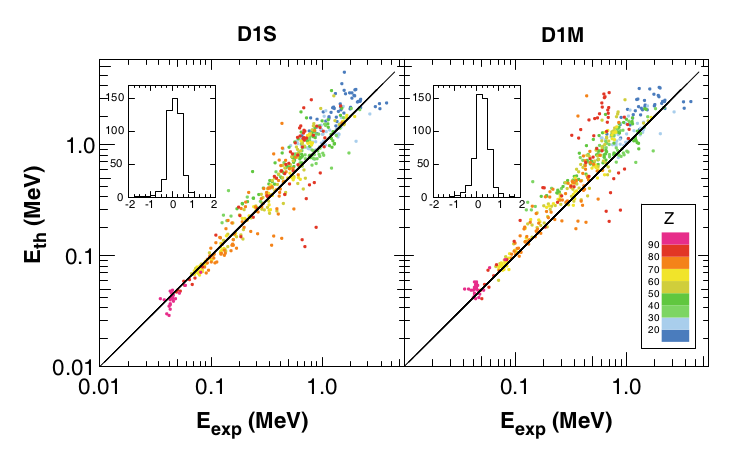}
\caption{Comparison between the theoretical predictions for the 
excitation energy (in MeV) of the lowest $2^{+}$ collective state 
obtained with the 5DCH and the corresponding experimental data. Results 
for both D1S and D1M are shown. Figure taken from from 
Ref.~\cite{Goriely2016a}. }
\label{Fig:D1SD1ND1M2}
\end{center}
\end{figure}
%%%%%%%%%%%%%%%%%%%%%%%%%%%%%%%%%%%%%%
 
In Fig \ref{Fig:D1SD1ND1M2} a comparison of the excitation energies of 
the collective $2^{+}$ states versus the experimental data is made both
for D1S and D1M~\cite{Goriely2016a}. The model used to obtain the $2^{+}$ excitation energies
is the five dimensional Collective Hamiltonian (5DCH) presented in Sec \ref{Sec:MF5D}
that makes use of quantities obtained solely at the HFB level like potential
energy surfaces, collective inertias, etc to be discussed below. In line
with the discussion above, the D1S and D1M results for this finite nucleus
observable are rather similar. The comparison with experimental data is
rather good, specially for those states with low excitation energy that
correspond to the first member of a rotational band. See ~\cite{Goriely2016a,Libert2016}
for further details.

% ------------------------------------------------------------------------------------------------------
%                                                                           D2
% ------------------------------------------------------------------------------------------------------

\subsection{D2}

The most recent addition to the family of Gogny forces is the one of 
Ref \cite{Chappert2015} where the D2 force was proposed. The main 
difference with respect to D1 and other parametrizations of the D1 
family is the inclusion of a finite range density dependent term. The 
new radial form implies that now the strongly repulsive density 
dependent term also contributes to the pairing field, a contribution 
that has to be canceled out in a delicate balance by the central 
potential contribution. Also, the previous dependence with the density 
at the center of mass coordinate has to be given up to facilitate the 
numerical implementation of the new force. The results of the few 
applications carried out so far with this new Gogny force seem to 
indicate a good reproduction of experimental data, at the level of 
other popular parametrizations like D1S or D1M. Also, other 
non-observable quantities like potential energy surfaces look very 
similar to the ones of D1S or D1M. So far, the main drawback of D2 is 
its heavy computer resources requirement \cite{Chappert2015} that make 
this force unsuited for large scale applications. In addition, the new density 
dependent term has still to be implemented in many of the computer 
codes for mean field and beyond calculations. These two facts together 
imply that D2 is not expected to be competitive with the D1 family of 
forces in the near future. 

% ------------------------------------------------------------------------------------------------------
%                                                                           Other parametrizations
% ------------------------------------------------------------------------------------------------------

\subsection{Other parametrizations}

In Ref \cite{Blaizot1995} several parametrizations of the Gogny force 
were proposed to study the relationship between the nuclear matter 
incompressibility parameter and the energy of breathing modes in 
spherical doubly magic nuclei like $^{40}$Ca and $^{208}$Pb. Those 
parametrizations, dubbed D250, D260, D280 and D300 according to their 
value of the incompressibility coefficient $K_{nm}$, have not been used 
apart from the mentioned study and therefore their reliability remains 
to be assessed.

In another study \cite{D1P}, Farine {\it et al} introduced a new type 
of Gogny force, denoted D1P where a new zero range density dependent 
term is added to the traditional Gogny force. The main merits of D1P 
with respect of D1 and D1S are: (i) the agreement with experimental 
data on the depth of optical potentials is improved (ii) sum rules of 
Landau parameters are better fulfilled and (iii) a more realistic 
behavior for the  equation of state of neutron matter is obtained. The 
new interaction has never been used in finite nuclei calculations and 
therefore their merits remain to be assessed. 

% ------------------------------------------------------------------ Tensor

The original Gogny force did not include tensor terms, which are known 
to modify the distribution of single particle levels around the Fermi 
surface. Attempts to include a tensor term have been made but the main 
difficulty is to find a fitting protocol to determine the parameters of 
the tensor part. Only the GT2 interaction by Otsuka {\it et al}  including a 
tensor isovector contribution of Gaussian form 
\cite{PhysRevLett.97.162501} has been fully fitted but at the HF level. 
All the other attempts so far to incorporate a tensor term to the Gogny 
force were carried out without modifying the other parameters of the 
interaction and therefore the results obtained can be considered as 
exploratory. Several sets of tensor parameters have been introduced in 
the literature: D1ST and D1MT including a radial part based on the one 
of Argonne V18 \cite{Anguiano11}, D1ST2a and D1ST2b including also a 
tensor isoscalar term but this time with a Gaussian shape 
\cite{Anguiano12} and also D1ST2c and D1MT2c in Ref \cite{Grasso13} 
where the spin orbit strength was modified alongside with the strengths 
of the different channels of the tensor term. With this latter 
improvement it is possible to successfully reproduce the 
$1f_{7/2}-2p_{3/2}$ gap in the $^{40-48}$Ca isotopic chain. All the 
previous results are obtained for spherical nuclei, but recently 
calculations in deformed nuclei with D1ST2a have been reported 
\cite{Bernard16}.

% ----------------------------------------------------------- Davesne three Gaussians

Another possibility to generalize the Gogny force is to increase the number 
of Gaussians in the central potential part to have more flexibility to 
adjust nuclear matter properties. This is the path taken in Ref 
\cite{Davesne17} where the extra freedom was used to adjust the four  
spin-isospin channels of the nuclear matter potential energy to the 
results of realistic calculations. As in the other cases, no finite 
nuclei calculations have been carried so far and therefore the merits 
of the new proposal for finite nuclei remain to be demonstrated.

% ---------------------------------------------------------------------- D1M*

Very recently, a new variant of D1M has been proposed \cite{Boquera2018} to cure one of its
most notorious drawbacks: its inability to reproduce the accepted value of
two solar masses for the mass of the heaviest neutron stars. In the new parametrization,
dubbed D1M$^{*}$, the slope of the symmetry energy coefficient in nuclear matter $L$ has
been fitted to a higher value (43 MeV) than the original rather low value of D1M (24.8 MeV). At the same
time, all other relevant combination of parameters have been kept as to
preserve the already outstanding properties of D1M, specially the properties of the pairing channel. It has
to be mentioned that the D2 parametrization also gives two solar masses for the
mass of neutron stars and could be considered as an alternative to D1M$^{*}$.
However, D2 is much more computationally demanding than D1M$^{*}$.

% ------------------------------------------------------------------------------ SEI
 
Although not a pure Gogny force, the recent proposal of Behera {\it et al} \cite{Beh16}
of a simple effective interaction (SEI) resembles very much the Gogny force. There
are two distinctive places where the two differ: one of the ranges of the Gogny
force is set to zero in SEI and the density dependence includes an additional denominator
to prevent supra-luminous effects in nuclear matter. The parameters of SEI
are fitted mostly to nuclear matter properties and only one (or two) are left
to fit the binding energies of finite nuclei. The results obtained in finite
nuclei, including binding energies, radii, deformation properties, etc are encouraging
and exploratory work in other aspects of the interaction is underway. 

% ------------------------------------------------------------------------------------------------------
%                                                  As pairing interaction
% ------------------------------------------------------------------------------------------------------

\subsection{The pairing channel of the Gogny force}

The pairing interaction coming from the central part of the Gogny force
has been used in many places as an alternative to zero range pairing interactions
\cite{GONZALEZLLARENA199613,PhysRevC.60.014310} due to its natural ultraviolet
cutoff. A systematic comparison with other alternatives in the relativistic
framework can be found in Ref \cite{Afanasjev2013}. Due to the non-local 
character of this pairing interaction its numerical cost represents a large
fraction of the total cost of the calculation and therefore other cheaper 
alternatives have been sought. In this respect we can mention the separable
expansion of the two-body Gaussian interaction proposed in \cite{TIAN200944}
that is widely used along with relativistic mean fields not only at 
the mean field level \cite{PhysRevC.81.054318,PhysRevC.80.024313} but
even as a cheap alternative to the full Quasiparticle Random Phase Approximation
(QRPA) \cite{PhysRevC.79.064301}. It is also interesting to mention 
another separable expansion \cite{Rob10} valid also for the Hartree 
Fock part and that could be used a simpler replacement of the Gaussian.

% ------------------------------------------------------------------------------------------------------
%                                                  Further improvements
% ------------------------------------------------------------------------------------------------------

\subsection{Future improvements}

It is not easy to forecast the future of the Gogny force, but there are 
a few things that are obvious and should be implemented in the short 
term. First of all, the release of the zero range radial dependence of 
the spin-orbit potential. This is a mostly aesthetic improvement and no 
big impact on any relevant observable is foreseen. A tensor term has 
already been introduced in the Gogny force and its impact analyzed in a 
variety of situations, but the tensor contribution has been introduced 
in a perturbative fashion, that is, no refitting of the core parameters 
of the force has been carried out. A version of the Gogny force with a 
finite range tensor term and a full refitting of the parameters at the 
HFB level would be highly welcome. Finally, the time-odd sector of the 
density dependent part of the interaction has never been explored. It 
is true that the time-odd fields obtained from the central, spin--orbit 
and Coulomb part lead to a nice reproduction of observables like 
moments of inertia \cite{Afa00,Delaroche2006,Dobaczewski2015}, spin and 
parities of the ground state of odd mass nuclei 
\cite{Rob12,Dobaczewski2015} or even the excitation energy of high-K 
isomeric states \cite{Rob15}. However, there are indications in the 
physics of odd-odd nuclei that additional time-odd fields could be 
required \cite{Rob14} in order to reproduce the rich phenomenology 
associated to these kind of nuclei. Finally, let us mention that 
in all the applications with the Gogny force proton-neutron pairing
has never been considered. It is important for nuclei near $N=Z$ and
considering it will require some additional constraints in the fitting
protocol.

%% file: MeanField.tex
% ------------------------------------------------------------------------------------------------------
%                               M e a n   F i e l d   
% ------------------------------------------------------------------------------------------------------

\section{Mean field  with the Gogny force}
\label{Sec:MF}

The mean field approximation can be considered as the simplest of all 
possible approximations  to the  fermion many body problem, as in 
the atomic nucleus \cite{Ring1980}. In nuclear physics, and as a consequence of the 
nuclear interaction properties, the mean field is required also to 
incorporate those short range correlations responsible for the 
existence of Cooper pairs in the atomic nucleus and also responsible 
for the related phenomenon of nuclear super-fluidity. These two aspects are 
implemented in the Hartree- Fock- Bogoliubov (HFB) mean field 
approximation that is a generalization encompassing both the Hartree 
Fock (HF) and the Bardeen Cooper Schriffer  (BCS) approximations into a single 
framework. A 
genuine aspect of the nuclear mean field is the ubiquitous  spontaneous 
breaking of symmetries. This is a direct manifestation of the 
properties of the nuclear force that lead to mean field wave 
functions that can eventually break all kind of spatial or internal 
symmetries.  Spontaneous symmetry breaking  leads to mean field 
solutions not preserving the  symmetries of the Hamiltonian like 
translational invariance, rotational invariance, reflection symmetry, 
etc. It is a consequence of the non-linear nature of the HFB 
equations and therefore is a consequence of the approximate mean field 
treatment of the problem. This aspect of the mean field could be 
considered as unphysical and undesirable but it turns out to be the 
other way around: it is a way to incorporate different kinds of 
correlations into a simple picture (the typical example being the BCS 
theory of superconductivity) and is behind the very 
successful concept of "intrinsic state" in nuclear physics and the 
associated grouping of levels in bands connected by strong electromagnetic
transitions and often found in the nuclear spectrum. 
Obviously, the whole idea of symmetry breaking  requires some further 
refinement in order to obtain the physical wave functions that are 
labeled with quantum numbers of the symmetries of the system (angular 
momentum, parity, etc). How to buid laboratory-frame wave functions 
with the proper quantum numbers of 
the Hamiltonian's symmetries out of the intrinsic states will the 
subject of the next section.

This characteristics leads, in a natural way, to a taxonomy of the 
different kind of mean field approximations based on the symmetries 
allowed to break in the calculations: it is common to talk about 
axially symmetric calculations, reflection asymmetric, triaxial etc 
depending on the symmetries preserved (or allowed to break) by the mean 
field (or the computer codes used to carry out the calculations). In 
the following we will make use of this terminology.

\subsection{Mean field calculations with the Gogny force}

General properties of the nuclear interaction require the treatment of 
both long range and short range correlations in the same footing. At 
the mean field level, this means that the traditional Hartree-Fock 
approximation has to be supplemented by the incorporation of short 
range correlations in the spirit of the BCS theory of superconductivity 
and super-fluidity. The incorporation  of these two effects requires the 
introduction of the so-called HFB quasiparticle annihilation and 
creation operators $\beta_\mu$ and $\beta^\dagger_\mu$ which are 
expressed as linear combinations (with amplitudes $U$ and $V$) of 
generic creation $c^\dagger_k$ and annihilation $c_k$ operators that 
correspond to a conveniently chosen basis
\begin{equation} \label{eq:Bogo}
\left(\begin{array}{c}
\beta\\
\beta^{\dagger}
\end{array}\right)=\left(\begin{array}{cc}
U^{+} & V^{+}\\
V^{T} & U^{T}
\end{array}\right)\left(\begin{array}{c}
c\\
c^{\dagger}
\end{array}\right)=W^{+}\left(\begin{array}{c}
c\\
c^{\dagger}
\end{array}\right).
\end{equation}
In order to alleviate the notation  we have introduced the block matrix $W$ 
encompassing both $U$ and $V$ in a convenient way. The associated 
single particle wave functions of the basis $\varphi_k(\vec{r}) = 
\langle \vec{r} | c_k^\dagger |0\rangle$ can in principle be anything 
we want provided they span the whole Hilbert space. However, practical 
limitations force the use a finite subset of single particle states 
that only generates a limited corner of the whole Hilbert space. As a 
consequence, the choice of the single particle states has to be adapted 
to the geometry of the problem at hand and the symmetries expected to 
be broken by the mean field. For instance, the physics of triaxial 
shapes is better described in terms of wave functions breaking 
spherical symmetry. A typical example are those Harmonic Oscillator (HO) wave 
functions which are tensor product of 1D HO wave functions with different 
oscillator lengths along each of the spatial directions. The oscillator 
lengths are adapted to the size of the major axis of the matter distribution 
of the triaxial configuration. There is 
an additional constraint in the choice of the $\varphi_k(\vec{r})$ which is related to 
the need to evaluate  billions of matrix 
elements of a two body interaction with those wave functions. In the case of the Gogny force, 
where the central part of the interaction is modeled in terms of a linear 
combination of Gaussians, the obvious choice for the basis is the set 
of eigen-states of the harmonic oscillator potential. Viewed from a 
mathematical perspective, the choice is very convenient as the 
Hermite polynomials entering the  HO wave functions are orthogonal with 
respect to a Gaussian weight and therefore the evaluation of the two 
body matrix elements can be carried out analytically. The other terms 
of the interaction are either zero range (spin-orbit and density 
dependent part of the interaction) or can be easily expressed in terms 
of Gaussians as it is the case with the Coulomb potential. In  appendix 
\ref{App:A} we discuss the general principles guiding the efficient 
evaluation of matrix elements of a Gaussian two body interaction, which 
is central to any mean field calculation with the Gogny force. The 
discussion is focused on 1D harmonic oscillator wave functions which 
are at the heart of the triaxial representation of the HO wave 
functions \cite{Gir83}. Other possibilities involving the 2D harmonic 
oscillator \cite{You09,Egido1997} or even the 3D one \cite{Gog75} rely 
on the same principles and will not be discussed in detail.

The Bogoliubov transformation of Eq (\ref{eq:Bogo}) has to preserve the commutation
relations of creation and annihilation quasiparticle operators. This requirement
restricts the form of the $W$ matrices to those satisfying some sort of unitarity 
constraint
\begin{equation} \label{Wconstr}
W^{\dagger}\sigma W^{*} = \sigma
\end{equation} 
where
\begin{equation}\label{sigma}
	\sigma = \left(\begin{array}{cc}
0 & \mathbb{I}\\ \mathbb{I} & 0 \end{array}\right)
\end{equation}
is a block matrix with the same block structure as $W$. Given a set of 
quasiparticle operators satisfying the canonical commutation relations, 
the associated mean field HFB wave function $|\varphi\rangle$ is defined 
by the condition  $\beta_{\mu} |\varphi\rangle = 0$ that is fulfilled by the
product state
\begin{equation}
|\varphi\rangle = \prod_{\mu}
\beta_{\mu} |-\rangle
\end{equation}
where the product extends to those labels $\mu$ for which the product 
is non-zero. Finally, the $U$ and $V$ amplitudes of the Bogoliubov 
transformation (or the $W$ amplitudes) are determined by the Ritz 
variational principle on the HFB energy $E_\mathrm{HFB} = \langle 
\varphi | \hat{H} | \varphi \rangle / \langle \varphi | \varphi \rangle 
$. The Hamiltonian is the sum of a one body kinetic energy term plus
a two body potential term that is written in second quantization form as
\begin{equation}
	\hat H = \sum_{ij} t_{ij}c^{\dagger}_{i}c_{j} + \frac{1}{4} \sum_{ijkl}
	\bar{\nu}_{ijkl}c^{\dagger}_{i}c^{\dagger}_{j}c_{l}c_{k}
\end{equation}
with $\bar{\nu}_{ijkl}$ the antisymmetrized two body matrix element
$\bar{\nu}_{ijkl}=\langle ij|v(1,2)|kl\rangle - \langle ij|v(1,2)|lk\rangle$.
 Very often the minimization of the energy is restricted to fulfill 
some constraints on the mean value of some operators $\hat{O}_{i}$ like 
the quadrupole or octupole moments of the mass distribution. These 
constraints have to be considered along with the traditional constraint 
on particle number $\langle N \rangle=N$ and $\langle Z \rangle=Z$ 
characteristic of the HFB theory. In order to handle this situation the 
introduction of Lagrange multipliers $\lambda_{i}$ is required. The 
quantity to be minimized becomes 
\begin{equation}
E_\mathrm{HFB}^{\prime}
= \langle \varphi | \hat{H}^{\prime} | \varphi \rangle / \langle \varphi | \varphi 
\rangle	
\end{equation}
where
\begin{equation}
	\hat{H}^{\prime}=\hat{H}-\sum_{i} \lambda_{i} \hat{O}_{i}
\end{equation}
It is now possible to carry out an unconstrained minimization of $E_\mathrm{HFB}^{\prime}(\lambda_{1},\lambda_{2},\ldots)$
but fixing the values of the  chemical potentials $\lambda_{i}$ by requiring  that 
the mean value of the constraining operators is equal to the desired value of
the constraints $\langle\varphi |\hat{O}_{i}|\varphi\rangle=o_{i}$.

Before proceeding with the application of the variational principle to $E_\mathrm{HFB}^{\prime}$ we
have to overcome an additional problem: the $U$ and $V$ amplitudes are not linearly independent due
to the constraint of Eq (\ref{Wconstr}) and they therefore cannot be used as
independent variational parameters. A set of variables which are linearly independent is
provided by the Thouless theorem \cite{Thouless1960,Ring1980} that gives
the most general form of an HFB state $|\varphi (Z) \rangle$ in terms of
some linearly independent complex parameters $Z_{\mu \mu'}$ (with $\mu^{\prime
}>\mu$ ) and a reference HFB state $|\varphi_{0}\rangle$
\begin{equation}  \label{eq:thoul}
|\varphi (Z)\rangle =\eta (Z,Z^{*})\exp \left( \sum_{\mu<\mu^{\prime
}}Z_{\mu\mu^{\prime }}\beta_\mu^{+}\beta_{\mu^{\prime }}^{+}\right) |\varphi
_0\rangle
\end{equation}
where $\eta (Z,Z^{*})$ is a normalization constant such that $\langle
\varphi (Z)|\varphi (Z)\rangle =1$. The only restriction on $|\varphi (Z) \rangle$ 
is that it must have a non-zero overlap with  $|\varphi
_0\rangle$. As it is customary, we will use as free
parameters $Z$ and $Z^{*}$ instead of $\Re(Z)$ and $\Im(Z)$. The
relation between the Bogoliubov wave functions $U(Z)$ and $V(Z)$
corresponding to $|\varphi (Z)\rangle$ and the $U_{0}$ and $V_{0}$ corresponding to
$|\varphi_0 \rangle$ is given by \cite{Egido95}
\begin{eqnarray}
U(Z) & = & (U_0+V_0^*Z^*)\left[L^{-1}\right]^\dagger \\
V(Z) & = & (V_0+U_0^*Z^*)\left[L^{-1}\right]^\dagger
\end{eqnarray}
where the matrix $L[Z,Z^{*}]$ is the Choleski decomposition of the
positive definite matrix $I+Z^T Z^*$, i.e.,
\begin{equation}
LL^\dagger = I+Z^TZ^*
\end{equation}
where $I$ is the unity matrix. The Choleski decomposition is nothing but
the ``square root" of the matrix $I+Z^TZ^*$.  Using the Thouless parametrization, the HFB
energy is given by a function of the linearly independent complex $Z$ and $Z^*$
parameters
\begin{equation}
E^{\prime} (Z,Z^*)=\frac{
\langle \varphi (Z)|H^{\prime} (Z,Z^{*})|\varphi (Z)\rangle}
{\langle \varphi (Z)|\varphi (Z)\rangle}
\end{equation}
The Gogny force is state dependent through the density dependent term 
that  depends on the mass density of the corresponding state $|\varphi (Z)\rangle$. This is
the reason why in the above expression we have considered that the Hamiltonian
explicitly depends on the $Z$ and $Z^{*}$ amplitudes and this dependence
has to be taken into account in the variational principle.
As the $Z$ and $Z^{*}$ amplitudes are independent variational parameter, 
the Ritz variational principle becomes 
\begin{equation}\label{eq:Ritz}
\frac{\partial E(Z,Z^{*})}{\partial Z_{\mu \mu^{\prime}}}=
\frac{\partial E(Z,Z^{*})}{\partial Z_{\mu \mu^{\prime}}^{*}}=0
\end{equation}
For practical reasons it is better to define the HFB amplitudes which are a solution of
the Ritz variational principle equation, as those corresponding to the 
reference HFB amplitude of the Thouless theorem
$|\varphi_{0}\rangle$ and therefore the derivatives in Eq (\ref{eq:Ritz})
are to be evaluated at $Z=Z^{*}=0$. Using the expression for the partial 
derivative of $|\varphi (Z)\rangle{}$ 
\begin{equation}
\frac{\partial}{\partial Z_{\mu \mu^{\prime}}} |\varphi (Z)\rangle =
\beta_\mu^{+} \beta_{\mu^{\prime }}^{+}| \varphi_0\rangle + O(Z,Z^{*})
\end{equation}
we obtain
\begin{equation}\label{eq:HFB1}
	\frac{\partial E^{\prime}(Z,Z^{*})}{\partial Z_{\mu \mu^{\prime}}}_{|Z=0}= 
	\langle \varphi_{0} |H^{\prime }(0,0) \beta_\mu^{+} \beta_{\mu^{\prime }}^{+}|\varphi_{0} \rangle{}
	+ \left\langle \frac{\delta
H^{\prime }}{\delta Z_{\mu\mu^{\prime }}}\right\rangle_{| Z=0}=0
\end{equation}
which is the HFB equation for density dependent forces. Traditionally,
the dependence on $Z$ and $Z^{*}$ of the Hamiltonian comes through a 
density dependent term depending on the spatial density 
\begin{equation}
\rho (\vec{R})=\langle \varphi (Z)|\hat{\rho }(%
\vec{R})|\varphi (Z)\rangle
\end{equation}
where $\hat{\rho }(\vec{R})=\sum_{i=1}^A\delta (%
\vec{r}_i-\vec{R})$ is the standard matter
density operator and $\vec{R}=\frac 12 (\vec{r}_1+\vec{r}_2)$ is the 
center of mass coordinate. Then
\begin{equation}
\frac{\partial \rho }{\partial Z_{\mu\mu^{\prime }}}_{| Z_{\mu
\nu}=0} =\langle \varphi _0|\hat{
\rho }(\vec{R})\beta _\mu^{+}\beta _{\mu^{\prime }}^{+}|\varphi
_0\rangle
\end{equation}
and therefore
\begin{equation}\label{eq:pHpZ}
\frac{\partial H^{\prime }}{\partial Z_{\mu\mu^{\prime }}}=\frac{\partial \rho }{%
\partial Z_{\mu\mu^{\prime }}}\frac{\partial H^{\prime }}{\partial \rho }=\langle
\varphi _0|\hat{\rho }(\vec{R})\beta _\mu^{+}\beta
_{\mu^{\prime }}^{+}|\varphi _0\rangle \frac{\partial H^{\prime }}{\partial \rho }
\end{equation}
In the above expression 
both $\langle
\varphi _0|\widehat{\rho }(\vec{R})\beta_\mu^{+}\beta
_{\mu^{\prime }}^{+}|\varphi _0\rangle $ and $\frac{\delta
H^\prime}{\delta \rho}$ have
to be understood as operators in the variable $\vec{R}$
and, therefore, $\frac{\delta H^{\prime }}{\delta Z_{\mu\mu^{\prime }}}$
has to be treated as a two body operator in the
evaluation of the mean values of Eq (\ref{eq:HFB1}).
Inserting the result of Eq (\ref{eq:pHpZ}) in Eq (\ref{eq:HFB1}) we finally
arrive to the HFB equation for density dependent forces in standard form
\begin{equation}  \label{gradef}
H^{\prime 20}_{\mu\mu^{\prime }}\equiv \langle \varphi _0|H^{\prime }\beta_\mu^{+}\beta
_{\mu^{\prime }}^{+}|\varphi _0\rangle +\left\langle \langle \varphi _0|%
\hat{\rho }(\vec{R})\beta _\mu^{+}\beta _{\mu^{\prime
}}^{+}|\varphi _0\rangle \frac{\partial H^{\prime }}{\partial \rho }%
\right\rangle =0
\end{equation}
In the present context, the Lagrange multipliers $\lambda_{i}$ are defined 
by the condition that the gradient of the density dependent Routhian has 
to be perpendicular to the gradient of  the constraints 
\begin{equation}
O^{20}_{j\, \mu\mu^{\prime}} = \langle \varphi _0|\hat{O}_{j}\beta_\mu^{+}\beta
_{\mu^{\prime }}^{+}|\varphi _0\rangle	
\end{equation}
That is
\begin{equation}
	\sum_{\mu\mu^{\prime}} H^{\prime 20}_{\mu\mu^{\prime }} O^{20}_{j\, \mu\mu^{\prime}} =
	\sum_{\mu\mu^{\prime}} H^{20}_{\mu\mu^{\prime }} O^{20}_{j\, \mu\mu^{\prime}} - \sum_{i} \lambda_{i}
    \sum_{\mu\mu^{\prime}} O^{20}_{i\, \mu\mu^{\prime }} O^{20}_{j\, \mu\mu^{\prime}}=0
\end{equation}
what represents a linear system of equations for the unknown $\lambda_{i}$.
The first term of the gradient in Eq (\ref{gradef}) can be easily computed by using the quasi-particle
representation of the Hamiltonian operator while the second needs further treatment.
Using the second quantization form of the density operator
\begin{equation}
\widehat{\rho }(\vec{R})=\sum_{ij}f_{ij}(\vec{R}
)c_i^{+}c_j
\end{equation}
where $f_{ij}(\vec{R})=\langle i|\delta (\vec{r}-
\vec{R})|j\rangle = \varphi^{*}_{i} (\vec{R})\varphi_{j} (\vec{R})$,
the last term of Eq. (\ref{gradef}) can be written as
\begin{equation}
\sum_{ij} \langle \varphi_0 | c_i^\dagger c_j \beta_\mu^\dagger
\beta_{\mu^\prime}^\dagger | \varphi_0 \rangle \left\langle f_{ij}
(\vec{R})
\frac{\delta H'}{\delta \rho} \right\rangle
\end{equation}
which suggests the definition of the one body operator
\begin{equation}
\widehat{\partial \Gamma} =\sum_{ij}\partial \Gamma _{ij}c_i^{+}c_j
\end{equation}
with matrix elements
\begin{eqnarray}
\partial \Gamma _{ij} & = & \left\langle \frac{\delta H^{\prime }}{\delta
\rho } f_{ij}(\vec{R})\right\rangle \\ \nonumber{}
& = & \frac{1}{4}\sum_{klmn}
\langle
kl|\frac{\delta H^\prime}{\delta \rho} f_{ij} (\vec{R})
|\widetilde{mn} \rangle
\left(\rho_{nl}\rho_{mk}-\rho_{ml} \rho_{nk}-\kappa_{mn}\kappa^*_{kl}
\right)
\end{eqnarray}
requiring antisymmetrized two body matrix elements of the rearrangement term.
With this definition, the last term of Eq. (\ref{gradef}) becomes
\begin{equation}
\langle \varphi _0|\widehat{\partial \Gamma} \beta _\mu^{+}\beta
_{\mu^{\prime}}^{+}|\varphi _0\rangle
\end{equation}
which shows that the calculation of the gradient of the Routhian for density
dependent forces proceeds in the same way as for standard forces except for
the fact that an additional density dependent one body operator
$\hat{\partial \Gamma}$ has to be
added to the Hamiltonian. The calculation of the gradient $G$ now follows the standard
procedure and we finally obtain
\begin{equation}
G_{\mu \mu^{\prime }}=H^{\prime 20}_{\mu \mu^{\prime }} = H^{20}_{\mu \mu^{\prime }} -
\sum_{i}\lambda_{i} O^{20}_{i\, \mu \mu^{\prime}}
\end{equation}
with
\begin{equation}
H^{20}_{\mu \mu^{\prime }} = \left( U^{+}hV^{*}-V^{+}h^TU^{*}+U^{+}\Delta
U^{*}-V^{+}\Delta ^{*}V^{*}\right) _{\mu \mu^{\prime }}
\end{equation}
given in terms of
\begin{eqnarray}
h_{ij} & = & t_{ij}+\partial \Gamma_{ij} + \Gamma_{ij} \\
\Gamma_{ij} & = & \sum_{qq^{\prime }}\widetilde{\upsilon }
_{iqjq^{\prime }}\rho _{q^{\prime }q} \\
\partial\Gamma_{ij} & = & \frac{1}{4}\sum_{klmn} \langle kl|\frac{\delta H^\prime}{\delta \rho} f_{ij} (\vec{R})|\widetilde{mn} \rangle\left(\rho_{nl}\rho_{mk}-\rho_{ml} \rho_{nk}-\kappa_{mn}\kappa^*_{kl}
\right) \\
\Delta _{ij} & = & \frac 12\sum_{qq^{\prime }}\widetilde{\upsilon
}_{ijq^{\prime}q}\kappa _{q^{\prime }q}
\label{hgd}
\end{eqnarray}

As the mean value of the HFB Routhian only depends upon the standard
density matrix $\rho_{ij}$ and pairing tensor $\kappa_{ij}$, it only depends
upon the two first transformations of the Bloch-Messiah decomposition of the 
$U$ and $V$ amplitudes 
\cite{Bloch1962,Zumino1962} (see
\cite{Ring1980} for a detailed discussion). As a consequence,
the HFB equation Eq. \ref{gradef} only determines the Bogoliubov
transformation up to an unitary transformation among the
quasiparticles (the third transformation of the Bloch-Messiah
theorem). To fix  this arbitrary unitary transformation it is customary to
introduce an additional imposition to the Bogoliubov transformation:
namely that the ${H^\prime}^{11}_{\mu \nu}$ matrix, defined as
\begin{equation}
H^{\prime 11}_{\mu \mu^{\prime }} = H^{11}_{\mu \mu^{\prime }} -
\sum_{i}\lambda_{i} O^{11}_{i\, \mu \mu^{\prime}}
\end{equation}
with 
\begin{equation}
H^{11}_{\mu \nu}=
\left( U^{+}hU-V^{+}h^TV+U^{+}\Delta
V-V^{+}\Delta ^{*}U\right) _{\mu \mu^{\prime }},
\end{equation}
has to be diagonal. The eigenvalues of this matrix are called quasiparticle
energies and denoted by $E_\mu$. For non-density dependent forces
they represent an approximation to the energy gain of the
odd system represented by the non-self-consistent wave function
$\beta^+_\mu |\varphi\rangle$
with respect to the corresponding even one $|\varphi \rangle${}
\begin{equation}
E_\mu \approx \langle \varphi |\beta_\mu H^\prime \beta^\dagger_\mu|
\varphi \rangle - \langle \varphi | H^\prime | \varphi \rangle
\end{equation}
in which the two quasiparticle interaction terms of the Hamiltonian
have been dropped. The first impression looking at the previous
equation is that
one should use $H$ instead of $H^\prime$ in it. However, the use of
$H^\prime$  in the definition of $E_\mu$ implies that,
at first order in perturbation theory, we are
correcting for the fact that the mean values of
the constraining operators for  the state $\beta^\dagger_\mu |\varphi \rangle$
are not the same as those of $|\varphi \rangle$ (the correct ones).

For density dependent forces it has to be taken into account
that the Hamiltonian used in the evaluation of the energy of
the odd system differs from the one used in the calculation of the
even one as the former depends on the density of the odd system
\begin{equation}
\rho^{\rm odd} (\vec{R})_\mu = \langle \varphi | \beta_\mu \hat{\rho}
\beta^\dagger_\mu | \varphi \rangle = \rho^{\rm even} + \delta
\rho_\mu
\end{equation}
where
\begin{equation}
\delta \rho (\vec{R})_\mu = \sum_{ij} f_{ij} (\vec{R}) (U^*_{i\mu}
U_{j\mu} -V^*_{j\mu} V_{i\mu})
\end{equation}
As $\delta \rho_\mu$ is the change on the density coming from the
addition of a particle it is small compared with the total density of
the system and therefore it is reasonable to expand the density
dependent Hamiltonian of the odd system around the one of the even
system as
\begin{equation}
\hat{H} (\rho^{\rm odd}_\mu) = \hat{H} (\rho_0) + \frac{\delta
H}{\delta \rho} \delta \rho_\mu + \ldots
\end{equation}
With this expansion the energy of the odd system can be evaluated as
\begin{eqnarray}
\langle \varphi_0 | \beta_\mu \hat{H} (\rho^{\rm odd}_\mu)
\beta^\dagger_\mu | \varphi_0 \rangle & = &
\langle \varphi_0 | \beta_\mu \hat{H} (\rho_0)
\beta^\dagger_\mu | \varphi_0 \rangle +
\langle \varphi_0 | \beta_\mu \frac{\delta \hat{H}}{\delta \rho}
\delta \rho_\mu
\beta^\dagger_\mu | \varphi_0 \rangle \\ \nonumber
& = &
\langle \varphi_0 |\hat{H} (\rho_0)| \varphi_0 \rangle +
H^{11}_{0_{\mu \mu}} +
\langle \varphi_0 | \beta_\mu \Delta \left[ \frac{\delta
\hat{H}}{\delta \rho} \delta \rho_\mu \right]
\beta^\dagger_\mu | \varphi_0 \rangle
\end{eqnarray}
where $H^{11}_{0_{\mu \mu}}$ is computed with the wave functions of
$|\varphi_0\rangle${}
\begin{equation}
H^{11}_{0_{\mu \mu}} = \left( U^+ h U - V^+h^T V + U^+ \Delta V -
V^+ \Delta^* U \right)_{\mu \mu}
\end{equation}
and contains in its definition the rearrangement potential. The
remaining term ( in which the definition $\Delta \hat{O} = \hat{O} -
\langle \varphi_0 | \hat{O} | \varphi_0 \rangle$ has been used)
represents the interaction of the quasiparticle with the change
induced by itself in the density. It is comparable with the magnitude
of the quadratic terms in $\delta \rho_\mu$ neglected in the expansion
of the Hamiltonian and therefore should not be considered.

The conditions $H^{\prime\,20}_{\mu \mu^{\prime}}=0$ and 
$H^{\prime\,11}_{\mu \mu^{\prime}}=E_{\mu}\delta_{\mu \mu^{\prime}}$
can be written in compact form as
\begin{equation}\label{HFBeq2}
	\left(\begin{array}{cc}
H^{\prime\,11} & H^{\prime\,20}\\ -H^{\prime\,20 *} & -H^{\prime\,11}  \end{array}\right){}
=
W
	\left(\begin{array}{cc}
h & \Delta\\ -\Delta^{*} & -h^{*}  \end{array}\right){}
W^{\dagger}
=	
\left(\begin{array}{cc}
E & 0 \\ 0 & -E  \end{array}\right){}
\end{equation}
Introducing the block Hamiltonian matrix
\begin{equation}\label{HFBH}
\mathcal{H} =	
	\left(\begin{array}{cc}
h & \Delta\\ -\Delta^{*} & -h^{*}  \end{array}\right){}
\end{equation}
and the block density 
\begin{equation}\label{HFBRo}
\mathcal{R} =	
	\left(\begin{array}{cc}
\rho & \kappa \\ -\kappa^{*} & -\rho^{*}  \end{array}\right){}
\end{equation}
the condition of Eq \ref{HFBeq2} is expressed as 
\begin{equation} \label{HFBeq3}
	\mathcal{H}W=W\left(\begin{array}{cc}
E & 0 \\ 0 & -E  \end{array}\right){}
\end{equation}
that is the traditional form of the HFB equation. Please note, that
the $W$ transformation also brings $\mathcal{R}$ to diagonal form
\begin{equation}
 W^{\dagger}\mathcal{R}W^{*}=\left(\begin{array}{cc}
0 & 0 \\ 0 & \mathbb{I} \end{array}\right){}
\end{equation}
and therefore Eq (\ref{HFBeq3}) implies that $\mathcal{H}$ must commute with
$\mathcal{R}$ at the self-consistent solution of the HFB equation.

The HFB equation is a non-linear equation where the matrix to be 
diagonalized $\mathcal{H}$ depends upon the eigenvectors $W$ through the 
density matrix and pairing tensor. Therefore, the HFB equation is not an 
standard eigenvalue problem. The best way to tackle the solution of the 
HFB equation is to solve it iteratively: starting with a reasonable 
guess for $U$ and $V$, the density $\rho$ and pairing tensor 
$\kappa$ and the corresponding HF $h$ and pairing $\Delta$ fields are computed
and then used to build the generalized Hamiltonian $\mathcal{H}$. The 
Hamiltonian is diagonalized to obtain a new set of $U$ and $V$ 
amplitudes. The process is repeated iteratively until convergence is 
achieved (i.e. the input $U$ and $V$ are the same as the output ones). 
This is the preferred approach by the Bruy\'eres-Le-Ch\^{a}tel group. To make 
it converge, some kind of ``slowing down" strategy has to be implemented 
in the iterative procedure. Traditionally, this ``slowing down" is 
implemented by mixing the $U^{i+1}$ and $V^{i+1}$ amplitudes with the 
previous iteration's ones $U^{i}$ and $V^{i}$ by means of some add-hoc 
mixing parameter $\eta$. The proper choice of $\eta$ requires some 
experience and the consideration of the type of calculation at hand. 
There is an alternative to this procedure based on the equivalence 
between the HFB equation and the variational principle over the HFB 
mean value of the Routhian: the HFB amplitudes solving the HFB equation 
are those that minimize the HFB mean value of the Routhian. Therefore, 
the HFB problem can be considered as a minimization problem with a very 
large set of variational parameters $Z$ and $Z^{*}$. This is one of the 
classical problems in numerical analysis and the usual numerical 
techniques used in this case can be invoked: the gradient method \cite{Mang1976},
the conjugate gradient method \cite{Egido95} and the 
(approximate) second order Newton-Rampson method \cite{Robledo2011}. 
The most notorious advantages of any variant of the gradient method 
versus the iterative one are the easy handling of multiple constraints 
that is often required in practical applications like fission or the 
determination of potential energy surfaces, the much lower iteration 
count and, finally, the guarantee that the method always converges to a 
solution (that might not be the optimal one).

In the case of the Gogny force and, at variance with similar type of 
calculations using the Skyrme EDF, the 
pairing field $\Delta$ is computed from the same interaction used in the 
central particle-hole (p-h) channel.
The finite range of the central potential 
makes unnecessary any kind of cut-off or restriction on the active 
configuration space. The pairing field gets contributions from the 
central potential and the spin-orbit term. The density dependent term 
does not contribute to the pairing channel in the traditional family of 
D1 like parametrizations (D1, D1S, D1N, etc) due to its zero range, the
specific spin structure (the $x_1$ parameter is one) and the fact that
the wave function is a product of independent proton and neutron wave 
functions. This is not the 
case for more recent parametrizations of the Gogny force with a finite 
range density dependent potential \cite{Chappert2015} that belong to 
the D2 family of next generation Gogny forces. The spin-orbit 
contribution to the pairing field is often neglected but the 
anti-pairing field coming from the two body kinetic energy correction 
included in the definition of the Gogny force is fully taken into 
account. Let us finally mention that the explicit central potential 
contribution to $\Delta$ is similar in structure to the HF exchange 
contribution to $h$ and shares its computational complexity.

In addition to the Gogny force, the interaction used to solve the HFB 
equation includes the Coulomb potential among protons and a two body 
kinetic energy correction. The Coulomb potential contributes in 
principle to the direct, exchange and pairing fields, but it is 
customary to treat exactly only the direct contribution whereas the 
exchange contribution is replaced by the Slater approximation and the 
Coulomb anti-pairing field is simply neglected. The relative importance 
of the exact Coulomb exchange and Coulomb anti-pairing effect has been 
discussed in detail in the context of particle number symmetry 
restoration and high spin physics at the HFB level in Ref 
\cite{Anguiano2001}. Although Coulomb exchange is usually well 
described by the Slater approximation, the impact of neglecting Coulomb 
anti-pairing can be rather dramatic in quantities like collective 
inertias or the moment of inertia of high spin states. On the other 
hand, the contribution of the two body kinetic energy correction to 
both the HF and pairing fields is fully taken into account. The two 
body kinetic energy correction contribution to pairing is repulsive and 
yields to a rather large anti-pairing effect that is compensated by the 
central potential contribution. 

For the implementation of all these techniques in the form of computer
codes see appendix \ref{App:B}.

%General principles in the solution of the HFB equation with the Gogny force:
%expansion in a harmonic oscillator basis, evaluation of matrix elements, solution
%of the selfconsistent and non-linear HFB problem, importance of the rearrangement in
%the mean field. Handling of the Coulomb force and the two body kinetic energy correction.

% ----------------------------------------------------------------------
%                      M e a n   F i e l d    A p p l i c a t t i o n s
% ----------------------------------------------------------------------

Mean field calculations with the Gogny force go back to the mid seventies. The 
first paper appearing in a regular journal was a HF calculation of deformation 
properties in the Sm isotopes \cite{Decharge1975} where it was shown 
that the recently proposed Gogny D1 interaction was able to reproduce 
quadrupole deformation properties of heavy nuclei. This early paper was 
followed by a beyond-mean field calculation of the charge density of 
$^{58}$Ni \cite{GIROD1976}. In both cases quadrupole deformation was 
allowed and an axially symmetric HO basis was used to expand the HF 
amplitudes. The full consideration of the HFB theory was performed in 
the seminal calculation  by J. Decharge and D. Gogny 
of spherical nuclei \cite{decharge1980}  where the D1 parametrization of the force was used 
to study semi-magic nuclei and their pairing properties along isotopic chains.
For mean field calculations using the Skyrme EDF or relativistic lagrangians
the reader is refered to the review paper of Ref \cite{bender2003}. 

%%%%%%%%%%%%%%%%%%%%%%%%%%%%%%%%%%%%%%%
\begin{figure}
\begin{center}
\includegraphics[width=0.5\textwidth]{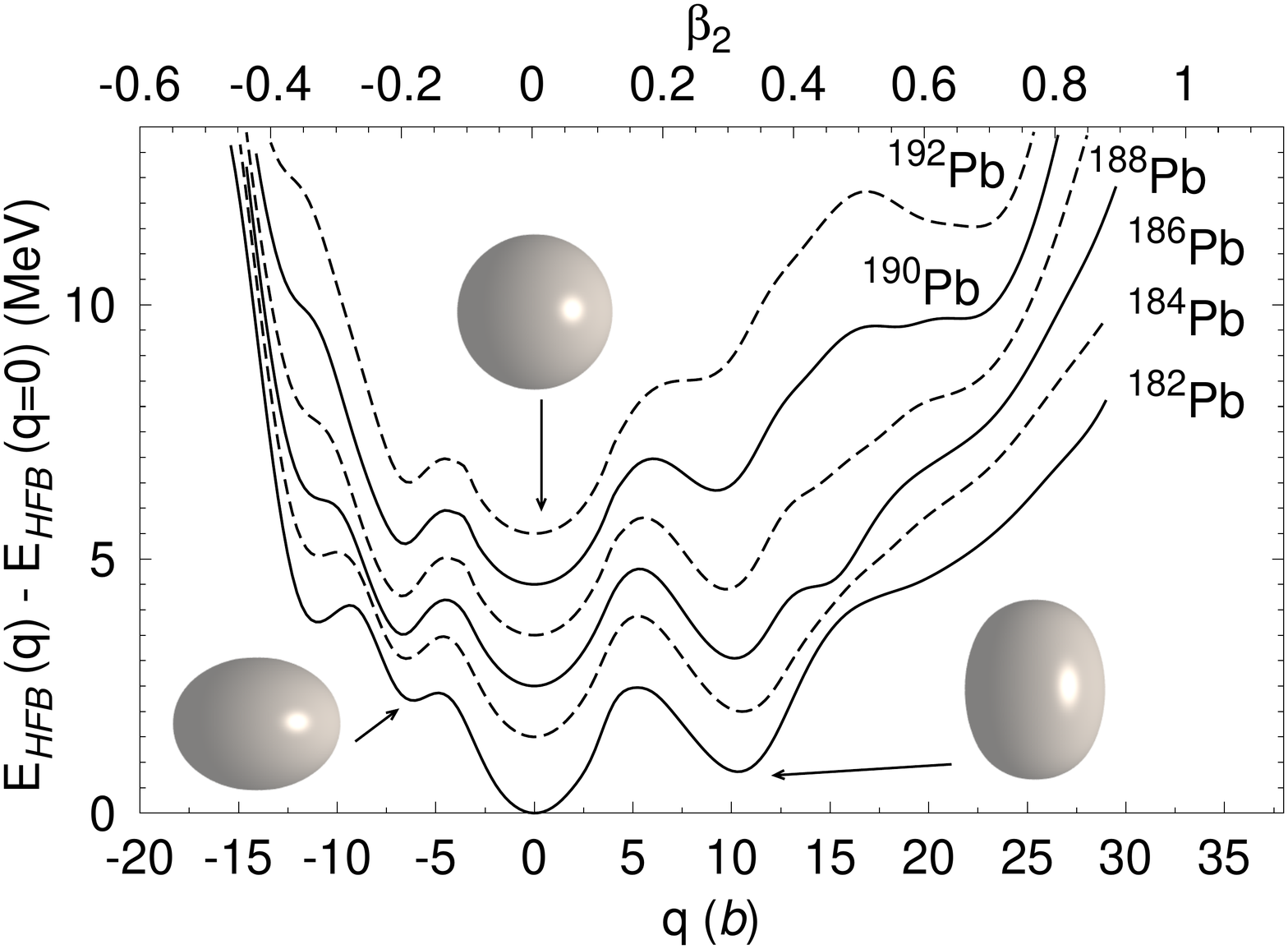}
\caption{ Potential energy surfaces given as a function of the axial quadrupole moment $q$ in barns
for several neutron deficient lead isotopes showing the phenomenon of 
triple shape coexistence. The real shape of the nucleus for each of the 
three minima is also shown. Figure adapted from Ref.~\cite{Egido2004}. }
\label{Fig:PbSC}
\end{center}
\end{figure}
%%%%%%%%%%%%%%%%%%%%%%%%%%%%%%%%%%%%%%

At the mean field level it is customary to carry out constrained 
calculations where the target wave function is required to produce 
specific values of the mean values of some observables like the 
quadrupole, octupole, etc mass moments. In this way, the so-called 
``potential energy surfaces" (PES) are obtained. They are linked to the 
dynamics of the associated collective degree of freedom (represented by 
the constrained operator: quadrupole, octupole, etc). Typical 
calculations of this kind are those studying the PES as a function of 
the axial quadrupole moment in order to identify the ground state's 
quadrupole deformation or the existence of shape coexistence between 
prolate or oblate minima (some times saddle points). A typical example is that of the triple shape 
coexistence in neutron deficient lead isotopes revealing three minima, 
one prolate, one oblate and other spherical \cite{Egido2004}. The PES corresponding to 
the relevant nuclei is shown in Fig \ref{Fig:PbSC} as a function of the
quadrupole moment. Triple shape coexistence is observed in most of the
nuclei displayed in the Figure and
can be connected with three low lying $0^{+}$ states, like the ones
experimentally identified in $^{186}$Pb \cite{Andreyev2000}.

These kind of PES calculations 
are also useful to identify the existence of super-deformed \cite{Valor1997} 
or even hyper-deformed intrinsic states \cite{Egido1997}. Quadrupole deformed
axially symmetric PESs are also very common in the description of fission 
as they allow for the description of the ``potential energy" felt by the 
nucleus in its way down to fission. In this case, reflection symmetry is
allowed to break  in order to describe asymmetric fission, where the mass of
the two resulting fragments is not equal.  

The calculation of PES using as collective
variable the axial octupole moment $Q_{30}$ has permitted to characterize
octupole correlations in nuclei. After some years of debate 
(see, for instance, \cite{egi92c}), the octupole 
deformed character of some nuclei like $^{224}$Ra \cite{Gaffney2013} or
$^{144-146}$Ba \cite{Bucher2017,PhysRevLett.116.112503} has
been unambiguously established experimentally. 
The coupling between the axially symmetric quadrupole and octupole degrees
of freedom has been analyzed at the mean field level in 
Refs \cite{Robledo2013,Rodriguez-Guzman2012}. A weak but not neglegible 
coupling is found in most of the cases studied.

%%%%%%%%%%%%%%%%%%%%%%%%%%%%%%%%%%%%%%%
\begin{figure}
\begin{center}
\includegraphics[width=0.5\textwidth]{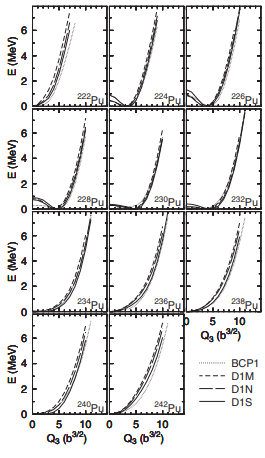}
\caption{HFB potential energy surfaces computed with three
variants of the Gogny force (D1S, D1N, D1M) and the BCP1 \cite{Rob10b}
energy density functional are plotted as a function of the
axial octupole moment for several isotopes of Pu.  Figure adapted from Ref.~\cite{Rob12l}. }
\label{Fig:Q3Pu}
\end{center}
\end{figure}
%%%%%%%%%%%%%%%%%%%%%%%%%%%%%%%%%%%%%%

Systematic mean field calculations exploring the existence of octupole deformation
in the ground state of even-even nuclei have been carried out with D1S, D1N and
D1M in Ref \cite{Robledo2011} in the context of a beyond mean field calculation.
Here, it was found that octupole deformation is only present in a
few actinides around Ra, a few rare earth around Ba and a few $A\approx 90$ nuclei
around Zr. In Ref \cite{Rob12l} another calculation in the actinide
region and taking into account other functionals came to the same conclusion.
In Fig \ref{Fig:Q3Pu} a sample of those calculations is given. Potential
energy curves are plotted as a function of the axial octupole moment $Q_{30}$ for
a few relevant Pu isotopes. It is found that irrespective of the interaction
used, the isotopes from $^{224}$Pu to $^{232}$Pu have an octupole deformed 
ground state. The deepest minimum occurs for $^{226}$Pu with a depth that
slightly depends on the force used but is of the order of 1 MeV.

\begin{figure}
\begin{center}
\includegraphics[width=0.65\textwidth]{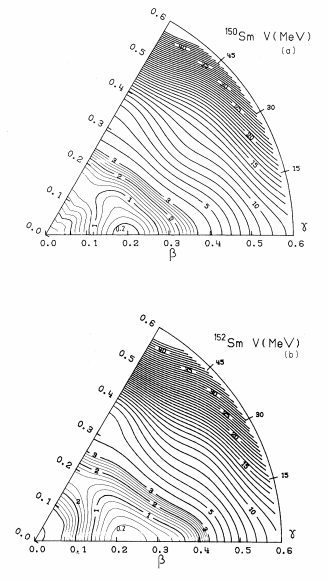}%
\caption{HFB $\beta-\gamma$  potential energy surfaces for the A=150 
and 152 isotopes of Sm computed with the Gogny D1S force. Contour lines 
are drawn every 0.2 MeV for energies between 0 and 3 MeV and every 1 
MeV for higher energies.  Figure taken from Ref.\cite{Gir83}.}
\label{Fig:BetGam}
\end{center}
\end{figure}

In order to look  for triaxial
deformed minima, another component of the 
quadrupole tensor has to be taken into account in addition to $Q_{20}$. 
The resulting "triaxial shapes" are characterized by the $\beta$ and $\gamma$ 
shape parameters. 
\begin{eqnarray}
	\beta & = & \sqrt{\frac{20\pi}{9}} \frac{\sqrt{Q_{20}^{2}+2Q_{22}^{2}}}{r_{0}^{2}A^{5/3}}   \\
	\tan \gamma & = & \frac{\sqrt{2} Q_{22} }{Q_{20}}   
\label{beta_gamma_definition}
\end{eqnarray}
The first calculation with the Gogny force including 
triaxiality was presented in Ref \cite{Gir83} where also details on how 
to compute the matrix elements of the Gogny force in a triaxial HO 
basis are given. In the triaxial case, PESs become the popular 
$\beta-\gamma$ planes that are very helpful in the interpretation of 
experimental results. This $\beta-\gamma$ planes are also essential 
ingredients in the 5DCH (see Sec \ref{Sec:MF5D} for details) to be 
discussed below as well as in the determination of the parameters of 
the interacting boson model (see Sec \ref{Sec:IBM}). Triaxial 
deformation also plays a relevant
role in the reduction, by a couple of MeV, of the first fission barrier
height in the Actinides  \cite{Gir83,Delaroche2006}. This reduction improves
substantially the agreement with experimental data and helps to reduce
the rather long predictions for the spontaneous fission half lives.
An example of $(\beta,\gamma)$ potential
energy surfaces is given in Fig \ref{Fig:BetGam} for the nuclei
$^{150}$Sm and $^{152}$Sm in the form of a polar contour plot. The two
nuclei show a prolate ground state with deformations $\beta=0.2$ and $\beta=0.23$, respectively
as well as oblate minima that are connected to the prolate minima through
a path that goes along triaxial shapes. The barrier for this path is, in
the two cases, much lower that the one corresponding to the path going 
through spherical shapes.

%%%%%%%%%%%%%%%%%%%%%%%%%%%%%%%%%%%%%%%
%%%%%%%%%%%%%%%%%%%%%%%%%%%%%%%%%%%%%%

% ----------------------------------------------------------------------
% 
%Applications:
%\begin{itemize}
%\item Spherical calculations (Decharge, Gogny) \cite{decharge1980}
%\item Deformed (axial) calculations (Girod, Berger Gogny): Potential energy surfaces and fission 
%(Berger, Gogny, Goutte, Robledo, Rodriguez)
%\item Triaxial calculations (Girod, Robledo, Rodriguez): $\beta$, $\gamma$ plane.
%\item Cranking calculations (time reversal breaking): high spin physics, 
%\item Odd mass nuclei: full blocking and the equal filling approximation
%\item Selfconsistent multiquasiparticle excitations
%\item The Gogny force in nuclear matter
%\item Finite temperature
%\end{itemize}

%                                                        Cranking

\begin{figure}
\begin{center}
\includegraphics[width=0.55\textwidth]{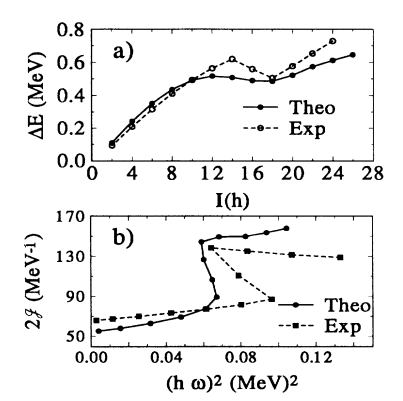}%
\includegraphics[width=0.40\textwidth]{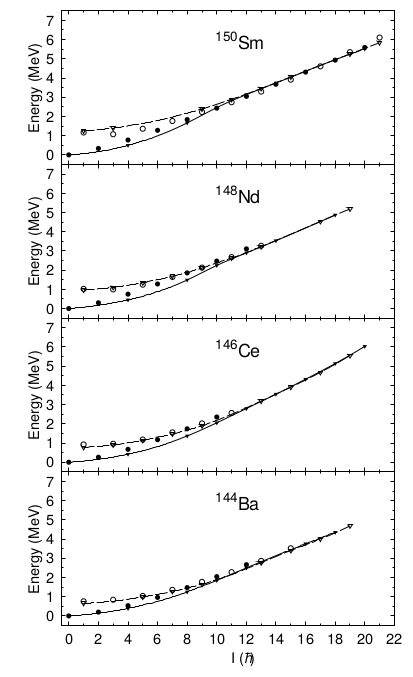}
\caption{ On the left hand side, upper panel the $\gamma$ ray energy 
$\Delta E_{\gamma}(I)=E(I)-E(I-2)$ is plotted vs the spin $I$ for the
rotational band of the nucleus $^{164}$Er.
In the lower panel, the static moment of inertia $\mathcal{J}$ for the same nucleus is
plotted vs the square of the angular velocity $\omega$ to show the phenomenon
of backbending. On the right hand side panel the $\gamma$ ray energy
of the positive and negative parity rotational bands of various rare earth isotopes is plotted as a
function of angular momentum $I$.   Figure adapted from Ref.\cite{EgidoRobledo93} ~\cite{Gar98}. }
\label{Fig:HS}
\end{center}
\end{figure}

High spin physics can also be described in the mean field framework by
using the Cranked HFB method where an additional constraint is introduced
on the mean value of the $\hat{J}_{x}$ operator \cite{Ring1970}
\begin{equation}
	\langle \hat{J}_{x} \rangle = \sqrt{J(J+1)-\langle \hat{J}_{z}^{2} \rangle}
	\label{cranking_constr}
\end{equation}
The last term $\langle \hat{J}_{z}^{2} \rangle$ is often neglected in practical
applications for even-even nuclei where it tends to be rather small compared to
$J(J+1)$.
By solving the HFB equation with this constraint a Lagrange multiplier $\omega$
has to be introduced. As the Lagrange multiplier corresponds to the derivative
of the energy with respect to the constraint, it can be interpreted as the
angular velocity of the rotating nucleus. The Cranked HFB can be derived
(see below) by starting with an angular momentum projected theory where
the intrinsic wave function is searched for  as to minimize the projected
energy (Variation After Projection, VAP) while assuming the strong deformation
limit for the intrinsic state \cite{Kamlah1968}. Another characteristic
feature of the Cranked HFB is the time-odd character of the constraining
operator that leads to the breaking of time reversal invariance even in
even-even nuclei. The first applications of the Cranking method with realistic
effective interactions go back to the early seventies when the
method was implemented with the Pairing+Quadrupole hamiltonian \cite{Ring1970}
and also with Skyrme interactions \cite{Fleckner1979}. 
The solution of the Cranked HFB equation
was first implemented in a computer code with the Gogny force a few years later 
in Ref \cite{EgidoRobledo93} in order to
study the evolution of rotational bands with spin and the backbending phenomenon.
It turns out that the moment of inertia of a rotational band as obtained with
the Cranked HFB (the Thouless-Valatin moment of inertia \cite{Thouless1962}) is a good
observable to test the pairing channel of the interaction and there have
been several thorough studies in this respect \cite{Afa00,Delaroche2006,Dobaczewski2015}.
An example of rotational band in a normal deformed nucleus $^{164}$Er showing
the phenomenon of backbending is shown in the left-hand side panels of Fig \ref{Fig:HS}
where the $\gamma$ ray energy  
$\Delta E_{\gamma}(I)=E(I)-E(I-2)$ is plotted vs the spin $I$. 
Also the static moment of inertia $\mathcal{J}$
is plotted as a function of the square of the angular velocity $\omega$. In both
plots the phenomenon of backbending (the crossing of two rotational bands
with different structures) is clearly seen as a sudden dip in the case of
$\Delta E(I)$ and as a back-bending curve in the case of $\mathcal{J}$.
Super-deformed intrinsic states also produce beautiful rotational bands
that have been the subject of systematic studies with the Gogny force
\cite{Valor1997,Valor2000}. The interplay between angular momentum and octupole
correlations leading to the concept of alternating parity rotational bands
has been analyzed in \cite{Gar97,Gar98}. An example of alternating parity
bands (requiring projection to good parity, see Sec \ref{Sec:SRPar} below) is
given in the right panel of Fig \ref{Fig:HS} where the energy of the members
of the positive and negative parity rotational bands are plotted as a 
function of the angular momentum for several rare earth isotopes. At low
spins the four nuclei are not octupole deformed and the positive and negative parity bands are 
well separated. However, as the spin increases, permanent octupole deformation develops
in the intrinsic state. As a consequence, the excitation energy of the negative parity 
state with respect to the positive parity one becomes very small (see Sec \ref{Sec:SRPar} below)
and the two rotational bands interleave. 

%%%%%%%%%%%%%%%%%%%%%%%%%%%%%%%%%%%%%%%
%%%%%%%%%%%%%%%%%%%%%%%%%%%%%%%%%%%%%%

% -----------------------------------------------------------------------
%                                                        Odd mass nuclei
% -----------------------------------------------------------------------

Another situation where the time-reversal symmetry of the wave function is
explicitly broken is in the description of odd mass, or odd-odd mass nuclei 
due to the presence of unpaired nucleons.
In the HFB framework, odd-A nuclei require the introduction of the
so-called ``blocked" HFB states
\begin{equation}\label{Block}
	|\Psi_{\mu}\rangle = \beta_{\mu}^{\dagger} | \Psi_{0}\rangle
\end{equation}
where $ |\Psi_{0}\rangle $ is the wave function of an even system and $\beta_{\mu}^{\dagger} $
is the quasiparticle creation operator on the quantum state characterized by the
label $\mu$.
In order to define in a more precise way the concepts just to be discussed, it
is convenient to introduce the concept of ``number parity". It can easily be proven 
by going to the BCS representation with the Bloch-Messiah theorem \cite{Bloch1962} that
any HFB wave function can be decomposed as a linear combination of wave
functions with definite number of particles. The decomposition is such that
wave functions with an even number of particles cannot be mixed with those
with an odd number of particles. As a consequence, we can catalog  any HFB intrinsic
state in terms of the ``number parity", even or odd, according to the parity 
of the number of particles entering into its decomposition. For instance, a 
HFB wave function has ``even number
parity" if it is decomposed as a linear combination of good particle number states
containing only even number of particles. In the same way ``odd number parity"
states are defined. A fully paired HFB state with a BCS like structure in the
canonical basis has even number parity. In principle, HFB states with even
number parity (both for protons and neutrons) could only be used to describe 
even-even nuclei. Using this language,  $ |\Psi_{0}\rangle $ in Eq (\ref{Block}) 
is an even number parity
state, whereas, by construction $|\Psi_{\mu}\rangle$ has odd number parity and
is only suited to describe odd mass systems.

The ``blocked" HFB state of Eq \ref{Block} is also a HFB state, as it is
the vacuum of the set of quasiparticle operators 
$\beta_{1},\ldots,\beta_{\mu}^{\dagger},\ldots$ \cite{Sugawara66}.
As $\beta_{\mu}^{\dagger}$ now plays the role of $\beta_{\mu}$ and both
states can be obtained from the other by a convenient exchange of the $\mu-th${}
column of the $U$ and $V$ amplitudes  it is not surprising that the 
``blocked" HFB method is essentially the same as the traditional, fully paired one,
but performing the $U$ and $V$ column exchange in the appropriate place \cite{Sugawara66}. 
For instance, the traditional form of the density matrix and pairing
tensor for ``even number parity states" $\rho_{kk'} = ( VV^{T})_{kk'}$ and 
$\kappa_{ij} =(V^{*}U^{T})_{kk'}$ now becomes
\begin{equation}
	\rho^{(\mu)}_{kk'} = 
\left(V^{*}V^{T}\right)_{kk'}+
\left(U_{k'\mu}^{*}U_{k\mu}-V_{k'\mu}V_{k\mu}^{*}\right)
\label{Roblock}
\end{equation}
and
\begin{equation}
\kappa_{kk'}^{(\mu)}=
\left(V^{*}U^{T}\right)_{kk'}+
\left(U_{k\mu}V_{k'\mu}^{*}-
U_{k'\mu}V_{k\mu}^{*}\right)\label{Kappablock}
\end{equation}
where $U$ and $V$ are the reference Bogoliubov amplitudes of the ``even number
parity" state. The ``blocked" density matrix, instead of being pairwise 
degenerate, has an eigenvalue 1 and
another 0 in the canonical basis.
The formal justification of this ``exchanging of columns" procedure \cite{Sugawara66} can be found, for instance in
\cite{Robledo}. The main consequence of using ``blocked" HFB states is
that both the density matrix $\rho$ and the pairing tensor $\kappa$ are not 
invariant under time reversal, that is they are given by the
sum of time-even and  time-odd terms. This forces to consider also 
time-even and time-odd contributions to the HF and pairing fields in the HFB
procedure. In order
to overcome the necessity to compute the time-odd fields, the so called
``equal (or uniform) filling approximation" (EFA) is used 
(see \cite{decharge1980} for an early use
of the EFA with the Gogny force). In this approximation
the blocked density matrix and pairing field of Eqs (\ref{Roblock}) and
(\ref{Kappablock}) are replaced by a ``weighted average"
\begin{equation}
\rho^{(\mu)}_{kk'} = 
\left(V^{*}V^{T}\right)_{kk'} + \frac{1}{2}
\left(
\left(U_{k'\mu}^{*}U_{k\mu}-V_{k'\mu}V_{k\mu}^{*}\right) + 
\left(U_{k'\bar{\mu}}^{*}U_{k\bar{\mu}}-V_{k'\bar{\mu}}V_{k\bar{\mu}}^{*}\right)  
\right)
\label{RoEFA}
\end{equation}
and
\begin{equation}
\kappa_{kk'}^{(\mu)}=
\left(V^{*}U^{T}\right)_{kk'}+\frac{1}{2}\left(
\left(U_{k\mu}V_{k'\mu}^{*}-U_{k'\mu}V_{k\mu}^{*}\right){}
\left(U_{k\bar{\mu}}V_{k'\bar{\mu}}^{*}-U_{k'\bar{\mu}}V_{k\bar{\mu}}^{*}\right){}
\right)
\label{KappaEFA}
\end{equation}
where both the contribution of the blocked state $\mu$ and its time reversed
counterpart $\bar{\mu}$ are considered with the same weight. The intuitive
justification  behind this approximation is that both $\mu$ and $\bar{\mu}$ have 
an occupancy of $1/2$ each. The drawback of this approximation is that
there is no single HFB wave function that leads to the density matrix and
pairing tensor of Eqs (\ref{RoEFA}) and (\ref{KappaEFA}). It took many years
since its first use to find a solid foundation of the EFA \cite{Per08}
in terms of statistical ensembles where both the blocked state $|\Psi_{\mu}\rangle${}
and its time reversed partner $|\Psi_{\bar{\mu}} \rangle$ are members of a
statistical ensemble with equal probability $p=1/2$.
In this way, the EFA formalism becomes the same as the one of finite temperature
HFB (discussed below) but with fixed probabilities. Also, it is clear that
the EFA is a variational approximation with all the associated advantages,
like the use of the gradient method for the
numerical solution of the EFA-HFB equation. 

From a practical perspective, both the full blocking and the EFA 
require the use of  starting one-quasiparticle configurations built on
the underlying even state. Due to self-consistency, the choice of the
quasiparticle with the lowest excitation energy within a given set of
quantum numbers does not represent a  guarantee for reaching the lowest
self-consistent solution with the same quantum numbers after solving the
self-consistent equation. Therefore, it is important to start
the calculation from several initial quasiparticle excitations
to make sure one is landing in the lowest energy solution for the
given set of quantum numbers. Typically, one needs to consider of
the order of ten starting quasiparticles to be sure to reach the
ground state and this is the reason why dealing with odd mass nuclei
is computationally more expensive
than dealing with even-even ones. 

\begin{figure}
\begin{center}
\includegraphics[width=0.40\textwidth]{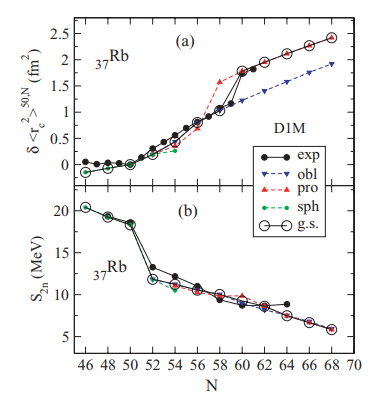}%
\includegraphics[width=0.55\textwidth]{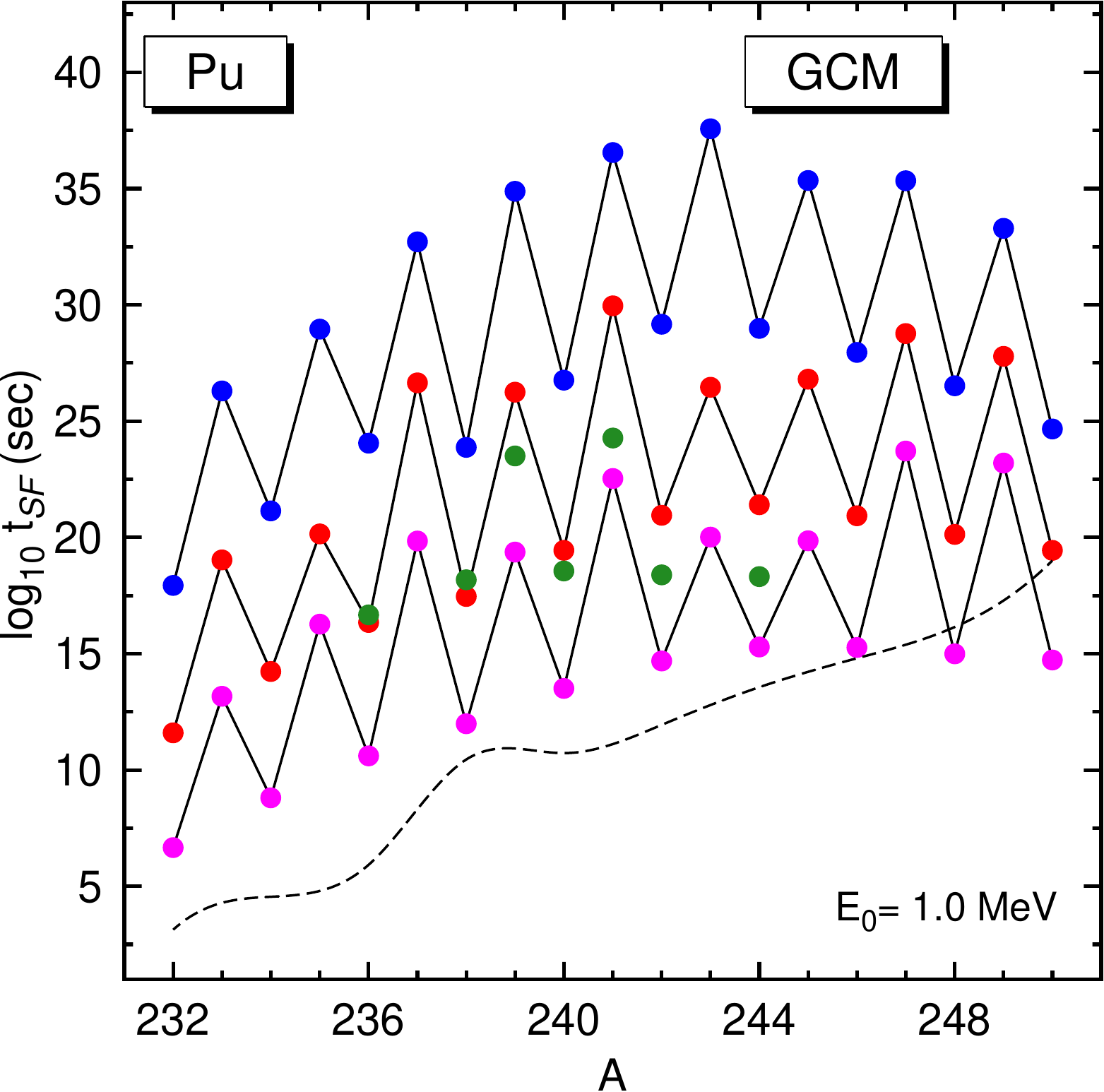}
\caption{On the left hand side, the evolution of the nuclear radius and
the two neutron separation energies is shown as a function of neutron
number for several Rb ($Z=37$) odd mass isotopes. The evolution of 
the ground state deformation as $N$ increases is observed to go from 
spherical to oblate to prolate. The changes induced in the radius as 
a consequence of the shape changes are clearly observed. On the right
hand side panel, the spontaneous fission half-live $t_\mathrm{SF}$ is
plotted as a function of mass number A for several isotopes of Pu including
even and odd number of neutrons. Experimental data is shown by green bullets,
whereas experimental predictions are shown by bullets connected by lines. 
The different theoretical predictions correspond to different levels of
enhancement of the pairing strength (blue, no enhancement; red, a 5\%{}
enhancement and purple, a 10 \% enhancement.
Figure adapted from Ref.\cite{Rod16y} ~\cite{Rodriguez-Guzman2010,Rod10}. }
\label{Fig:OddEFA}
\end{center}
\end{figure}

Odd mass systems have been mostly described in the framework of the EFA 
as, for instance in the seminal work of Ref \cite{decharge1980}, the fitting
protocol of D1M \cite{goriely2009}, or other applications like 
the study of shape evolution in some isotopic chains \cite{Rodriguez-Guzman2010,Rod10}. Special
attention deserves also the seminal evaluation of fission properties in odd systems
that has shown that the most relevant factor in the description of those
systems is the quenching of pairing correlations induced by the presence
of the unpaired nucleon \cite{Rod16y,RodG17}. As a consequence of the severe
quenching of pairing, the collective inertia governing spontaneous fission
half-lives ($t_\mathrm{SF}$) increases enormously leading to a huge odd-even
staggering of $t_\mathrm{SF}$. The staggering is reduced to a level 
comparable with the experimental data if the pairing strength is artificially
increased by 10 \% \cite{Rod16y}. In Fig \ref{Fig:OddEFA} a few examples
of results obtained with the EFA are presented. They range from the 
evolution of properties (like radii or $S_{2n}$ separation energies)
of some odd mass Rb isotopes to the aforementioned description of 
the staggering of $t_\mathrm{SF}$ in the Pu isotopic chain.

%%%%%%%%%%%%%%%%%%%%%%%%%%%%%%%%%%%%%%%
%%%%%%%%%%%%%%%%%%%%%%%%%%%%%%%%%%%%%%

\begin{figure}
\begin{center}
\includegraphics[width=0.45\textwidth]{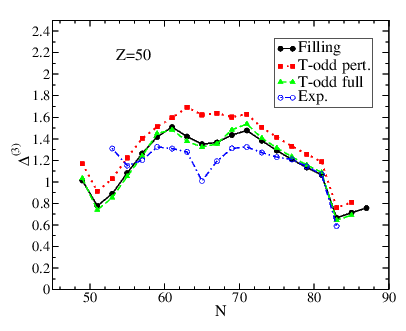}%
\caption{The pairing gap $\Delta^{(3)}$ is plotted as a function of neutron
number for the tin isotopic chain for different approximations in the
description of the unpaired odd neutron (EFA, full blocking and perturbative
blocking). The theoretical results are compared to the experimental data.
Figure adapted from Ref.\cite{Rob12}. }
\label{Fig:Odd}
\end{center}
\end{figure}

Calculations with full blocking using the Gogny force are scarce: there is
the calculation of the properties of super-deformed rotational bands in
$^{191}$Hg \cite{VILLAFRANCA199735} or the recent proposal to do full blocking
calculations but preserving axial symmetry \cite{Rob12}. The advantage of
this proposal is that the $K$ quantum number is preserved and the
assignment of the spin and parity of the ground state  and excited band-heads
is greatly facilitated. This 
approach has been used in a recent study of the properties of super-heavy 
nuclei \cite{Dobaczewski2015}. In Fig \ref{Fig:Odd} a comparison of the
pairing gap $\Delta^{(3)}$ obtained with the EFA, full blocking and the
simple Perturbative Quasiparticle approximation (see \cite{Rob12} for
details) is presented for the Sn isotopic chain. From the plot we conclude
that the time odd fields of the full blocking method have little impact
on the gap $\Delta^{(3)}$ as the results obtained with this method are
very similar to the ones of the EFA. Obviously, the perturbative approximation
fails to quench pairing correlations as much as the other two approaches
and therefore the pairing gap is significantly larger. 

%%%%%%%%%%%%%%%%%%%%%%%%%%%%%%%%%%%%%%%
%%%%%%%%%%%%%%%%%%%%%%%%%%%%%%%%%%%%%%

% -----------------------------------------------------------------------
%                          Selfconsistent multiquasiparticle excitations
% -----------------------------------------------------------------------

\begin{figure}
\begin{center}
\includegraphics[width=0.45\textwidth]{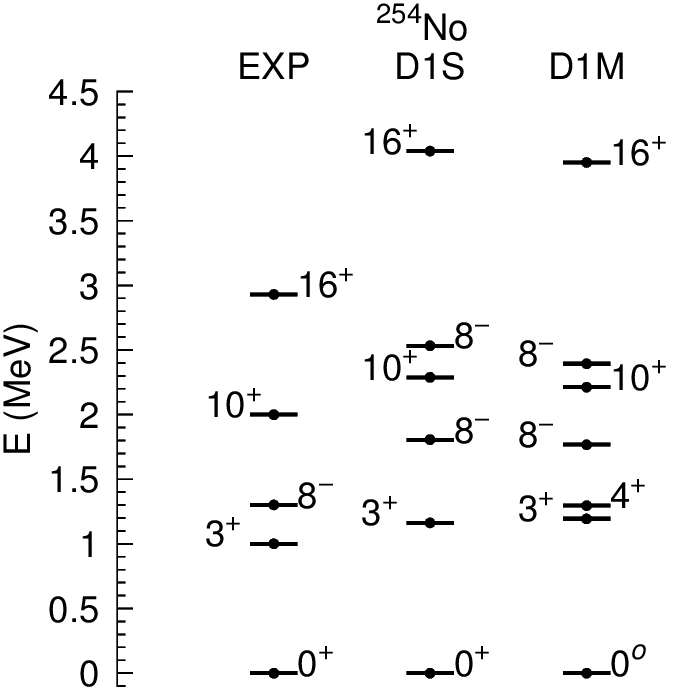}%
\caption{Excitation spectrum of high-K isomers computed self-consistently
using the fully blocked HFB procedure with the D1S and D1M parametrizations
of the Gogny force. On the left hand side of the plot the experimental
spectrum is shown for comparison. Figure adapted from Ref.\cite{Rob15}. }
\label{Fig:HighK}
\end{center}
\end{figure}

Within the mean field HFB formalism it is also possible to study 
excited states that are given by multi-quasiparticle excitations. 
Multi-quasiparticle excited states can be considered perturbatively as 
the standard quasiparticle excitations built on top of a given HFB 
reference state \cite{Delaroche2006}. The advantages of this approach 
are evident: the results are already available after a HFB calculation 
and the excited states are orthogonal by construction. On the other 
hand, self-consistency can play a very relevant role as 
multi-quasiparticle excitations tend to severely quench the pairing 
correlations present in the ground state. The drawbacks are multiple, 
first, as time reversal symmetry can be broken, the induced time odd 
fields (also present in the description of odd-A nuclei) have to be 
considered. Second, the iterative solution of the non linear problem is 
not easy to achieve even using standard gradient method techniques. 
Finally, the issue of orthogonality between the self-consistent 
multi-quasiparticle excitations and the ground state and among 
themselves becomes relevant. Some of these issues have been addressed 
in Ref. \cite{Rob15} where high $K$ isomers in $^{254}$No have been studied
with both the D1S and D1M parametrizations of the Gogny force. The effects
of self-consistency and the quenching of pairing correlations substantially
reduce the excitation energy of the two-quasiparticle and four-quasiparticle
isomers with respect to the naive sum of quasiparticle excitation energies
built on top of the reference HFB ground state. In Fig \ref{Fig:HighK} we
have plotted the excitation energy of those high-$K$ isomers with known
experimental excitation energies and compared them with our results. 
It is remarkable the good reproduction of the excitation energies given
the rather universal scope of the Gogny force and its fitting protocol.

%%%%%%%%%%%%%%%%%%%%%%%%%%%%%%%%%%%%%%%
%%%%%%%%%%%%%%%%%%%%%%%%%%%%%%%%%%%%%%

%---------------------------------------------------------------------------
%                                                       Nuclear Matter
%---------------------------------------------------------------------------

Another interesting field of application of the Gogny force  is the 
study of nuclear matter properties at the mean field level. This studies
allows to obtain, among other things, the Equation of State (EOS) of
nuclear matter, of great relevance in astrophysical environments like
the interior of neutron stars. In addition, the nuclear matter results 
can be compared to the ones of more sophisticated realistic interactions obtained 
with more elaborated  many body techniques. In this respect, the 
evaluation of the Landau parameters of the different Gogny interactions 
has become a useful tool \cite{Gogny1977,DAVESNE2016288}. The 
analysis of  isovector properties like the symmetry energy or its slope 
give hints on the expected performance of the force in astrophysical 
environments or in very neutron rich scenarios \cite{Sellahewa2014}.
Another interesting application of nuclear matter calculations is the study
of the pairing gap and its comparison with realistic forces as analyzed
in Refs \cite{PhysRevC.60.064312,KUCHAREK1989249}.

%---------------------------------------------------------------------------
%                                                       Finite temperature
%---------------------------------------------------------------------------

To end  this section of applications of the HFB method, we will discuss the use of the finite 
temperature HFB (FT-HFB) formalism to describe the physics of highly 
excited nuclei using a grand-canonical ensemble formalism at fixed temperature T. 
Although the nucleus is a finite, isolated system, the ideas of quantum 
statistical mechanics have been used to describe situations where the 
intrinsic excitation energy of the nucleus is very high and therefore its wave function can be 
any among those in a bunch of excited states with a very large level density. Given 
the description of pairing correlations in terms of a mean field theory 
with no definite number of particles, the statistical ensemble to be 
used in nuclear physics is the grand canonical one, which allows both 
the exchange of energy and particles with a fictitious external 
reservoir. The quantity determining the density operator through a 
minimization principle is the free energy $F=H-TS$ that depends not 
only on the energy but also on the entropy $S$ and temperature $T$ of the system. The 
minimization of the free energy under the assumption that the density 
matrix is the exponential of a one-body operator and that the statistical 
trace has to be taken for all possible multi-quasiparticle excitations of 
a HFB ground state, leads to the FT-HFB equation. Its form is the same
as in the zero temperature case and only the definition of the density 
and pairing tensor has to be replaced by the appropriate one
\begin{equation}
	\mathcal{R}= W \left(\begin{array}{cc}
f & 0\\ 0 & 1-f \end{array}\right) W^{\dagger}
\end{equation}
where the $f_{\mu}=1/(1+\exp(E_{\mu}/(K_{B}T))$ are the Fermi statistical
occupation factors depending on the quasiparticle energies $E_{\mu}$ and
the temperature. The FT-HFB equation 
has been solved with the Gogny force in order to study the  phase 
transitions from super-fluid to normal-fluid systems with temperature as 
well as the transition from deformed to spherical driven also by 
temperature \cite{Egi00}. In the same reference, level densities are 
also evaluated using the FT-HFB formalism. Thermal fluctuations and 
their effect of washing out the abrupt phase transitions observed at 
the mean field level are analyzed in \cite{Mar03} in a variety of 
systems. Finally, the evolution of the fission barrier heights with 
temperature is studied in the case of $^{240}$Pu in Ref  \cite{Mar09} where
the decrease with temperature of the fission barrier heights is observed.
As mentioned before, increasing the temperature means a quenching of pairing
correlations that yield to an increase in the collective inertia \cite{Mar09}. 

%%%%%%%%%%%%%%%%%%%%%%%%%%%%%%%%%%%%%%%
\begin{figure}
\begin{center}
\includegraphics[width=0.5\textwidth]{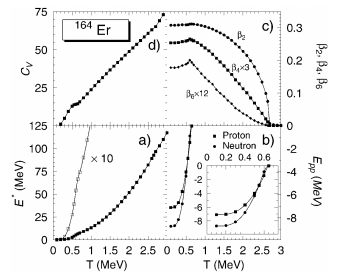}%
\includegraphics[width=0.45\textwidth]{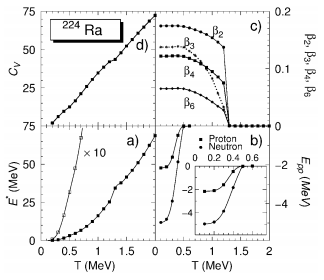}%
\caption{On the left hand side panels, the evolution with
temperature ($T$ in MeV) of different quantities for the 
nucleus $^{164}$Er. In panel a) the excitation energy 
$E^{*}$ in MeV, in panel b) the particle-particle energy,
in panel c) the $\beta_{2}$, $\beta_{4}$ and $\beta_{6}$ deformation
parameters and in panel d) the specific heat at constant
volume $C_{V}$. In the right hand side panels, the same
quantities are plotted for the octupole deformed nucleus $^{224}$Ra.
The results were obtained with the Gogny D1S interaction.
 Figures taken from Ref.\cite{Egi00}. }
\label{Fig:FTbeta}
\end{center}
\end{figure}
%%%%%%%%%%%%%%%%%%%%%%%%%%%%%%%%%%%%%%

In Fig \ref{Fig:FTbeta} the behavior with temperature of different quantities
are shown for two different types of nuclei, the quadrupole deformed $^{164}$Er
and the quadrupole, octupole deformed $^{224}$Ra. The calculations are
carried out with the Gogny D1S force in the context of FTHFB. In panels b) and c) in the
two cases we observe (panel c)) the behavior of the deformation parameters
$\beta_{\lambda}$ with temperature. A phase transition to an spherical
regime is observed at $T=2.6$ MeV in the $^{164}$Er case and at $T=1.3$ MeV
in the octupole deformed $^{224}$Ra nucleus. Also, in panels b) a  
phase transition from a paired regime to an unpaired
one at $T=0.6$ MeV. Both phase transitions produce a kink in the behavior
of the specific heat at constant volume $C_{V}$ shown in panels d) and
some discontinuity in the excitation energy $E^{*}$ of the system. The
observed phase transitions are washed out when thermal fluctuations are
considered \cite{Mar03}.

In passing, let us mention that level densities can also be computed 
using combinatorial techniques and the uncorrelated quasiparticle 
spectrum obtained from a mean field calculation with the Gogny force 
\cite{Robledo2012}. The advantages and drawbacks of this method over the finite
temperature formalism have to be still assessed but it is clear that any
method based on the HFB ground state should have more difficulties to 
take into account the effects of finite temperature driven phase transitions 
like the transition to a normal fluid regime or from a deformed intrinsic
state to a spherical one \cite{Egi00}.

\begin{figure}
\begin{center}
\includegraphics[width=0.45\textwidth]{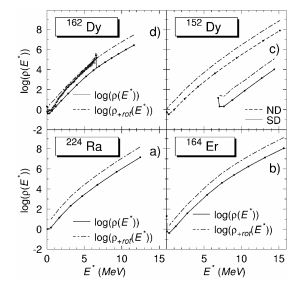}%
\caption{Logarithm of the level density $\rho (E^{*})$ plotted as a 
function of the excitation energy of the nucleus $E^{*}$ for four 
different nuclei $^{152}$Dy, $^{162}$Dy, $^{164}$Er and $^{224}$Ra. The 
(full) dashed lines correspond to the calculation (without) with the 
rotational correction.  Experimental data is also presented in the 
$^{162}$Dy case. The calculations are carried out with the D1S force. 
Figures taken from Ref.\cite{Egi00}. }
\label{Fig:FTLev}
\end{center}
\end{figure}

In Fig \ref{Fig:FTLev} the behavior of level densities as a function of 
the excitation energy are shown for the four nuclei considered in Ref \cite{Egi00},
In the $^{162}$Dy case it is compared with the experimental data. A rather
good agreement with the experiment is observed when a phenomenological correction
to take into account rotational bands is introduced. 
 
%%%%%%%%%%%%%%%%%%%%%%%%%%%%%%%%%%%%%%%
%%%%%%%%%%%%%%%%%%%%%%%%%%%%%%%%%%%%%%

% -----------------------------------------------------------------------------------------------------------------
%                                                                                             TDHFB
% -----------------------------------------------------------------------------------------------------------------

\subsection{Time dependent Hartree-Fock-Bogoliubov}

The time dependent Hartree-Fock-Bogoliubov (TDHFB) method is the 
natural extension to treat dynamical aspects within the HFB theory. 
The generalized density matrix of HFB $\mathcal{R}$ is no longer static
and its time evolution is governed by the TDHFB equation
\begin{equation}\label{eq:TDHFB}
i\hbar \frac{\partial \mathcal{R}}{\partial t}=[\mathcal{H},\mathcal{R}]
\end{equation}
The TDHFB equation preserves the energy of the initial state (defined by $\mathcal{R}(t=0)$) 
and therefore cannot be used to describe tunneling through barriers, i.e. it
is not possible to describe spontaneous fission. On the other hand,
it can be used to study the time evolution of a wave packet confined in a potential well 
and in this way extract several properties of this kind of motion (elementary excitation
energies, strengths, etc). The TDHFB method with the Gogny force
has been recently implemented  by Hashimoto \cite{Hashimoto2012} in a 
harmonic oscillator basis
and preserving axial symmetry. The idea is to replace Eq (\ref{eq:TDHFB}) by
the equivalent in terms of the Bogoliubov amplitudes $U$ and $V$
\begin{equation}
	i\hbar \frac{\partial }{\partial t} 
\left(\begin{array}{c}
\bar{U}\\
\bar{V} \end{array}\right)={}
\mathcal{H}
\left(\begin{array}{c}
\bar{U}\\
\bar{V} \end{array}\right)
\end{equation}
This is a non-linear system of equations because the HFB Hamiltonian 
$\mathcal{H}$ itself depends on $U$ and $V$ through its dependency in 
the density matrix $\rho=V^{*}V^{T}$ and the pairing tensor $\kappa=UV^{T}$. 
The solution to these coupled multidimensional non-linear differential 
equations is carried out using predictor-corrector techniques. The 
calculations focus on the isoscalar quadrupole and/or isovector dipole 
vibrations, in the linear (small amplitude) region, in some oxygen, 
neon, magnesium and titanium  isotopes. The isoscalar quadrupole and 
isovector dipole strength functions are calculated from the expectation 
values of the isoscalar quadrupole and isovector dipole moments. This 
kind of calculations represent an alternative to the QRPA type of calculations 
mentioned in the next section. A more recent implementation, also preserving axial 
symmetry, replaces the harmonic oscillator basis along the $z$ 
direction by a Lagrange mesh. The reason for such replacement is to have 
far more  flexibility in describing the time dependent density along 
the symmetry axis. This new implementation has been successfully 
applied to the description of head-on collisions in light systems 
\cite{Hashimoto2016}.

%% file: BMF1.tex
% ------------------------------------------------------------------------------------------------------
%                               M e a n   F i e l d   
% ------------------------------------------------------------------------------------------------------

\section{Beyond mean field with the Gogny force: first applications}
\label{Sec:BMF1}

The mean-field approximation \cite{Ring1980} allows the description of several 
basic nuclear properties all over the nuclear chart. However, in order 
to access the spectroscopy of atomic nuclei, one needs to go beyond the 
mean-field level. Several routes have emerged in recent years in order 
to afford such a task. On the one hand, the symmetry-projected 
Generator Coordinate Method (GCM)  has already provided a wealth of 
results in different regions of the nuclear chart (see, for example, 
Refs.~\cite{bender2003,niksic2011,Rodriguez-Guzman2002a} and references therein). On 
the other hand, calculations are very demanding from a computational 
point of view, specially in the case of medium and heavy nuclei and/or 
when several collective coordinates should be included in the 
symmetry-projected GCM ansatz. This certainly limits, at least for the 
moment, the applicability of the approach from a computational point of 
view though progress is growing due to the new generation of 
computational facilities. All these methods along with some approximations
to them will be discussed below in Sec \ref{Sec:LAM}.
Another route is represented by the 
Quasiparticle Random Phase Approximation (QRPA) which can be viewed
as a small-amplitude approximation to the GCM or the TDHFB methods \cite{Ring1980}. 
A different alternative 
route is  represented by fermion-to-boson mapping procedures 
described in Sec \ref{Sec:IBM} below. 

% -----------------------------------------------------------------------------------------------------------------
%                                                                                             QRPA
% -----------------------------------------------------------------------------------------------------------------

% \cite{Libert1999,Peru2011,Goriely2016,Martini2016}

\subsection{The QRPA}
\label{Sec:QRPA}

The quasiparticle random phase approximation (QRPA) is nothing but the small amplitude limit of arbitrary 
vibrations around the HFB equilibrium configuration~\cite{Thouless1961,Ring1980}. In the QRPA it is 
possible to study within the same framework not only collective 
excitations of isoscalar and isovector character, but also single 
particle excitations. The solution of the QRPA equation involves the 
construction and subsequent diagonalization of a huge matrix in the 
space of two quasiparticle excitations and therefore, applications with 
the Gogny force require a huge computational effort. Nevertheless, the 
advent of powerful computers allows for large scale calculations based 
on this method.
QRPA calculations with Gogny interactions have been recently reviewed by P\'eru and Martini in Ref.~\cite{Peru2014}. 
Other QRPA implementations and applications of the most widely used energy density functionals like 
Skyrme~\cite{Sarriguren1996,Engel1999,Sarchi2004,Terasaki2005,Terasaki2006,Terasaki2010} and/or 
RMF~\cite{Paar2003,Paar2004,Arteaga2008} are out of the scope of the present review.
Therefore, here we will only summarize the most important aspects and applications of the method, and update the list of most recent works 
done within the QRPA framework with Gogny interactions.

Derivation of QRPA equations and its formalism can be looked up in 
several textbooks (e.g. Ref.~\cite{Ring1980}). It is based on building 
two-quasiparticle excitations on top of a QRPA vacuum, 
$|\mathrm{QRPA}\rangle$, defined as:
\begin{equation}
\hat{\theta}_{\nu}|\mathrm{QRPA}\rangle=0;\hat{\theta}^{\dagger}_{\nu}=\frac{1}{2}\sum_{k<k'}X^{\nu}_{kk'}\beta^{\dagger}_{k}\beta^{\dagger}_{k'}-Y^{\nu}_{kk'}\beta_{k'}\beta_{k}
\label{QRPA_ansatz}
\end{equation}
where $(\beta^{\dagger}_{k},\beta_{k})$ are HFB quasiparticle creation and annihilation operators (see Eq.~\ref{eq:Bogo}) and the amplitudes $(X^{\nu},Y^{\nu})$ are obtained from the QRPA equation given by~\cite{Ring1980}:
\begin{eqnarray}
\left(\begin{array}{cc}
  A & B  \\
  B^{*} &  A^{*}
\end{array}
\right)\left(\begin{array}{c}
  X^{\nu}  \\
  Y^{\nu}
\end{array}
\right)=E_{\nu}\left(\begin{array}{c}
  X^{\nu}  \\
  -Y^{\nu}
\end{array}
\right)\label{QRPA_eq}
\end{eqnarray}
Here, $E_{\nu}$ are the excitation energies and the matrices $(A,B)$ 
are determined by the HFB state and the interaction through:
\begin{eqnarray}
A_{kk'll'}&=&\langle \mathrm{HFB}|\left[\beta_{k'}\beta_{k},[\hat{H},\beta^{\dagger}_{l}\beta^{\dagger}_{l'}]\right]|\mathrm{HFB}\rangle\nonumber\\
B_{kk'll'}&=&-\langle \mathrm{HFB}|\left[\beta_{k'}\beta_{k},[\hat{H},\beta_{l}\beta_{l'}]\right]|\mathrm{HFB}\rangle
\end{eqnarray}
In the implementations performed with Gogny forces, the same 
interaction is considered in the HFB and QRPA parts (normally 
neglecting two-body center-of-mass corrections and, in some cases, 
Coulomb exchange and pairing terms~\cite{Peru2014}).

Apart from the spectrum, $E_{\mu}$, the QRPA approach is widely used to 
study the response of the system to external fields. In the first 
implementations with the Gogny interaction, only spherical RPA (without 
pairing) were 
considered~\cite{Blaizot1976,Blaizot1977,Decharge1981,Decharge1983,Blaizot1995,Peru2005,DeDonno2011,Co2013}. 
In these works, nuclear compressibility, giant and pygmy resonances and 
excitations in closed-shell nuclei were studied. The inclusion of 
pairing (QRPA) and axial quadrupole deformation in this kind of 
calculations was performed for the first time by P\'eru et al. in 
Refs.~\cite{Peru2007,Peru2008}. The assumption of axial and parity 
symmetry conservation allows for the classification of the QRPA states 
in terms of $K$ (the projection of the angular momentum $J$ along the 
symmetry axis) and $\pi$ (parity) quantum numbers, i.e., the QRPA 
states are given by~\cite{Peru2014}:
\begin{equation}
|\theta_{\nu},K\rangle=\hat{\theta}_{\nu,K}|0_{\mathrm{def}},(K=0)\rangle
\end{equation} 
where $|0_{\mathrm{def}}\rangle$ is the HFB ground state obtained with 
axially deformed calculations. The final evaluation of transition 
matrix elements of external fields given by the corresponding multipole 
operator, $\hat{Q}_{\lambda\mu}$, is obtained after projecting onto 
good angular momentum (see Sec \ref{Sec:SR}, Eq \ref{Eq:AMPOP}) both the ground and the QRPA excited states:
\begin{eqnarray}
|0^{+}_{\mathrm{g.s.}}\rangle&=&\frac{1}{8\pi^{2}}\int \mathcal{D}^{0*}_{00}(\Omega)\hat{R}(\Omega)|0_{\mathrm{def}}\rangle d\Omega\nonumber\\
|JM(K)_{\nu}\rangle&=&\frac{2J+1}{8\pi^{2}}\int \mathcal{D}^{J*}_{MK}(\Omega)\hat{R}(\Omega)|\theta_{\nu},K\rangle d\Omega\nonumber\\
\end{eqnarray}
Hence, the transition matrix elements are given by 
$\langle0^{+}_{\mathrm{g.s.}}|\hat{Q}_{\lambda\mu}|JM(K)_{\nu}\rangle$ 
and the sum rules and moments by:
\begin{equation}
M_{k}\left(\hat{Q}_{\lambda\mu}\right)=\sum_{\nu}E_{\nu}^{k}|\langle0^{+}_{\mathrm{g.s.}}|\hat{Q}_{\lambda\mu}|JM(K)_{\nu}\rangle|^{2}\nonumber\\
\end{equation}
For example, the energy weighted sum rule 
($\mathrm{EWSM}(\hat{Q}_{\lambda\mu})$) is given by $M_{k=1}$ in the 
expression above.

Many calculations with the deformed QRPA method with Gogny interactions 
have been performed in the recent years, some of them with direct 
applications to nuclear astrophysics. For example, low-lying excitation 
energies, pygmy and giant resonances, electromagnetic multipole 
excitations and response functions, gamma-ray strength functions, 
reaction cross-sections, etc., in nuclei all along the nuclear chart 
have been studied and compared with experimental data where 
available~\cite{Peru2007,Peru2008,Martini2011,Peru2011,Utsunomiya2013,Martini2014,Filipescu2014,Peru2014,Nyhus2015,Corsi2015,Dupuis2015,Mayer2016,Martini2016,Goriely2016,Versteegen2016}. 
In addition, Gamow-Teller response has been also computed with a 
proton-neutron QRPA method in Refs.~\cite{Martini2014a,Deloncle2017}. 
Finally, the role of continuum~\cite{DeDonno2011,DeDonno2016}, tensor 
interactions~\cite{DePace2016} and correlations beyond 
two-quasiparticle excitations~\cite{Martini2016} has been also analyzed 
recently. A thorough discussion of the performance of the method with 
several examples of its applications is found in the review of P\'eru 
and Martini and we refer the reader to Ref.~\cite{Peru2014} for further 
details.

% -----------------------------------------------------------------------------------------------------------------
%                                                                                    IBM mapping
% -----------------------------------------------------------------------------------------------------------------

\subsection{IBM mapping}
\label{Sec:IBM}

In order to extend the realm of the mean field to deal with spectroscopy 
in a simple way a novel method  has been introduced in recent years \cite{Nomura2008}. It essentially 
maps the fermionic energy surfaces, obtained within the constrained 
mean-field approximation, onto the bosonic ones computed as the 
expectation value of a chosen Interacting Boson Model (IBM) \cite{IBM} 
Hamiltonian in the boson coherent state. The fermion-to-boson mapping 
procedure determines the parameters of the chosen IBM Hamiltonian for 
each nuclear system and, therefore, no phenomenological adjustment of 
those parameters to the experimental data is required. It only relies on 
microscopic mean-field energy surfaces as the key input and, therefore, 
has the potential to provide predictions in those regions of the 
nuclear chart where experimental data are rather scarce or not even 
available as it is the case, for example, of exotic neutron-rich 
nuclei. The IBM Hamiltonian, obtained via the mapping procedure, is 
then diagonalized and the resulting wave functions are used to compute 
spectroscopic properties and transition rates. The fermion-to-boson 
mapping procedure has already allowed an accurate computationally 
economic and systematic description of basic properties in several 
regions of the nuclear chart.

\begin{figure}
\begin{center}
\includegraphics[angle=0,width=0.8\columnwidth]{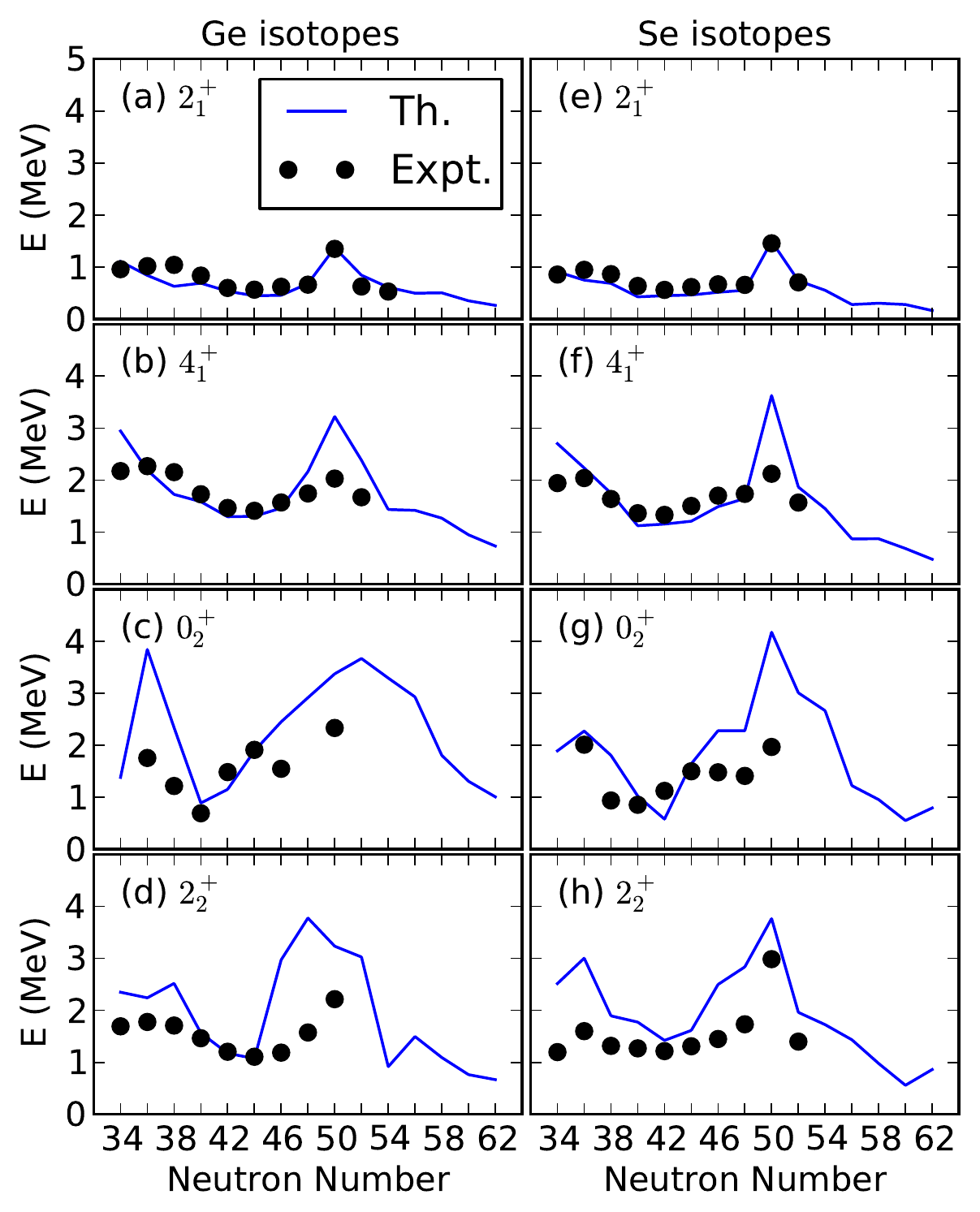} 
\end{center}
\caption{(Color online) The $2^+_1$, $4^+_1$, $0^+_2$ and $2^+_2$ 
excitation energies obtained in the diagonalization of the
mapped IBM Hamiltonian are plotted as functions of the neutron number, for 
the Ge and Se nuclei, along with 
the available experimental data \cite{data}. Taken from Ref.~\cite{Nomura2017c}.}
\label{fig:energies}
\end{figure}
%%%%%%%%%%%%%%%%%%%%%%%%%%%%%%%%%%%%%%%%%%%%%%%%%%%%%%%%%%%%%%%%%%%%%%%%%%%%%%%%%%%%%%%%%%%%%%%%%

As an example of application  we will consider a sample of results obtained for 
Ge and Se nuclei as they belong to one of the most challenging 
regions of the nuclear chart. Their structure and decay patterns have 
received  considerable experimental 
\cite{Guerdal2013,Corsi2013,Toh2013,Sun2014} and theoretical  
\cite{Yoshinaga2008,Honma2009,Kaneko2015,Gaudefroy2009,Niksic2014,Wang2015,%
Sarriguren2015,Barea2009,Petrovici1988,Petrovici1989,Petrovici1990,Petrovici1992,Petrovici2002} 
attention in recent years.
 In particular, the shape transitions around 
the neutron sub-shell closure $N = 40$ have already been studied in 
detail. Moreover, Ge and Se nuclei exhibit a pronounced competition 
between different configurations associated with a variety of intrinsic 
shapes, i.e., shape coexistence. The corresponding  energy spectra 
display  low-lying excited 0$^{+}$ energy levels which could be linked 
to proton intruder excitations across the $Z=28$ shell gap.

The nuclei $^{66-94}$Ge and $^{68-96}$Se have been studied in 
Ref.~\cite{Nomura2017c}. The employed IBM model space consisted of collective 
nucleon pairs in the valence space with spins and parity $J^{\pi}$ = 
0$^{+}$ (monopole $S$ pair) and 2$^{+}$ (quadrupole $D$ pair). They are 
associated with the $J^{\pi}$ = 0$^{+}$ (s) and  2$^{+}$ (d) bosons, 
respectively \cite{Otsuka1978}. The microscopic input for the mapping, 
i.e., the mean-field energy surfaces as functions of the $\beta$ and 
$\gamma$ deformation parameters, have been obtained via 
Hartree-Fock-Bogoliubov (HFB) calculations with the parametrization D1M 
\cite{goriely2009} of the Gogny interaction. Configuration mixing has 
been included in the corresponding (mapped) IBM calculations. To this 
end, the boson model space has been extended following a method, 
proposed by Duval and Barret \cite{Duval1981}, that incorporates the 
intruder configurations by introducing several independent IBM 
Hamiltonians. The intruder configurations correspond to proton $2p-2h$ 
excitations across the Z=28  shell gap and the different boson spaces 
are allowed to mix via certain mixing interaction. The criterion to 
include the configuration mixing for a given nucleus is that the second 
lowest-energy minimum  in the mean-field  energy surface is clear 
enough so as to constrain the corresponding (unperturbed) Hamiltonian 
for the intruder configuration. Therefore, configuration mixing has 
been taken into account for $^{66,70-74,90-94}$Ge and 
$^{68-76,90-96}$Se. All the required equations as well as the fitting 
procedure to obtain the parameters of the (mapped) IBM model can be 
found in Ref.~\cite{Nomura2011b}. Once the IBM parameters are determined for 
a given system, the Hamiltonian is diagonalized and the resulting wave 
functions are used to compute electromagnetic properties that could be 
considered as signatures of shape coexistence and/or shape transitions.

\begin{figure}
\begin{center}
\includegraphics[angle=0,width=0.8\columnwidth]{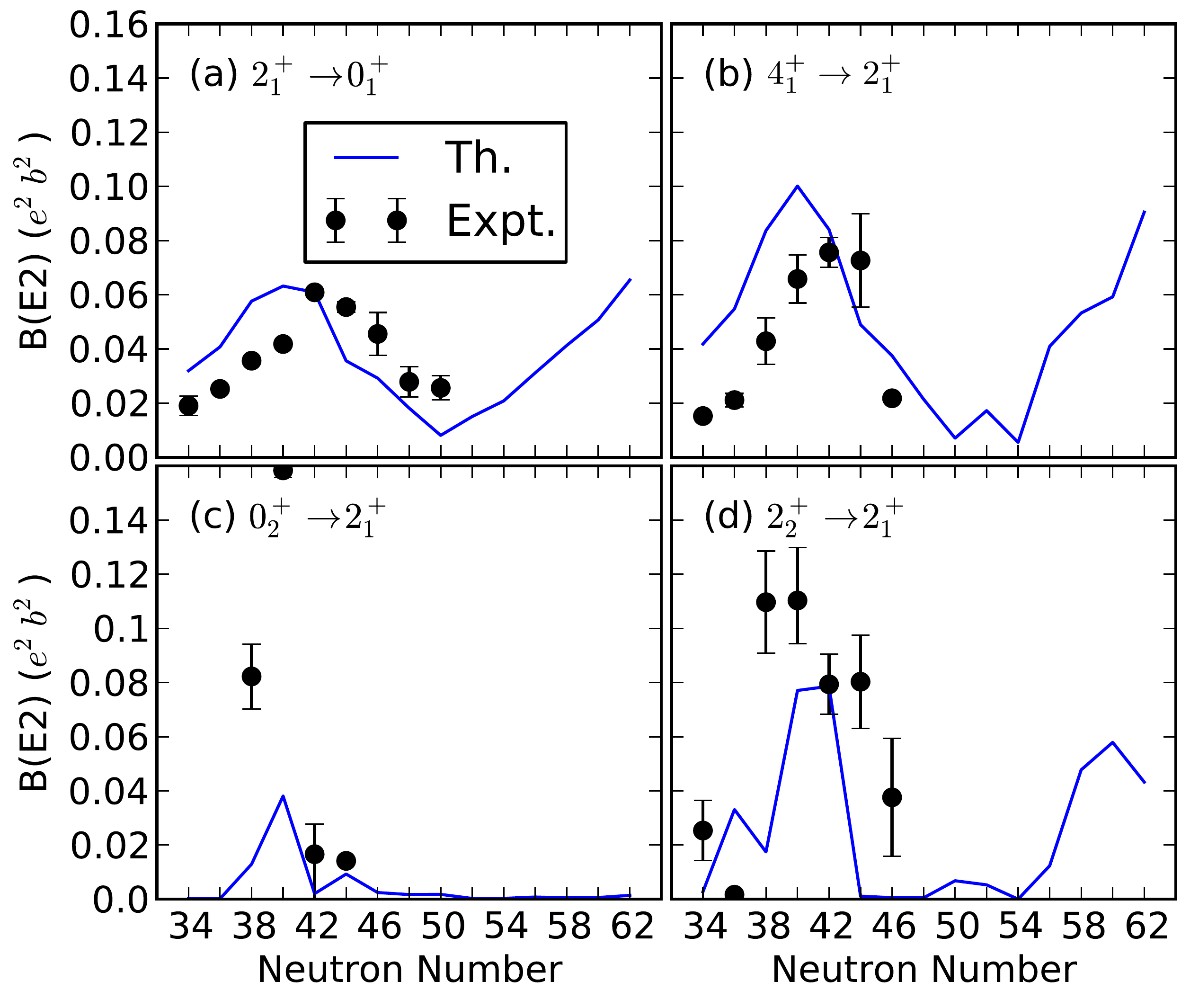} 
\end{center}
\caption{(Color online) The $B(E2; 2^+_1\rightarrow 0^+_1)$, 
$B(E2; 4^+_1\rightarrow 2^+_1)$, $B(E2; 0^+_2\rightarrow 2^+_1)$ 
and $B(E2; 2^+_2\rightarrow 2^+_1)$ transition probabilities 
obtained for Ge isotopes are plotted as functions 
of the neutron number. Experimental data have been 
taken from Ref.~\cite{data}. Taken from Ref.~\cite{Nomura2017c}.}
\label{fig:ge_e2}
\end{figure}

The excitation energies of the $2^+_1$, $4^+_1$, $0^+_2$ and $2^+_2$ 
states, obtained via the diagonalization of the
mapped IBM Hamiltonian, are displayed in Fig.~\ref{fig:energies} 
as functions of the neutron number. They are compared with the 
available experimental data \cite{data}. The 
calculations provide a reasonable agreement with the experimental 
systematics, especially for the yrast states. For both Ge and Se nuclei, the computed $E(2^+_1)$ 
energies decrease as one approaches $N=40$. For the former, this is at
variance with the experiment, a  discrepancy that could be 
attributed to the  $N=40$ neutron sub-shell closure  not  explicitly 
taken into account  in the  calculations \cite{Nomura2017c}. 
The $E(2^+_1)$ 
values display a pronounced peak at $N=50$. 
The  experimental
$E(4^+_1)$ excitation energies [panels (b) and (f)]
are overestimated around $N=50$. 
This could be linked 
to the limited IBM configuration space comprising only $s$ and $d$ 
bosons. Within this context, the inclusion of the $J=4^+$ ($G$) pair in the IBM model could 
improve the agreement with the experiment.

The predicted $E(0^+_2)$ energies are plotted in 
panels (c) and (g). They display a pronounced decrease towards 
$N\approx 40$. This correlates well with the  shape coexistence 
observed in the underlying Gogny-D1M energy surfaces around this 
neutron number \cite{Nomura2017c}. The overestimation of the $E(0^+_2)$ energy
in the case 
of $^{68}$Ge is due to the fact that a configuration mixing calculation 
has not been carried out in this case. For the considered 
neutron-rich nuclei, several 
examples of low-lying $0^+_2$ states are found beyond the $N=50$ 
shell closure [panels (c) and (g)].  As can be seen from
 panels (d) and (h), the calculations  provide a reasonable description of 
the energies of the $2^+_2$ states which are, either interpreted as 
band-heads of the quasi-$\gamma$ bands or as members of the $0^+_2$ 
bands \cite{Nomura2017c}. 

For both Ge and Se nuclei,  the predicted excitation energies
of the non-yrast states $E(0^+_2)$ and $E(2^+_2)$ are generally higher
than the experimental ones especially for $46\leq N\leq 50$. 
This discrepancy is commonly observed in  previous calculations
for other mass regions 
(see, e.g., Ref.~\cite{Nomura2016}) and could be, in most cases,
attributed to the restricted model space of the IBM when the shell
closure is approached. 

The transition probabilities $B(E2; 2^+_1\rightarrow 0^+_1)$, $B(E2; 
4^+_1\rightarrow 2^+_1)$, $B(E2; 0^+_2\rightarrow 2^+_1)$ and $B(E2; 
2^+_2\rightarrow 2^+_1)$ are depicted in Fig.~\ref{fig:ge_e2} for Ge
isotopes. Similar results have been obtained for Se nuclei. The 
maximum $B(E2; 
2^+_1\rightarrow 0^+_1)$ value is reached around $N=40$ where the 
deformation is the largest \cite{Nomura2017c}. The agreement with the experimental 
data is fairly good. A similar trend is also 
found for the $B(E2; 4^+_1\rightarrow 2^+_1)$ transition rates shown in panel 
(b). The quantity $B(E2; 
0^+_2\rightarrow 2^+_1)$, shown in panel (c), can be regarded 
as a measure of the mixing between 
different intrinsic configurations. The experimental $B(E2;
0^+_2\rightarrow 2^+_1)$ value is very large around $N=38$ or 40 where,  
a pronounced configuration mixing could be expected. 
Such a large 
value is not reproduced by the  calculations. The origin 
of such a discrepancy between the theoretical  predictions 
and the experiment
could be  
associated to a weak mixing between the $2^+_1$ and $0^+_2$ states 
in the model. Some  discrepancies between the predicted and experimental $B(E2;
2^+_2\rightarrow 2^+_1)$ values can be seen from panel (d).
Nevertheless, the experimental trend, i.e., the $B(E2;
2^+_2\rightarrow 2^+_1)$ transition probability reaches its largest
value at around $N=40$, being almost of the same order of magnitude as
$B(E2; 2^+_1\rightarrow 0^+_1)$, is reproduced rather well by the
calculations.

Several applications of the fermion-to-boson mapping procedure, with 
different  levels of sophistication and intrinsic degrees of freedom, 
i.e., deformations, have already been discussed in the literature 
including the recent extensions of the model to describe the properties 
of odd-mass nuclear systems  
\cite{Nomura2011b,Nomura2011a,Nomura2011,Albers2012,Albers2013,Nomura2013,Rudigier2015,%
Nomura2015,Nomura2016,Nomura2017c,Nomura2016a,nomura2016odd,Nomura2017,%
Nomura2014,Nomura2013}.

%% file: SR.tex
% ------------------------------------------------------------------------------------------------------
%                                                                                   Symmetry restoration
% ------------------------------------------------------------------------------------------------------
\section{Beyond the mean field: symmetry restoration}
\label{Sec:SR}

The spontaneous breaking of symmetries by the nuclear mean field is one 
of its most salient features and it is a direct consequence of the 
properties of the nuclear interaction. An important aspect of the mean 
field, namely its non-linearity, is also responsible for the 
spontaneous breaking of symmetries. By breaking symmetries at the mean 
field level, a lot of correlations can be encoded into the wave 
function in a simple way. However, those broken symmetries have to be 
restored by considering appropriate linear combinations of mean field 
wave functions. In this way, the resulting wave functions recover the 
quantum number characteristic of the eigenstates of a symmetry 
invariant interaction/Hamiltonian~\cite{Ring1980,Duguet2010}. The mechanism of symmetry breaking 
along with the one of symmetry restoration leads to the fruitful 
concepts of  "intrinsic" and "laboratory" frames of reference. The 
intrinsic frame corresponds to the symmetry breaking "intrinsic" mean 
field state with its characteristic "deformations" (like the prolate 
quadrupole deformation typical of many mid-shell nuclear ground states, 
or the diffuseness of the Fermi level typical of BCS wave functions 
present when the symmetry of the number of particles is broken). On the 
other hand the "laboratory" system corresponds to the symmetry restored 
wave functions obtained as linear combinations of the intrinsic frame 
ones. The laboratory frame can be visualized as the frame of reference 
obtained after adding to the intrinsic frame the quantum fluctuations 
of "orientation"  in the appropriate variables \footnote{Think of a 
rugby ball, when it is at rest it looks like a prolate deformed nuclei, 
however, if its orientation is changed randomly and "very quickly" it 
"looks on the average" as an spherical object}. The impact of symmetry 
restoration in the absolute value of the binding energy of a typical 
nucleus is not too high, of a few per thousand. This is also the case 
if the observable is associated to a scalar quantity (i.e. invariant 
under the symmetry operation). However, the values of the 
electromagnetic transition strengths involving overlaps of tensor 
operators are very sensitive to the effects of symmetry restoration. 
For instance, to reproduce the selection rules typical of those 
strengths, the wave functions involved must have the proper quantum 
numbers, a property that is only obtained after symmetry restoration.

In this section, we will develop the main ideas of symmetry restoration 
and apply them to the study of the most relevant cases in nuclear 
physics: (a) reflection symmetry (b) particle number, (c) rotational 
symmetry and (d) translational invariance. Prominent examples of the 
application of such symmetry restorations will be briefly discussed. We 
will also discuss the approximate evaluation of projected quantities 
under the assumptions of strong deformation of the intrinsic state not 
only because this is a cheap, and in many cases, sufficient way to 
incorporate the effects of symmetry restoration but also because it is 
the basis of the rotational model of Bohr and Mottelson \cite{bohr1975} 
that has proven to be so successful in explaining lots of phenomenology 
in nuclear structure.

% -----------------------------------------------------------------------------------------------------------------
%                                                                       General principles
% -----------------------------------------------------------------------------------------------------------------

\subsection{General principles}
\label{Sec:SRGP}

The starting point of symmetry restoration is, obviously, the existence of
a spontaneous symmetry breaking intrinsic mean field wave function. This 
wave function is obtained from some interaction or EDF by solving the 
non-linear HF or HFB equations. The symmetry restored (or projected) wave
functions are obtained by considering appropriate linear combinations
of the "rotated" intrinsic states (that is, the intrinsic state after
applying to it a member of the symmetry group). With these wave functions
(there are as many as the number of irreducible representations of the 
underlying symmetry group) we can compute physical observables like the
projected energy, radius, electromagnetic moments, etc. This procedure
is known under the name of projection after variation (PAV) because
the "intrinsic wave function" is obtained by minimizing the intrinsic
energy, and therefore knows nothing about its subsequent projection~\cite{Peierls1957,Yoccoz1957,Ring1980}. 
At this point one might wonder about the usefulness of considering the
projected energies (one for each irrep of the symmetry group) as the
quantities to apply the variational principle in order to obtain the
"intrinsic" states. The most immediate consequence of this procedure
is that different intrinsic states are obtained for each projected
energy considered. Another consequence is that the projected wave
function is richer than the set of intrinsic ones and therefore the
procedure is variationally effective, that is, a lower energy than the
intrinsic one is always going to be obtained at least for one of the projected
energies. This procedure is known as variation after projection (VAP)~\cite{
Zeh1965,Rouhaninejad1966,Yoccoz1966,Ring1980} 
and is formally superior to PAV as shown in many examples and test 
studies in simple models. The main disadvantage of VAP over PAV
is that the projected energies are far more involved to compute than
the intrinsic (mean field) one. Not to mention also the fact that
for each quantum number a minimization has to be carried out. 

% -----------------------------------------------------------------------------------------------------------------
%                                                                       Group theory arguments
% -----------------------------------------------------------------------------------------------------------------

\subsubsection{Symmetry restoration, general group theory arguments}
\label{Sec:SRGTA}

The procedure to restore symmetries is deeply rooted in the underlying group
structure of the symmetry operations~\cite{Ring1980,Duguet2010}. The set of symmetry operations is endowed with 
the mathematical structure of a group. For instance, spatial rotations are associated to the 
continuous Lie group $SU(2)$, reflection symmetry is associated to
a discrete, idempotent group made of two elements, the identity and
the parity operator $\Pi$ satisfying $\Pi^{2}=\mathbb{I}$. The quantum
numbers of the restored symmetries are the labels of the irreducible
representations (irreps) of the group, or, in the case of Lie groups of the irreps
of the associated algebra. For instance, in the case of $SU(2)$ the 
corresponding algebra $su(2)$ is composed of the angular momentum operators $J_{x}$, $J_{y}$ and $J_{z}$.
The irreps of the angular momentum operators correspond to the eigenstates
of $J^{2}=J_{x}^{2}+J_{y}^{2}+J_{z}^{2}$ and $J_{z}$ and are labeled by the typical $J$ and $M$ quantum numbers.
The basis states of the irreps are the eigenstates $|JM\rangle$. The subspace spanned by each
irrep is invariant under the operator realizing the symmetry. For instance, consider the
rotation operator which can be expressed in terms of the Euler angles
$\Omega = (\alpha,\beta,\gamma)$
\begin{equation}
	\hat{R} (\Omega) = e^{-i/\hbar \alpha J_{z}} e^{-i/\hbar \beta J_{y}} e^{-i/\hbar \gamma J_{z}}
\end{equation}
The matrix element of the rotation operator between basis states of two
different irreps gives
\begin{equation}
	\langle J'M' | \hat{R} (\Omega) | J M \rangle = \delta_{JJ'}\mathcal{D}^{J}_{M'M} (\Omega)
\end{equation}
where the Wigner rotation matrix $\mathcal{D}^{J}_{M'M}(\Omega)$ has 
been introduced. The $\delta_{JJ'}$ in the above expression shows that 
the subspace spanned by $|JM\rangle$ is indeed invariant under the 
symmetry operator $\hat R (\Omega)$. By taking linear combinations of 
the rotation operator with the Wigner functions and using the 
appropriate integration measure (de Haar measure) we end up with the 
projection operator 
\begin{equation}\label{Eq:AMPOP}
	\hat{P}^{J}_{MK} = \frac{2J+1}{8\pi^{2}} \int d\Omega \mathcal{D}^{J\, *}_{MK} (\Omega) \hat{R}(\Omega)
\end{equation}
which is labeled with the indices of the irreps of the corresponding group.
In the $SU(2)$ case it has the properties
\begin{equation}
	P^{J}_{MK} P^{J}_{M'K'} = \delta_{JJ'}\delta_{M'K} \hat{P}^{J}_{MK'}
\end{equation}
and 
\begin{equation}
	\left(\hat{P}^{J}_{MK}\right)^{\dagger}=\hat{P}^{J}_{KM}
\end{equation}
Due to the special structure of the projector, the action of the 
symmetry operator on the left (or the right) of the projector leads to 
a linear combination of the same projector with weights which are 
nothing but the representation of the symmetry operator in the linear 
space of the irreps
\begin{equation}
	\hat{R}(\Omega) \hat{P}^{J}_{MK} = \sum_{K'} \mathcal{D}^{J}_{MK'}(\Omega)\hat{P}^{J}_{K'M}
\end{equation}
which suggests the following form of the projector
\begin{equation}
 \hat{P}^{J}_{MK} = \sum_{\alpha} |JM \alpha \rangle \langle JM \alpha|
\end{equation}
where $\alpha$ represents an additional set of quantum numbers required
to fully characterize the quantum states. 

% -----------------------------------------------------------------------------------------------------------------
%                                                                       Rotational bands
% -----------------------------------------------------------------------------------------------------------------

\subsubsection{Symmetry breaking as generator of symmetry bands (rotational, parity, etc)}
\label{Sec:SRRB}

For each intrinsic (deformed) state $|\Phi\rangle$ we can generate a 
whole set of states with the appropriate symmetry quantum numbers. 
Using again the angular momentum as an example, for each deformed 
states $|\Phi\rangle$ we can generate infinite (in principle) states
\begin{equation}
	|\Psi_{JM} \rangle = \sum_{K} g_{K}^{J}\hat{P}^{J}_{MK} |\Phi\rangle
\end{equation}
eigenstates of $\hat J^{2}$ and $\hat J_{z}$ labeled by the $J$ and $M$ 
quantum numbers. We have generated, in this way, a "rotational band" 
which is nothing but the set of $|\Psi_{JM}\rangle$ states. The 
physical properties of those states will strongly depend on the degree 
of deformation of the intrinsic state $|\Phi\rangle$ as discussed 
below. We can compute the energy for each member of the band
\begin{equation}
	E^{J} = \frac{\langle \Psi_{JM} | \hat{H} |\Psi_{JM}\rangle}{\langle \Psi_{JM} |\Psi_{JM}\rangle}
\end{equation} 
which is independent of $M$ as a consequence of the scalar nature of $\hat{H}$.
The dependence of $E^{J}$ with $J$ depends on the structure of the intrinsic
state $|\Phi\rangle$ although it is possible to extract general properties
in the case in which the intrinsic state is strongly deformed. As it will 
be shown below, in the strong deformation limit the energy is given by
\begin{equation}
	E^{J} = \langle \Phi | \hat{H} | \Phi \rangle - \frac{\langle \Delta \vec{J}^{2}\rangle}{2\mathcal{J}_\mathrm{Y}}
	+ \frac{\hbar^{2} J(J+1)}{2\mathcal{J}_\mathrm{Y}}
\end{equation}
that is the characteristic expression for the energy of a rotational 
band with the characteristic $J(J+1)$ growing of the energy with 
angular momentum. In this expression $\mathcal{J}_\mathrm{Y}$ is the Yoccoz 
moment of inertia defined, for instance in \cite{Peierls1957,Yoccoz1957,Mang1975325,Ring1980}.  
Along the same line of reasoning, the breaking of reflection symmetry 
followed by parity restoration requires a projector which is 
proportional to $\mathbb{I}+\pi\hat\Pi$ with the parity quantum number 
$\pi=\pm 1$. In the strong deformation limit $\langle \hat\Pi \rangle 
\rightarrow 0$ the projected energy $E^{\pi}=(\langle \hat{H} \rangle + 
\pi \langle \hat{H}\hat{\Pi} \rangle)/(1+\pi \langle \hat{\Pi} \rangle$ 
is degenerate leading to a "band" composed of two elements which are 
degenerated in energy.

Breaking of several symmetries at the same time is also possible and its
restoration leads to an "admixture" of bands, like the alternating parity
rotational bands characteristic of octupole deformation at high spins. In
this case, rotational and reflection symmetries are broken at the same time.

% -----------------------------------------------------------------------------------------------------------------
%                                                                       GWT and Pfaffian
% -----------------------------------------------------------------------------------------------------------------

\subsubsection{Evaluation of overlaps: generalized Wick theorem and Pfaffian}
\label{Sec:SROVER}

The evaluation of the "projected" quantities, like energies or transition
strengths, requires the evaluation of an integral over the parameters of
the symmetry group of the corresponding operator's overlaps. To be more specific,
and using again the example of $SU(2)$ we need to evaluate integrals involving
quantities like
\begin{eqnarray}
	H(\Omega) & = & \langle \Phi | \hat{H} \hat{R}(\Omega) |\Phi\rangle \\ \nonumber{}
	N(\Omega) & = & \langle \Phi |         \hat{R}(\Omega) |\Phi\rangle 
\end{eqnarray}
which are overlaps of the corresponding operators between the intrinsic
HFB wave function $|\Phi\rangle$ and the "rotated" one $\hat{R}(\Omega) |\Phi\rangle$.
Fortunately, all the symmetry groups considered in nuclear physics are
either discrete or compact Lie groups. In the later case, the symmetry operators 
are written in terms of exponentials of the corresponding Lie algebra that 
is represented in Fock space as linear combinations of one body operators ($J_{i}$, $P_{i}$, etc).
As a consequence, we can use Thouless theorem to show that $\hat{R}(\Omega) |\Phi\rangle${}
is again a HFB wave function and therefore, we are dealing with overlaps
of operators between different HFB wave functions (the arguments below can
be extended straightforwardly to the case where $|\Phi\rangle$ is replaced
by a more general HFB wave function $|\Phi '\rangle$ ). The evaluation
of the overlaps is carried out using the generalized Wick theorem and
the overlap formula as discussed in Appendix \ref{App:C}. The expressions
derived in the Appendix are obtained under the assumption that the original
single particle basis and the rotated one span the same subspace of the
total Hilbert space. This is not the case if the basis breaks the symmetry
under consideration. Typical examples are the use of deformed harmonic
oscillator basis in the case of rotations or basis not translational invariant
in the case of Galilei invariance restoration. This difficulty leads to
either the use of symmetry restricted basis (for instance, HO basis with
equal oscillator lengths along all possible spatial directions - see below) or a careful
evaluation of the impact of truncation errors as is the case in the
restoration of Galilei invariance (see below). 

The overlaps can be evaluated in terms of linear combinations of 
products of contractions with the use of the generalized Wick's theorem. For
instance, the ratio $h(\Omega)=H(\Omega)/N(\Omega)$ is given by
\begin{equation} \label{Hoverlap}
	h(\Omega) = \sum_{ab} t_{ab} \rho_{ba} (\Omega) + 
	\frac{1}{4} 
	\sum_{abcd} \bar{\nu}_{abcd} [ \rho_{db} (\Omega)\rho_{ca} (\Omega)
	                              -\rho_{da} (\Omega)\rho_{cb} (\Omega){}
	                              +\kappa_{dc}(\Omega) \bar{\kappa}_{ba}(\Omega) ]
\end{equation}
where the "elemental" contractions are given by quantities that resemble
very much the standard HFB density matrix and pairing tensor
\begin{eqnarray}
	\rho_{db}   (\Omega) & = & \frac{\langle \Phi | c^{\dagger}_{b} c_{d}           \hat{R}(\Omega) |\Phi\rangle}{\langle \Phi |         \hat{R}(\Omega) |\Phi\rangle } \\
	\kappa_{dc} (\Omega) & = & \frac{\langle \Phi | c_{c}           c_{d}           \hat{R}(\Omega) |\Phi\rangle}{\langle \Phi |         \hat{R}(\Omega) |\Phi\rangle } \\
\bar\kappa_{ba} (\Omega) & = & \frac{\langle \Phi | c^{\dagger}_{a} c^{\dagger}_{b} \hat{R}(\Omega) |\Phi\rangle}{\langle \Phi |         \hat{R}(\Omega) |\Phi\rangle } 
\end{eqnarray}
Once the quantity $h(\Omega)$ is obtained, the evaluation of $H(\Omega)${}
is straightforward (however, see below for the case where $N(\Omega)=0$ for
some $\Omega$). The remaining term, $N(\Omega)$ is given in terms of 
determinants or Pfaffians of the appropriate matrices (see Appendix \ref{App:C}).

% -----------------------------------------------------------------------------------------------------------------
%                                                                       Pauli exclusion principle
% -----------------------------------------------------------------------------------------------------------------

\subsubsection{Difficulties associated with the Pauli exclusion principle and self-energies}
\label{Sec:SRDIFF1}

In the implementation of symmetry restoration and configuration mixing,
to be discussed below, one has to face technical difficulties associated
with the evaluation of the overlaps in Eq \ref{Hoverlap} and having to do with
the vanishing of the overlap between the two HFB states. Let us
define the overlap 
\begin{equation}
	h(q,q')=\frac{\langle \varphi_{q}|\hat{H}|\varphi_{q'}\rangle}%
                 {\langle \varphi_{q}|        \varphi_{q'}\rangle}
\end{equation}
where $|\varphi_{q}\rangle$ is a HFB wave function parametrized in terms
of the labels denoted collectively as $q$. The generalized Wick's theorem
allows to express the above overlap in terms of the contractions
\begin{eqnarray}
	\rho_{db}   (q,q') & = & \frac{\langle \varphi_{q} | c^{\dagger}_{b} c_{d}           |\varphi_{q'}\rangle}%
	                              {\langle \varphi_{q} |                                  \varphi_{q'}\rangle } \\
	\kappa_{dc} (q,q') & = & \frac{\langle \varphi_{q} | c_{c}           c_{d}           |\varphi_{q'}\rangle}%
	                              {\langle \varphi_{q} |                                  \varphi_{q'}\rangle } \\
\bar\kappa_{ba} (q,q') & = & \frac{\langle \varphi_{q} | c^{\dagger}_{a} c^{\dagger}_{b} |\varphi_{q'}\rangle}%
                                  {\langle \varphi_{q} |                                  \varphi_{q'}\rangle } 
\end{eqnarray}
In general, the quantities $\langle \varphi_{q} | c^{\dagger}_{b} c_{d} |\varphi_{q'}\rangle $, 
$\langle \varphi_{q} | c_{c}           c_{d}|\varphi_{q'}\rangle $ or 
$\langle \varphi_{q} | c^{\dagger}_{a} c^{\dagger}_{b} |\varphi_{q'}\rangle $
are, by construction, finite. In the case when $\langle \varphi_{q} | \varphi_{q'}\rangle$ is zero,
this implies that $\rho_{db}   (q,q') $, $\kappa_{dc} (q,q') $ and $\bar\kappa_{ba} (q,q')$
must be divergent. If the overlap goes to zero like a small parameter $\epsilon$, then
the contractions must diverge like $1/\epsilon$. As a consequence, the overlap
of one body operators $\langle \varphi_{q}|\hat{T}|\varphi_{q'}\rangle $ is 
manifestly finite as it only involves the contraction $\rho_{db}   (q,q') $ times the 
overlap $\langle \varphi_{q}|\varphi_{q'}\rangle$. The difficulties arise
in the evaluation of two (or higher order) operators as in this case we have
products of two contractions that behave like $1/\epsilon^{2}$. A simplistic
analysis would lead to the conclusion that $ \langle \varphi_{q}|\hat{H}|\varphi_{q'}\rangle$ 
must diverge when $\langle \varphi_{q}|\varphi_{q'}\rangle$ goes to zero.
However, a more careful analysis reveals that, due to the properties of
the fermion creation and annihilation operator algebra (that lead to the Pauli exclusion principle), the sum
of the products of two contractions 
\begin{equation}\label{soc}
	\rho_{db}   (q,q')\rho_{ac}(q,q') - \rho_{da}   (q,q')\rho_{bc}(q,q') + 
	\kappa_{dc} (q,q')\bar\kappa_{ba} (q,q')
\end{equation}
cancel one of the inverse powers of $\epsilon$ as to render $ \langle 
\varphi_{q}|\hat{H}|\varphi_{q'}\rangle$ finite. This discussion might 
look a bit of an academic one as it is very unlikely that given a 
discrete set of $q$ values one is going to find a zero overlap. 
However, the important point is that if the overlap is small because it 
is close to a zero, then the contractions are going to take very large 
values that still need large cancellations in order to get the proper 
value of the overlap $ \langle 
\varphi_{q}|\hat{H}|\varphi_{q'}\rangle$. Very often, both in the 
evaluation of the energy or the Hamiltonian overlap several 
contributions ,like Coulomb exchange or Coulomb anti-pairing are 
neglected. This means that not all the terms in Eq. \ref{soc} are 
considered and therefore in the event of being close to a zero of the 
overlap the cancellation mentioned above does not take place and the 
overlap might become unnaturally and unphysically large. This also 
happens if two different interactions are used in the ph and pp 
channels as only the first two terms in the sum of Eq. \ref{soc} are 
considered in the ph channel and the last one in the pp channel. This 
problem is dubbed in the literature as the "self-energy" and/or 
"self-pairing" problem \cite{Lacroix2009}. 
The reason for that name is that it shares the 
same origin with another typical problem encountered in the evaluation 
of two body operators and consequence of neglecting the exchange or 
pairing terms: the mean value of the Hamiltonian  computed with a mean 
field wave function corresponding to just one particle is different 
from zero in spite of not having other particle to interact with. In 
the case of the Gogny force calculations, where the pairing field is 
obtained from the same interaction used in the ph channel, the best 
solution is to fully consider all the neglected terms, namely the 
Coulomb exchange, Coulomb anti-pairing and spin-orbit pairing fields. 
Unfortunately, considering the Coulomb contributions increase by almost 
and order of magnitude the computational requirements to compute mean 
values of overlaps. Another, more important, consequence is that all 
the Gogny forces incepted so far do not consider exact Coulomb exchange 
and Coulomb anti-pairing and therefore their parameters are not fully 
adapted to their consideration. The "self-energy" problem is ubiquitous 
in Particle number projection calculations and seems to be less 
dangerous in applications of the generator coordinate method without 
projection. This issue has been discussed at length in the literature 
\cite{Don98,Anguiano2001,Dobaczewski2007,Lacroix2009,Egido2016a} and the reader is 
referred to the mentioned literature for further details. 

%----------------------------------------------------------------------------------------------------------------- 
%                                                                        Non integer powers of the density 
% -----------------------------------------------------------------------------------------------------------------

\subsubsection{Difficulties associated with non-integer powers of the density}
\label{Sec:SRDIFF2}

Density dependent forces, like Gogny, are state dependent, i.e. the 
interaction depends upon the matter density of the corresponding state. 
Nevertheless, they are unambiguously defined in the calculation of the 
mean value of the energy as the matter density to be used in the density
dependent part of the interaction is just the mean value of the density
operator. This is not the case, however, in the calculation of
overlaps of the Hamiltonian between 
different HFB states that is required, for instance, in the implementation of 
symmetry restoration or large amplitude motion discussed
below. In those situations, a prescription for the density dependent
part of the interaction is required. The most obvious one is to replace
the density in the density dependent part of the interaction by the
{\it mixed} density \cite{Bonche1990}
\begin{equation}
	\rho_{q,q'} = \frac{\langle \varphi (q) | \hat\rho (\vec{R}) | \varphi (q') \rangle}%
	{\langle \varphi (q)  | \varphi (q') \rangle}
\end{equation}
which is the overlap of the matter density operator $\hat\rho (\vec{R})=\sum_{i=1}^{A} \delta(\vec{r}_{i}-\vec{R})$ between
the corresponding HFB sates $|\varphi(q)\rangle$ and $|\varphi(q')\rangle$.
This prescription fulfills a series of requirements
like leading to overlaps of scalar operators invariant under symmetry operations or other properties
essential to ensure that observables are real quantities \cite{Rodriguez-Guzman2002a,ROBLEDO2007}.
The main objection to this prescription is that, in the general case, the overlap density  is
a complex quantity that has to be raised to a non integer power $\alpha$, as it is the case in
all  of the popular phenomenological effective interactions of the Gogny
or Skyrme type. Raising complex numbers $z=|z|e^{i\varphi}$ to non-integer
powers is a delicate operation that leads to multi-valued solutions (in general 
infinite ones) of the form $|z|^{\alpha}e^{i \alpha\varphi + 2\pi \alpha i n }$
where $n$ represent all positive integers $n=0,1,2,\ldots$. Those multi-valued
powers (or roots) have to be located in Riemann sheets and special attention has to
be paid to continuity issues in going from one Riemann sheet to another.
If those discontinuities are not treated adequately they might lead to 
jumps in the integrals leading to the projected energy. 
This issue has been discussed in detail in Refs \cite{Dobaczewski2007,Duguet2009}
but so far no evident general solution to this problem seems to exist. However,
wise choices of phases lead to very reasonable results which seem to be
free from the artifacts of discontinuous integrands. 
A possible solution to this problem would be to use symmetry invariant
densities which are given by mean values with projected wave functions or
linear combinations of mean values evaluated with rotated intrinsic quantities.
In any of the two possibilities, they are real quantities and therefore free from the uncertainties
in the determination of the non-integer power. However, it has been shown
that this prescription is inconsistent with the general philosophy of the
density dependent term in the case of spatial symmetries like parity or
angular momentum \cite{Robledo10} leading to catastrophic results.
Nevertheless, this prescription, when used only in the particle number restoration
leads to reasonable results as it will be shown below. 

% -----------------------------------------------------------------------------------------------------------------
%                                                                                             Parity projection
% -----------------------------------------------------------------------------------------------------------------

\subsection{Parity Projection}
\label{Sec:SRPar}

%%%%%%%%%%%%%%%%%%%%%%%%%%%%%%%%%%%%%%%
\begin{figure}
\begin{center}
\includegraphics[width=0.45\textwidth]{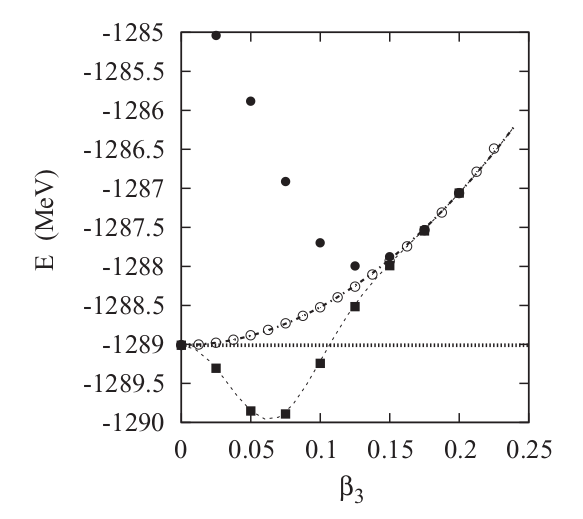}%
\includegraphics[width=0.45\textwidth]{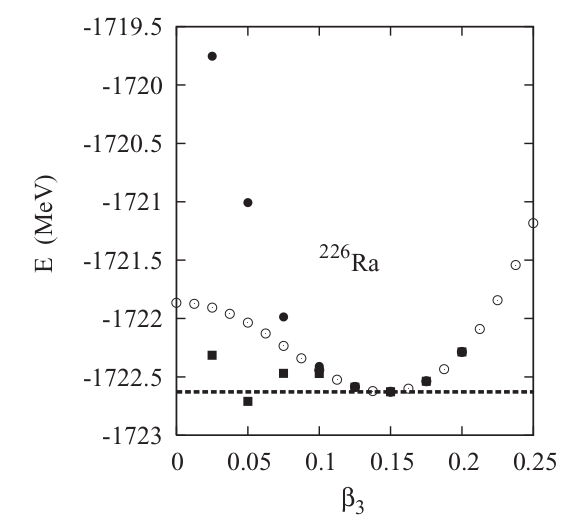}%
\caption{HFB  (open circles), positive parity (full squares)
and negative parity (bullets) energies as a function
of the $\beta_{3}$ deformation parameter for the $^{158}$Gd (left)
and $^{226}$Ra (right) nuclei. The Gogny D1S parameterization is used.
Figure adapted from Ref.~\cite{Robledo2011}. }
\label{Fig:PPRVAP}
\end{center}
\end{figure}

Parity projection is the simplest possible example of symmetry 
restoration because the symmetry operators belong to a discrete set 
with only two elements, the identity $\mathbb{I}$ and the parity 
operator $\hat\Pi$. As the parity operator satisfies 
$\hat\Pi^2=\mathbb{I}$ the parity quantum number can only take two values 
$\pi=\pm 1$. The expression of the projection operator $\hat{P}_\pi$ 
and the projected energy are specially simple in this case
\begin{equation}
\hat{P}_\pi = \frac{1}{2} (\mathbb{I}+\pi \hat\Pi)
\end{equation}
and 
\begin{equation}
E_\pi = \frac{\langle \hat{H} \rangle + \pi \langle \hat{H} \hat{\Pi}\rangle}{1 + \pi \langle \hat{\Pi}\rangle}
\end{equation}
When the degree of symmetry breaking of the intrinsic wave function is 
large $\langle \hat{\Pi}\rangle \approx 0$, the Hamiltonian overlap 
$\langle \hat{H} \hat{\Pi}\rangle$ can be neglected with respect to 
$\langle \hat{H} \rangle$ and therefore we obtain a degenerate energy 
for both parities (a parity doublet) that corresponds to the intrinsic 
energy. The parity doublet  is nothing but the "band" generated by the 
restoration of the parity symmetry of an intrinsic state which is 
strongly deformed.

Parity (or reflection symmetry) breaking is associated with the lowest 
relevant negative parity multipole moment of the mass distribution, 
namely the octupole moment $Q_{3\mu}$. Octupole deformation of the 
ground state of atomic nuclei is not as common  as quadrupole 
deformation and its presence requires that proton and neutron numbers 
are close or equal to specific values such that the Fermi level is 
close to opposite parity orbitals with $\Delta l=3$. Those specific 
numbers are $N=40, 56, 88, 136$, etc. Nuclei with proton and neutron 
number close to them   correspond to paradigmatic octupole deformed 
systems like $^{144}$Ba \cite{PhysRevLett.116.112503,Bucher2017} or 
$^{224}$Ra \cite{Gaffney2013}. When the nucleus is octupole deformed, 
the parity splitting $\Delta E_\pm = E_{\pi=+1}-E_{\pi=-1}$ is very 
small and parity doublets appear characterized by strong $B(E3)$ 
transition probabilities.  This description is not very accurate as 
fluctuations in the octupole degree of freedom (taken into 
consideration, for instance, with the GCM, see Sec \ref{Sec:LAM}) 
modify quantitatively the picture. Octupole effects are, however, 
pervasive and appear all over the periodic table when parity projection 
is considered in the VAP or restricted variation after projection 
(RVAP) framework. In the latter case, we consider a set of HFB 
configurations constrained on the value of the octupole moment $Q_{30}$ 
and compute the projected energies $E_\pi (Q_{30})$. The corresponding 
minima for the positive and negative parity curves are in general 
different (they only coincide in octupole deformed nuclei) leading to 
two intrinsic states associated to the two possible parities. In this 
way, it is possible to compute the parity splitting $\Delta E$ for all 
nuclei and the corresponding $B(E3)$ strengths. In addition, the 
absolute minimum of the two curves very often lies at an excitation 
energy lower than the one of the HFB minimum being the energy gain the 
octupole correlation energy. This procedure has been carried out with 
the D1S, D1N and D1M variants of the Gogny force in Refs 
\cite{Robledo2011,Robledo2015} for a large set of even-even nuclei 
spanning a wide subset of the nuclide chart. In Fig \ref{Fig:PPRVAP} an 
example of the RVAP procedure taken from \cite{Robledo2011} is shown 
for an octupole soft nucleus $^{158}$Gd and for a well deformed one 
$^{224}$Ra. Both parity projected energies are plotted as a function of 
the $\beta_{3}$ deformation parameter,  along with the HFB energy. In 
the octupole soft case we observe two well differentiated minima, one 
for positive parity the other for negative parity, at different 
excitation energies. On the other hand, for the nucleus $^{224}$~Ra the 
three minima (HFB, positive parity and negative parity) are located at 
the same $\beta_{3}$ value and have the same energy. Therefore, the 
parity splitting between positive and negative parity states is 
essentially zero.

Another useful application of parity projection is to show some of the 
difficulties associated to symmetry restoration in the presence of 
density dependent forces like Gogny: in Ref \cite{Robledo10} it was 
shown that the use of the symmetry restored density in the density 
dependent term of the Gogny interaction results in catastrophic 
consequences. The only consistent density dependent prescription in the 
restoration of spatial symmetries seems to be the "mixed density" (see 
above, Sec \ref{Sec:SRDIFF2}). This result can be straightforwardly 
extended to any of the flavors of the Skyrme EDF.

Parity projection has also been applied with intrinsic HFB cranking 
states to describe the stabilization of octupole deformation with 
increasing angular momentum and the corresponding emergence of 
"alternating parity rotational bands" \cite{Gar97,Gar98}.

%\begin{itemize}
%\item Discrete symmetry
%\item Parity doublets
%\item Static octupole deformation is rare, but dynamical effects are pervasive.
%\item Projected density versus overlap density and parity projection
%\end{itemize}

% -----------------------------------------------------------------------------------------------------------------
%                                                                                             PN projection
% -----------------------------------------------------------------------------------------------------------------
%%\subsection{Particle number projection}

%%Abelian symmetry

%%Intimately connected with pairing correlations: great impact on
%%collective inertias

\subsection{Particle number projection (PNP)}\label{PNR}
%%%%%%%%%%%%%%%%%%%%%%
As it was mentioned above, pairing correlations are allowed to be 
present in the mean-field approximation through the Bogoliubov 
transformation. This implies that the HFB wave functions are not (in 
general) eigenstates of the proton and neutron number operators. Such a 
symmetry breaking mechanism is very useful to explore some relevant 
terms of the nuclear interaction with relatively simple many-body wave 
functions. However, it is obvious that the physical nuclear states of a 
specific nucleus must have a definite number of protons and neutrons, 
i.e., they have to be eigenstates of those operators. Additionally, in 
cases where the single-particle level density around the Fermi level is 
small, the HFB approach itself is unable to capture pairing 
correlations at all and the method collapses into a pure HF state. 
These and other drawbacks can be corrected by restoring the 
particle-number symmetry of the nuclear system. The restoration relies 
on projection techniques that produce many-body states as linear 
combinations of mean-field wave-functions with the coefficients 
dictated by the symmetry group U(1) (see discussion in Sec \ref{Sec:SRGTA}).
Hence, the particle-number projected (PNP) wave functions are defined 
as~\cite{Ring1980}:
\begin{equation}
|NZ\rangle=\hat{P}^{N}\hat{P}^{Z}|\Phi\rangle
\label{pnp_wf}
\end{equation}
where $|\Phi\rangle$ is a HFB state and $\hat{P}^{N(Z)}$ is the 
projection operator onto a good number of neutrons (protons):
\begin{equation}
\hat{P}^{N}=\frac{1}{2\pi}\int_{0}^{2\pi}e^{i\varphi(\hat{N}-N)}d\varphi
\label{pnp_op}
\end{equation}
Once the PNP wave functions are defined, the projection itself can be 
performed before or after the variational procedure~\cite{Ring1980}. In 
both cases the variational space is made of HFB (intrinsic) states and 
the difference comes from the energy functional that is minimized. 

In the projection after variation (PAV) approach the HFB energy is 
minimized first and the resulting intrinsic wave function is projected 
afterwards:
\begin{eqnarray}
\left.\delta E^{'\mathrm{HFB}}\left[|\Phi\rangle\right]\right|_{|\Phi\rangle=|\mathrm{HFB}\rangle}=0\Rightarrow\nonumber\\ \Rightarrow|NZ\rangle_{\mathrm{PN-PAV}}=\hat{P}^{N}\hat{P}^{Z}|\mathrm{HFB}\rangle
\label{HFB_variational}
\end{eqnarray}
where $E^{'\mathrm{HFB}}\left[|\Phi\rangle\right]$ is the energy 
density functional (EDF) computed within the HFB approximation:
\begin{equation}
E^{'\mathrm{HFB}}\left[|\Phi\rangle\right]=\langle\hat{H}\rangle+\varepsilon^{\mathrm{DD}}\left[|\Phi\rangle\right]-\lambda_{Z}\langle\hat{Z}\rangle-\lambda_{N}\langle\hat{N}\rangle
\label{HFB_functional}
\end{equation}
Here $|\rangle\equiv|\Phi\rangle$, $\lambda_{Z(N)}$ are the Lagrange 
multipliers to ensure the correct mean value of the number of protons 
(neutrons) in the intrinsic wave function, and $\hat{H}$ and 
$\varepsilon^{\mathrm{DD}}\left[|\Phi\rangle\right]$ are the 
Hamiltonian piece and the explicit density-dependent part of the 
nuclear interaction, respectively. 

This method is computationally cheap since it only requires one PN 
projection at the end of the process. However, the collapse of the 
pairing correlations in the HFB approach in weak pairing regimes 
remains unsolved using the PAV method. In these situations the HFB 
states are pure HF wave functions that do not break the particle number 
symmetry and PN-PAV does not have any effect. 

The natural way to improve the PN-PAV method is the so-called variation 
after projection (PN-VAP) technique where the projected energy instead 
of the HFB energy is minimized. 
\begin{eqnarray}
\left.\delta E^{\mathrm{PNP}}\left[|\Phi\rangle\right]\right|_{|\Phi\rangle=|\mathrm{PN-VAP}\rangle}=0\Rightarrow\nonumber\\ \Rightarrow|NZ\rangle_{\mathrm{PN-VAP}}=\hat{P}^{N}\hat{P}^{Z}|\mathrm{PN-VAP}\rangle
\label{PN_VAP_variational}
\end{eqnarray}
In this case $E^{\mathrm{PNP}}\left[|\Phi\rangle\right]$ is the EDF that includes the particle number restoration:
\begin{equation}
E^{\mathrm{PNP}}\left[|\Phi\rangle\right]=\frac{\langle\hat{H}\hat{P}^{N}\hat{P}^{Z}\rangle}{\langle\hat{P}^{N}\hat{P}^{Z}\rangle}+\varepsilon^{\mathrm{DD,PNP}}\left[|\Phi\rangle\right]
\label{PNP_functional}
\end{equation}
It is important to notice that the EDF coming from the explicit 
density-dependent part of the interaction, 
$\varepsilon^{\mathrm{DD,PNP}}\left[|\Phi\rangle\right]$, requires a 
more detailed explanation that will be given below (see also Secs \ref{Sec:SRDIFF1} and
\ref{Sec:SRDIFF2}).

The amount of pairing correlations included by using this method is 
larger than the PN-PAV and, moreover, the collapse of the pairing 
correlations is avoided. The problem is the higher computational cost 
of the PN-VAP method since the symmetry restoration is performed in 
every step of the resolution of the variational equations. 
Nevertheless, this is not a serious limitation of the method with the 
present computing capabilities.

\begin{figure}
\begin{center}
\includegraphics[width=0.4\textwidth]{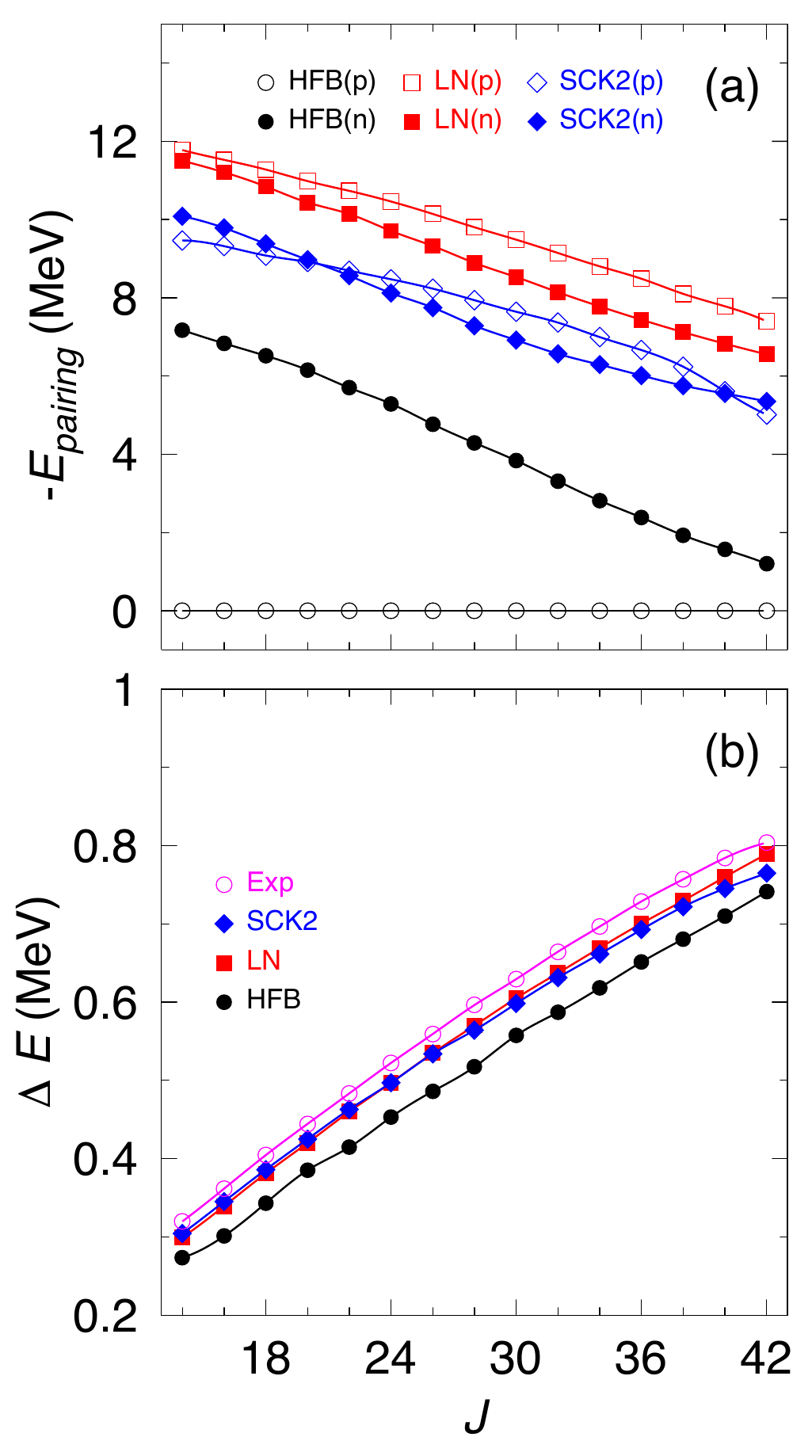}
\caption{(a) Pairing energies (protons and neutrons) and (b) transition 
energies as a function of the cranking angular momentum, $J$, 
calculated with HFB, LN and SCK2 methods for $^{190}$Hg with the Gogny 
D1S interaction. Experimental energies are also plotted in panel (b). 
Figure adapted from Refs.~\cite{Valor2000,Valor2000a}. }
\label{HFB_LN_SCK2_190Hg}
\end{center}
\end{figure}

Both PN-PAV and PN-VAP approaches are exact projections onto good 
number of protons and neutrons. However, the first implementations of 
the PNP with Gogny interactions were made in the self-consistent 
second-order Kamlah (SCK2)~\cite{Kamlah1968} and Lipkin-Nogami (LN)~\cite{Lipkin1960,Nogami1964} approximations, being 
the latter an approximation itself to the 
former~\cite{Valor1996,Valor1997,Valor2000,Valor2000a} (see Sec \ref{Sec:SRApp} below). 
One of the first applications of these approximate methods was the study of deformed and super-deformed bands at high spin in rare earth nuclei and in Hg isotopes around $A=190$~\cite{Valor1996,Valor1997,Valor2000,Valor2000a}. In Fig.~\ref{HFB_LN_SCK2_190Hg} we show an example of the performance of HFB, LN and SCK2 approaches in cranking calculations of the structure of the nucleus $^{190}$Hg. Pairing energies (protons and neutrons) and transition energies for the super-deformed yrast band are plotted as a function of the cranking angular momentum in the top and bottom panels, respectively. We observe that, in this particular case, proton pairing correlations are zero in the HFB approximation while this drawback is corrected with the LN and SCK2 methods. In these cases, pairing correlations decrease with increasing the angular momentum due to the Coriolis anti-pairing effect. More interestingly, the increase of pairing correlations with LN and SCK2 approaches lowers the moment of inertia of the band and the transition energies are larger (and closer to the experimental results) than the HFB ones.
%%%%%%%%%%%%%%%%%%%%%%%%%%%%%%%%%%%%%%%
%%%%%%%%%%%%%%%%%%%%%%%%%%%%%%%%%%%%%%
Lipkin-Nogami method (instead of plain HFB) is still routinely used 
with Skyrme and RMF energy density functionals to find the intrinsic 
mean-field-like wave functions. This is mainly due to the problems that 
arise when the particle number symmetry restoration is implemented in 
EDFs that deal differently with the HF and pairing parts of 
functional~\cite{Anguiano2001,Bender2009,Duguet2009,Lacroix2009} (see Sec \ref{Sec:SRDIFF1} for 
a discussion of this issue). A 
closer approach to PN-VAP energies can be obtained by projecting LN 
intrinsic mean-field states onto good number of particles exactly. This 
is the so-called projected Lipkin-Nogami approach (PLN). This method, 
compared to the full PN-VAP, is only able to approximate the minima of 
the PN projected energy surfaces defined along the direction of $\Delta 
N^{2}$~\cite{Rodriguez2005a}.

\begin{figure}
\begin{center}
\includegraphics[width=0.5\textwidth]{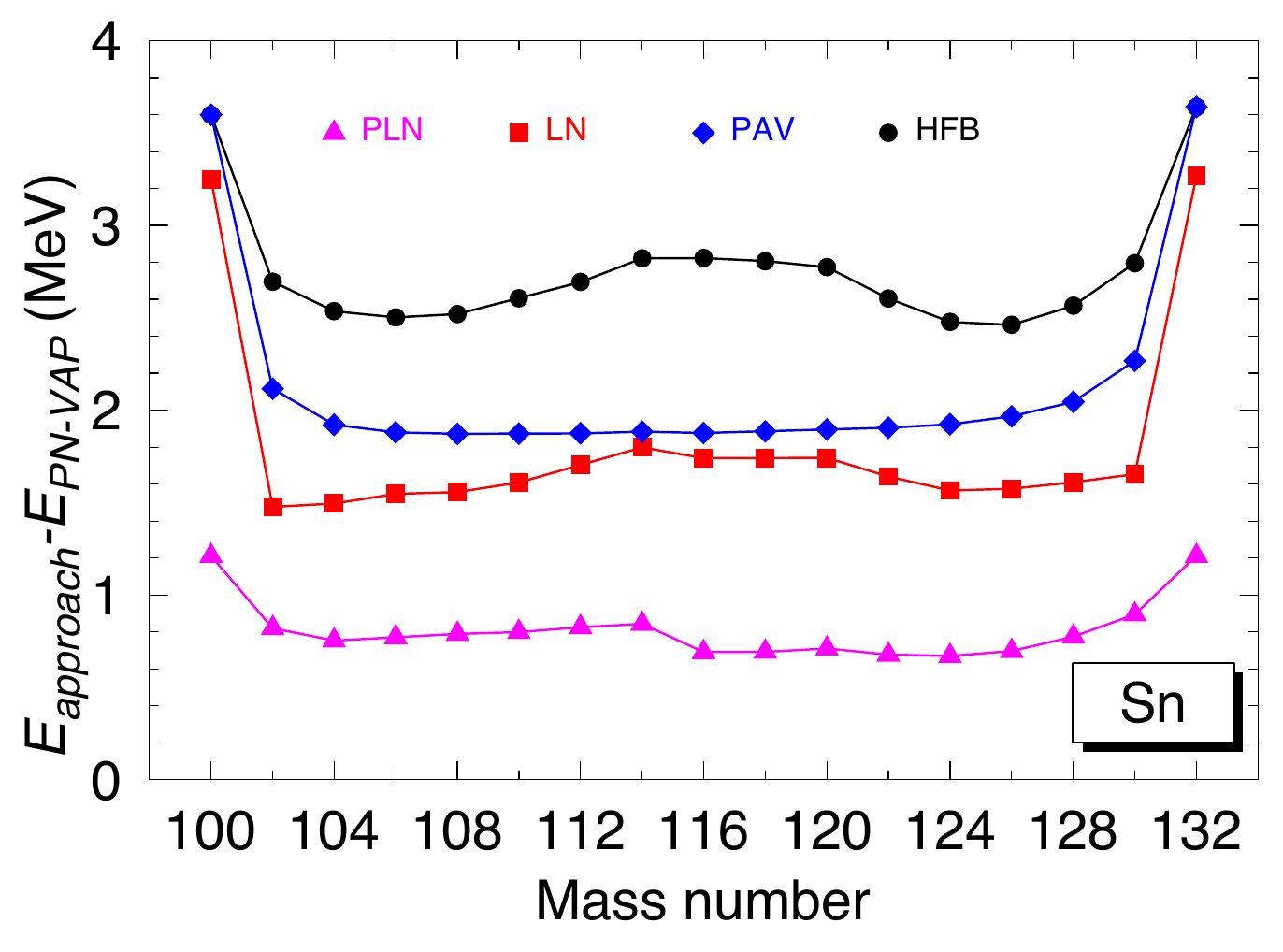}
\caption{Difference between the ground state energies obtained with the 
PN-VAP method and HFB, LN, PN-PAV and PLN approaches in the Sn isotopic 
chain. Gogny D1S interaction is used. Figure adapted from 
Ref.~\cite{Anguiano2002}. }
\label{Sn_PNVAP}
\end{center}
\end{figure}

Nevertheless, Kamlah and Lipkin-Nogami methods with Gogny EDFs were 
quickly improved with the full implementation of the exact particle 
number projection both within the PN-PAV and PN-VAP 
approximations~\cite{Anguiano2001}. In Fig.~\ref{Sn_PNVAP} the energy 
difference between the ground-state energy provided by the PN-VAP 
method and HFB, PN-PAV, LN and PLN methods in the Sn isotopic chain is 
shown. We observe that the lowest ground-state energies are obtained 
with the PN-VAP method, i.e., the energy difference is always positive. 
Since PN-VAP, PN-PAV, HFB, and PLN are variational methods, the PN-VAP 
approach is the best one in that respect. In addition, the largest 
energy differences are obtained in the doubly-magic isotopes 
$^{100,132}$Sn. In those nuclei, not only the proton but also the 
neutron pairing correlations collapse in the HFB approach and, 
subsequently, in the PN-PAV one. LN and PLN are able to attain some 
pairing correlations in the whole isotopic chain, but these shell 
effects are still present. 
%%%%%%%%%%%%%%%%%%%%%%%%%%%%%%%%%%%%%%%
%%%%%%%%%%%%%%%%%%%%%%%%%%%%%%%%%%%%%%

\begin{figure}
\begin{center}
\includegraphics[width=0.5\textwidth]{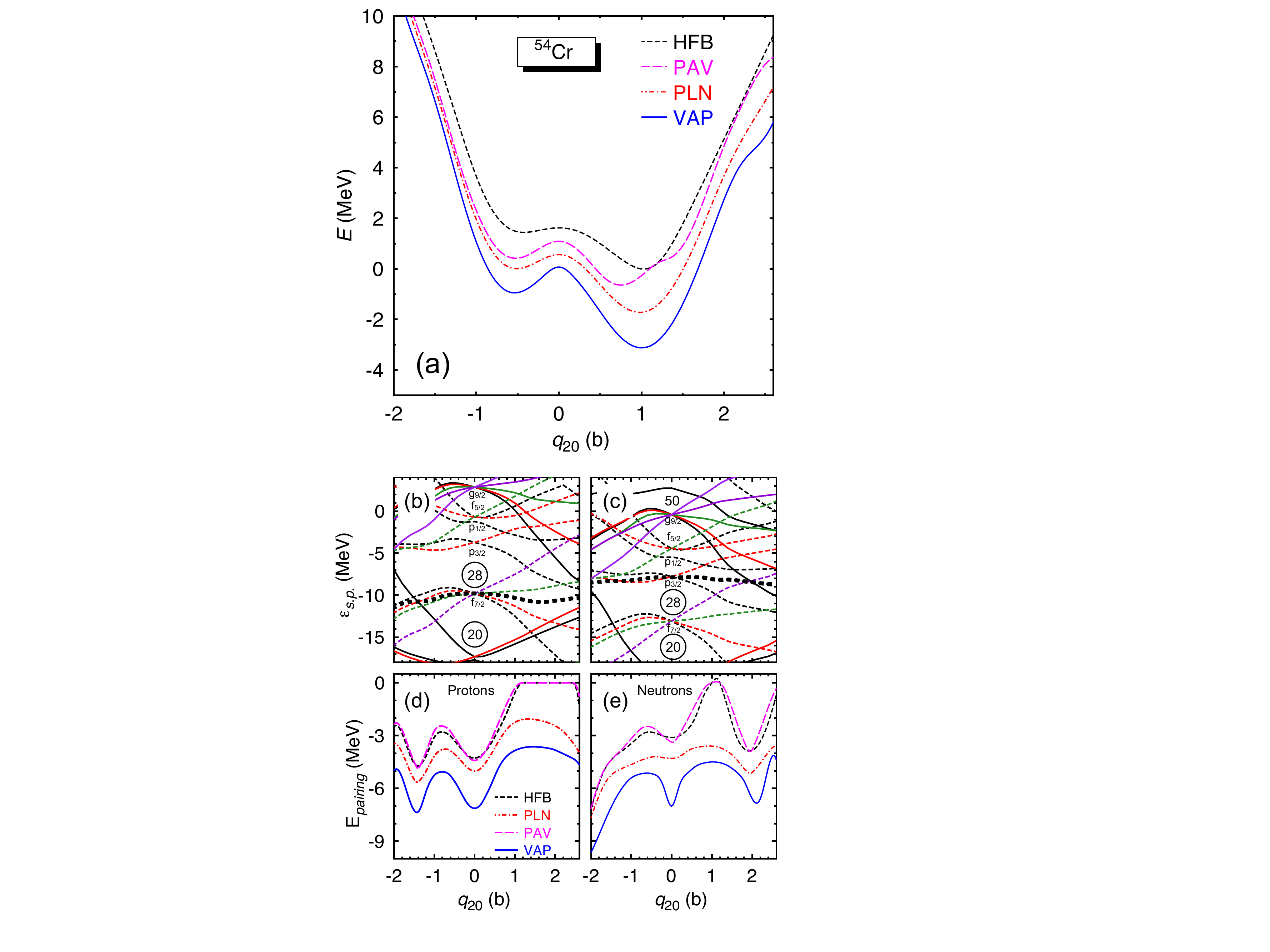}
\caption{(a) Energy (normalized to the minimum of the HFB solution) as 
a function of the axial quadrupole deformation in $^{54}$Cr calculated 
with HFB, PN-PAV, PNL and PN-VAP approximations. (b)-(c) 
Single-particle-energies and (d)-(b) pairing energies as a function of 
the axial quadrupole deformation for protons (left panel) and neutrons 
(right panel), respectively. Gogny D1S interaction is used. Dash thick 
lines in (b) and (c) represent the Fermi level. Figure adapted from 
Ref.~\cite{Rodriguez2007}. }
\label{54Cr_PNVAP}
\end{center}
\end{figure}

Another way of studying the performance of these approaches with the 
Gogny force is represented in Fig.~\ref{54Cr_PNVAP}. Here, several 
potential energy surfaces of $^{54}$Cr as a function of the quadrupole 
deformation are plotted (Fig.~\ref{54Cr_PNVAP}(a)), depending on the 
variational many-body method used to obtain the total energy of the 
nucleus. As previously, the best approach is given by the PN-VAP method 
and the PLN approach, although gives a similar qualitative result, is 
still above the PN-VAP result in the whole range of quadrupole 
deformations. Moreover, if we analyze the pairing energies, both for 
protons and neutrons, in connection with the corresponding 
single-particle-energies, we observe again that the largest amount of 
pairing correlations is obtained with the PN-VAP approach. The HFB and 
PN-PAV solutions show spurious super-fluid normal-fluid phase 
transitions for deformations for which the level density is small. This 
is not the case for the PLN method but this approximation is unable to 
capture enough pairing correlations in regions with small and large 
level density around the Fermi level. 
%%%%%%%%%%%%%%%%%%%%%%%%%%%%%%%%%%%%%%%
%%%%%%%%%%%%%%%%%%%%%%%%%%%%%%%%%%%%%%

Despite the PN-VAP method is widely used in current EDF calculations, 
there is still one problem left when beyond-mean-field methods with 
density-dependent interactions are implemented, i.e., the definition of 
the spatial density that enters in the interaction. The 
density-dependent term in the Gogny interaction is 
$\propto\rho^{1/3}(\vec{R})$ (where $\vec{R}$ is the position of the 
center-of-mass of the two nucleon system) and in the HFB approximation 
the spatial density is defined unambiguously as 
$\rho(\vec{r})=\langle\Phi|\hat{\rho}(\vec{r})|\Phi\rangle$. 
However, it is not obvious how to extend the definition of this density 
in the case of particle number projection, or, more generally, in the 
generator coordinate method where matrix elements of operators between 
different HFB states on the left (bra) and right (ket) are used. Hence, 
two prescriptions have been used mainly in PNP implementations with 
Gogny interactions (see Sec \ref{Sec:SRDIFF2} for a discussion), namely, the \textit{mixed} density, 
$(\rho(\vec{r},\varphi))$, and the \textit{PN-projected} density, 
$(\rho^{N}(\vec{r}))$:
\begin{eqnarray}
\rho(\vec{r},\varphi)&=&\frac{\langle\Phi|\hat{\rho}(\vec{r})e^{i\varphi\hat{N}}|\Phi\rangle}{\langle\Phi|e^{i\varphi\hat{N}}|\Phi\rangle}\label{mix_dens}\\
\rho^{N}(\vec{r})&=&\frac{\langle\Phi|\hat{\rho}(\vec{r})\hat{P}^{N}|\Phi\rangle}{\langle\Phi|\hat{P}^{N}|\Phi\rangle}\label{proj_dens}
\end{eqnarray}
In both cases rearrangement terms appear in the variational equations 
(exact and approximate as LN and/or SCK2) that have to be properly 
taken into account~\cite{Valor2000,Anguiano2001}. Additionally, when 
the mixed prescription is used to compute the PNP energy obtained from 
the density-dependent term of the Gogny force (proportional to a 
non-integer power of the density), such a term is not well 
defined~\cite{Bender2009,Duguet2009,Lacroix2009}. Nevertheless, the 
PN-projected density prescription has been used in practically all of 
the PNP calculations performed so far with the Gogny interaction. This 
prescription is free from these problems although it has a more 
phenomenological character than the \textit{mixed} prescription.
 
Let us mention two other important implementations of exact 
particle-number restorations with Gogny EDFs. The first one is the 
study of the impact of the PNP on nuclear halos, skins and drip-lines 
including the continuum~\cite{Schunck2008,Schunck2008a}. These 
calculations were performed using a spherical Woods-Saxon basis and a 
restricted PN-VAP (R-PN-VAP) method as an approximation to the full 
PN-VAP one. Such a R-PN-VAP procedure consisted in: a) finding a set of 
HFB wave functions, $\lbrace|\Phi(x)\rangle\rbrace$, that minimize the 
HFB energy whose pairing field matrix, $\Delta(x)=x\Delta$, is 
multiplied by a pairing enhancement factor, $x$. Obviously, the HFB 
solution will be directly given by $x=1$; b) computing the PNP energy 
surface along the pairing enhancement factor, 
$\langle\Phi(x)|\hat{H}\hat{P}^{N}\hat{P}^{Z}|\Phi(x)\rangle$; and, c) 
selecting the minimum of the PNP energy surface as the total energy. It 
is shown in Refs.~\cite{Schunck2008,Schunck2008a} that the two-neutron 
drip line is extended by two neutrons in a half a dozen cases with 
respect to the drip line predicted with continuum HFB calculations. 
These cases correspond to closed-shell nuclei where, contrary to the 
HFB method, R-PN-VAP can still get pairing correlations and, therefore, 
a lower total energy that produces a positive two-neutron separation 
energy. 

On the other hand, pairing fluctuations have been also included in the 
study of nuclear structure properties with Gogny 
EDFs~\cite{Vaquero2011,Vaquero2013a}. In this case, axial symmetric 
PN-VAP calculations have been performed along the quadrupole 
deformation and particle-number fluctuation ($\Delta N^{2}$) degrees of 
freedom on an equal footing. The inclusion of the latter degree of 
freedom allows, among other aspects, the study of pairing vibrations 
because states with the same spatial deformation but with different 
pairing gaps ("pairing deformations") can be mixed. This mixing can 
generate states that "vibrate" around an average pairing gap. Since 
these calculations serve as the starting point of more involved 
beyond-mean-field techniques (including not only particle number, but 
also angular momentum restoration, and configuration mixing) we will 
discuss this implementation in a subsequent Section (see 
Sec.~\ref{SCCM_section}).

%% -----------------------------------------------------------------------------------------------------------------
%%                                                                                    Angular momentum projection
%% -----------------------------------------------------------------------------------------------------------------

%%%%%%%%%%%%%%%%%%%%%%
\subsection{Angular Momentum Projection (AMP)}\label{AMP}
%%%%%%%%%%%%%%%%%%%%%%
Similarly to the particle-number symmetry breaking mechanism, the HFB 
wave functions can also break the rotational symmetry of the system by 
definition. This procedure enlarges the HFB variational space and 
allows the inclusion of correlations related to spatial multipole 
deformations of the Hamiltonian. However, the mean-field many-body wave 
functions (in general) are not eigenstates of the total nor the third 
component of the angular momentum operators, 
($\hat{J}^{2},\hat{J}_{z}$), but a linear combination of them:
\begin{equation}
|\phi\rangle=\sum_{\alpha,JM}a_{\alpha,JM}|\alpha JM\rangle
\label{mf_lin_com_JM}
\end{equation}
where $|\alpha JM\rangle$ are eigenstates of 
($\hat{J}^{2},\hat{J}_{z}$) and $\alpha$ refers to other quantum 
numbers of the system. This drawback can be corrected through symmetry 
restoration techniques (angular momentum projection, AMP) in a similar 
way as the particle-number symmetry discussed in Sec.~\ref{PNR}. In 
this case, the projection methods are more involved since the symmetry 
group associated to rotations is more complex. Due to the large 
computational cost of the AMP, this kind of symmetry restoration has 
been only carried out in a PAV approach with the most sophisticated 
EDFs (Skyrme, Gogny, RMF). Thus, we can build an eigenstate of the 
angular momentum operators, ($\hat{J}^{2},\hat{J}_{z}$), applying a 
projection operator to a symmetry-breaking many-body state, 
$|\phi\rangle$~\cite{Ring1980}:
\begin{equation}
|JM\sigma\rangle=\sum_{K=-J}^{J}g^{J\sigma}_{K}|JMK\rangle=\sum_{K=-J}^{J}g^{J\sigma}_{K}\hat{P}^{J}_{MK}|\phi\rangle
\label{proj_state}
\end{equation}
where $J$, $M$ and $K$ are the total angular momentum and the 
projection of $\vec{J}$ on the laboratory and intrinsic $z$-axes, 
respectively; $\hat{P}^{J}_{MK}$ is the angular momentum projector 
operator written in terms of an integral over the Euler angles 
($\Omega=(\alpha,\beta,\gamma)$)~\cite{Ring1980}.
\begin{equation}
\hat{P}^{J}_{MK}=\frac{2J+1}{16\pi^{2}}\int^{4\pi}_{0}d\alpha\int^{\pi}_{0}d\beta\sin\beta\int^{2\pi}_{0}d\gamma\mathcal{D}^{J*}_{MK}(\Omega)\hat{R}(\Omega)
\label{proj_operator}
\end{equation}
with $\mathcal{D}^{J}_{MK}(\Omega)$ are the Wigner matrices and 
$\hat{R}(\Omega)=e^{-i\alpha\hat{J}_{z}}e^{-i\beta\hat{J}_{y}}e^{-i\gamma\hat{J}_{z}}$ 
is the rotation operator defined with the Euler angles\footnote{From now on, units of $\hbar=1$ are used.}. The interval of 
integration over the angle $\alpha$ can be reduced to $[0,2\pi]$ for 
even-even nuclei, multiplying Eq.~\ref{proj_operator} by a factor 2.

Furthermore, the coefficients $g^{J\sigma}_{K}$, with 
$\sigma=1,2,3,...$ labeling the different states for a given value of 
the angular momentum, are obtained by minimizing the angular momentum 
projected energy within the $K$-space of dimension 
$(2J+1)\times(2J+1)$~\cite{Ring1980}. This is equivalent to solve a 
Hill-Wheeler-Griffin (HWG) equation (see Sec.~\ref{SCCM_section}) 
defined in such a space:
\begin{equation}
\sum_{K'=-J}^{J}\left(\mathcal{H}^{J}_{KK'}-E^{J\sigma}\mathcal{N}^{J}_{KK'}\right)g^{J\sigma}_{K'}=0
\label{HWG_AMP}
\end{equation}
The projected norm and Hamiltonian overlaps are defined as: 
\begin{eqnarray}
\mathcal{N}^{J}_{KK'}&=&\langle\phi|\hat{P}^{J}_{KM}\hat{P}^{J}_{MK'}|\phi\rangle=\langle\phi|\hat{P}^{J}_{KK'}|\phi\rangle\nonumber\\
\mathcal{H}^{J}_{KK'}&=&\langle\phi|\hat{P}^{J}_{KM}\hat{H}\hat{P}^{J}_{MK'}|\phi\rangle=\langle\phi|\hat{H}\hat{P}^{J}_{KK'}|\phi\rangle
\label{AMP_overlaps}
\end{eqnarray}
These equations are trivially obtained from the commutation properties, 
$[\hat{\mathcal{I}},\hat{R}(\Omega)]=0,[\hat{H},\hat{R}(\Omega)]=0$, 
and the property of the angular momentum projector 
$\hat{P}^{J}_{KM}\hat{P}^{J}_{MK'}=\hat{P}^{J}_{KK'}$. However, the 
last expression in Eq.~\ref{AMP_overlaps} must be taken with care when 
we deal with density-dependent interactions since the density-dependent 
term is not, in general, rotational invariant. As discussed thoroughly 
in Ref.~\cite{Rodriguez-Guzman2000a}, the \textit{mixed}-density 
prescription fulfills two basic requirements to provide a meaningful 
AMP energy, i.e., this term should produce: a) a scalar quantity (it 
should not carry angular momentum); and, b) a real (non-complex) 
quantity. The mixed-density prescription is defined in the AMP case as:
\begin{equation}
\rho_{\Omega}(\vec{r})=\frac{\langle\phi|\hat{\rho}(\vec{r})\hat{R}(\Omega)|\phi\rangle}{\langle\phi|\hat{R}(\Omega)|\phi\rangle}
\label{mixed_dd_amp}
\end{equation} 
Again, as discussed in Secs.~\ref{Sec:SRDIFF2} and \ref{PNR}, this term could be ill-defined 
whenever a non-integer power of the density is used to evaluate the 
energy (e.g., 1/3 as in Gogny 
EDFs)~\cite{Dobaczewski2007,Bender2009,Duguet2009,Lacroix2009}. However, the 
suitability of this prescription has been only tested numerically (see 
Ref.~\cite{Vaquero2013a}) and a more detailed work along the lines of 
Refs.~\cite{Duguet2009,Dobaczewski2007} but in the AMP context should 
be performed in the future. From now on, all the results that refer to 
angular momentum projected energies will be obtained by using the 
mixed-density prescription given by Eq.~\ref{mixed_dd_amp}.
 
The calculation of the overlaps (Eq.~\ref{AMP_overlaps}) requires a 
three dimensional integration over the Euler angles:
\begin{eqnarray}
\mathcal{N}^{J}_{KK'}&=&\frac{2J+1}{16\pi^{2}}\int \mathcal{D}^{J*}_{KK'}(\Omega)\langle\phi|\hat{R}(\Omega)|\phi\rangle d\Omega\nonumber\\
\mathcal{H}^{J}_{KK'}&=&\frac{2J+1}{16\pi^{2}}\int \mathcal{D}^{J*}_{KK'}(\Omega)\langle\phi|\hat{H}\hat{R}(\Omega)|\phi\rangle d\Omega
\label{AMP_overlaps2}
\end{eqnarray}
with $d^{J}_{KM}(\beta)$ the reduced Wigner matrices.
Such a calculation can be largely simplified if the intrinsic many-body 
states, $|\phi\rangle$, are axially-symmetric, i.e., 
$\hat{J}_{z}|\phi\rangle=K|\phi\rangle$. For the even-even case, we can 
choose the $z$-axis along the symmetry axis, having $K=0$. Therefore, 
Eq.~\ref{AMP_overlaps2} is reduced to the evaluation of only one 
integral:
\begin{eqnarray}
\mathcal{N}^{J}_{00}&=&\frac{2J+1}{2}\int_{0}^{\pi} d^{J*}_{00}(\beta)\langle\phi|e^{-i\beta\hat{J}_{y}}|\phi\rangle \sin\beta d\beta\nonumber\\
\mathcal{H}^{J}_{00}&=&\frac{2J+1}{2}\int_{0}^{\pi} d^{J*}_{00}(\beta)\langle\phi|\hat{H}e^{-i\beta\hat{J}_{y}}|\phi\rangle \sin\beta d\beta
\label{AMP_overlaps3}
\end{eqnarray}
Furthermore, the HWG equation (Eq.~\ref{HWG_AMP}) is trivially solved 
because it just expresses the single value obtained for the AMP energy 
in the axial case.

Most of the implementations of the AMP with Gogny interactions has been 
considered as an intermediate step from constrained HFB or PN-VAP 
towards configuration mixing (within the generator coordinate method) 
calculations. Nevertheless, we present some relevant properties of the 
AMP itself and the main differences/similarities between the 
calculations that have been published so far. 

First developments of the AMP with Gogny interactions were applied to 
the study of super-deformed bands and collectivity around $N=20$ and 
$N=28$ magic numbers in light nuclei, as well as the description of 
shape coexistence in the neutron-deficient lead 
region~\cite{Rodriguez-Guzman2000,Rodriguez-Guzman2000a,Rodriguez-Guzman2002,Rodriguez-Guzman2004}. 
These calculations assumed HFB many-body states with axial, parity and 
time-reversal symmetry conservation. These intrinsic HFB states were 
obtained from constrained-HFB calculations along the axial quadrupole 
degree of freedom ($q_{20}$), although some other directions 
-respecting axial, parity and time-reversal symmetries- were also 
explored~\cite{Rodriguez2005}. Particle number projection was not 
performed in these calculations either. Therefore, the AMP energy 
surfaces were evaluated as:
\begin{equation}
E^{J}(q_{20})=\frac{\mathcal{H}^{J}_{00}(q_{20})}{\mathcal{N}^{J}_{00}(q_{20})}-\sum_{\tau=p,n}\lambda_{N_{\tau}}(q_{20})\left(\frac{(\mathcal{N}_{\tau})^{J}_{00}(q_{20})}{\mathcal{N}^{J}_{00}(q_{20})}-N_{\tau}\right)
\label{AMP_PES_axial}
\end{equation}
where 
$\frac{(\mathcal{N}_{\tau})^{J}_{00}(q_{20})}{\mathcal{N}^{J}_{00}(q_{20})}$ 
is the AMP expectation value of the number of protons ($\tau=p$) and 
neutrons ($\tau=n$), $\lambda_{N_{\tau=p,n}}$ are the Lagrange 
multipliers obtained in the constrained-HFB calculation and 
$N_{\tau=p,n}$ are the actual number of protons and neutrons of the 
nucleus under study.
%%%%%%%%%%%%%%%%%%%%%%%%%%%%%%%%%%%%%%%
\begin{figure}
\begin{center}
\includegraphics[width=0.5\textwidth]{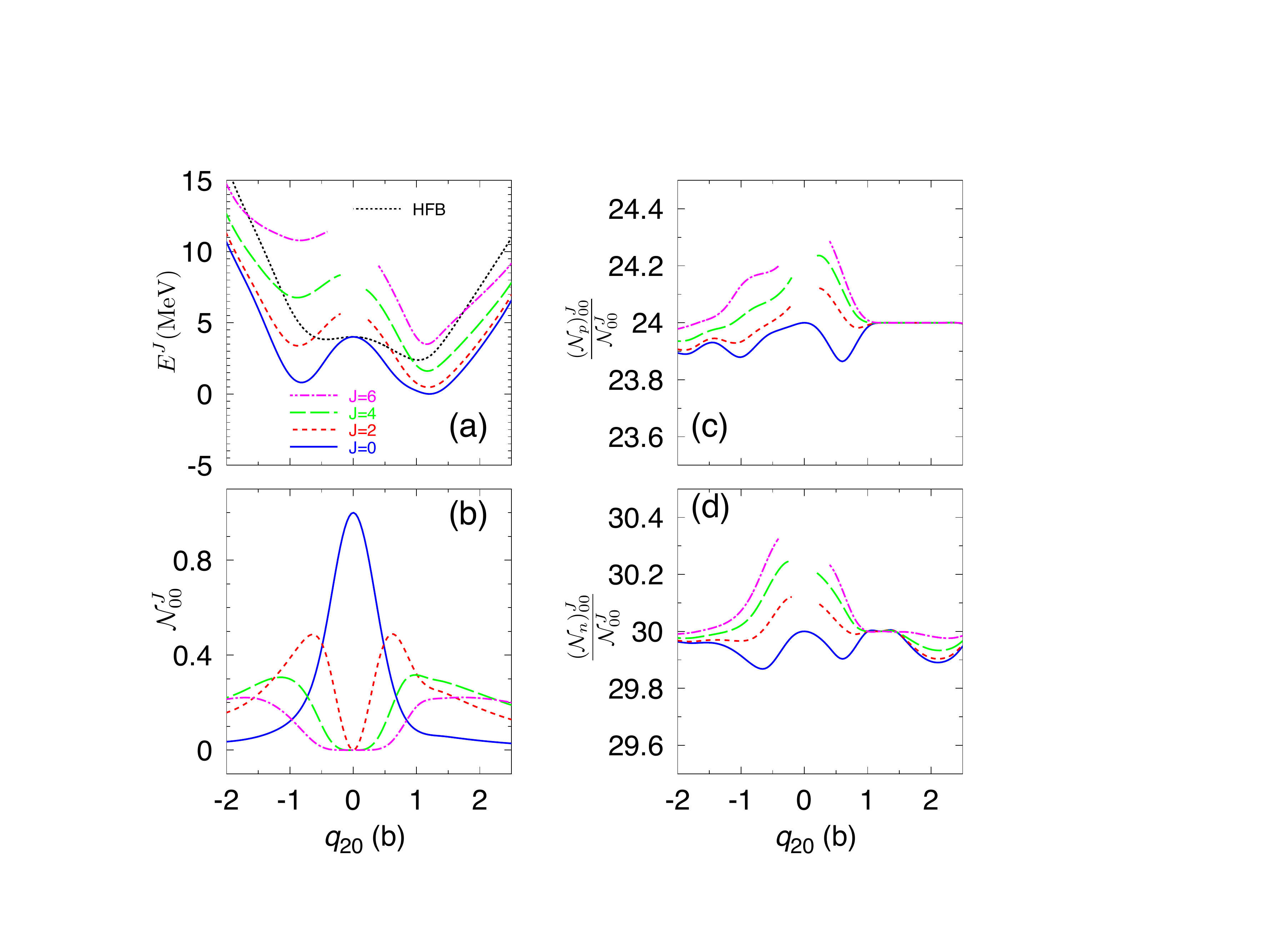}
\caption{(a) Energy (normalized to the minimum of the $J=0$ solution) 
as a function of the axial quadrupole deformation in $^{54}$Cr 
calculated with HFB and HFB+AMP with $J=0^{+}$, $2^{+}$, $4^{+}$ and 
$6^{+}$ methods with Gogny D1S. (b) AMP norm overlap and expectation 
values of the number of (c) protons and (d) neutrons as a function of 
the axial quadrupole deformation for $J=0^{+}$, $2^{+}$, $4^{+}$ and 
$6^{+}$.}
\label{54Cr_AMP_PES}
\end{center}
\end{figure}
%%%%%%%%%%%%%%%%%%%%%%%%%%%%%%%%%%%%%%   

\begin{figure}
\begin{center}
\includegraphics[width=0.5\columnwidth]{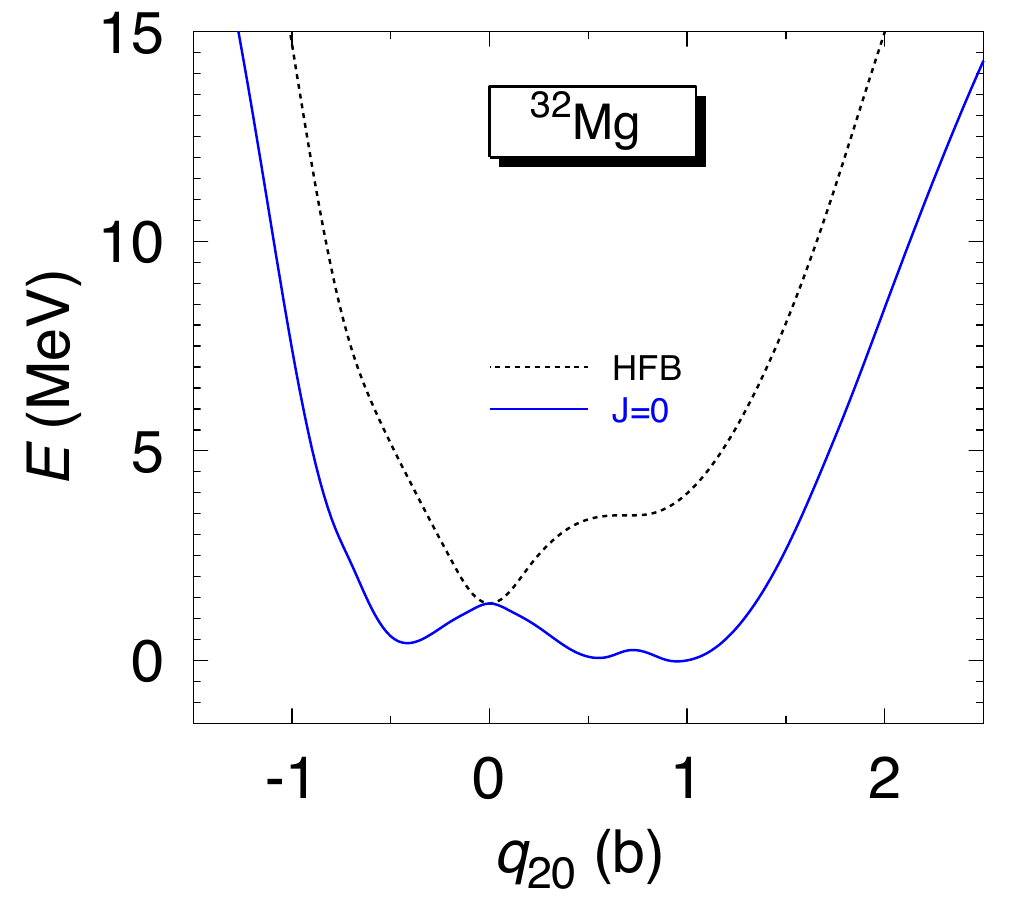}
\caption{(a) Energy (normalized to the minimum of the $J=0$ solution) 
as a function of the axial quadrupole deformation in $^{32}$Mg 
calculated with HFB and HFB+AMP with $J=0^{+}$ methods. Gogny D1S is 
used. Adapted from Ref.~\cite{Rodriguez-Guzman2002a}.}
\label{32Mg_AMP_PES}
\end{center}
\end{figure}

We show the main aspects of a HFB+AMP calculation in 
Fig.~\ref{54Cr_AMP_PES} along the axial quadrupole moment, where the 
nucleus $^{54}$Cr is used as an example. In Fig.~\ref{54Cr_AMP_PES}(a) 
the potential energy surfaces (PES) are represented for the results 
obtained with the constrained-HFB method and the subsequent AMP with 
$J=0^{+}$, $2^{+}$, $4^{+}$ and $6^{+}$. For $J=0^{+}$ we observe an 
energy gain with respect to the HFB solution, except for the spherical 
point ($q_{20}=0$). Therefore, we obtain a correlation energy due to 
the rotational symmetry restoration of deformed nuclei. In addition, 
the two minima (one prolate and one oblate) found in the HFB 
calculation are also obtained in the AMP-PES but shifted to larger 
deformations.  Thus, the PES without AMP are always modified by the 
inclusion of the angular momentum projection. This modification can 
significantly change the character of the nucleus under study in those 
cases where two minima, one less deformed (or spherical) than the 
other, are obtained in PES without AMP. A paradigmatic example of this 
effect is the nucleus $^{32}$Mg. Since this nucleus has $N=20$ neutrons 
(magic number), the HFB-PES has its absolute minimum in the spherical 
point and a shoulder related to the crossing of neutron $f_{7/2}$ and 
$d_{3/2}$ orbits appears at $q_{20}\approx0.8$ b  (see 
Fig.~\ref{32Mg_AMP_PES} and Ref.~\cite{Rodriguez-Guzman2002a}). 
However, once the angular momentum projection is performed, the 
absolute minimum corresponds now to a prolate deformed state, in 
agreement with the structure inferred from experiments. 
%%%%%%%%%%%%%%%%%%%%%%%%%%%%%%%%%%%%%%%
%%%%%%%%%%%%%%%%%%%%%%%%%%%%%%%%%%%%%%   

Coming back to the $^{54}$Cr example, we observe in 
Fig.~\ref{54Cr_AMP_PES}(a) that the AMP-PES with $J\neq0^{+}$ present 
discontinuities around $q_{20}=0$. Their origin is the same as the zero 
energy gain in the spherical point for $J=0^{+}$ and the absence of 
negative parity and odd-$J$ states. The projected norm overlap shown in 
Fig.~\ref{54Cr_AMP_PES}(b) represents the probability of finding a 
given eigenstate of the angular momentum in a intrinsic state. Using 
the decomposition given in Eq.~\ref{mf_lin_com_JM}, we see that such a 
probability is:
\begin{equation}
P(JM)=\sum_{\alpha}|\langle\alpha JM|\phi\rangle|^{2}=\sum_{\alpha}|a_{\alpha,JM}|^{2}=\mathcal{N}^{J}_{00}
\end{equation}
Since the underlying HFB states are axial and parity symmetric, only 
positive parity and even-$J$ angular momentum projected states can be 
obtained. In other words, the coefficients of the linear combination 
given in Eq.~\ref{mf_lin_com_JM} for negative parity and odd-$J$, i.e., 
their projected norm overlaps, are strictly zero in this case. In 
addition, the spherical intrinsic state, $|\phi(q_{20}=0)\rangle$, is 
already an eigenstate of the angular momentum operators with $J=0$. 
Therefore, the projected norm overlap for $J=0$ is one, and zero for 
other values of $J$ (see Fig.~\ref{54Cr_AMP_PES}(b)). This is the 
reason why there is not an energy gain in this special point by 
projecting onto $J=0$ and the appearance of the gap in the $J\neq0$ PES 
around this deformation. 

Finally, on the right panel of Fig.~\ref{54Cr_AMP_PES} we plot the 
projected expectation values of the number of protons and neutrons. 
Here we observe a non-negligible deviation from the nominal values for 
$^{54}$Cr. In fact, the deformations for which the number of 
protons/neutrons is correct correspond to regions where the HFB method 
presents a phase transition to HF solutions (without pairing). From 
this plot it is already obvious that if we want to perform more 
reliable beyond-mean-field calculations, the PNP must be also carried 
out. 

This simultaneous particle number and angular momentum projection 
(PNAMP) was first implemented with Gogny EDFs in 
Ref.~\cite{Rodriguez2007} as part of a more general beyond-mean-field 
method to study the potential $N=32$ and $N=34$ shell closures in 
neutron rich calcium, titanium and chromium isotopes. One of the main 
advantages introduced here was that the PNAMP was performed onto 
intrinsic HFB-like wave functions that are obtained from constrained 
PN-VAP instead of plain constrained-HFB calculations. This fact 
improved the pairing properties of the method. Therefore, the intrinsic 
many-body states in Eq.~\ref{proj_state} have the form of 
Eq.~\ref{pnp_wf}. As in the previous case, in the first applications of 
the PNAMP with Gogny interactions, axial, parity and time-reversal 
symmetries were conserved to reduce the computational burden. 
Nevertheless, these restrictions have been overcome in the most recent 
developments as we discuss below. 

A major improvement of PNAMP calculations with Gogny EDFs was the 
implementation of the triaxial angular momentum projection for 
even-even nuclei in Ref.~\cite{Rodriguez2010}. Again, parity and 
time-reversal symmetries were not broken, or, more specifically, 
$D^{T}_{2h}$ symmetry was 
conserved~\cite{Dobaczewski2000,Dobaczewski2000a,Frauendorf2001}. 
Therefore, the $(\beta_{2},\gamma)$ plane \footnote{We use indistinctly $\beta$ and
$\beta_{2}$ to refer to the quadrupole deformation parameter} can be reduced to values of 
$\gamma\in[0^{\circ},60^{\circ}]$. Many details such as the convergence 
of the triaxial AMP with respect to the number of Euler angles used to 
discretize the three-dimensional integral (Eq.~\ref{proj_operator}) or 
the best choice of the mesh in the $(\beta_{2},\gamma)$ plane are 
discussed thoroughly in Ref.~\cite{Rodriguez2010}. The inclusion of 
triaxiality opened several possibilities, for example: the study of 
shape evolution, mixing and/or coexistence in a more appropriate way; 
the study of $\gamma$-bands and $J^{+}$-odd states; a better 
description of $J\neq0$ excited states through $K$-mixing, etc.. These 
applications will be reviewed in Sec.~\ref{SCCM_section}. 

\begin{figure*}[t]
\begin{center}
\includegraphics[width=0.8\textwidth]{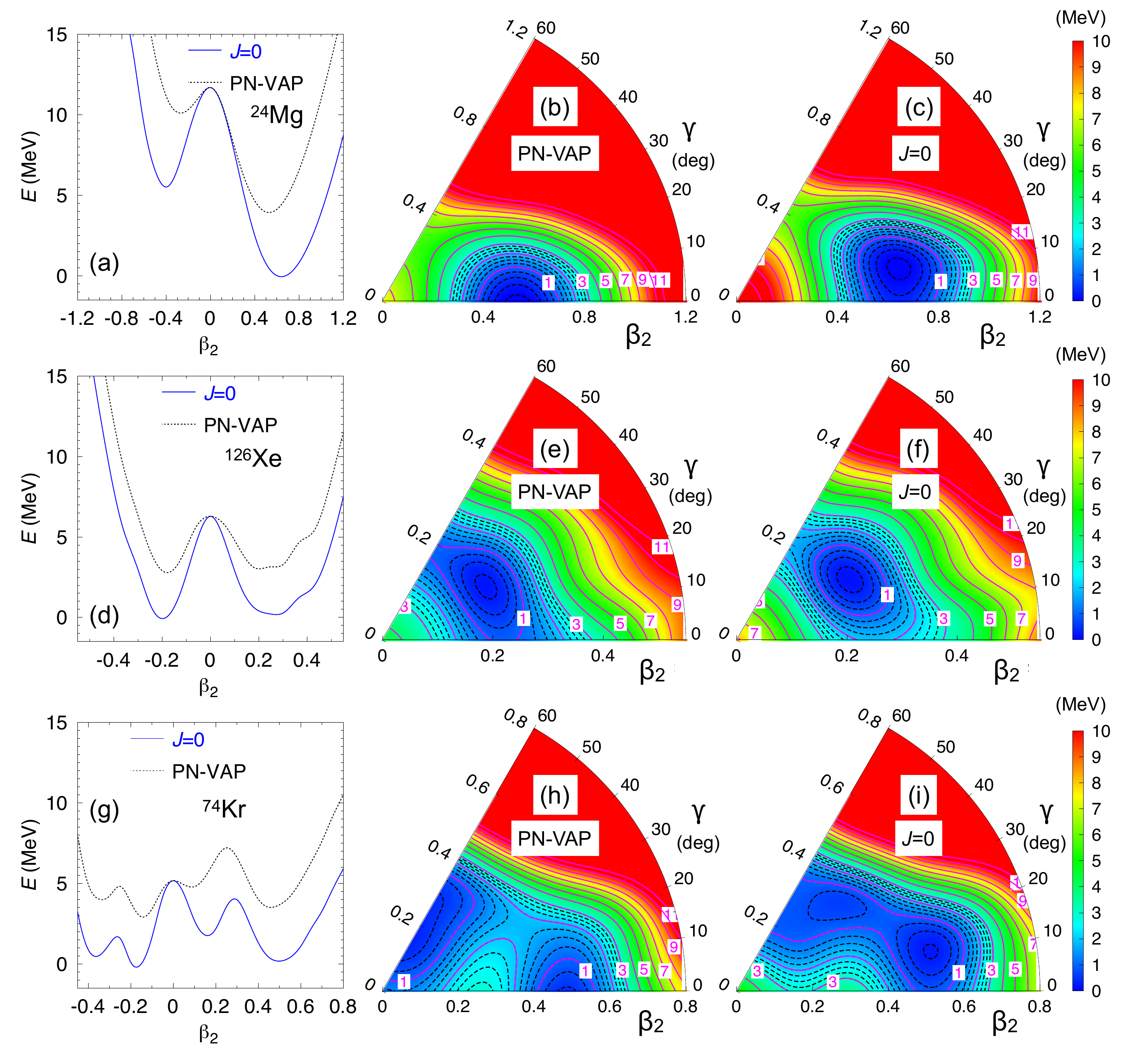}
\caption{Left panel: PN-VAP and PNAMP ($J=0$) potential energy surfaces 
(normalized to the absolute minimum of the $J=0$ PES) calculated with 
axially symmetric intrinsic wave functions. Middle panel and right 
panels: PN-VAP and PNAMP ($J=0$) potential energy surfaces (normalized 
to their minima) in the triaxial $(\beta_{2},\gamma)$ plane, 
respectively. Calculations are performed with Gogny D1S EDF for (a)-(c) 
$^{24}$Mg, (d)-(f) $^{126}$Xe and (g)-(i) $^{74}$Kr isotopes.}
\label{Axial_triaxial_PES_AMP}
\end{center}
\end{figure*}

Let us show in three examples how the inclusion of the triaxial degree 
of freedom could change the interpretation of the potential energy 
surfaces obtained with an axial calculation. In 
Fig.~\ref{Axial_triaxial_PES_AMP}(a)-(c) PN-VAP and PNAMP ($J=0$) 
results for the nucleus $^{24}$Mg are plotted. We observe that, in the 
axial case (Fig.~\ref{Axial_triaxial_PES_AMP}(a)), both the PN-VAP and 
$J=0$ PES have two well-defined minima, one prolate (the absolute one) 
and one oblate. These minima are shifted to slightly larger values in 
the AMP case. If we now explore additionally the triaxial degree of 
freedom we see that there is only one minimum, i.e., the prolate one in 
the PN-VAP case (Fig.~\ref{Axial_triaxial_PES_AMP}(b)) that is 
displaced towards larger and more triaxial deformations when AMP is 
performed (Fig.~\ref{Axial_triaxial_PES_AMP}(c)). Therefore, the axial 
oblate minimum is actually a saddle point in the $(\beta_{2},\gamma)$ 
plane. The situation is even worse in nuclei where a well-defined 
triaxial minimum is found in the triaxial-PES as in the isotope 
$^{126}$Xe Fig.~\ref{Axial_triaxial_PES_AMP}(d)-(f). Two minima (oblate 
and prolate) very close in energy are obtained in the axial PN-VAP and 
$J=0$ PES. This result could be interpreted as a possible signature of 
shape coexistence and/or shape mixing. However, if we study the 
triaxial PES, we see that these minima correspond to the saddle points 
produced by the absolute (and single) triaxial minimum. Finally, in 
Fig.~\ref{Axial_triaxial_PES_AMP}(g)-(i) we show an example ($^{74}$Kr) 
of a more plausible case of shape coexistence. Here, actual axial 
oblate and prolate minima are obtained in the PN-VAP axial and triaxial 
calculations. However, the $J=0$ surfaces show again differences 
between the axial and triaxial results exploring the triaxial degree of 
freedom. Hence, the barrier between prolate and oblate configurations 
is much smaller through pure triaxial configurations and the absolute 
minimum is located at a more prolate deformation in the triaxial case 
than in the axial calculation, that corresponds to an oblate shape. 
%%%%%%%%%%%%%%%%%%%%%%%%%%%%%%%%%%%%%%%
%%%%%%%%%%%%%%%%%%%%%%%%%%%%%%%%%%%%%% 

Apart from the inclusion of triaxial shapes, the octupole degree of 
freedom has been recently explored in PNAMP calculations with Gogny 
EDFs~\cite{Tagami2015,Bernard2016}. These states break the parity 
symmetry. Therefore, a simultaneous parity, particle number and angular 
momentum projection (PPNAMP) is also carried out. The intrinsic 
many-body states in Eq.~\ref{proj_state} are thus parity and particle 
number projected HFB-like wave functions, 
$|\phi\rangle=\hat{P}^{N}\hat{P}^{Z}\hat{P}^{\pi}|\psi\rangle$ 
($\hat{P}^{\pi}$ is the parity projection operator~\cite{Ring1980}). This 
method is used to study regions of the nuclear chart where the octupole 
correlations are expected to play an important role. In addition, both 
positive and negative parity states can be computed within this 
approach. 

\begin{figure*}[t]
\begin{center}
\includegraphics[width=0.8\textwidth]{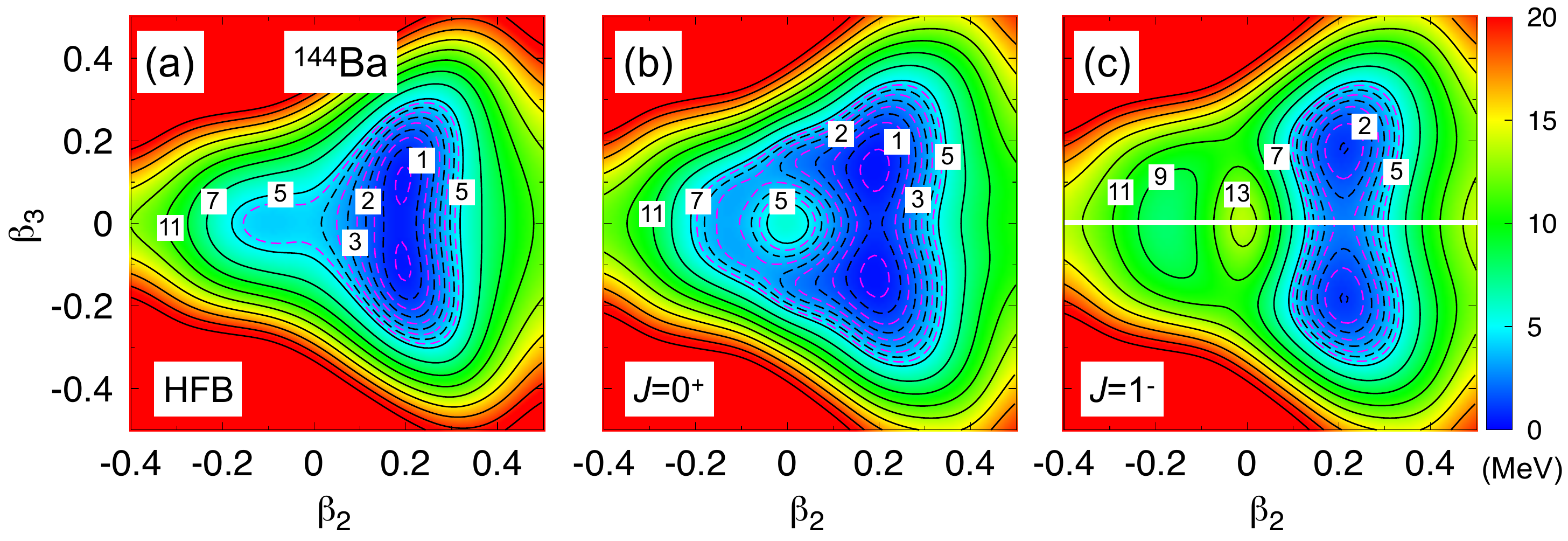}
\caption{Potential energy surfaces in the $(\beta_{2},\beta_{3})$ plane 
for (a) HFB, (b) PPNAMP-$J=0^{+}$ and (c) PPNAMP-$J=1^{-}$ 
approximations computed for $^{144}$Ba with the Gogny D1S 
parametrization. PES are normalized to the energy of the minimum of (a) 
HFB and (b)-(c) PPNAMP-$J=0^{+}$ surfaces. Contour lines are separated 
by 0.5 MeV (dashed lines) and 2.0 MeV (continuous lines) respectively. 
Adapted from Ref.~\cite{Bernard2016}.}
\label{Axial_octupole_PES_AMP}
\end{center}
\end{figure*}

The possible appearance of tetrahedral shapes in the region of Zr 
isotopes has been examined analyzing PPNAMP potential energy surfaces 
along axial and non-axial deformations~\cite{Tagami2015}. In this case, 
a full triaxial angular momentum projection is also included. On the 
other hand, PPNAMP potential energy surfaces have been studied in the 
Ba region as the intermediate step of configuration mixing 
calculations~\cite{Bernard2016}. Here, only axial-symmetric HFB wave 
functions have been considered. We show in 
Fig.~\ref{Axial_octupole_PES_AMP} an example of the performance of the 
PPNAMP method in the nucleus $^{144}$Ba. First of all, the PES along 
the $(\beta_{2},\beta_{3})$ plane are symmetric by exchanging 
$\beta_{3}\rightarrow-\beta_{3}$ due to the parity conservation of the 
nuclear interaction. This isotope is found to be both quadrupole and 
octupole deformed already within the mean-field approximation. The main 
effects of the simultaneous PPNAM projection are: a) the widening of 
the potential wells around the absolute minima; b) the connection 
between prolate and oblate configurations through the octupole degree 
of freedom direction for $J=0^{+}$; and, c) an energy gain of $\sim4.5$ 
MeV of correlation energy (difference between the HFB and $J=0^{+}$ 
absolute minima). Additionally, projection to odd-angular momenta and 
negative parity is not possible for the intrinsic states with 
$\beta_{3}=0$ since the projected norm overlaps for those parity- and 
axially-symmetric wave functions are zero.
%%%%%%%%%%%%%%%%%%%%%%%%%%%%%%%%%%%%%%%
%%%%%%%%%%%%%%%%%%%%%%%%%%%%%%%%%%%%%% 

\begin{figure}
\begin{center}
\includegraphics[width=0.7\textwidth]{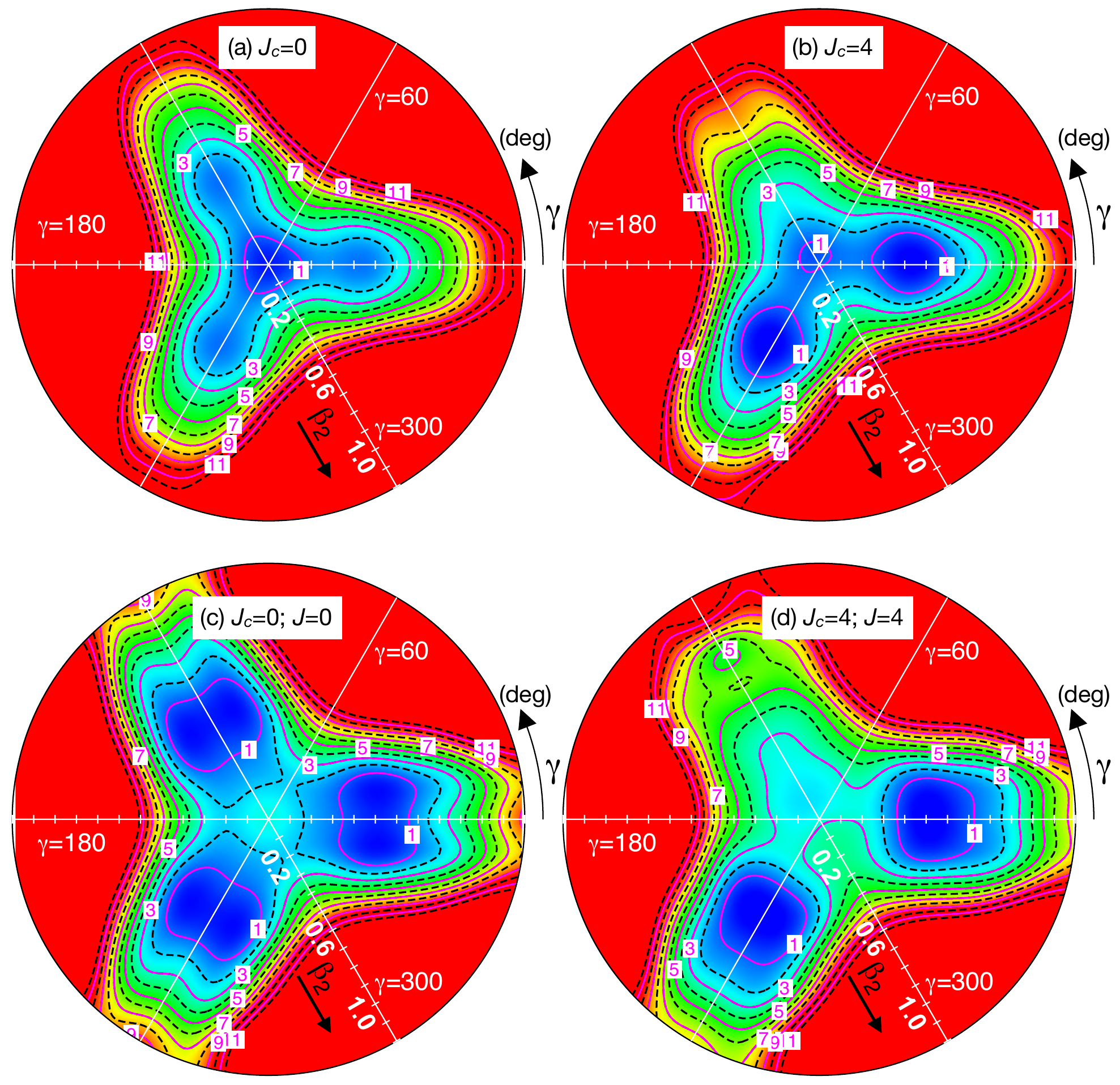}
\caption{(a) Energy as a function of the quadrupole 
$(\beta_{2},\gamma)$ deformation in $^{32}$Mg calculated with (a)-(b) 
PN-VAP and (c)-(d) PNAMP methods. In (a) and (c) the cranking angular 
momentum is $J_{c}=0$ while in (b) and (d) it is $J_{c}=4$. Gogny D1S 
is used. Adapted from Ref.~\cite{Borrajo2015}.}
\label{32Mg_AMP_PES_CRANK}
\end{center}
\end{figure}

The last implementation of simultaneous particle number and angular 
momentum projections with Gogny EDFs has been carried out onto 
intrinsic wave functions that break the time-reversal symmetry. These 
developments have been applied to study both even-even nuclei -with 
intrinsically rotating wave functions 
(cranking)~\cite{Borrajo2015,Egido2016,Tagami2016,Shimada2016,Rodriguez2016,Egido2016a}- 
and odd-even nuclei -with blocking~\cite{Borrajo2016,Borrajo2017}. As a 
consequence of the time-reversal symmetry breaking, the equivalence of 
the six sextants in the $(\beta_{2},\gamma)$ plane is lost 
($D^{T}_{2h}$ symmetry is 
broken~\cite{Dobaczewski2000,Dobaczewski2000a,Frauendorf2001}). In 
cranking calculations, where the intrinsic wave functions are 
constrained to have, for example, $\langle 
\hat{J}_{x}\rangle=\sqrt{J_{c}(J_{c}+1)}$, it is usual to keep the 
simplex as a self-consistent symmetry ($\hat{\Pi} 
e^{-i\pi\hat{J}_{x}}|\rangle=|\rangle$, with $\hat{\Pi}$ the 
parity operator). Therefore, the $(\beta_{2},\gamma)$ plane is divided 
now in two equivalent sextants that are symmetric with respect to the 
$\gamma=(120^{\circ},300^{\circ})$ direction (because the cranking 
direction is chosen to be the $x$-axis). This property can be used to 
perform consistency checks of the PNAMP with cranking wave functions as 
it is shown in Ref.~\cite{Borrajo2015}. In 
Fig.~\ref{32Mg_AMP_PES_CRANK}(a)-(b) we show energy surfaces in the 
whole $(\beta_{2},\gamma)$ computed for the $^{32}$Mg isotope with the 
PN-VAP method with cranking. Two different values of the cranking 
angular momentum, $J_{c}=0$ and 4, are chosen to discuss the effect of 
time-reversal symmetry breaking in this kind of calculations. 
Furthermore, these intrinsic states are also angular momentum projected 
and the corresponding PES for $(J_{c}=0,J=0)$ and $(J_{c}=4,J=4)$ are 
plotted in Fig.~\ref{32Mg_AMP_PES_CRANK}(c)-(d). We observe (left 
panel) that, as anticipated, the six sextants of PES computed with 
$J_{c}=0$ are equivalent. This is not the case for the PES calculated 
with $J_{c}=4$ (right panel). Since the cranking method produces 
intrinsically rotating deformed states about the $x$-axis, thus, the 
energies are not independent on the orientation of the coordinate 
system. Therefore, a better description of the nucleus is obtained by 
exploring three sextants of the $(\beta_{2},\gamma)$ plane. This has 
been already implemented with Gogny EDFs to study isomeric states in 
$^{44}$S~\cite{Egido2016} and $^{42}$Si~\cite{Egido2016a} since, due to 
the intrinsic rotations, single-particle excitations associated to 
alignments can be obtained within the PNAMP plus cranking approach. 
%%%%%%%%%%%%%%%%%%%%%%%%%%%%%%%%%%%%%%%
%%%%%%%%%%%%%%%%%%%%%%%%%%%%%%%%%%%%%%  

\begin{figure*}[t]
\begin{center}
\includegraphics[width=0.8\textwidth]{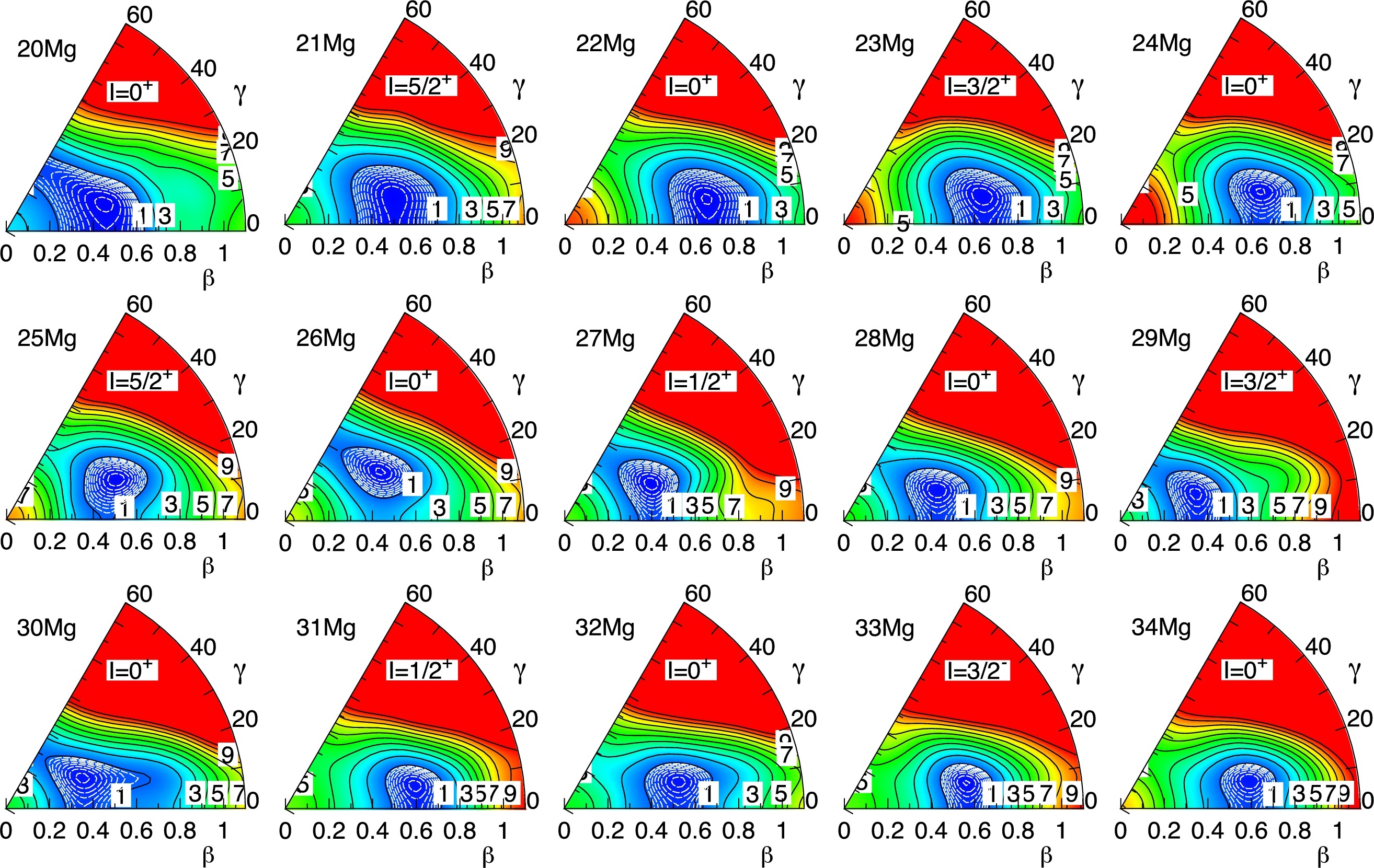}
\caption{Particle number and angular momentum projected potential 
energy surfaces (normalized to each minimum) as a function of 
$(\beta_{2},\gamma)$ Mg isotopic chain including both even-even and 
odd-even nuclei with the Gogny D1S interaction. The angular momentum is 
chosen to be the one that gives the ground state for each isotope. 
Figure taken from Ref.~\cite{Borrajo2017}.}
\label{Mg_odd_AMP}
\end{center}
\end{figure*}

As mentioned above, PNAMP energy surfaces have been also calculated 
with blocking (one-quasiparticle) PN-VAP 
states~\cite{Borrajo2016,Borrajo2017}. These works constitute the first 
implementations of PNAMP for odd-nuclei with Gogny EDF. In particular, 
bulk properties and electromagnetic moments were computed for the 
magnesium isotopic chain, obtaining a good agreement with the 
experimental data. Figure~\ref{Mg_odd_AMP} shows the PNAMP-PES 
corresponding to the angular momentum that provides the ground state 
energy for each isotope. Obviously, this value corresponds to $J=0^{+}$ 
for even systems while for odd systems, several blocking configurations 
(only neutrons of both positive and negative parity) were explored to 
find the predicted ground state.  
%%%%%%%%%%%%%%%%%%%%%%%%%%%%%%%%%%%%%%%
%%%%%%%%%%%%%%%%%%%%%%%%%%%%%%%%%%%%%%  

Let us finish this section with two important comments. The first one 
is that the variation after angular momentum projection (AM-VAP) is far 
from being reached with sophisticated EDFs as Gogny. However,  the 
exploration of AM-PAV energy surfaces defined along different degrees 
of freedom (quadrupole, octupole, cranking rotation, etc.) can be 
considered as a meaningful approach to the full VAP. In fact, a 
restricted VAP (RVAP) can be performed by choosing the minima of the 
AM-PAV energy surfaces. The second comment concerns the prescription 
for the spatial density in PNAMP calculations with Gogny EDFs. This was 
phenomenologically taken as a mixed version of projected (in the 
particle number projection) and \textit{mixed} (in the angular momentum 
projection) prescriptions:
\begin{equation}
\rho^{NZ}_{\Omega}(\vec{r})=\frac{\langle\phi|\hat{\rho}(\vec{r})\hat{R}(\Omega)\hat{P}^{N}\hat{P}^{Z}|\phi\rangle}{\langle\phi|\hat{R}(\Omega)\hat{P}^{N}\hat{P}^{Z}|\phi\rangle}
\label{mixed_dd_ampPNP}
\end{equation}
This prescription has been used in all of the applications of PNAMP 
approaches presented above and also in the symmetry conserving configuration 
mixing (SCCM) approaches that will be 
presented in Sec.~\ref{SCCM_section}.

% ------------------------------------------------------------------------------------------------
%                                                                       Linear momentum projection
% ------------------------------------------------------------------------------------------------

\subsection{Linear momentum projection (Brink-Boecker force)}

Although the Gogny force is not invariant under Galilei transformations
owing to the phenomenological density dependent term, the consequences 
of the restoration of Galilei invariance on physical observables are 
important and therefore we have discussed them with a simplified nuclear 
interaction that shares its central potential with the Gogny force, the 
Brink-Boecker interaction \cite{Brink19671}.

The restoration of Galilei invariance is 
considered one of the most challenging symmetry 
restorations within the nuclear many-body problem.  The 
homogeneity of space requires that the total linear momentum of the
nucleus, considered as a closed system of interacting, non-relativistic 
nucleons, is conserved. Therefore, the Hamiltonian cannot depend on the 
center of mass (COM) coordinate of its constituents, but only on 
relative coordinates and momenta. The dependence on the total linear 
momentum just accounts for the free motion of the system as a whole 
and can always be transformed away by considering the COM rest frame. 
In principle, the remaining (internal) Schr\"odinger equation can be 
solved by writing the Hamiltonian in Jacobi coordinates. However, since 
nucleons obey the Pauli principle and the Jacobi coordinates depend on 
all the nucleon coordinates, an explicit anti-symmetrization of the 
wave function should be carried out. Such an explicit 
anti-symmetrization, thought still feasible in few-body systems, 
becomes impossible for systems with a large number of nucleons. In this case,
anti-symmetrization is then taken into account implicitly by 
considering (mean-field) product trial wave functions \cite{Ring1980} 
that automatically incorporate Pauli's principle. Nevertheless, those 
product wave functions depend on 3A instead of the allowed 3A-3 
coordinates and thus contain spurious contaminations due to the motion 
of the system as a whole. As a result, Galilei invariance is broken.
This fact was already recognized \cite{Elliott61} as one of the problems of
the nuclear  shell model. It also was shown 
\cite{Giraud65} that for pure harmonic oscillator configurations the 
problem can be treated by diagonalizing the (oscillator) COM 
Hamiltonian and projecting out of the spectrum all states not corresponding to the ground 
state of this operator. However, 
this requires the use of complete $n\hbar\omega$ spaces since only then 
COM and internal excitations decouple exactly. 

A more general solution is obtained via symmetry restoration 
\cite{Peierls1957,Peierls1962,Yoccoz1966,Ring1980}, i.e., by projecting 
the considered wave functions into the COM rest frame. Note, that the 
projection techniques only ensure translational invariance and in order 
to recover the full Galilei invariance, the projection into the COM 
rest frame should be carried out before solving the corresponding 
mean-field (variational) equations \cite{Ring1980}, i.e., full Galilei 
invariance can only be recovered if the projection into the COM rest 
frame is carried out before the variation.

%%%%%%%%%%%%%%%%%%%%%%%%%%%%%%%%%%%%%%%%%%%%%%%%%%%%%%%%%%%%%%%%%%%%%%%%%%%%%%%%%%%%%%%%%%%%%%%%%
\begin{figure}
\begin{center}
\includegraphics[width=8cm]{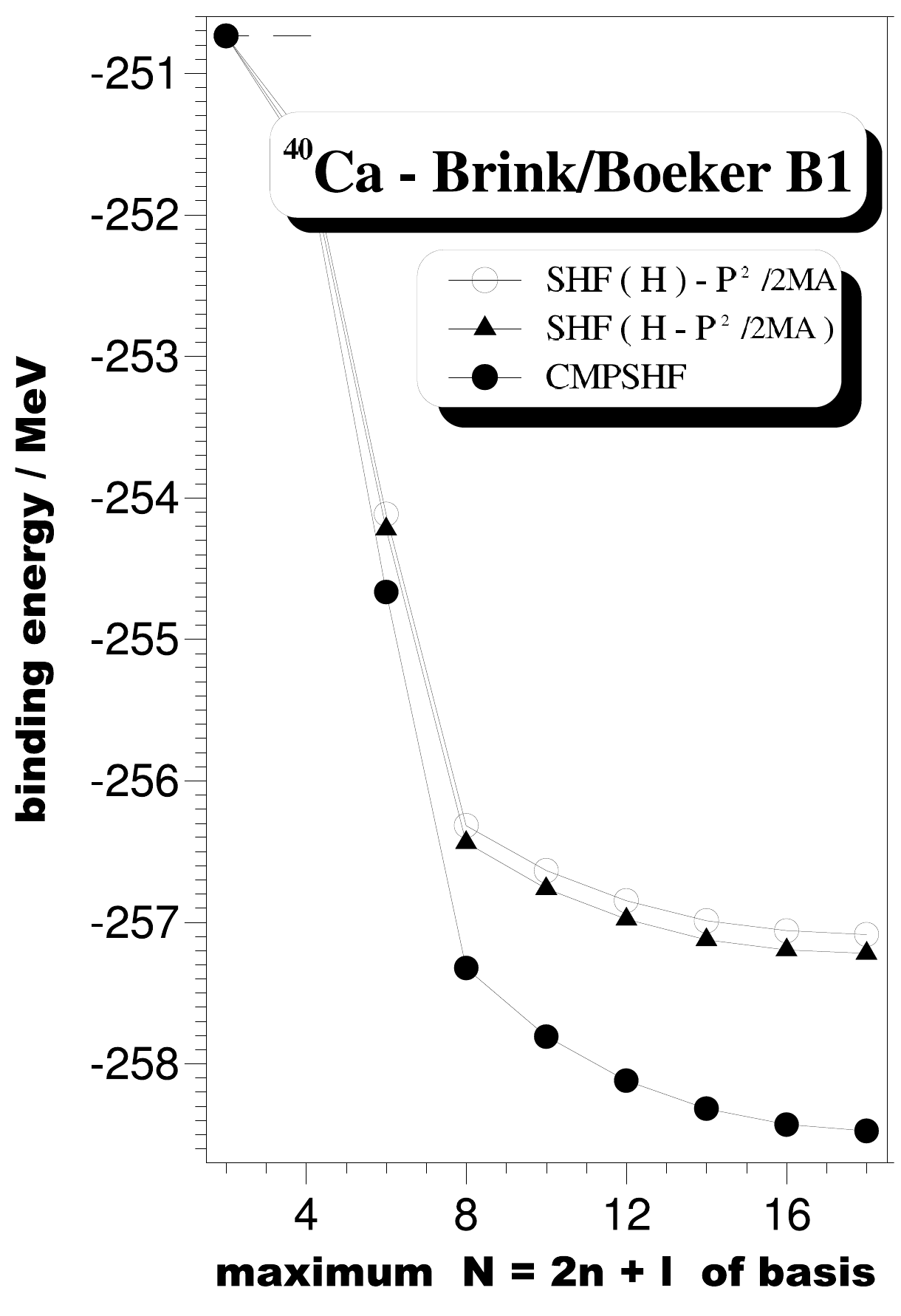} 
\end{center}
\caption{Total binding energy of $^{40}$Ca plotted as a function of the size of the single particle basis. Three curves 
are shown in the plot: the open circles correspond to a normal spherical Hartree-Fock calculation 
with the expectation value of ${\hat{P}}/2MA$ subtracted after convergence is achieved, the full 
triangles correspond to a calculation in which this COM correction is included in the variational 
procedure and the full circles correspond to a spherical Hartree-Fock calculation with projection 
 into the COM rest frame before the variation. Taken from Ref.~\cite{Rod04a}.
}
\label{fig1-COM-Rayner} 
\end{figure}
%%%%%%%%%%%%%%%%%%%%%%%%%%%%%%%%%%%%%%%%%%%%%%%%%%%%%%%%%%%%%%%%%%%%%%%%%%%%%%%%%%%%%%%%%%%%%%%%%
The COM projection method works in general model spaces as well as for 
general wave functions. However, it also entails challenging technical 
difficulties. For example, the associated projection operator links states 
in the model space to those in the core because any change of the momentum of the 
valence nucleons should be accompanied by a change in the momentum of 
the core to ensure zero total linear momentum. In other words, unlike 
other nuclear symmetries, linear momentum is a true A-body correlation 
and hence more difficult to afford than other symmetry restorations. 

The previous difficulties might be the main reasons why the exact 
restoration of Galilei invariance in the nuclear many-body problem has 
received much less attention in comparison with other symmetries. 
Instead, approximations like subtracting the kinetic energy of the COM 
from the original Hamiltonian which is in the definition of the Gogny 
force or the Tassie-Barker  \cite{Tassie1958} corrections to form 
factors are commonly employed. However, previous Hartree-Fock 
calculations with projection onto the COM rest frame in the case of 
$^{4}$He \cite{Schmid1990} as well as for form factors and charge 
densities in spherical nuclei \cite{Schmid1990a,Schmid1991} have shown, 
that a correct treatment of Galilei invariance leads to effects far 
beyond the usually assumed 1/A dependence. Considerable effects have 
also been found for scattering states in $^{4}$He \cite{Schmid1995} as 
well as for spectral functions, spectroscopic factors, transition form 
factors and densities, energies of hole states, Coulomb sum rules and 
response functions in 
Refs.~\cite{Schmid2001,Schmid2002,Schmid2002a,Schmid2003}.
%%%%%%%%%%%%%%%%%%%%%%%%%%%%%%%%%%%%%%%%%%%%%%%%%%%%%%%%%%%%%%%%%%%%%%%%%%%%%%%%%%%%%%%%%%%%%%%%%
\begin{figure}
\begin{center}
\includegraphics[width=13.05cm]{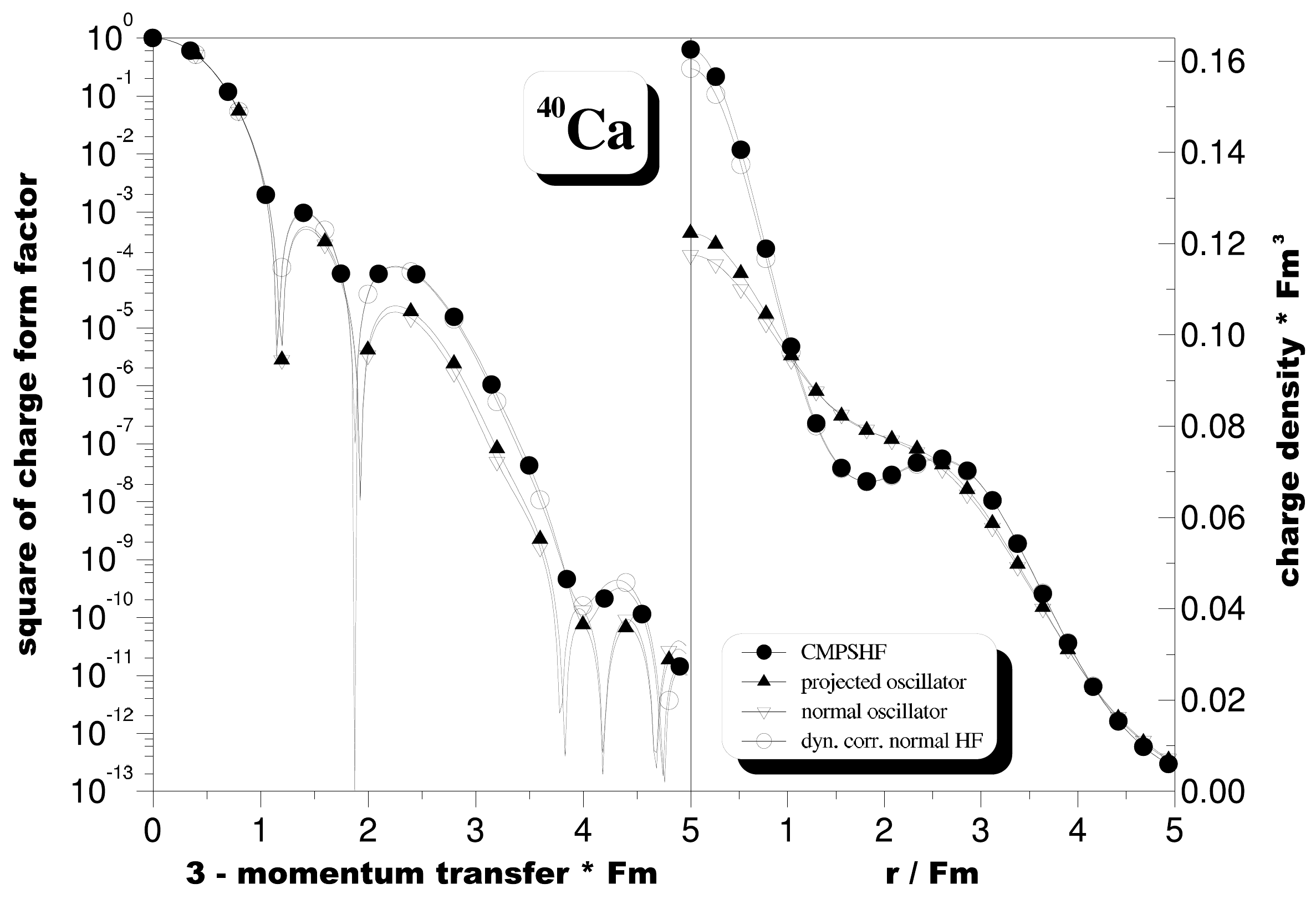}
\end{center}
\caption{In the left panel of the figure the square of the charge form factor for $^{40}$Ca is 
displayed as a function of the 3-momentum transfer. Open inverted triangles correspond to an 
oscillator occupation with no COM correction included, full triangles  
give the oscillator result including the Tassie-Barker factor (projected oscillator), open circles
display the form factor obtained with the normal Hartree-Fock (including COM correction in the 
Hamiltonian during the variation) taking into account the dynamic correction Eq.(\ref{Tassie})
in the main text. Full circles display the result of the full VAP-projected calculation. The right part
of the figure shows the corresponding charge densities. Taken from Ref.~\cite{Rod04a}.
}
\label{fig2-COM-Rayner} 
\end{figure}
%%%%%%%%%%%%%%%%%%%%%%%%%%%%%%%%%%%%%%%%%%%%%%%%%%%%%%%%%%%%%%%%%%%%%%%%%%%%%%%%%%%%%%%%%%%%%%%%%

In order to obtain a Galilei invariant wave function, the following ansatz
is used 
\begin{equation} \label{eqno5}
| \Phi\,;\;0\rangle\,\equiv\,{{\hat C(0)| \Phi\rangle}\over
{\sqrt{\langle \Phi|\hat C(0)| \Phi\rangle}}}
\end{equation}
as a trial variational state. The operator
\begin{equation}
\hat C(0)\,\equiv\,\int d^3\,\vec a\,\exp\{i\vec a\cdot\hat P\}
\end{equation}
projects into the COM rest frame by superposing all shifted (by $\vec{a}$) states 
$\exp\{i\vec a\cdot\hat P\} | \Phi\rangle$ with identical weights. The
operator $\hat P$ in the exponent is the operator of the total linear
momentum. The state $| \Phi\rangle$
represents a Slater determinant that is determined as the one minimizing 
\cite{Brodlie1977} the projected energy
\begin{equation} \label{eqno8}
E_\mathrm{proj}\;=  {{\langle \Phi |  \hat H \hat C(0)| \Phi \rangle}\over
{\langle \Phi|\hat C(0)| 	\Phi\rangle}}
\end{equation}

As an example of the performance of the COM projection, we analyze the results of spherical HF calculations
with a Gogny-like interaction without the density-dependent term. The effect of the symmetry restoration
on the total energy of the nucleus $^{40}$Ca is shown in Fig.~\ref{fig1-COM-Rayner}. Here, the total energy
as a function of the size of the single particle basis is represented for three different calculations , namely:
a) HF calculations with the subtraction of the expectation value of the kinetic energy of the COM motion afterwards;
b) HF calculations where the minimization of the energy is carried out with a modified Hamiltonian given by 
$\hat{H'}=\hat{H}-\frac{\hat{P}^{2}}{2MA}$; and, c) projection onto the COM rest frame before the variation.
As can be seen from the figure, in the case of pure oscillator 
occupations (i.e., the smaller basis) the three curves coincide, as 
expected, since these are non spurious configurations. With increasing 
basis size, however, they display a rather different major shell 
mixing. The two curves for the unprojected approaches, i.e., solutions a) and b), run 
almost parallel with the latter providing a larger binding energy than 
the former. Furthermore, the energy gain of the projected approach is the largest among all of the methods and,
as we observe in the figure, this contribution to the total energy is not negligible even in a not-so-light system as $^{40}$Ca. 

%%%%%%%%%%%%%%%%%%%%%%%%%%%%%%%%%%%%%%%%%%%%%%%%%%%%%%%%%%%%%%%%%%%%%%%%%%%%%%%%%%%%%%%%%%%%%%%%%
\begin{figure}
\begin{center}
\includegraphics[width=13cm]{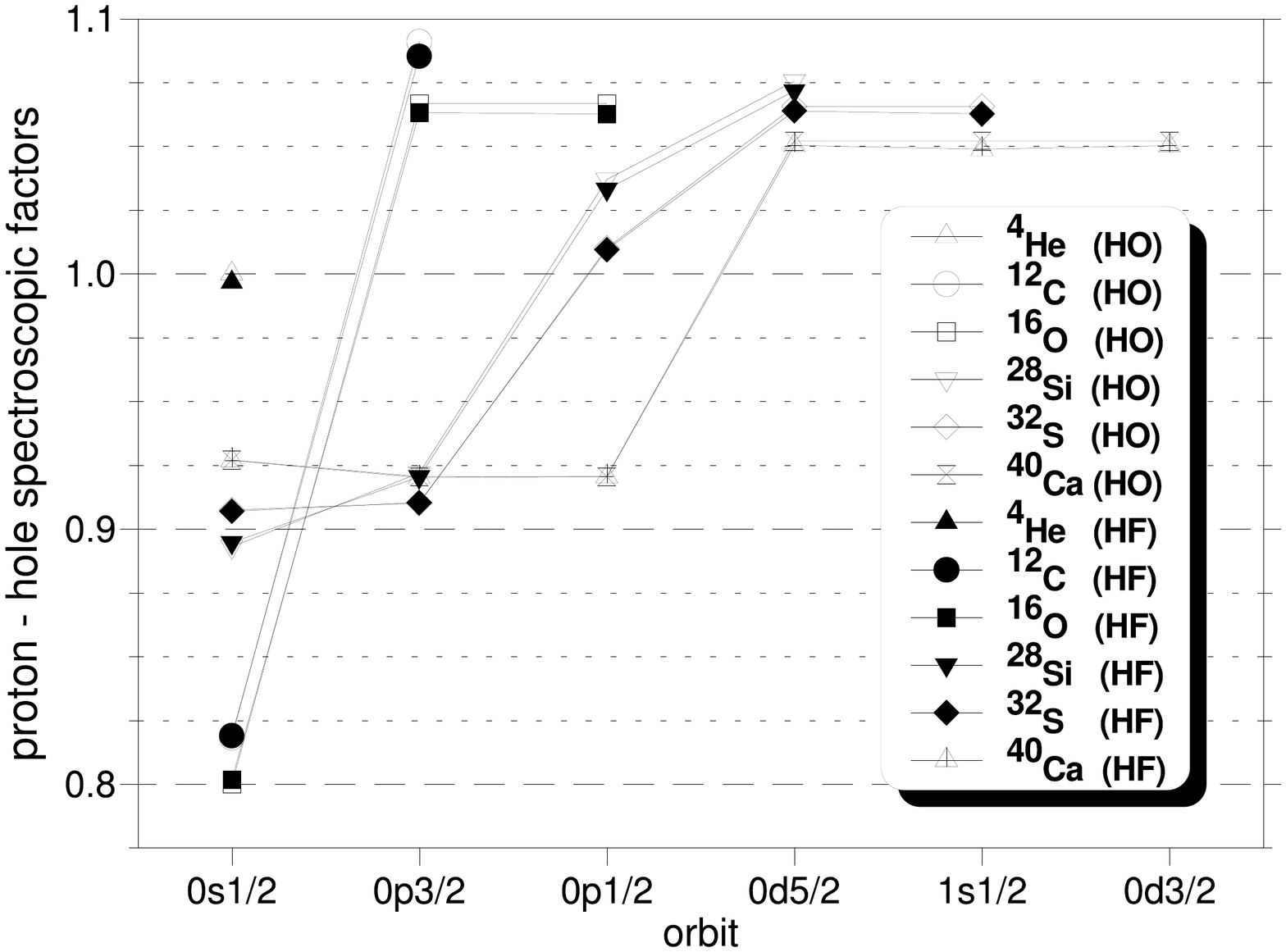}
\end{center}
\caption{Proton-hole-spectroscopic factors for the various spherical 
hole-orbits  in $^{4}$He, $^{12}$C, $^{16}$O, $^{28}$Si, $^{32}$ and 
$^{40}$Ca. Open symbols refer to the projected results using pure 
harmonic oscillator occupations. For $^{32}$S and $^{40}$Ca
the $0s_{1/2}$ denotes orthonormalized states (with respect to the $1s_{1/2}$ orbit). 
Full symbols refer to results obtained with 
projection into the COM rest frame before the variation. In this case $0s_{1/2}$ denotes the lowest
$s_{1/2}$ solution, $1s_{1/2}$ the second lowest $s_{1/2}$ solution and for the other orbits $0lj$ always 
the lowest solution is meant. Note, that in the usual approximation all the numbers displayed 
in the figure should be 1 regardless of whether pure oscillator or projected Slater determinants
are used.
Taken from Ref.~\cite{Rod04b}
}
\label{fig3-COM-Rayner} 
\end{figure}
%%%%%%%%%%%%%%%%%%%%%%%%%%%%%%%%%%%%%%%%%%%%%%%%%%%%%%%%%%%%%%%%%%%%%%%%%%%%%%%%%%%%%%%%%%%%%%%%%

Other interesting observables that are sensitive to Galilei invariance restoration are the charge 
form factors and the charge densities. The 
charge density in momentum space \cite{Schmid2002,Donnelly1979} is
given by
\begin{equation}
\hat\rho_n\,\equiv\,\sum_{\tau =p,n} f_{\tau}(Q^2)\,\sum_{i=1}^{N_{\tau}}\,
\exp\{i\vec q\cdot\vec r_i\} 
\end{equation}
where $f_{\tau}(Q^2)$ are the nucleon charge form factors as functions 
of the (negative) square of the four-momentum transfer $Q^2$ \cite{Rod04a}.
In the case of a Slater determinant $| \Phi \rangle$, the charge 
form factor in this approximation reads 
\begin{equation} \label{KAKA}
F_{ch}^n(Q^2)\,=\,\langle \Phi|\hat\rho_n| \Phi\rangle
\end{equation}
and the corresponding charge density can be obtained via the Fourier transform
of Eq.(\ref{KAKA}). On the other hand, the translationally invariant charge density reads 
\cite{Schmid2002,Rod04a}
\begin{equation} \label{KAKA-INV}
\hat\rho_{inv}\,\equiv\,\hat\rho_n\,\exp\{-i\vec q\cdot\vec R\}.
\end{equation} 
and, using Eq.(\ref{KAKA-INV}), the Galilei-invariant charge form factor takes 
the form 
\begin{equation} \label{CHRGE_FORM_FC_INV}
F_{ch}^{pr}(Q^2)\,=\,{{\langle \Phi|\hat\rho_{inv}\hat C(0)|
\Phi\rangle}\over{\langle \Phi|\hat C(0)| \Phi\rangle}}
\end{equation}
Obviously, the charge density corresponding to this form factor can be 
obtained through its Fourier transform. On the other hand, if the Gaussian overlap
approximation (GOA) \cite{Ring1980} is applied to both the shift operator
and the operator of Eq.(\ref{KAKA-INV}), one obtains the
dynamically corrected charge form factor 
\begin{equation} \label{Tassie}
F_{ch}^{dy}(Q^2)\,=\,F_{ch}^n(Q^2)\,\exp\left\{{3\over 8}\,{{{\vec q\,}^2}
\over{\langle \Phi|{\hat P\,}^2| \Phi\rangle}}\right\}
\end{equation}
Note that when the Slater determinant corresponds 
to a non spurious state, then the exponent in Eq.(\ref{Tassie}) takes 
the form $\exp\{(\vec q\,b/2)^2/A\}$ which is the 
Tassie-Barker correction \cite{Tassie1958,Rod04a}. 

These quantities are represented in Fig.~\ref{fig2-COM-Rayner} again for $^{40}$Ca. 
In particular, Eq.~\ref{KAKA} and Eq.~\ref{CHRGE_FORM_FC_INV} are evaluated with 
the spherical HF solution denoted by a) above (inverted open and full triangles respectively). 
The latter result is nothing but a PAV approach. Additionally, Eq.~\ref{Tassie} is evaluated with the HF solution
with the corrected Hamiltonian, referred as b) above. Finally, Eq.~\ref{CHRGE_FORM_FC_INV} has been 
computed with the VAP (Eq.~\ref{eqno8}) approximation. Charge form factors and their corresponding
charge densities (Fourier transformations) are given on the left and right side of Fig.~\ref{fig2-COM-Rayner}
respectively. Here we observe clearly the difference between pure mean-field and PAV approaches with the 
full COM-VAP restoration. Interestingly, the dynamically corrected form factor provides a very good approximation
to the projection before the variation results, at least in the present nucleus. 

Finally, let us 
consider the spectroscopic factors. They play a key role in one-nucleon 
transfer reactions where they are used to study nucleon-nucleon 
correlations. The normal hole-spectral functions $f_{h\tau\sigma}^{nor}(\vec k\,)$ are essentially given 
by the complex conjugate of the Fourier transform of the corresponding
single particle wave functions \cite{Rod04b}.
The normal hole-spectroscopic factor is defined as
\begin{equation}
S_h^{nor}\,\equiv\,\sum\limits_{\sigma}\int d^3\vec k\,
| f_{h\tau\sigma}^{nor}(\vec k\,)
|^2
\end{equation}
and satisfies the sum rule 
\begin{equation} \label{sum_rule}
\sum\limits_{h}S_h^{nor}\,=\,A
\end{equation}

In the usual picture, the hole-spectroscopic 
factors are one for all the occupied states and zero for the
unoccupied ones. However, such a picture is not Galilei-invariant
and therefore requires a reformulation~\cite{Rod04b} where Galilei-invariant hole states 
$\overline{h}$ have to be considered in the  projected hole-spectral functions 
$f_{\tilde h\tau\sigma}^{proj}(\vec k\,)$ and forms factors $S_{\tilde h}^{proj}$
The projected hole-spectroscopic factors satisfy a new sum rule
\begin{equation} \label{new_sum_rule}
\sum\limits_{\tilde h}\,S^{proj}_{\tilde h}\,=\,A\,-\,\epsilon
\end{equation}
where the ratio $\epsilon/A$ varies between 0.12 and 0.35 percent
for the cases discussed in Ref.~\cite{Rod04b}. Therefore, the violation of the sum rule Eq.(\ref{sum_rule}) due to
the correlations induced by the projection operator on the 
uncorrelated Hartree-Fock system, is rather small.

The hole-spectroscopic factors for several $N=Z$ nuclei
are depicted in Fig.~\ref{fig3-COM-Rayner}. In the non-projected HF
approximation they are all equal to one. Therefore, in the 
figure only projected hole-spectroscopic factors, both for the oscillator occupations~\cite{Schmid2001}
and the Hartree-Fock ground states are shown. Moreover, the proton and neutron spectroscopic 
factors are identical in the oscillator approach and even quite similar in the 
Hartree-Fock approach. Therefore, only results for protons are shown in the figure.
As can be seen from the figure the oscillator and Hartree-Fock results are quite similar 
suggesting that the size of the single particle basis is not relevant in 
the Galilei-invariant prescription at least as long as only uncorrelated system 
are considered.

A considerable depletion of the strengths of the
hole-states with excitation energies larger or equal to $1\hbar\omega$ and an
enhancement of the strengths of the hole-orbits near the Fermi energy is observed. 
The oscillator results fulfill the sum rule Eq.(\ref{sum_rule})
exactly, while for the HF determinants this is only approximately true
Eq.(\ref{new_sum_rule}). However,  the violation
$\epsilon$ of the sum rule is rather small. 

The results presented in Fig.~\ref{fig3-COM-Rayner} suggest that the 
usual picture of an uncorrelated system has to be modified 
considerably. In particular, those results indicate that the 
interpretation of experiments in which deviations of the 
hole-spectroscopic factors from one are usually regarded  as 
fingerprints of nucleon--nucleon correlations, should be taken with 
care. In a Galilei-invariant description, the spectroscopic factors 
even of an uncorrelated system differ from one and only deviations from 
the COM-projected results should be related to non-trivial 
nucleon-nucleon correlations.

The results discussed in this section illustrate that a correct treatment 
of Galilei invariance is possible, via projection techniques, in the 
case of finite range forces and large configuration spaces. They also 
suggest that, in the long run, the up to now almost neglected 
restoration of Galilei invariance should be incorporated in more 
sophisticated approaches like the shell model \cite{Bro01}, the quantum 
Monte Carlo diagonalization method \cite{ots01} and in 
symmetry-projected (mean-field based) methods \cite{hje02}.
For a more detailed account of these aspects, the reader is referred to 
Refs.~\cite{Rod04a,Rod04b,Schmid2001,Schmid2002,Schmid2002a,Schmid2003}.
To the best of our knowledge, those COM-projected results still 
represent the only ones of their kind available in the literature.

% -----------------------------------------------------------------------------------------------------------------
%                                                                       Approximate versus exact projection
% -----------------------------------------------------------------------------------------------------------------
\subsection{Approximate versus exact projection}
\label{Sec:SRApp}

Symmetry restoration requires the evaluation of multidimensional integrals
where the integrand is an overlap of operators that is expensive to compute (see
Ref \cite{Rodriguez2016} for a discussion on computational costs). As this 
has to be repeated for many overlaps in the GCM case it is important
to reduce the computational burden. An approach is to benefit from the
fact that often many of the intrinsic configurations are strongly 
deformed, the overlaps are strongly peaked and therefore the integrands 
contribute to the integral in a limited subset of the integration interval.
In addition, the assumption of an overlap with Gaussian form can be often 
made allowing for an analytic approach to the problem. To illustrate the
procedure the PNP case will be discussed in detail below. 

The main assumption of the method is  the approximate Gaussian form of 
the rotated overlap 
\begin{equation}
	\langle \Phi | e^{i\varphi \hat{N}} | \Phi \rangle = e^{-\langle \Delta N^{2}\rangle \varphi^{2}}
\end{equation}
with the width $ \langle \Delta N^{2}\rangle$  large enough as to make
the overlap strongly peaked around $\varphi=0$. The next assumption is
that the overlap ratio of any relevant operator 
$\langle\Phi|\hat{H}e^{i\varphi\hat{N}}|\Phi\rangle / \langle\Phi|e^{i\varphi\hat{N}}|\Phi\rangle$ 
is a smooth function of $\varphi$ and its expansion to second order
around $\varphi=0$ is enough as to faithfully represent that quantity
in the relevant subset of the integration interval. Instead of 
expanding this quantity we follow the method suggested by Kamlah \cite{Kamlah1968}
where the Hamiltonian kernel 
$h(\varphi)=\langle\Phi|\hat{H}e^{i\varphi\hat{N}}|\Phi\rangle$ (for 
the sake of simplicity we omit the double projection onto protons and 
neutrons and we assume for the moment density independent interactions 
provided by a Hamiltonian, $\hat{H}$) is expanded as:
\begin{equation}
h(\varphi)=\sum_{m=0}^{M}h_{m}\hat{K}^{m}n(\varphi)
\label{kamlah_expansion}
\end{equation}
where $\hat{K}=\frac{1}{i}\frac{\partial}{\partial\varphi}-\langle\hat{N}\rangle$ 
is a representation of the particle number operator in the space 
parametrized by the gauge angle $\varphi$ and 
$n(\varphi)=\langle\Phi|e^{i\varphi\hat{N}}|\Phi\rangle$ is the norm 
overlap kernel. Eq.~\ref{kamlah_expansion} is exact when 
$M\rightarrow\infty$ but for situations with a relatively strong 
symmetry breaking the expansion can be reduced to the lowest orders, 
typically $M=2$. The coefficients $h_{m}$ are thus found by solving a 
system of equations obtained by the application of the operators 
$\hat{K}^{0}$, $\hat{K}^{1}$, ..., $\hat{K}^{M}$ to 
Eq.~\ref{kamlah_expansion} and taking the limit $\varphi\rightarrow0$:
\begin{equation}
\langle\hat{H}(\Delta\hat{N})^{n}\rangle=\sum_{m=0}^{M}h_{m}\langle(\Delta\hat{N})^{n+m}\rangle
\end{equation}
where $\Delta\hat{N}=\hat{N}-\langle\hat{N}\rangle$.

The PNP energy is calculated as:
\begin{equation}
E^{\mathrm{PNP}}=\frac{\frac{1}{2\pi}\int_{0}^{2\pi}e^{-i\varphi N}h(\varphi)d\varphi}{\frac{1}{2\pi}\int_{0}^{2\pi}e^{-i\varphi N}n(\varphi)d\varphi}
\label{pnp_ener}
\end{equation}
Substituting the Kamlah expansion (Eq.~\ref{kamlah_expansion}) in the 
previous expression we obtain the PNP energy at order $M$:
\begin{equation}
E^{\mathrm{PNP}}_{(M)}=\sum_{m=0}^{M}h_{m}\left(N-\langle\hat{N}\rangle\right)^{m}
\end{equation}
If we keep only the terms up to $M=2$, the PNP energy is written as:
\begin{equation}
E^{\mathrm{PNP}}_{(2)}=h_{0}+h_{1}\left(N-\langle\hat{N}\rangle\right)+h_{2}\left(N-\langle\hat{N}\rangle\right)^{2}
\label{Kamlah_2nd}
\end{equation}
with
\begin{eqnarray}
h_{0}&=&\langle\hat{H}\rangle-h_{2}\langle(\Delta\hat{N})^{2}\rangle\nonumber\\
h_{1}&=&\frac{\langle\hat{H}\Delta\hat{N}\rangle-\langle(\Delta\hat{N})^{3}\rangle}{\langle(\Delta\hat{N})^{2}\rangle}\nonumber\\
h_{2}&=&\frac{\langle\Delta\hat{H}(\Delta\hat{N})^2\rangle-\langle\hat{H}\Delta\hat{N}\rangle\langle(\Delta\hat{N})^{3}\rangle/\langle(\Delta\hat{N})^{2}\rangle}{\langle(\Delta\hat{N})^{4}\rangle-\langle(\Delta\hat{N})^{2}\rangle^{2}-\langle(\Delta\hat{N})^{3}\rangle^2/\langle(\Delta\hat{N})^{2}\rangle}\nonumber\\
\end{eqnarray}
Now the intrinsic HFB wave function are obtained by solving the 
variational equations extracted from the minimization of the projected 
energy at order $M=2$ in the Kamlah expansion (Eq.~\ref{Kamlah_2nd}). 
The full variation of such an energy functional gives the 
self-consistent second-order Kamlah (SCK2) approach to the PN-VAP 
energy whose equation is written as:
\begin{equation}
\delta\langle\hat{H}\rangle-h_{2}\delta\langle(\Delta\hat{N})^{2}\rangle-\langle(\Delta\hat{N})^{2}\rangle\delta h_{2}=0,
\end{equation}
with the additional condition of having $h_{1}=\lambda$ as a Lagrange 
multiplier that ensures the constraint $\langle\hat{N}\rangle=N$. 
However, it is also usual to perform a further approach that consists 
in ignoring the variation of the coefficient $h_{2}$. This is the 
Lipkin-Nogami (LN) approximation~\cite{Lipkin1960,Nogami1964} that produces the equation:
\begin{equation}
\delta\langle\hat{H}\rangle-h_{2}\delta\langle(\Delta\hat{N})^{2}\rangle=0,
\end{equation}  
with the same constraint in $\langle\hat{N}\rangle=N$. Both the variational
Kamlah as well as the Lipkin-Nogami have been used to describe super-deformed
high spin bands as discussed in Sec \ref{PNR}.

The same ideas can be used, for instance, with angular momentum projection.
In this case, however, the non-abelian nature of the underlying symmetry
group SU(2) brings additional complications. A nice derivation is given
in Ref \cite{Mang1975325} which is complemented by the discussion on the
approximate form of transition matrix elements in \cite{Islam1979} (see
also \cite{Robledo2012} for a discussion of the rotational formula
in the near spherical limit ). The main conclusion is that if a strongly
deformed intrinsic state is projected onto good angular momentum (PAV) the
approximate projected energy is given by  
\begin{equation}\label{Eq:EJapp}
	E^{J} = \langle \Phi | \hat{H} | \Phi \rangle - \frac{\langle \Delta \vec{J}^{2}\rangle}{2\mathcal{J}_\mathrm{Y}}
	+ \frac{\hbar^{2} J(J+1)}{2\mathcal{J}_\mathrm{Y}}
\end{equation}
as discussed in Sec \ref{Sec:SRRB}. The typical rotational band pattern
with the $J(J+1)$ dependence is obtained with the Yoccoz moment of inertia
$\mathcal{J}_ \mathrm{Y}$~\cite{Peierls1957,Yoccoz1957}. Also, the ground state rotational energy correction 
$\frac{\langle \Delta \vec{J}^{2}\rangle}{2\mathcal{J}_\mathrm{Y}}$ is
obtained: it represents the energy gained by the ground state as a consequence
of the quantum correlations associated to the symmetry restoration. 
On the other hand, if the intrinsic state is varied as to minimize 
the projected energy $E^{J}$ the same approximate angular momentum projection
leads to the cranking model where the intrinsic states are determined
by solving the HFB equation with a constraint in $\langle J_{x}\rangle=\sqrt{J(J+1)}$
in the spirit of the VAP method. As the intrinsic state is now $J$ dependent
the projected energies follow a somehow distorted rotational pattern that 
can take into account typical high spin effects like the Coriolis anti-pairing
effect or the angular momentum dependent intrinsic deformation parameters, see
Sec \ref{Sec:MF} for details and calculations with the Gogny force. As argued
in \cite{Rodriguez-Guzman2002} and discussed in \cite{Friedman1970,Villars1971} the 
projected angular momentum energy of the cranked states can be well represented
by a formula similar to Eq \ref{Eq:EJapp} but replacing the Yoccoz moment
of inertia by the Thouless-Valatin one~\cite{Thouless1962} in the $J(J+1)$ term. 

The application of the large deformation approximation to the evaluation
of transition probabilities leads to the well known rotational formula
connecting those quantities with mean values of the associated multipole
moments \cite{Mang1975325,Islam1979}. Unfortunately, the assumption of
large deformation made to derive the rotational formula is often overlooked
and the formula is improperly applied to spherical or near spherical nuclei.
In Refs \cite{Robledo2012,Robledo2016} a detailed comparison between
transition probabilities computed with projected wave functions or computed
with the rotational formula for near spherical nuclei is made. The conclusion
is that the rotational formula can be wrong by a factor that can be as
large as $J(J+1)$ with $J$ being the multipolarity of the transition.

%% file: GCM.tex
\section{Beyond the mean field: large amplitude collective motion}
\label{Sec:LAM}

\def\bq{\mathbf{q}}
\def\bQ{\mathbf{Q}}

The mean field approximation is tailored to yield the ground state of
the nucleus although it is also possible to describe rotational bands in 
the cranking model or single particle excitations in the form of multi-quasiparticle
excitations. Collective excitations can be handled with the QRPA if they
are of small amplitude and for the more general case approximations like the IBM mapping procedure is available
as shown in previous sections.
However, for a general description of large amplitude collective motion
a more general theory is required: the generator coordinate method (GCM).
This framework has been widely used with Gogny interactions within an exact implementation (see 
Sec.~\ref{SCCM_section}) or by assuming approximations to deduce collective Pfaffians like the collective 
Schr\"odinger equation (CSE) (see Sec.~\ref{Sec:CSE}) and the five-dimensional collective Hamiltonian (5DCH) 
(see Sec.~\ref{Sec:MF5D}).
 
The first step in the application of the GCM is the selection of a collective
manifold of many-body states $|\Phi (\bq)\rangle$ where the symbol $\bq$
stands for a set of collective coordinates of any kind (shape parameters,
pairing correlations, etc). These states are normally either HFB or symmetry-restored HFB states.
Correlations are introduced by considering
general linear combinations of those "generating states"
\begin{equation}
    |\Psi_\sigma\rangle = \int f_\sigma(\bq) |\Phi (\bq)\rangle d\bq
\end{equation} 
As the $|\Phi (\bq) \rangle$ are, in general, non-orthogonal states the
$f_\sigma$ amplitudes cannot be interpreted as probability distributions.
In addition, in order to represent the most general collective states,
the collective amplitudes are not restricted to continuous functions and
in general they have to be treated as distributions. However, it is 
customary to use a discrete version of the GCM where the continuous
variables are replaced by discrete ones, the collective amplitudes 
become plain numbers and the integrals become sums (see Sec.~\ref{SCCM_section}). 

The energy of each of the correlated states is simply given by the 
double integral
\begin{equation}
   E_\sigma = \frac{\int\int f_\sigma(\bq)^* f_\sigma (\bq ') \mathcal{H}(\bq,\bq')d\bq d\bq '}
                   {\int\int  f_\sigma(\bq)^* f_\sigma (\bq ') \mathcal{N}(\bq,\bq')d\bq d\bq '}
\end{equation}
involving the norm and Hamiltonian overlaps $\mathcal{N}(\bq,\bq')$ and 
$\mathcal{H}(\bq,\bq')$ given by
\begin{equation}
\mathcal{N}(\bq,\bq')= \langle \Phi (\bq)|        \Phi(\bq ')\rangle
\end{equation}
and
\begin{equation}
\mathcal{H}(\bq,\bq')= \langle \Phi (\bq)|\hat{H}|\Phi(\bq ')\rangle
\end{equation}
To simplify the evaluation of $\mathcal{H}(\bq,\bq')$ one introduces the
ratio 
\begin{equation}\label{Eq:hs}
h(\bq,\bq') = \frac{\mathcal{H}(\bq,\bq')}{\mathcal{N}(\bq,\bq')}
\end{equation}
that can be easily computed using the generalized Wick's theorem (GWT) 
discussed in Sec. \ref{Sec:SROVER} and in Appendix \ref{App:C}. In the evaluation of
the Hamiltonian kernel for density dependent forces  the same
precautions regarding the definition of the density dependent term as the
ones discussed in Sec. \ref{Sec:SRDIFF1} have to be taken into account. Just mention
that the overlap density is the quantity to be used in the density dependent
part of the interaction.

The collective amplitudes $f_\sigma (\bq)$ are determined through the
Ritz variational principle that leads to the so-called Hill-Wheeler-Griffin (HWG) integral equation~\cite{Hill1953,Griffin1957,Ring1980}:
\begin{equation}
\int [ \mathcal{H} (\bq,\bq') - E_\sigma \mathcal{N} (\bq,\bq') ] f_\sigma (\bq')d\bq'=0
\end{equation}
For discrete collective variables the above equation becomes a generalized eigenvalue
problem with a positive definite norm overlap matrix $\mathcal{N}_{ij}$. Hence, $\sigma$ labels the different 
energies and states that can be obtained from solving the HWG equation.
The procedure to reduce it to standard form is to introduce the square
root of the norm (by means of the Cholesky decomposition) $\mathcal{N}^{1/2} (\bq,\bq')$
and the collective amplitude
\begin{equation}
  G_\sigma (\bq) = \int  \mathcal{N}^{1/2} (\bq,\bq') f_\sigma (\bq') d \bq'
\end{equation}
to reduce the HWG equation to a standard eigenvalue problem
\begin{equation}
\int \tilde{\mathcal{H}} (\bq,\bq') G_\sigma (\bq')d\bq'  = E_\sigma  G_\sigma (\bq)
\end{equation}
In practical applications, this reduction is performed through the 
definition of the so-called natural basis. Thus, the basis of 
eigenvectors of the norm $u_\Lambda (\bq)$ and the corresponding 
eigenvalues $n_\Lambda$ satisfying
\begin{equation}
\int \mathcal{N} (\bq,\bq') u_\Lambda (\bq')d\bq' = n_\Lambda  u_\Lambda (\bq)
\end{equation}
are used to define a new set of many-body states
\begin{equation}
|\Lambda\rangle=\int\frac{u_\Lambda (\bq)}{\sqrt{n_\Lambda}}|\Phi (\bq)\rangle d\bq
\end{equation}
These states are orthonormal by construction but only states with 
$n_\Lambda>0$ are well-defined. In fact, the zeros of the eigenvalues 
of the norm overlap matrix reflect the linear dependencies of the 
original set of states, $|\Phi(\bq)\rangle$, and the condition 
$n_\Lambda>0$ is a very effective way of removing such linear 
dependencies. Therefore, the GCM ansatz can be written now as:
\begin{equation}
|\Psi_\sigma\rangle=\sum_{\Lambda}g_{\sigma}(\Lambda)|\Lambda\rangle
\end{equation}
and the HWG equation as:
\begin{equation}
\sum_{\Lambda'}\langle\Lambda|\hat{H}|\Lambda'\rangle g_{\sigma}(\Lambda')=E_{\sigma}g_{\sigma}(\Lambda)
\end{equation}
The latter equation can be solved with standard diagonalization 
techniques and the spectrum is directly given by $E_{\sigma}$. 
Furthermore, expectation values and transition probabilities are 
computed as:
\begin{equation}
\langle\Psi_{\sigma_{1}}|\hat{O}|\Psi_{\sigma_{2}}\rangle=
\int\int f^{*}_{\sigma_{1}}(\bq_{1})\mathcal{O}(\bq_{1},\bq_{2})f_{\sigma_{2}}(\bq_{2})d\bq_{1}d\bq_{2}
\end{equation}
with $\mathcal{O}(\bq_{1},\bq_{2})$ the overlap of the operator 
$\hat{O}$, which is not necessarily a scalar operator (e.g., 
electromagnetic transitions, electroweak decays, etc.).

As it stands, the GCM method is rather simple to implement, apart from 
the evaluation of the norm and Hamiltonian overlap. The only difficulty 
with the method is the choice of collective variables. Obvious choices 
are shape deformation parameters like the multipole moments 
$Q_{\lambda\mu}$, pairing degrees of freedom (associated to particle number
fluctuations $\langle \Delta N^{2}\rangle$ or even discrete sets of 
multi-quasiparticle excitations. However, it is not easy beforehand to
know which are the relevant degrees of freedom
and unfortunately the computational cost grows exponentially with the 
number of collective variables. In addition, there is the issue of 
symmetry restoration: it can be treated as a subclass of the GCM method 
where the manifold of HFB states is generated from a given one by 
applying the symmetry operators. Fortunately, as the algebra of the 
symmetry operators is composed of one body operators, the "rotated" HFB 
states are again HFB states and the whole machinery used to compute 
overlaps can be used verbatim. Therefore, the GCM plus symmetry 
restoration, i.e., the so-called symmetry conserving configuration 
mixing method (SCCM) is formally the same as the traditional GCM but 
adding the parameters of the symmetry groups to the set of collective 
coordinates. The SCCM method, its performance and several examples of 
applications are presented in Sec.~\ref{SCCM_section}. Just mention 
that the pure GCM has been mostly used in the context of octupole 
deformation, with global calculations for even-even nuclei of the 
$3^{-}$ excitation energies, $B(E1)$ and $B(E1)$ transition strengths 
using the three most popular parametrizations of 
Gogny~\cite{Robledo2011} and the axial octupole moment as generating 
coordinate. As the GCM with the octupole moment also restores the parity
symmetry this and other \cite{Robledo2015} associated results will be 
discussed in Sec.~\ref{SCCM_section}.

Here we just mention that as angular momentum symmetry was not restored, a systematic 
deviation of the predicted $B(E3)$ as compared to the experimental data 
was observed for near spherical nuclei. This was understood as a 
consequence of the deficiencies of the rotational formula used to 
relate transition strengths and intrinsic multipole moments  that can 
only be fixed by considering angular momentum projected wave 
functions~\cite{Robledo2012,Robledo2016}. The coupling of the 
quadrupole and octupole degree of freedom has also been considered in  
two dimensional GCM calculations in several regions of the periodic 
table~ \cite{Rodriguez-Guzman2012,Robledo2013,Robledo2014} (see Sec.~\ref{SCCM_section}). 
In an  attempt to understand the physics and emergence of alternating parity 
rotational bands, GCM calculations with the octupole moment as 
generating coordinate with cranking wave functions was carried out in 
Refs.~\cite{Gar97,Gar98}. Finally, the GCM with zero and two quasiparticle 
configurations has been applied in Ref.~\cite{Bernard2011} to the 
description of non-adiabatic fission.

Another form of the GCM is just to consider linear combinations of 
multi particle-hole excitations built from a given Slater determinant.
This method is known in the literature as Configuration Interaction 
method (see \cite{Pople1987} for a discussion of the method in Quantum
Chemistry). The application of this method to the nuclear physics
case requires the consideration of the density dependent term present
in the Gogny force with its associated rearrangement terms. This 
was formulated in Refs. \cite{Pillet2006,Pillet2008}. One of the
advantages of the method is that particle number is conserved while 
pairing correlations are accounted for by the mixing of different
mp-mh excitations. Applications of the method to the spectroscopic 
description of nuclei and using the Gogny D1S force
was presented in \cite{Pillet2011,Pillet2012,LeBloas2014}. The method
has been further extended as to determine the orbitals used to construct 
the underlying Slater determinant self-consistently \cite{Robin2016,Pillet2017a}.
So far the method has been restricted to light nuclei. 

The Gogny force has also been used in beyond mean field calculations
using Slater determinant built from triaxially deformed Gaussian wave
packets (the deformed basis Antisymmetrized Molecular Dynamics method).
The parameters of the Gaussian wave packet are optimized as to minimize
the parity projected energy and subsequently projected to good angular
momentum. The laboratory frame wave functions are then combined in a
GCM like study with the deformation parameters $\beta$ and $\gamma$ used
as a generating coordinates. Using this framework, shape coexistence
in $^{43}$S and the connection with the loss of magicity at N=28 has 
been studied in \cite{PhysRevC.87.011301}. Triaxial superdeformed structures
in $^{40}$Ca have been also considered in \cite{PhysRevC.82.011302} as
well as the clustering properties of $^{20}$Ne in \cite{PhysRevC.69.044319}.

% ----------------------------------------------------------------------
%                                              C o l l e c t i v e   S E
% ----------------------------------------------------------------------

\subsection{Approximate solutions: The collective Schr\"odinger 
equation and collective inertias.}
\label{Sec:CSE}

There is an approximation to the GCM that avoids the evaluation of all
the Hamiltonian overlaps by assuming that the norm overlaps behaves approximately
as a Gaussian
\begin{equation}
	\mathcal{N} (\bq,\bq') = \exp [-\Gamma(\bq,\bq') \cdot (\bq-\bq')^{2}]
\end{equation}
with a width $\Gamma(\bq,\bq') $ that is a smooth tensor function of 
the collective variables with components $\Gamma_{ij}$ running over all 
the collective degrees of freedom. The width tensor is often taken as a 
constant all over the range of the collective variables. This 
approximation is supplemented by the assumption that the ratio 
$h(\bq,\bq')= \mathcal{H} (\bq,\bq')/\mathcal{N} (\bq,\bq')$ of 
overlaps is again a smooth function of the collective coordinates and 
therefore can be expanded around the mid point $\bQ = 
\frac{1}{2}(\bq+\bq')$ up to second order with respect to non-locality 
$(\bq-\bq')$. Using this two approximations it is possible to reduce 
the integral equation form of the HWG equation to a differential 
equation (see \cite{Ring1980} for detail) that is referred to as the 
Collective Schr\"odinger Equation (CSE). In order to simplify the 
notation we will from now on describe the situation corresponding to a 
single collective degree of freedom, that will be denoted as $q$.
\begin{equation} \label{Eq:CSE}
	\left[ -\frac{\hbar^{2}}{2\sqrt{\gamma(q)}} \frac{\partial}{\partial q} \sqrt{\gamma(q)} \frac{1}{M(q)}\frac{\partial}{\partial q}
	+ V(q) - \epsilon_{0} (q)\right]g_{\sigma} (q) = \epsilon_{\sigma} g_{\sigma} (q)
\end{equation}
that determine the collective amplitudes $g_{\sigma}(q)$ as well as the 
collective energies $\epsilon_{\sigma}$. In the above equation $\gamma(q)$ stands
for $\Gamma (q,q)$ and the quantity $M(q)$ in the 
collective kinetic energy term is the collective mass  defined in terms 
of derivatives of the Hamiltonian overlap of Eq \ref{Eq:hs} as 
\begin{equation}\label{Eq:CSEM}
M(q) = 	\frac{-1}{\gamma^{2}(q)}\left[ \frac{\partial^{2}}{\partial q^{2}} h(q,q') - 
\frac{\partial^{2}}{\partial q'^{2}} h(q,q') \right]
\end{equation}
The potential energy $V(q)$ is nothing but the mean field energy for the
members of the collective manifold $|\Phi (q)\rangle$ and $\epsilon_{0}(q)$
is the zero point energy correction (see \cite{Ring1980} for details).
In order to compute mean values of observables it
is also required to approximate the ratios
$o(q,q')= \mathcal{O} (q,q')/\mathcal{N} (q,q')$ with respect to the
non-locality parameter $q-q'$. For most of the observables it is enough
to restrict to zero order and approximate $o(q,q')$ by $o(Q)$ resulting
in the general expression for overlaps
\begin{equation}
	\langle \Psi_{\sigma} | \hat{O} | \Psi_{\sigma'} \rangle = 
	\int dq \sqrt{\gamma}\, g_{\sigma}^{*}(q) o(q) g_{\sigma'}(q)
\end{equation} 

When the method is applied to the $\beta$ and $\gamma$ quadrupole 
deformation parameters and the Euler angle variables for rotations are 
added phenomenologically, one ends up with the 5D Bohr Hamiltonian 
discussed in Sec \ref{Sec:MF5D}. For other degrees of freedom, the CSE 
has been used together with the Gogny force mostly to describe octupole 
properties from the early studies of \cite{egi92c} to the most 
systematic ones of \cite{Rob12l}. 

The collective kinetic energy of Eq \ref{Eq:CSE} depends upon the so 
called GCM collective inertia $M(q)$ given by Eq \ref{Eq:CSEM}. 
Collective inertias are also defined in the Adiabatic Time dependent 
Hartree-Fock-Bogoliubov (ATDHFB) theory of collective motion. The 
expression differ from the one of the GCM and it is not clear which one 
of the two should be used in the CSE. The situation is similar to the 
differences between the Yoccoz (Y)~\cite{Peierls1957,Yoccoz1957} 
and the Thouless-Valatin (TV) \cite{Thouless1960,Thouless1962} 
moments of inertia (see \cite{Ring1980} and the next Section): the Y 
moment of inertia is the equivalent of the GCM inertia (it is derived 
in a pure quantum-mechanic fashion) whereas the TV one corresponds to 
the ATDHFB inertia. There are arguments that favor the TV versus the Y 
moment of inertia: TV comes from a VAP approach to the problem where 
the intrinsic states depend on the quantum numbers of the laboratory 
wave function whereas Y corresponds to the PAV case where the intrinsic 
states is given and there is no feedback between the laboratory frame 
quantities. In addition, in the case of translational invariance, where 
the equivalent of the moment of inertia is the nuclear mass it is only 
the VAP theory the one that provides the correct mass. These are strong 
arguments that favor the use of TV like inertias in the CSE but so far 
there is no founded justification in this case. The collective inertias 
are also used in fission in the determination of the spontaneous 
fission half-lives using the Wenzel Kramers Brillouin (WKB) approach to 
the tunneling through the fission barrier. The evaluation of both the 
collective inertias require the inversion of the linear response matrix 
of the HFB theory which is a matrix whose dimension is the number of 
two-quasiparticle excitations. As this is an enormous number for the 
configuration spaces typically used with phenomenological effective 
interactions like Gogny, approximations are used to reduce the 
computational cost. The most usual approximation is the use of the 
diagonal matrix elements of the linear response matrix in all the 
instances where this huge matrix appears. This is the so called 
perturbative inertia approximation and also yields to the 
Inglis-Belyaev \cite{Inglis1956,BELIAEV1961322} approximate form of 
the TV moment of inertia \cite{Thouless1960,Thouless1962}. Attempts 
to improve this approximation relay on the numerical evaluation of the 
collective momentum operator plus the diagonal approximation for the 
linear response matrix: this is the "non-perturbative" approach to the 
collective inertias. 

% ----------------------------------------------------------------------
%                                                    5D Collective hamiltonian
% ----------------------------------------------------------------------

\subsection{The five-dimensional collective Hamiltonian Equation}
\label{Sec:MF5D}

The most widely used beyond-mean-field approximation with Gogny 
interactions based on an approximation to the GCM method is the 
five-dimensional collective Hamiltonian (5DCH)~\cite{bohr1975,Kumar1967,Peru2014}. 
This method has been implemented in a similar fashion and with the same applicability 
as explained below with Skyrme~\cite{Deloncle1989} and RMF~\cite{Niksic2009} density functionals.
As mentioned in the 
previous section, a collective Hamiltonian can be extracted from the more 
general HWG equation if a Hamiltonian overlap approximation is assumed for 
the norm overlaps and the Hamiltonian overlaps behave smoothly with the 
collective coordinates. In the 5DCH, the collective coordinates are the 
quadrupole deformations $\bq=(\beta,\gamma)$ (see 
Eq.~\ref{beta_gamma_definition}) and the collective Bohr Hamiltonian is 
given by~\cite{Peru2014}:
\begin{equation}
\mathcal{H}_{\mathrm{coll}}=-\frac{\hbar^{2}}{2}\sum_{m,n=0,2}D^{-\frac{1}{2}}\frac{\partial}{\partial a_{m}}D^{\frac{1}{2}}B_{mn}^{-1}\frac{\partial}{\partial a_{n}}+\frac{\hbar^{2}}{2}\sum_{k=1}^{3}\frac{\hat{J}^{2}_{k}}{\mathcal{J}_{k}}+\mathcal{V}(\bq)-\mathrm{ZPE}(\bq)\,\,
\label{5DCH_ham}
\end{equation}
The first two terms of this Hamiltonian describe the kinetic energy 
associated to quadrupole vibrations (with $B_{mn}$ being the collective 
quadrupole vibrational inertia) and rotations (with $\mathcal{J}_{k}$ 
being the moments of inertia), respectively. The potential energy is 
given by 
$\mathcal{V}(\bq)=\langle\Phi(\beta,\gamma)|\hat{H}|\Phi(\beta,\gamma)\rangle$, 
where $| \Phi(\beta,\gamma)\rangle$ are obtained by solving 
constrained-HFB calculations with Gogny interactions. The zero-point 
energy correction, $\mathrm{ZPE}(\bq)$, takes into account the 
fluctuations in the quadrupole coordinates. Furthermore, the parameters 
$a_{0}=\beta\cos\gamma$ and $a_{2}=\beta\sin\gamma$ are convenient 
redefinitions of the quadrupole deformations and $\hat{J}_{k}$ is the 
$k$-component of the angular momentum operator. Finally, the metric is 
given by
\begin{equation} 
D=(B_{00}B_{22}-B^{2}_{02})\prod_{k}\mathcal{J}_{k}
\end{equation} 
In Eq.~\ref{5DCH_ham}, there are three rotational inertia and three 
quadrupole mass parameters. They are all computed from the local 
properties of mean-field solutions at the 
($\beta$,$\gamma$)-grid~\cite{Libert1999}. Hence, the vibrational 
inertia are computed as~\cite{Peru2014}:
\begin{equation}
B_{\mu\nu}(\bq)=\frac{\hbar^{2}}{2}\left[\mathcal{M}_{-1,\mu\nu}(\bq)\right]^{-1}\mathcal{M}_{-3,\mu\nu}(\bq)
\left[\mathcal{M}_{-1,\mu\nu}(\bq)\right]^{-1}
\end{equation}
where the moments are obtained by (cranking 
formula)~\cite{Girod1979,giannoni1980,giannoni1980-a,Libert1999}:
\begin{equation}
\mathcal{M}_{-n,\mu\nu}(\bq)=\sum_{ij}\frac{|\langle\Phi(\bq)|\beta_{j}\beta_{i}\hat{Q}_{2\mu}|\Phi(\bq)\rangle
\langle\Phi(\bq)|\beta_{j}\beta_{i}\hat{Q}_{2\nu}|\Phi(\bq)\rangle|}{(E_{i}+E_{j})^{n}}
\label{5DCH_moments}
\end{equation}
Here, $\beta^{\dagger}_{i}$ and $E_{i}$ is the quasiparticle creator operator and the quasiparticle energy, respectively.

Concerning the moments of inertia along the three axis ($k=x,y,z$), two choices have been proposed, namely, the Inglis-Belyaev and cranking formulae:
\begin{eqnarray}
\mathcal{J}^{IB}_{k}(\bq)&=&\hbar^{2}\sum_{ij}\frac{|\langle\Phi(\bq)|\beta_{j}\beta_{i}\hat{J}_{k}|\Phi(\bq)\rangle|^{2}
}{E_{i}+E_{j}}\\
\mathcal{J}^{I}_{k}(\bq)&=&\frac{\langle\Phi^{I}_{\omega_{k}}(\bq)|\hat{J}_{k}|\Phi^{I}_{\omega_{k}}(\bq)\rangle}{\omega_{k}}
\end{eqnarray} 
The latter is computed with HFB wave functions obtained with the cranking method (see Eq.~\ref{cranking_constr} in Sec.~\ref{Sec:MF}) 
and takes into account rearrangement to rotations for each value of the cranking angular momentum, $\langle\Phi^{I}_{\omega_{z}}|\hat{J}_{z}|
\Phi^{I}_{\omega_{z}}\rangle=\sqrt{I(I+1)}$. Moreover, in the limit $\omega_{k} \rightarrow 0$, this moment of inertia is equivalent to the 
Thouless-Valatin inertia. In practice, one takes a small value (for example, $\omega$ = 0.002 MeV) to
approximate the limit \cite{Delaroche2010}.

The zero-point quantum energy corrections, $\mathrm{ZPE}(\bq)$ are 
associated with the rotational and vibrational motions, i.e., 
$\mathrm{ZPE}(\bq)=\mathrm{ZPE}_{\mathrm{rot}}(\bq) + 
\mathrm{ZPE}_{\mathrm{vib}}(\bq)$ and both are computed as a 
combination of the moments given in 
Eq.~\ref{5DCH_moments}~\cite{Libert1999}. In practical implementations 
the ZPE corrections only contain the part arising from the kinetic 
energy operator while the part arising from the potential is neglected. 
This approximation is valid in typical situations with shallow minima 
in the corresponding potential energy surfaces. However, such a piece 
might become significant close to magic numbers where the curvature of 
the potential energy surface tends to be higher 
\cite{Libert1999,Delaroche2010}. Moreover, for nuclei near the magic 
ones, the correlation energy $E_{corr} = E_{HFB,min} - E_{5DCH}$ may 
even come out negative, which is unphysical. In particular, the 
rotational ZPE correction provided by the standard GOA is known to lead 
to difficulties for configurations close to the spherical ones as, in 
such a case, it does not scan the rotational degrees of freedom 
properly. A more realistic approximation for those configurations, 
though still far from quantitative when compared to the exact 
restoration of the rotational symmetry, is offered by the Topologically 
Invariant GOA (TopGOA) 
\cite{Reinhard1978,Reinhard1987,Rodriguez-Guzman2000b}.

The eigenstates $|J M \rangle$ and energies $E(J)$ are obtained by 
solving the equation
\begin{equation}
{\cal{H}}_{coll} |J M \rangle = E(J) |J M \rangle
\end{equation}
where the orthonormal $|J M \rangle$ states are expanded as
\begin{equation}
|J M \rangle = \sum_{K} g_{K}^{J}(\beta,\gamma) |J M K\rangle
\end{equation}
Here, $|J M K\rangle$ represents a linear combination of Wigner functions \cite{Libert1999}. The 
probability of a given $K$ value in the wave function reads 
\begin{equation}
P_{K} = \int da_{0}da_{2} |g_{K}^{J}(\beta,\gamma)|^{2}
\end{equation}
The 5DCH formalism has already been applied to describe low-lying 
energy spectra and shed light on a wide variety of nuclear phenomena, 
in particular, in the study of rotational bands, shell closures, and 
shape evolution, shape mixing and shape 
coexistence~\cite{Girod1988,Delaroche1989,Delaroche1994,Libert1999,Peru2000,Obertelli2005,Bertsch2007,Clement2007,Ljungvall2008,Girod2009,Gaudefroy2009,Obertelli2009,Zielinska2009,Ljungvall2010,Obertelli2011,Corsi2013,Louchart2013,BelloGarrote2015,Peru2014,Libert2016,Clement2016,Flavigny2017,Paul2017}. 
In addition, the 5DCH model has been used in the fitting protocol of 
the parametrization D1M of the Gogny interaction \cite{goriely2009}, as 
it was mentioned in Sec.~\ref{Sec:Gogny_force}. A thorough discussion 
of the performance of the method with several examples of its 
applications is found in the review of P\'eru and Martini and we refer 
the reader to Ref.~\cite{Peru2014} for further details.
%

% -----------------------------------------------------------------------------------------------------------------
%                                                                                            SCCM
% -----------------------------------------------------------------------------------------------------------------
%%%%%%%%%%%%%%%%%%%%%%
\subsection{SCCM methods and applications with Gogny EDF}\label{SCCM_section}
%%%%%%%%%%%%%%%%%%%%%%
The most advanced method currently used to solve the nuclear 
many-problem with Gogny EDF is based on the combination of the GCM 
method with symmetry-restored HFB-like wave functions, both discussed 
in previous Sections. These are the so-called symmetry conserving 
configuration mixing (SCCM) approximations and are rooted in the 
variational principle. Similar implementations as those described below have been carried out 
with the other two most popular energy density functionals, namely, Skyrme 
(see Refs.~\cite{Bender2008,Bally2014} and references therein) and RMF 
(see Refs.~\cite{Niksic2006,Niksic2006a,Yao2009,Yao2010,Yao2015} and references therein).
A vast amount of applications of the SCCM method with these EDF has been published in the last twenty years. 
We will only report here in more detail the implementations performed with Gogny interactions since they are aim 
of the present review. We refer the reader to explore the references given above. 
 
The nuclear states are defined in the SCCM method through the 
realization of the GCM ansatz as: 
\begin{equation}
|\Psi^{J\pi}_{\sigma}\rangle=\sum_{\mathbf{q}}f^{J\pi}_{\sigma} (\mathbf{q})|\Phi^{J\pi} (\mathbf{q})\rangle
\label{GCM_ansatz} 
\end{equation}
where $\sigma=1,2,...$ labels the different quantum states for a given 
angular momentum and parity, $J^{\pi}$, and $|\Phi^{J\pi} 
(\mathbf{q})\rangle$ are the projected intrinsic states
\begin{equation}
|\Phi^{J\pi} (\mathbf{q})\rangle=\hat{P}^{J}\hat{P}^{\pi}\hat{P}^{N}\hat{P}^{Z}|\mathbf{q}\rangle
\label{PPNAMP_state}
\end{equation}
Here, $\hat{P}^{J}$ is a shortening of the angular momentum projector 
and  $\hat{P}^{\pi}$, $\hat{P}^{N}$ and $\hat{P}^{Z}$ are again the 
projectors onto good parity, neutron number and proton number 
respectively. Furthermore, the intrinsic states, $|\mathbf{q}\rangle$, 
are obtained by solving HFB or PN-VAP equations, imposing the 
constraints on the corresponding collective coordinates 
$\mathbf{q}=\{q_i,i=1,\ldots,N_c\}$. Since in practical applications 
the number of projected intrinsic states entering Eq.~\ref{GCM_ansatz} 
is finite, we have discretized the collective variables and substituted 
the integrals by sums in the general GCM expressions.

As mentioned above, the coefficients of the linear combination given in 
Eq.~\ref{GCM_ansatz} are found by solving the HWG equations, now one for each value of the angular 
momentum and parity
\begin{equation}
\sum_{\mathbf{q}'} \left(\mathcal{H}^{J\pi} (\mathbf{q},\mathbf{q} ')-E^{J\pi}_{\sigma}\mathcal{N}^{J\pi} (\mathbf{q}, \mathbf{q}')\right) f^{J\pi}_{\sigma} (\mathbf{q}')=0\label{HWG_1}
\end{equation}
with the norm $\mathcal{N}^{J\pi} (\mathbf{q}, 
\mathbf{q}')=\langle\Phi^{J\pi} (\mathbf{q})|\Phi^{J\pi} 
(\mathbf{q}')\rangle$ and Hamiltonian $\mathcal{H}^{J\pi} 
(\mathbf{q},\mathbf{q} ')=\langle\Phi^{J\pi} 
(\mathbf{q})|\hat{H}|\Phi^{J\pi} (\mathbf{q} ')\rangle$ overlaps. These 
are the generalization of Eqs.~\ref{AMP_overlaps3} to the non-diagonal 
case.

To solve the HWG generalized eigenvalue problem, the natural basis approach is usually adopted.
Thus, the norm overlap matrix is diagonalized first:
\begin{equation}
\sum_{\mathbf{q}'}\mathcal{N}^{J\pi} (\mathbf{q}, \mathbf{q}')u^{J\pi}_{\Lambda}(\mathbf{q}')=n^{J\pi}_{\Lambda}u^{J\pi}_{\Lambda}(\mathbf{q})
\label{norm_overlap_ev}
\end{equation}
Then, we use the eigenvalues and eigenvectors ($n^{J\pi}_{\Lambda}$ and 
$u^{J\pi}_{\Lambda}(\mathbf{q})$) to define the states of the natural basis as:
\begin{equation}
|\Lambda^{J\pi}\rangle=\sum_{\mathbf{q}}\frac{u^{J\pi}_{\Lambda}(\mathbf{q})}{\sqrt{n^{J\pi}_{\Lambda}}}|\Phi^{J\pi} (\mathbf{q})\rangle
\label{natural_basis}
\end{equation}
To ensure that the linear dependencies of the original set 
of states have been removed only the eigenvalues 
$n^{J\pi}_{\Lambda}\neq0$ are chosen. In numerical applications, such a condition 
is substituted by $n^{J\pi}_{\Lambda}>\varepsilon$, being $\varepsilon$ 
a threshold value. Therefore, we can express the nuclear states in this 
basis as:
\begin{equation}
|\Psi^{J\pi}_{\sigma}\rangle=\sum_{\Lambda}g^{J\pi}_{\sigma}(\Lambda)|\Lambda^{J\pi}\rangle
\label{GCM_ansatz2} 
\end{equation}
and the HWG equations (Eq.~\ref{HWG_1}) as:
\begin{equation}
\sum_{\Lambda'} \langle\Lambda^{J\pi}|\hat{H}|\Lambda'^{J\pi}\rangle g^{J\pi}_{\sigma}(\Lambda')=E^{J\pi}_{\sigma}g^{J\pi}_{\sigma}(\Lambda)
\label{HWG_2}
\end{equation}
The solution of the latter equations give us the spectrum, 
$E^{J\pi}_{\sigma}$, and the coefficients $f^{J\pi}_{\sigma} 
(\mathbf{q})$ that can be used to compute expectation values and/or 
transition probabilities and moments, choosing the proper operator 
$\hat{O}$:
\begin{eqnarray}
\langle\Psi^{J_{1}\pi_{1}}_{\sigma_{1}}|\hat{O}|\Psi^{J_{2}\pi_{2}}_{\sigma_{2}}\rangle=&\nonumber\\
\sum_{\mathbf{q}_{1},\mathbf{q}_{2}}\left(f^{J_{1}\pi_{1}}_{\sigma_{1}} (\mathbf{q}_{1})\right)^{*}
\mathcal{O}^{J_{1}\pi_{1}, J_{2}\pi_{2}} (\mathbf{q}_{1}, \mathbf{q}_{2})\left(f^{J_{2}\pi_{2}}_{\sigma_{2}} (\mathbf{q}_{2})\right)&
\label{GCM_Expec_Val}
\end{eqnarray}
with $\mathcal{O}^{J_{1}\pi_{1}, J_{2}\pi_{2}} (\mathbf{q}_{1}, \mathbf{q}_{2})=\langle\Phi^{J_{1}\pi_{1}} (\mathbf{q}_{1})|\hat{O}|\Phi^{J_{2}\pi_{2}} (\mathbf{q}_{2})\rangle$. 

Finally,  the weights of the different collective coordinate in a given 
GCM wave function~\cite{Ring1980}
\begin{eqnarray}
G^{J\pi}_{\sigma} (\mathbf{q})&\equiv& \sum_{\Lambda} g^{J\pi}_{\sigma}(\Lambda)u^{J\pi}_{\Lambda}(\mathbf{q})\nonumber\\
&=&\sum_{\mathbf{q}'} \langle\Phi^{J\pi} (\mathbf{q})|\Phi^{J\pi} (\mathbf{q}\,')\rangle^{1/2}f^{J\pi}_{\sigma} (\mathbf{q}\,')\label{coll_wf}
\end{eqnarray}
are very useful quantities to analyze the character of the GCM states. 
The square of these quantities are the so-called collective wave 
functions.

As it is mentioned in previous sections, a prescription is required for 
the evaluation of Hamiltonian overlaps coming from the 
density-dependent term of the Gogny EDF. In every application shown in 
this section, the particle number projected spatial density (in those 
cases where PNP is performed) combined with the mixed prescription for 
the angular momentum projection, parity (if performed) and GCM parts is 
used. This is a generalization to non-diagonal kernels of 
Eq.~\ref{mixed_dd_ampPNP}.

It is important to note that the amount of correlations that the SCCM 
method can include in the nuclear states depends on three interrelated 
factors: a) the relevance and the number of the degrees of freedom 
explored by the GCM ansatz, $\mathbf{q}$; b) the method used to build 
the set of intrinsic wave functions (e.g., HFB or a more sophisticated 
version as PN-VAP); and, c) the number of broken and subsequently 
restored symmetries of the system. In addition, the quality of the 
approach is different depending on the specific nuclear state and also 
varies nucleus by nucleus. For example, the ground state of a 
well-deformed axial symmetric even-even nucleus can be accurately 
described by mixing PNAMP states with different values of the the axial 
quadrupole deformation. However, a more elaborated approximation would 
be required to describe, for instance, the first negative parity state 
of such a nucleus. 

We summarize in Table~\ref{Table1_GCM} the different implementations of 
the GCM with Gogny EDF depending on: the collective coordinates 
explored; the underlying method used to find the HFB-like intrinsic 
wave functions that are subsequently projected and mixed; the 
self-consistent symmetries imposed to such HFB-like intrinsic wave 
functions; and, finally, the symmetry restorations that are performed. 
In addition, we give the references where those calculations were 
reported for the first time with Gogny EDF.
%%%%%%%%%%%%%%%%%%%%%%%%%%%%%%%%%%%%%%%%%%%%%
\begin{table}
\caption{Different implementations of SCCM calculations with Gogny 
EDFs. The acronyms refer to: Hartree-Fock-Bogoliubov (HFB), particle 
number variation after projection (PN-VAP), parity (P), simplex (S), 
time-reversal (TR), axial (Ax), angular momentum projection (AMP), 
particle number and angular momentum projection (PNAMP) and parity, 
particle number and angular momentum projection (PPNAMP). Additionally, 
$q_{20}$, $q_{22}$, $q_{30}$ and $\Delta N^{2}$ are quadrupole, 
octupole and particle number fluctuations respectively.}
\begin{center}
\begin{indented}
\item[]
\begin{tabular}{c|c|c|c|c}\hline\hline
Intrinsic w.f. & $\mathbf{q}$ & Self-consistent symm. & Symm. restoration & Ref. \\\hline
HFB & $q_{20}$ & P, S, TR, Ax & AMP & \cite{Rodriguez-Guzman2000}\\
PN-VAP & $q_{20}$ & P, S, TR, Ax & PNAMP & \cite{Rodriguez2007}\\
PN-VAP & $q_{20}$, $\Delta N^{2}$ & P, S, TR, Ax & PNAMP & \cite{Vaquero2011}\\
PN-VAP & $q_{20}$, $q_{22}$ & P, S, TR & PNAMP & \cite{Rodriguez2010}\\
HFB & $q_{30}$ & S, TR, Ax & P & \cite{Robledo2011}\\
HFB & $q_{20}$, $q_{30}$ & S, TR, Ax & P & \cite{Rodriguez-Guzman2012}\\
PN-VAP & $q_{20}$, $q_{22}$, $J_{c}$ or $\omega$ & P, S & PNAMP & \cite{Borrajo2015}\\
HFB & $q_{20}$,$q_{30}$ & TR, S, Ax & PPNAMP & \cite{Bernard2016}\\\hline\hline
\end{tabular}
\end{indented}
\end{center}
\label{Table1_GCM}
\end{table}
%%%%%%%%%%%%%%%%%%%%%%%%%%%%%%%%%%%%%%%%%%%%%%

\begin{figure*}[t]
\begin{center}
\includegraphics[width=0.8\textwidth]{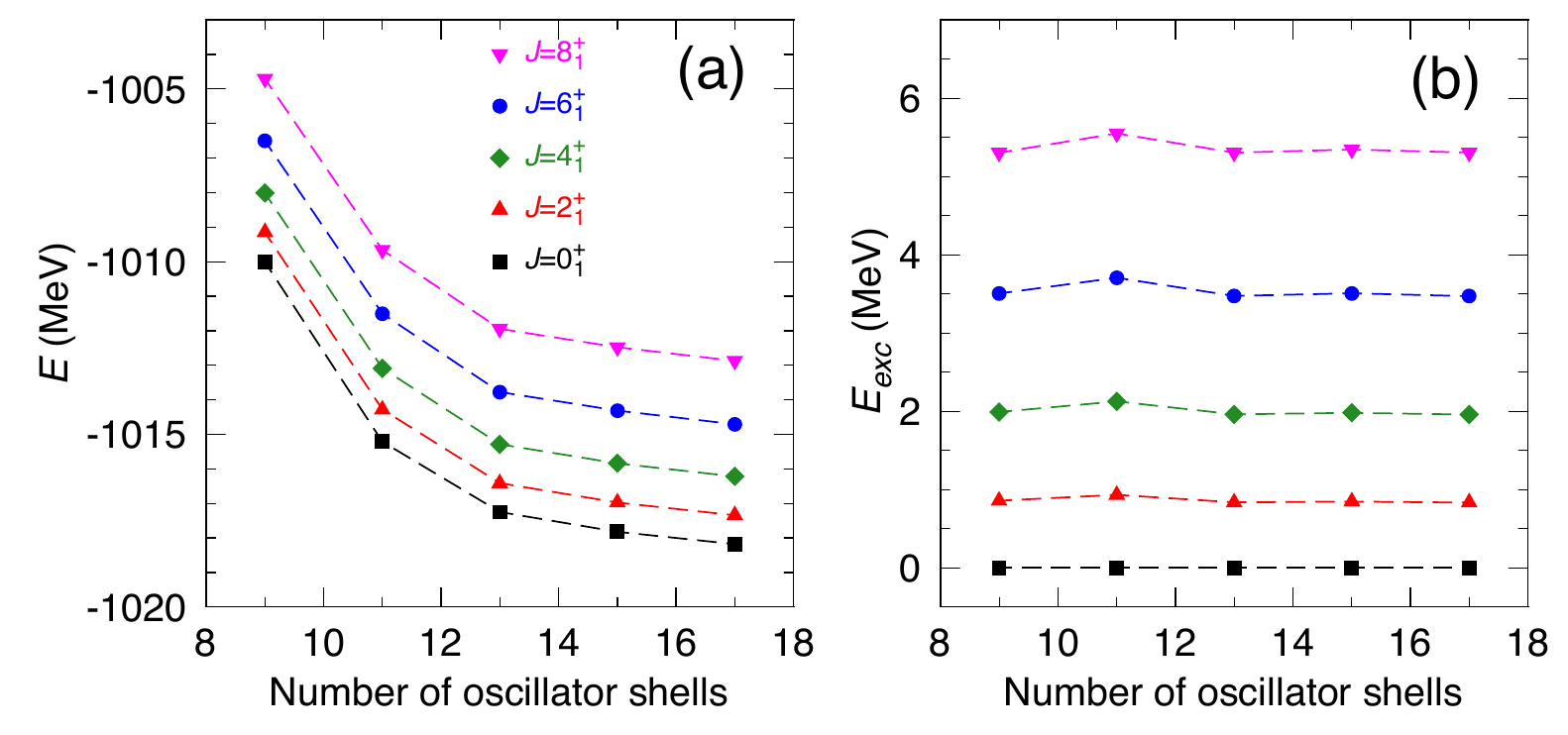}
\caption{(a) Absolute and (b) excitation energies of the yrast band of 
$^{120}$Cd calculated with a SCCM method -that includes PN-VAP+PNAMP 
wave functions along the axial quadrupole degree of freedom 
($\beta_{2}$), using Gogny D1S- as a function of the number of major 
spherical harmonic oscillator shells included in the working basis.}
\label{120Cd_convergence_GCM}
\end{center}
\end{figure*}

\begin{figure}
\begin{center}
\includegraphics[width=0.5\textwidth]{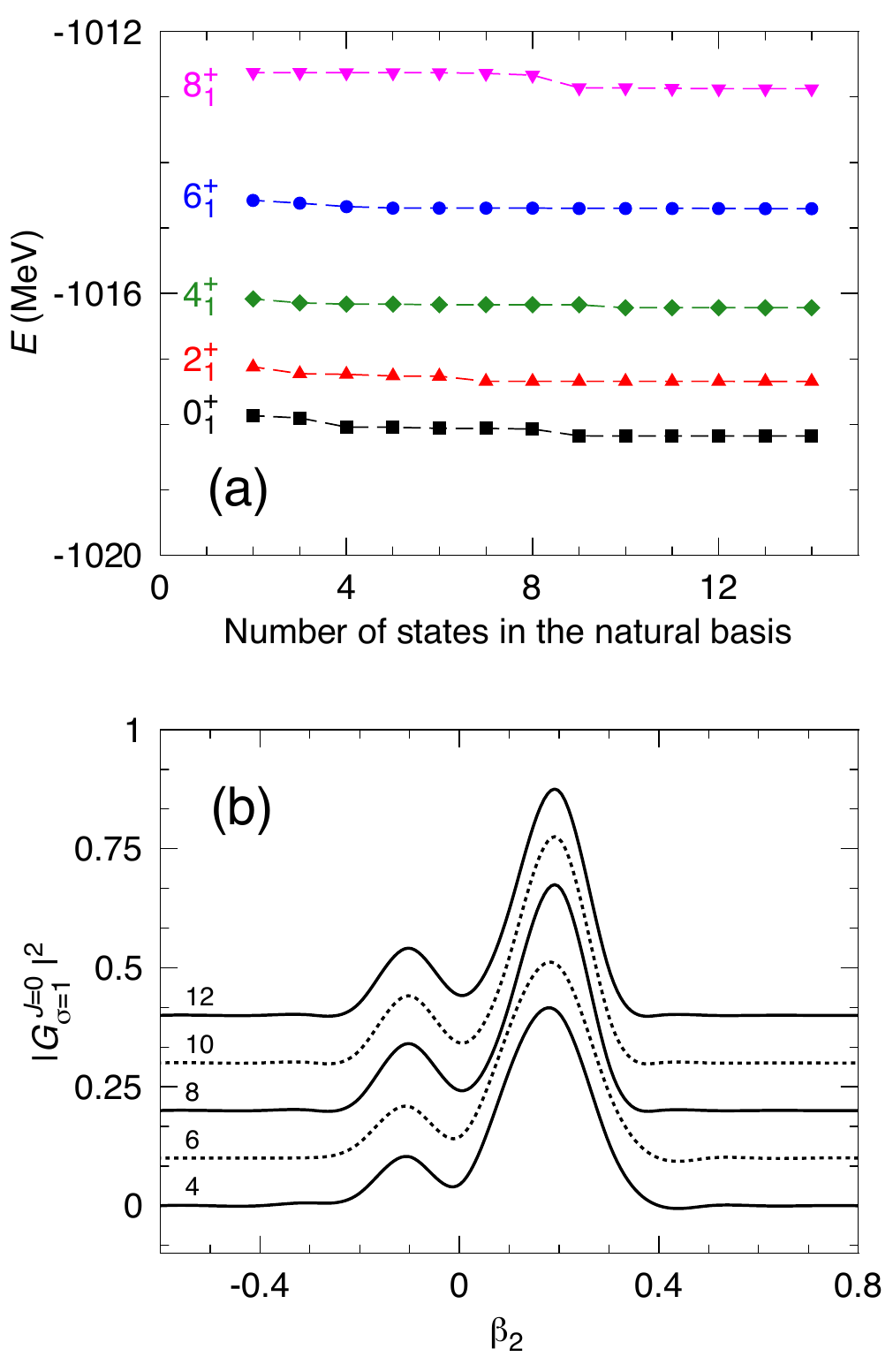}
\caption{(a) Absolute energies of the yrast band of $^{120}$Cd 
calculated with a SCCM method -that includes PN-VAP+PNAMP wave 
functions along the axial quadrupole degree of freedom ($\beta_{2}$), 
using Gogny D1S- as a function of the number states in the natural 
basis. (b) Collective wave function of the ground state, $0^{+}_{1}$, 
computed for different sizes of the natural basis (4, 6, 8, 10, 12). 
Curves are shifted for a better visualization.}
\label{120Cd_convergence_GCM_natural_basis}
\end{center}
\end{figure}

Before reviewing the many applications of the SCCM method performed 
until now, we briefly discuss some technical aspects, in particular, 
the convergence of the results. Such a convergence is a manifold 
problem. On the one hand, we have to assume first the amount of degrees 
of freedom (collective coordinates) explored, the self-consistent 
symmetries imposed in the intrinsic wave functions, and, consequently, 
the kind of symmetry restorations that must be carried out. Once the 
problem is defined in these terms, SCCM calculations for the low-lying 
states of the nucleus should have to converge with respect to: 
\begin{itemize}
\item The size of the intervals in which the collective coordinates are 
defined and the number of intrinsic states included in such intervals. 
These intervals are chosen in such a way that the energy difference between 
the boundaries and the minimum of the multidimensional energy surface 
is around 20 MeV. Furthermore, the collective wave functions should 
decay to zero in the boundaries. The number of points within these 
intervals should be sufficiently large to include all minima and 
relevant points in the surface. However, including too many points 
increases significantly the computational time and, in most of the 
cases, only introduces linear dependencies that must be subsequently 
removed out of the natural basis.    
\item The number of integration points chosen to perform the 
corresponding symmetry restorations (e.g., gauge and/or Euler angles). 
The suitability of this number is set to reproduce (up to 
$\sim10^{-4}$) the nominal expectation values, using projected wave 
functions, of the operators related to the symmetries (number of 
protons/neutrons ($\hat{N},\hat{Z}$), fluctuations of the number of 
protons/neutrons ($\hat{\Delta N^{2}},\hat{\Delta Z^{2}}$), angular 
momentum operators ($\hat{J}^{2}$, $\hat{J}_{z}$), etc.). These 
expectation values are also checked after performing the GCM calculations using 
the full SCCM states.
\item The number of major harmonic oscillator shells included in the 
working basis. This is a critical point because the computational time 
of the SCCM methods grows roughly exponentially with the number of 
oscillator shells (see, e.g., Table 1 of Ref.~\cite{Rodriguez2016}). 
Therefore, these aspects are not studied in detail except for axial 
calculations~\cite{Rodriguez2010,Rodriguez2015}. In 
Fig.~\ref{120Cd_convergence_GCM} we show the results for SCCM 
calculations for different sizes of the spherical harmonic oscillator 
working basis. This implementation includes both particle number and 
angular momentum restoration of PN-VAP parity and axially symmetric 
intrinsic wave functions for $^{120}$Cd. Only one generating 
coordinate, namely, the axial quadrupole degree of freedom is taken 
into account. In Fig.~\ref{120Cd_convergence_GCM}(a) the absolute 
energies for the yrast states, $J^{+}_{1}$, are plotted. These energies 
decrease whenever the number of states in the working basis is 
increased, but this energy gain is getting smaller and a convergence 
pattern can be easily identified. How to extrapolate these results to 
infinity is still under debate in the Gogny EDF 
context~\cite{Arzhanov2016}. Nevertheless, energy differences like 
particle separation energies, correlation energies, etc., converge much 
faster with respect to the size of the working 
basis~\cite{Rodriguez2015}, as we can see in 
Fig.~\ref{120Cd_convergence_GCM}(b) with the excitation energies of the 
yrast states.
\item The removal of the linear dependencies of the original set of 
wave functions, i.e., the size of the natural basis. As mentioned 
above, the HWG equation is transformed into a regular eigenvalue 
problem by defining a set of orthonormal states given by 
Eq.~\ref{natural_basis}. The number of states in this set depends on 
the choice of the smallest eigenvalue of the norm overlap matrix used 
in Eq.~\ref{natural_basis}. If this number is sufficiently large, the 
solution of the HWG should not vary. This is the so-called 
\textit{plateau} condition and its appearance is a signature of the 
convergence of the GCM method~\cite{Bonche1990}. Therefore, we can 
solve the HWG equations for different sizes of the natural basis and 
represent the results as a function of the size. This is represented in 
Fig.~\ref{120Cd_convergence_GCM_natural_basis}(a) where the same SCCM 
calculation with 17 major harmonic oscillator shells of 
Fig.~\ref{120Cd_convergence_GCM} is analyzed. In this example, the GCM 
is performed with 15 states in the original basis along the axial 
quadrupole deformation. We observe that the \textit{plateau} condition 
is nicely obtained for the yrast states. This procedure is normally 
complemented by examining the shape of the collective wave
functions obtained for different sizes of the natural basis. In 
Fig.~\ref{120Cd_convergence_GCM_natural_basis}(b) we plot the 
$J^{+}=0^{+}_{1}$ wave functions for several dimensions of the natural 
basis (4, 6, 8, 10, 12). Again, we observe that the collective wave 
functions are almost constant, proving a good convergence of the SCCM 
calculation in this respect.
\end{itemize}
%%%%%%%%%%%%%%%%%%%%%%%%%%%%%%%%%%%%%%%
%%%%%%%%%%%%%%%%%%%%%%%%%%%%%%%%%%%%%%  
%%%%%%%%%%%%%%%%%%%%%%%%%%%%%%%%%%%%%%%
%%%%%%%%%%%%%%%%%%%%%%%%%%%%%%%%%%%%%%  

We now review the multiple applications of the SCCM methods with Gogny 
EDFs that have been reported in recent years. Most of these studies 
have been focused on the calculation of bulk properties (masses, radii, 
etc.), excitation energies and electromagnetic properties (transition 
probabilities and moments) at low excitation energy. These quantities 
can be directly compared to experimental data and/or can give actual 
predictions for not-yet-measured nuclei. Moreover, apart from these 
observables, potential energy surfaces, Nilsson-like levels, occupation 
numbers and/or collective wave functions can be computed. These 
non-observable quantities provide a meaningful interpretation of the 
data in terms of the underlying shell structure, shapes, etc.. 
Most of these applications have their equivalent versions with Skyrme and RMF 
energy density functionals. In most of the cases, the three
EDF provide similar results and global trends. As mentioned several times throughout the
paper, we will restrict ourselves to review only the results obtained with Gogny interactions.
%%%%%%%%%%%%%%%%%%%%%%%%%%%%%%%%%%%%%% 
\subsubsection{Global and local studies of BMF correlation energies}
\begin{figure}
\begin{center}
\includegraphics[width=0.5\textwidth]{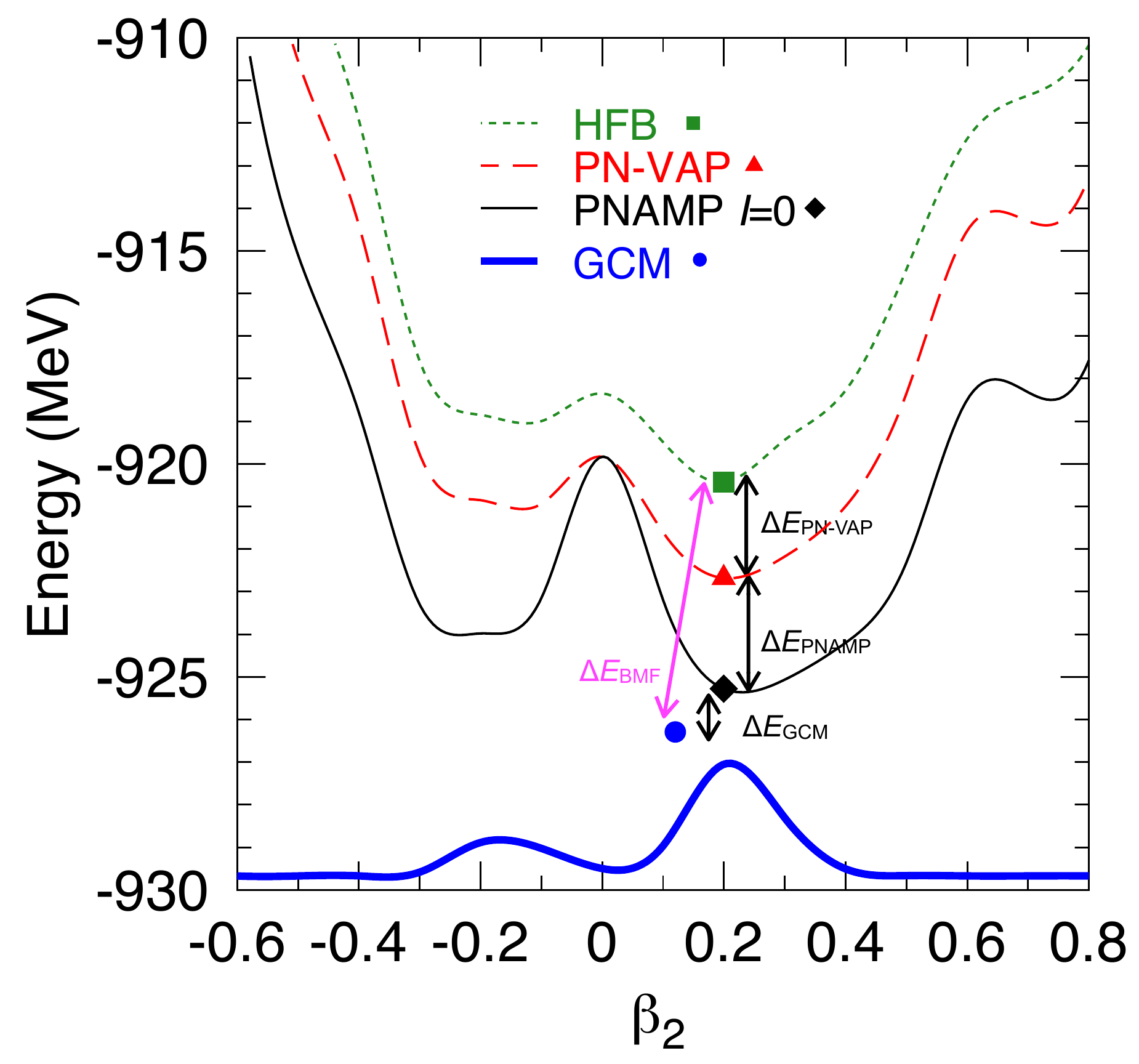}
\caption{ Potential energy surfaces as a function of the axial 
quadrupole deformation calculated with HFB (green dotted line), PN-VAP 
(red dashed line), and PNAMP (thin black continuous line) 
approximations for 108Pd with the Gogny D1S parametrization. The 
square, triangle, and diamond represent the minima of each surface. The 
blue dot corresponds to the full SCCM energy and the blue line 
represent the ground state collective wave function normalize to 1. The 
arrows point out the energy gain between the different BMF approaches.}
\label{bmf_gs_corr_ener_q2_ind}
\end{center}
\end{figure}

\begin{figure}
\begin{center}
\includegraphics[width=0.5\textwidth]{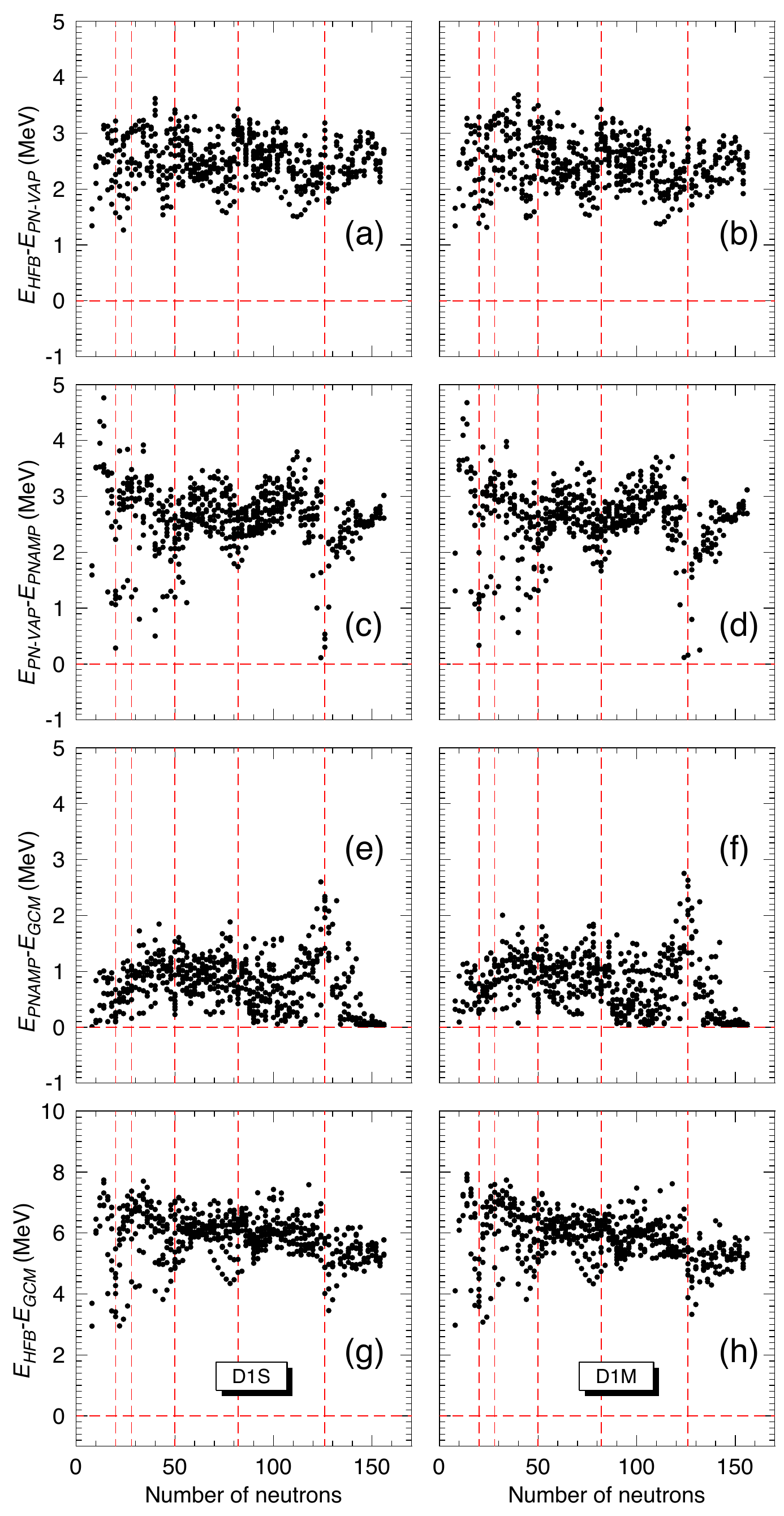}
\caption{Gain in total energy as a function of the number of neutrons 
obtained by including BMF effects depicted in 
Fig.~\ref{bmf_gs_corr_ener_q2_ind}. Dashed vertical lines represent the 
neutron magic numbers 20, 28, 50, 82, and 126. Left and right columns 
correspond to Gogny D1S and D1M parametrizations, respectively. Figure 
taken from Ref.~\cite{Rodriguez2015}}
\label{bmf_gs_corr_ener_q2_glob}
\end{center}
\end{figure}

SCCM methods are based on the variational principle. The increase of 
complexity of the many-body states from purely mean-field (HFB-like) 
wave functions to symmetry restored and configuration mixing states is 
a way of getting closer and closer to the exact solution of the 
many-body problem. Therefore, including BMF correlations is essential 
to give a reliable theoretical description of the system. These 
correlations depend on the breaking and restoration of the symmetries 
of mean-field states, and the amount and relevance of the collective 
coordinates explored with the GCM method. Primarily, SCCM methods are 
designed to provide the best approach to the ground state energy. In 
fact, ground state correlations energies have been globally studied in 
Gogny EDFs with two kind of axial SCCM calculations, namely, parity 
projection plus GCM along the octupole degree of 
freedom~\cite{Robledo2011,Robledo2015}, and particle number and angular 
momentum projection plus GCM along the quadrupole degree of 
freedom~\cite{Rodriguez2015}. In the latter case, the ground state 
correlation energies provided by the different BMF approaches used in a 
SCCM calculation are depicted for the nucleus $^{108}$Pd in 
Fig.~\ref{bmf_gs_corr_ener_q2_ind} as an example. This calculation has 
been extended to a wide range of even-even nuclei and D1S and D1M 
parametrizations as in 
Fig.~\ref{bmf_gs_corr_ener_q2_glob}~\cite{Rodriguez2015}. Here, we 
notice first the similar BMF correlation energies obtained with the two 
parametrizations. Moreover, we identify the order of magnitude of the 
different BMF correlation energies and the shell effects. These shell 
effects are rather prominent in PNAMP and GCM approximations. Hence, 
the energy gain is larger in mid-shell nuclei and negligible near magic 
nuclei when the rotational symmetry is restored (PNAMP), and the 
opposite happens when the quadrupole shape mixing is performed (GCM). 
Total BMF ground state correlation energies range from 4-8 MeV, having 
$\sim6$ MeV in most of the cases for this kind of SCCM calculation.   
%%%%%%%%%%%%%%%%%%%%%%%%%%%%%%%%%%%%%%%
%%%%%%%%%%%%%%%%%%%%%%%%%%%%%%%%%%%%%% 
%%%%%%%%%%%%%%%%%%%%%%%%%%%%%%%%%%%%%%%
%%%%%%%%%%%%%%%%%%%%%%%%%%%%%%%%%%%%%% 
%

\begin{figure}
\begin{center}
\includegraphics[width=0.5\textwidth]{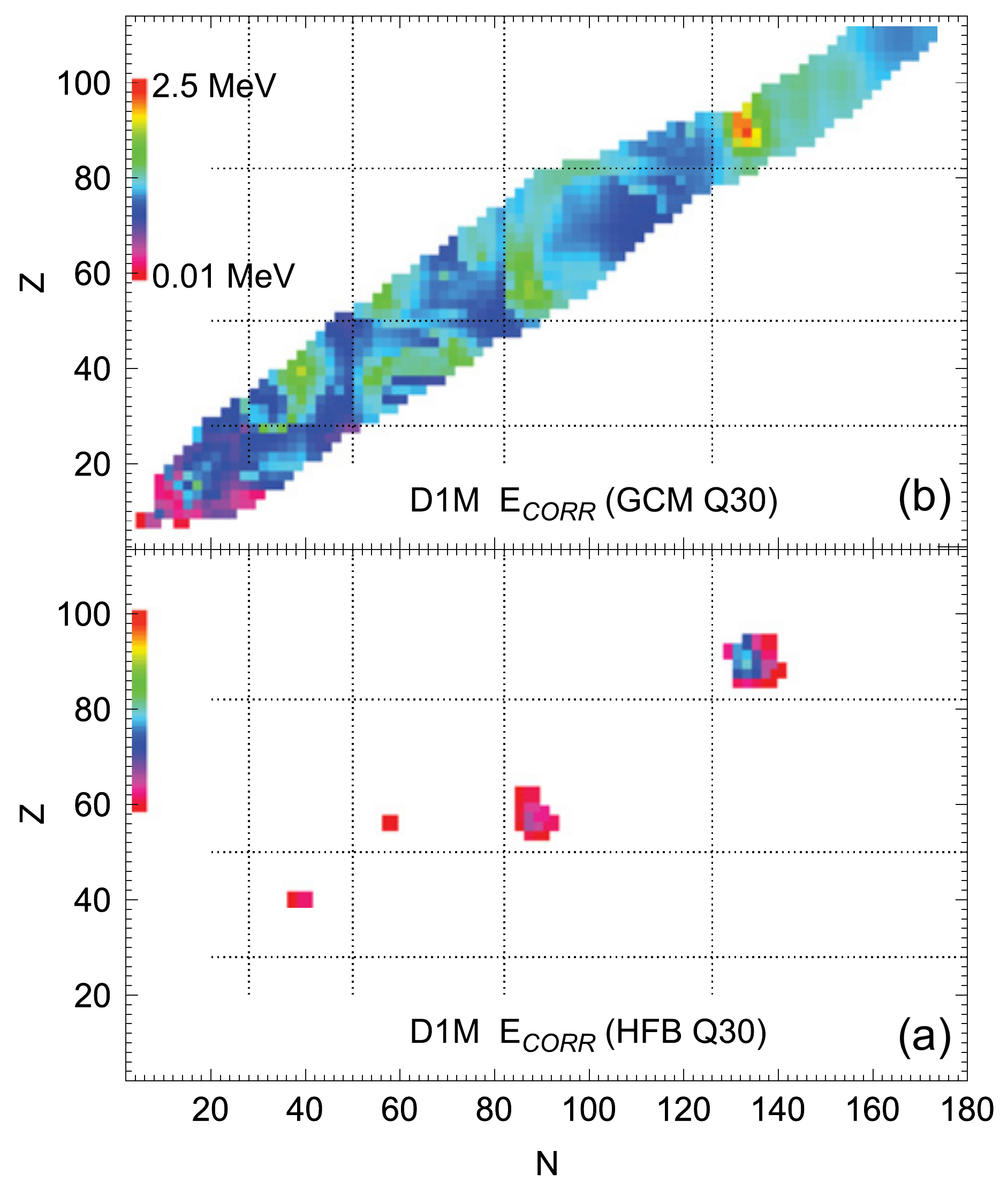}
\caption{Octupole correlation energy gain as compared to HFB results 
preserving reflection symmetry. (a) the HFB correlation energy gained 
by breaking reflection symmetry; and, (b) the parity projection plus 
octupole GCM correlation energy. Horizontal and vertical dotted lines 
correspond to magic proton and neutron numbers. Figure adapted from 
Ref.~\cite{Robledo2015}.}
\label{bmf_gs_corr_ener_q3_glob}
\end{center}
\end{figure}

\begin{figure}
\begin{center}
\includegraphics[width=0.5\textwidth]{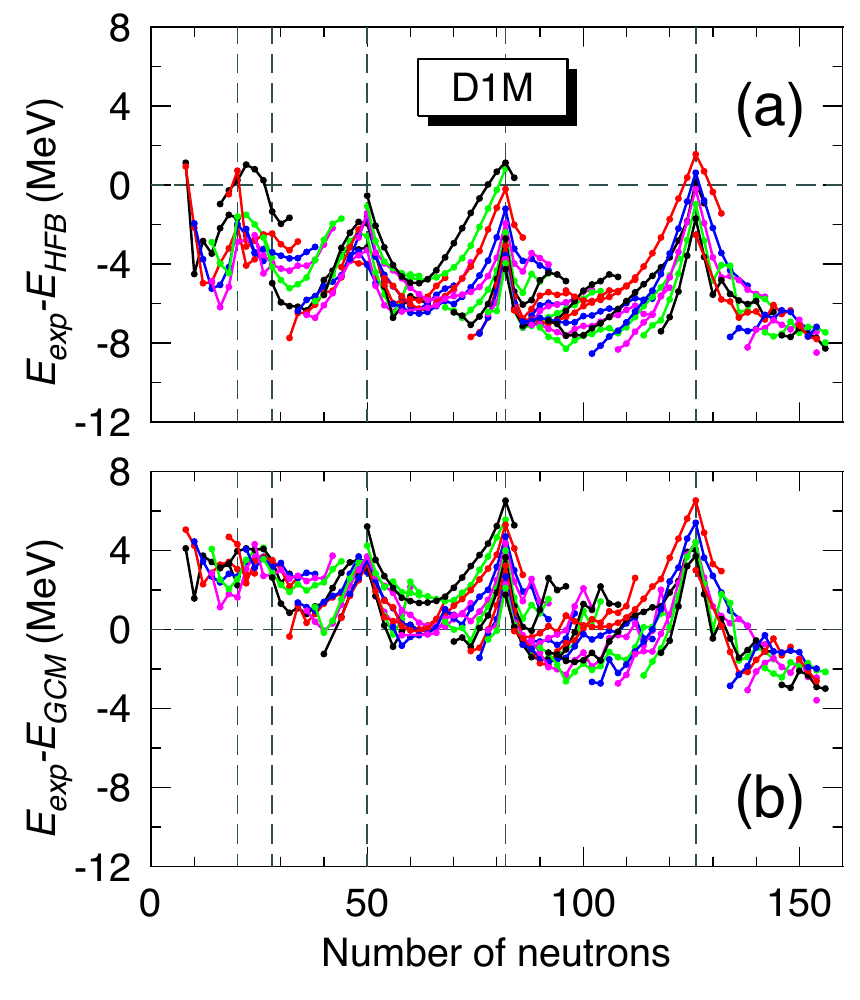}
\caption{ Difference between the experimental total energies (taken 
from Ref.~\cite{Wang2012}) and (a) HFB; and, (b) GCM total energies 
calculated with an axial SCCM method (including PNAMP and axial 
quadrupole shape mixing) with the D1M parametrization. Lines connect 
isotopic chains starting from $Z=10$. Black, red, blue, magenta, and 
green lines represent isotopic chains with $Z$ = x0, x2, x4, x6, and 
x8, where x = 1,2,..., etc. Dashed vertical lines mark the neutron 
magic numbers 20, 28, 50, 82, and 126. Figure adapted from 
Ref.~\cite{Rodriguez2015}.}
\label{d1m_global_masses}
\end{center}
\end{figure}

Similarly, BMF ground state correlation energies obtained by parity 
breaking, and parity projection plus octupole shape mixing are shown in 
Fig.~\ref{bmf_gs_corr_ener_q3_glob} for Gogny D1M, although D1S and D1N 
parametrizations give almost the same results~\cite{Robledo2015}. These 
energy gains are all defined with respect to the parity symmetric HFB 
ground state. In panel (a) only a few nuclei in the Ra, Ba and Zr 
regions show a non-zero correlation energy that never exceeds 1.2 MeV. 
These are the regions where a larger energy gain is also obtained when 
parity projection and octupole shape mixing is performed (panel (b)). 
The GCM correlation energy can be as large as 2.5 MeV. In the rest of 
the nuclei studied, the correlation energy is smaller, typically of the 
order of 1 MeV. Even though the computational cost compared to more 
traditional fits is much larger, BMF ground state correlation energies 
computed with SCCM methods should be taken into account in the future 
as it is clearly shown in 
Fig.~\ref{d1m_global_masses}~\cite{Rodriguez2015}. Here, the difference 
between the experimental and theoretical values for total energies are 
plotted both for mean-field (HFB) and the SCCM calculations with the 
Gogny D1M EDF. This parametrization already takes into account 
quadrupole shape mixings within the 5DCH in the fitting protocol. 
Therefore, the mean-field results are under-bound except for very few 
closed-shell nuclei (Fig.~\ref{d1m_global_masses}(a)). This effect is 
partially corrected with the inclusion of BMF ground state correlation 
energies (Fig.~\ref{d1m_global_masses}(b)). However, these corrections 
tend to overestimate (in average) the total energies. In addition, 
shell effects are very noticeable around the magic numbers in the 
mean-field approach and this drawback is not totally washed out with 
these SCCM calculations.    
%%%%%%%%%%%%%%%%%%%%%%%%%%%%%%%%%%%%%%%
%%%%%%%%%%%%%%%%%%%%%%%%%%%%%%%%%%%%%%
%%%%%%%%%%%%%%%%%%%%%%%%%%%%%%%%%%%%%%%
%%%%%%%%%%%%%%%%%%%%%%%%%%%%%%%%%%%%%%   

\begin{figure}
\begin{center}
\includegraphics[width=0.5\textwidth]{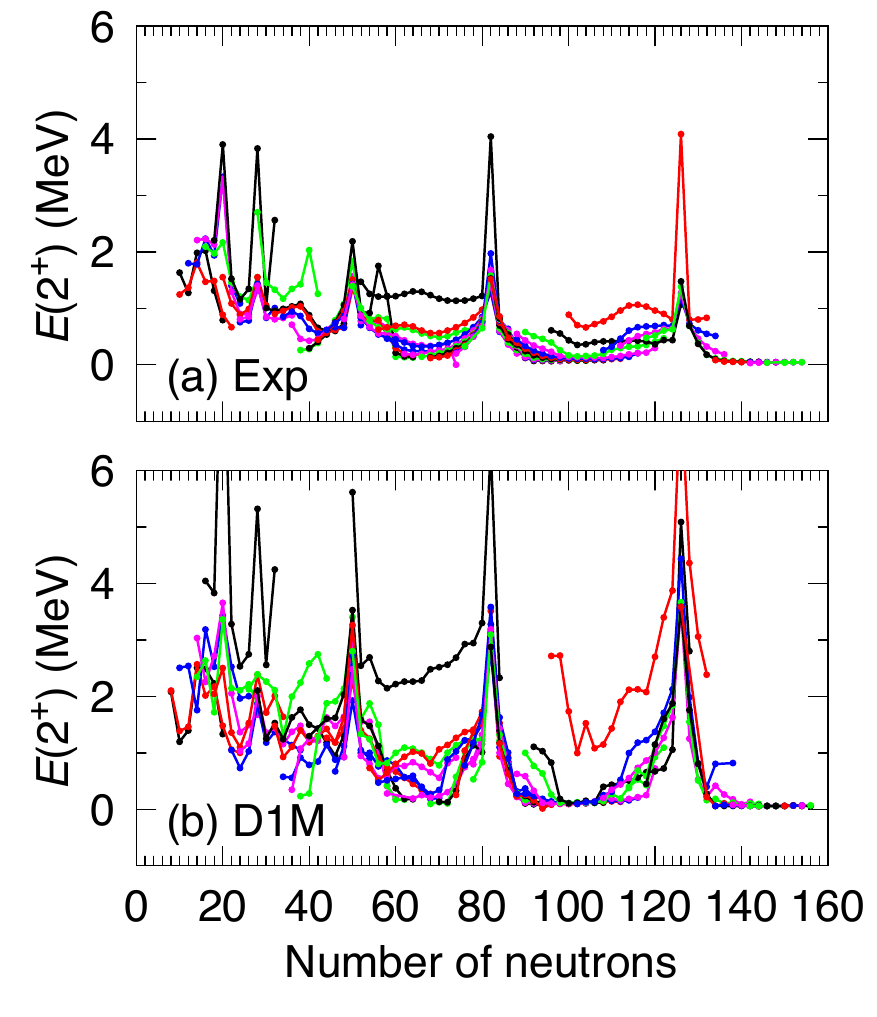}
\caption{$2^{+}_{
1}$ excitation energies for (a) experimental
data (taken from Ref.~\cite{Raman2001}); and (b) an axial SCCM method 
(including PNAMP and axial quadrupole shape mixing) with the D1M 
parametrization. Lines connect isotopic chains starting from $Z=10$. 
The color code is the same as in Fig.~\ref{d1m_global_masses}. Figure 
adapted from Ref.~\cite{Rodriguez2015}.}
\label{d1m_global_E2}
\end{center}
\end{figure}

\begin{figure}
\begin{center}
\includegraphics[width=0.5\textwidth]{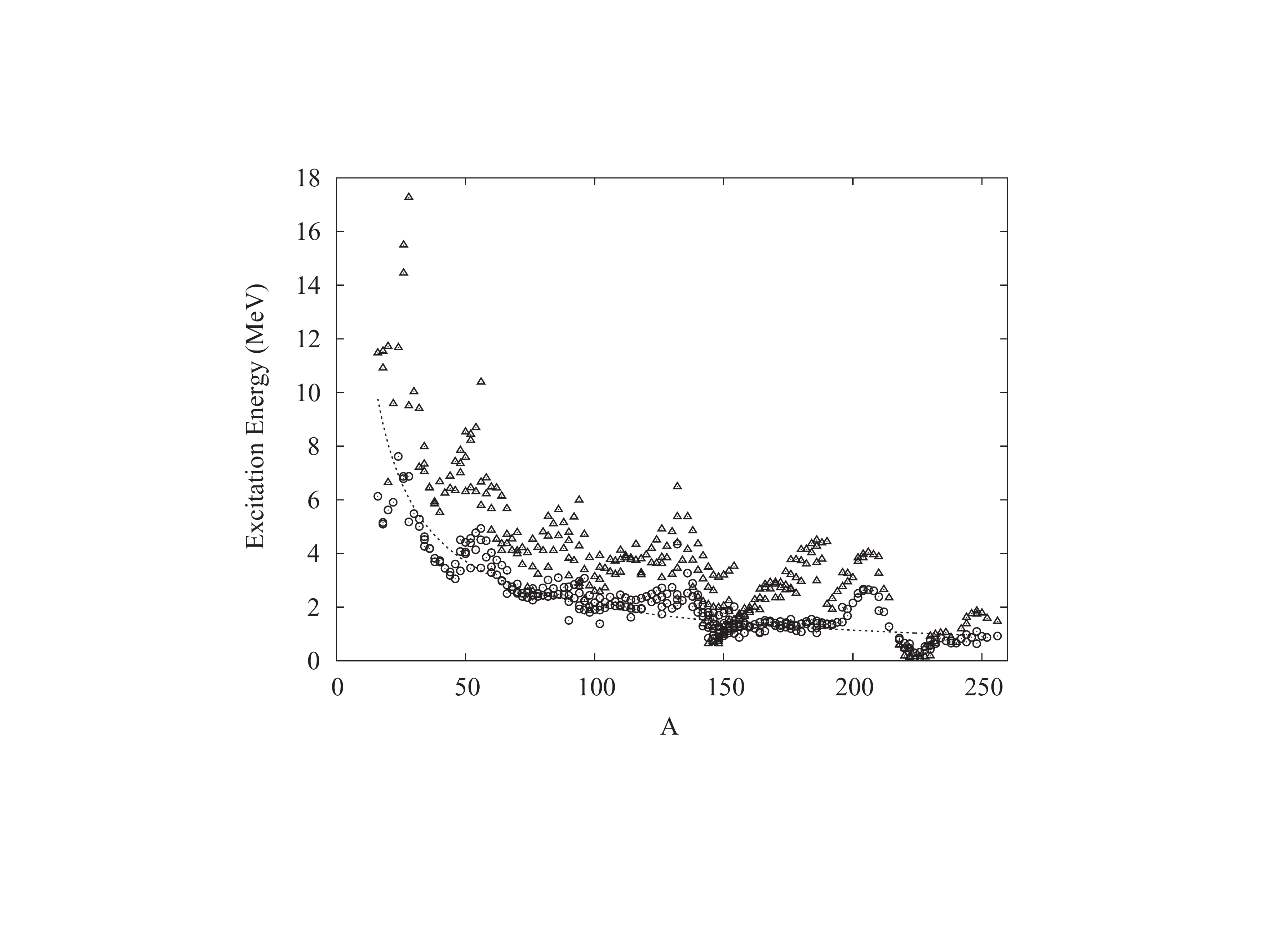}
\caption{Octupole excitation energies as a function of mass number $A$ 
calculated with an axial SCCM method (including parity projection and 
axial octupole shape mixing) with the D1S parametrization. Circles: 
experiment; triangles: theory. Figure taken from 
Ref.~\cite{Robledo2011}.}
\label{d1s_global_octupole_exc}
\end{center}
\end{figure}

Apart from the effect on total energies, axial SCCM calculations are 
well suited to study globally the performance of the method to 
reproduce other spectroscopic observables like $2^{+}$ 
(Fig.~\ref{d1m_global_E2}) or octupole excitation energies 
(Fig.~\ref{d1s_global_octupole_exc}). The overall trends of the 
experimental excitation energies are well reproduced with this kind of 
calculations although there are local discrepancies, and most 
importantly, the theoretical energies are too high in general. The 
origin of this stretching of the excitation energies could be mainly 
the lack of BFM correlations in the excited states. SCCM methods using 
axial and time-reversal symmetry conserving (TRSC) intrinsic wave 
functions to build the GCM basis explore variationally better ground 
states than excited states. Therefore, ground states calculated in this 
manner are closer (better converged from the variational point of view) 
to their exact values than the excited states that are still too above 
their exact values. These differences provoke the stretching of the 
theoretical spectra and the poor quantitative agreement with the 
experimental data shown in Figs.~\ref{d1m_global_E2} 
and~\ref{d1s_global_octupole_exc}. 
%%%%%%%%%%%%%%%%%%%%%%%%%%%%%%%%%%%%%%%
%%%%%%%%%%%%%%%%%%%%%%%%%%%%%%%%%%%%%%   
%%%%%%%%%%%%%%%%%%%%%%%%%%%%%%%%%%%%%%%
%%%%%%%%%%%%%%%%%%%%%%%%%%%%%%%%%%%%%%  

\begin{figure}
\begin{center}
\includegraphics[width=0.5\textwidth]{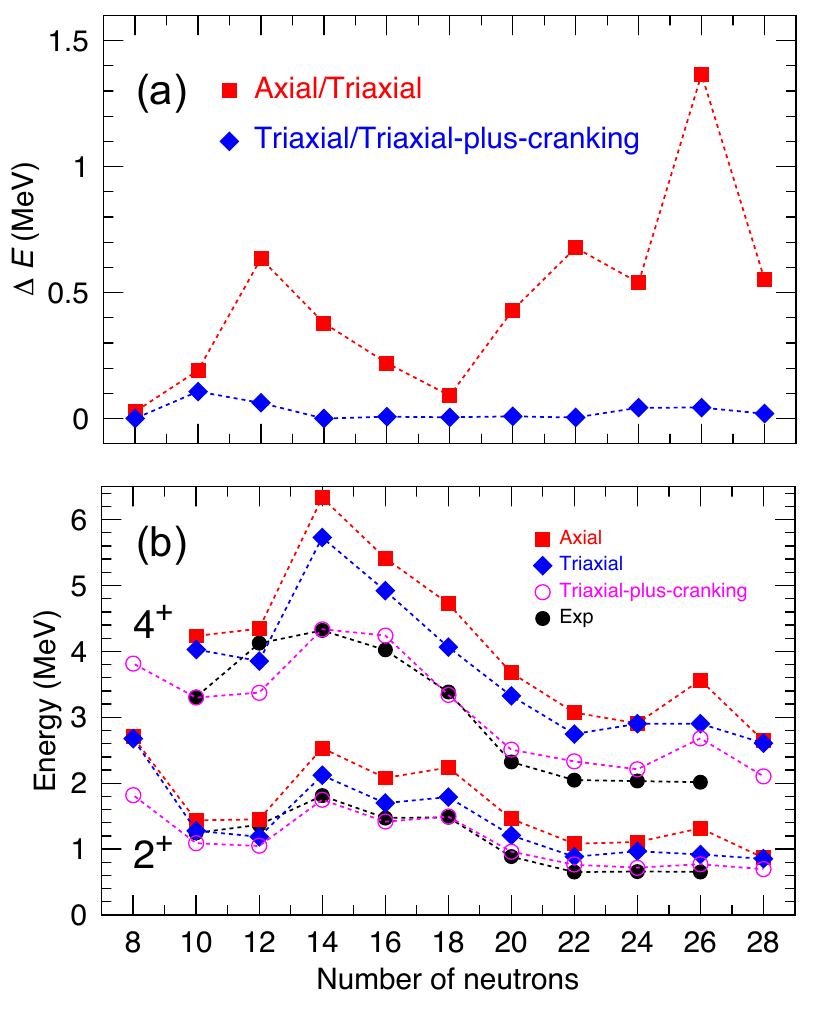}
\caption{(a) Energy differences between the SCCM ground state energies 
in the magnesium isotopic chain computed with: axial and triaxial 
$J_{c}=0$ shapes (red squares); triaxial $J_{c}=0$ and 
triaxial-plus-cranking $J_{c}=0,2$ states (blue diamonds). (b) 
Experimental and SCCM excitation energies for the first $2^{+}$ and 
$4^{+}$  states in the magnesium isotopic chain. Data points are taken 
from 
Refs.~\cite{Detraz1979,Gade2007,Doornenbal2013,Yoneda2001,DataBase,Motobayashi1995,Doornenbal2016}. 
Theoretical values are obtained with the D1S parametrization. Figure 
adapted from Ref.~\cite{Rodriguez2016}.}
\label{Mg_cranking}
\end{center}
\end{figure}

Several improvements can be proposed to correct this drawback and have 
more predictive calculations at a quantitative level, for example, the 
inclusion of multi-quasiparticle intrinsic states. However, within the 
SCCM framework, where the symmetry restorations and configuration 
mixings are performed on top of quasiparticle vacua, the solution is to 
explore more collective degrees of freedom, in particular, triaxial 
deformations and intrinsically rotating states (cranking). In fact, 
excited states are particularly sensitive to the addition of cranking 
states~\cite{Borrajo2015,Rodriguez2016,Shimada2016,Shimada2016a}. This 
procedure has also the convenient property of leaving unaltered the 
ground state. These effects are analyzed in Fig.~\ref{Mg_cranking}. 
Here, the results of three different SCCM calculations with increasing 
complexity in the magnesium isotopic chain are shown. The simplest case 
corresponds to an axial SCCM method that includes PNAMP and axial 
quadrupole shape mixing, the intermediate case includes additionally 
static triaxial shapes, and cranking states (time-reversal symmetry 
breaking) are added to the previous ones in the most involved 
calculation. Therefore, the collective coordinates are the quadrupole 
deformation parameters, $(\beta_{2},\gamma)$, and the cranking 
intrinsic angular momentum, $J_{c}$ (see Fig.~\ref{32Mg_AMP_PES_CRANK} 
for an example of states included in $^{32}$Mg). Obviously, the 
cranking angular momentum is $J_{c}=0$ in the two first approaches, and 
the angle is only $\gamma=0^{\circ},180^{\circ}$ in the axial case. 
Fig.~\ref{Mg_cranking}(a) shows the energy gain in the ground state by 
including static triaxial shapes to the ones used in the axial 
calculation. Since the method is based on the variational principle, 
this gain is always positive. Depending on the nucleus, the energy 
difference can be as large (small) as 1.4 MeV (0.1 MeV). More 
interestingly, the ground state correlations obtained by including the 
cranking states is very small ($< 0.1$ MeV) revealing that the ground 
state energy is almost insensitive to this degree of freedom. However, 
we observe in Fig.~\ref{Mg_cranking}(b) that this is not the case for 
excited states. Here, a similar qualitative behavior along the isotopic 
chain is obtained for the three calculations and also for the 
experimental data. However, the axial results are systematically above 
the triaxial values and the latter above the triaxial-plus-cranking. 
The stretching of the spectrum found in the axial and triaxial 
approximations is due to a privileged variational exploration of the 
ground state when the time-reversal symmetry is conserved. Then, the 
small compression of the energies in the triaxial approach is explained 
mainly by the possibility of $K$-mixing for states with $J\neq0$. 
The larger compression of the spectrum given by the 
triaxial-plus-cranking method is due to a better variational 
exploration of the excited states since the ground state energies 
remain practically the same as the ones obtained with the triaxial 
approach (see Fig.~\ref{Mg_cranking}(a)). Finally, the quantitative 
agreement with the experimental data reached with the most 
sophisticated SCCM method is excellent in this particular example.

%%%%%%%%%%%%%%%%%%%%%%%%%%%%%%%%%%%%%%%
\begin{figure*}[t]
\begin{center}
\includegraphics[width=1.0\textwidth]{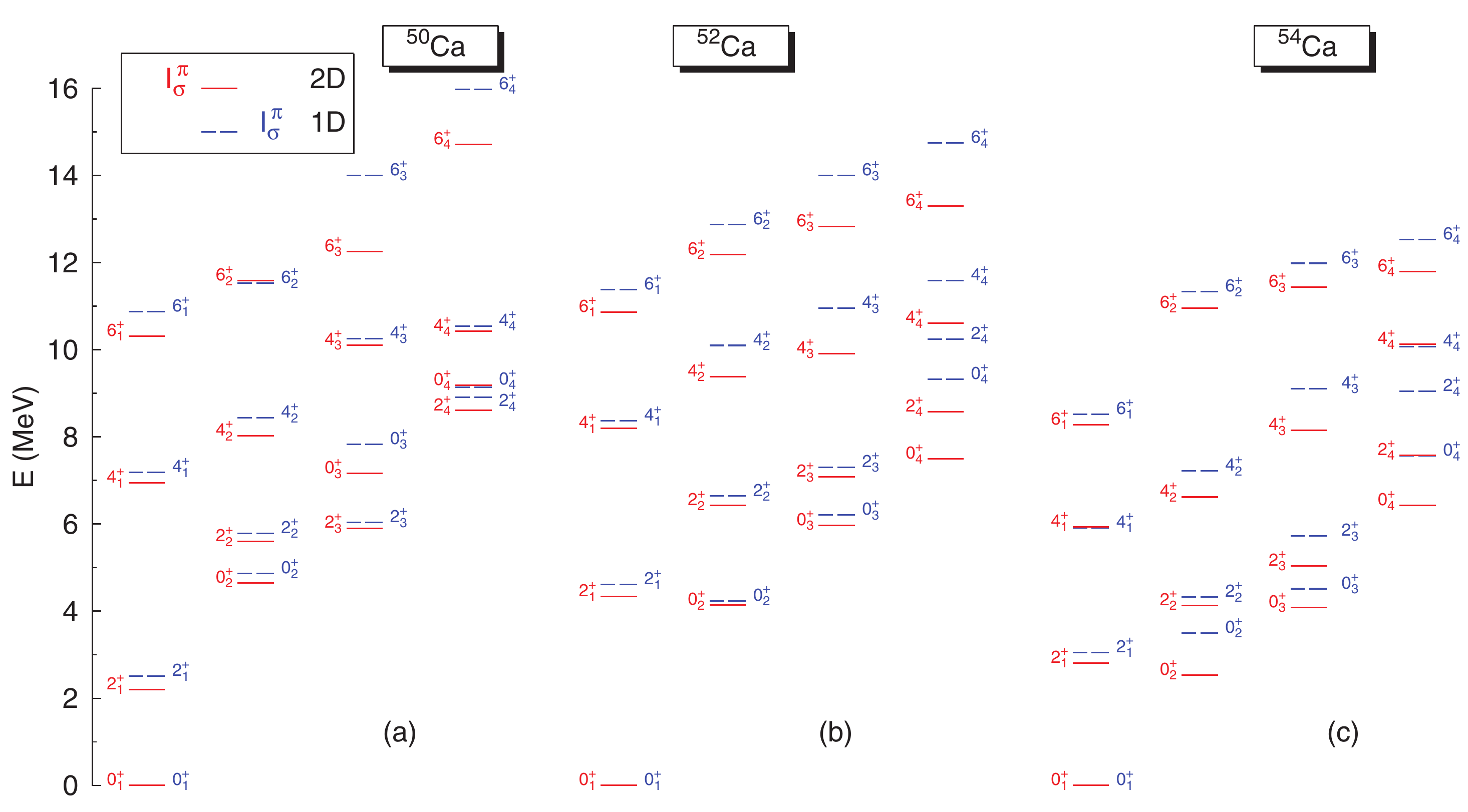}
\caption{Spectra of $^{50}$Ca, $^{52}$Ca, and $^{54}$Ca computed with 
two SCCM approximations, namely, including axial quadrupole shape 
mixing (1D) and adding particle number fluctuations (2D). Gogny In both 
cases . Figure taken from Ref.~\cite{Vaquero2013a}.}
\label{Ca_isotopes_dn2}
\end{center}
\end{figure*}

\begin{figure}
\begin{center}
\includegraphics[width=0.4\textwidth]{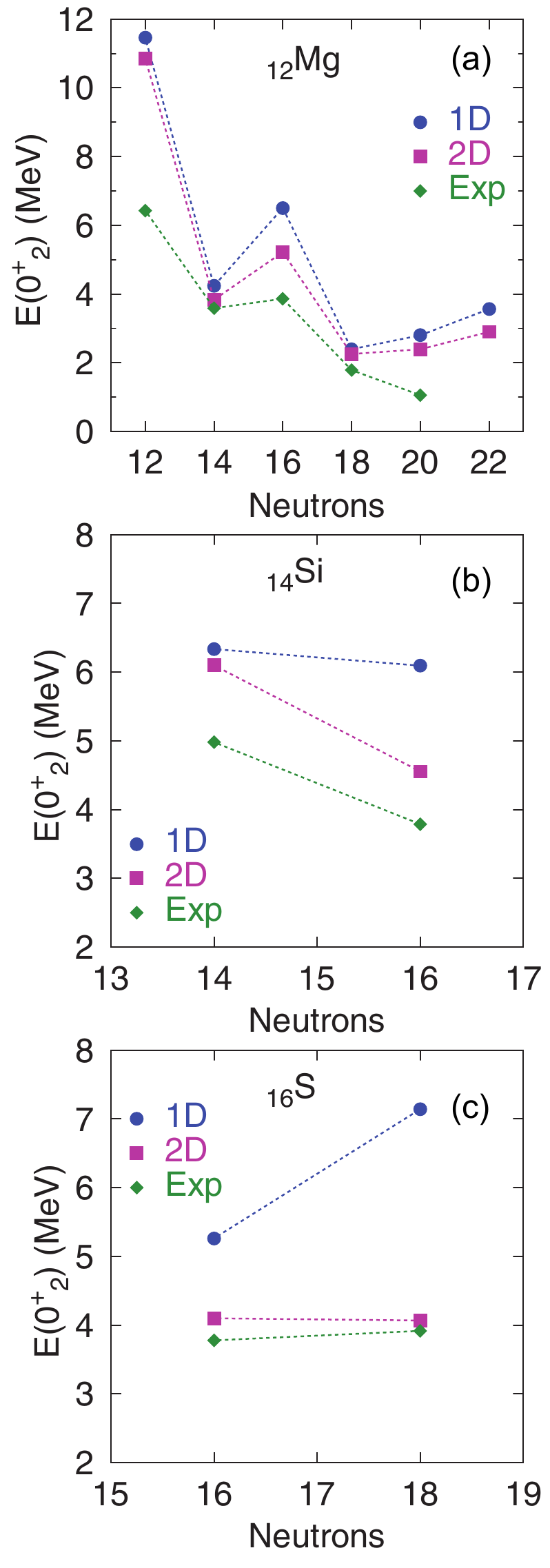}
\caption{Excitation energies of the $0^{+}_{2}$ states in (a) Mg, (b) 
Si, and (c) S isotopes. Theoretical values are calculated as in 
Fig.~\ref{Ca_isotopes_dn2} and experimental data are taken from 
Ref.~\cite{DataBase}. Figure taken from Ref.~\cite{Vaquero2013a}.}
\label{0+_dn2}
\end{center}
\end{figure}

%%%%%%%%%%%%%%%%%%%%%%%%%%%%%%%%%%%%%%  
This method cannot be used to improve the description of $0^{+}$ 
excited states within the SCCM framework. For these cases, simultaneous 
quadrupole shape and pairing fluctuations mixing can be performed. This 
SCCM method has been implemented in 
Refs.~\cite{Vaquero2011,Vaquero2013a} with axial symmetric intrinsic 
wave functions and the quadrupole deformation ($\beta_{2}$) and the 
particle number fluctuations $(\Delta N^{2})$ as collective 
coordinates. This degree of freedom introduces mean-field states with 
different values of the pairing gap. The comparison between the results 
obtained only with quadrupole shape mixing (one-dimensional GCM 
calculations, 1D) and those obtained with adding $(\Delta N^{2})$ (2D 
calculations) reveal that the latter method produces in general a more 
compressed spectrum than the former. In Fig.~\ref{Ca_isotopes_dn2} 1D 
and 2D calculations for $^{50-54}$Ca isotopes are shown as an example. 
Again, this effect is produced by a better variational exploration of 
the excited states. Particularly, the $0^{+}$ excitation energies can 
be pushed down improving their comparison with the experimental data as 
it is plotted in Fig.~\ref{0+_dn2}. Nevertheless, this compression is 
not large enough to reproduce the experimental values. This is an 
indication that some of the low-lying $0^{+}$ excited states correspond 
to explicit quasiparticle excitations rather than having a collective 
character. However, these states are out of the present SCCM methods 
with Gogny EDFs.

%%%%%%%%%%%%%%%%%%%%%%%%%%%%%%%%%%%%%%%
%%%%%%%%%%%%%%%%%%%%%%%%%%%%%%%%%%%%%%  
%%%%%%%%%%%%%%%%%%%%%%%%%%%%%%%%%%%%%%%
%%%%%%%%%%%%%%%%%%%%%%%%%%%%%%%%%%%%%%  
%%%%%%%%%%%%%%%%%%%%%%%%%%%%%%%%%%%%%% 
\subsubsection{Shape evolution/mixing/coexistence}
Most of SCCM methods considers the lowest multipole (quadrupole and, to 
a lesser extent, octupole) intrinsic deformations as the basic 
collective coordinates. Therefore, this is the perfect framework to 
study, within a microscopic theory, the nuclear shape and its related 
phenomena (shape evolution in isotopic/isotonic chains, shape mixing 
and shape coexistence in a single nucleus). A complete analysis of the 
shape of a nucleus usually starts with the evaluation of mean-field 
(HFB) and/or particle number projected potential energy surfaces (PES) 
defined along the deformations of the system. The structure of the PES, 
i.e., the number of minima, their location, depth and width, already 
provides an overall view of the character of the nucleus in question. 
Moreover, the shape of the PES can be related to an underlying shell 
structure given by self-consistent Nilsson-like orbits. For example, 
the minima appear at deformations where the Fermi level crosses a gap 
in these single particle energies. However, SCCM methods go beyond 
these mean-field analyses and the theoretical predictions for 
excitation energies and electromagnetic properties can be directly 
compared to experimental data. Moreover, the most probable intrinsic 
shapes can be also obtained for each individual state within the 
nucleus by computing the collective wave functions. 

%%%%%%%%%%%%%%%%%%%%%%%%%%%%%%%%%%%%%%%
\begin{figure}
\begin{center}
\includegraphics[width=0.5\textwidth]{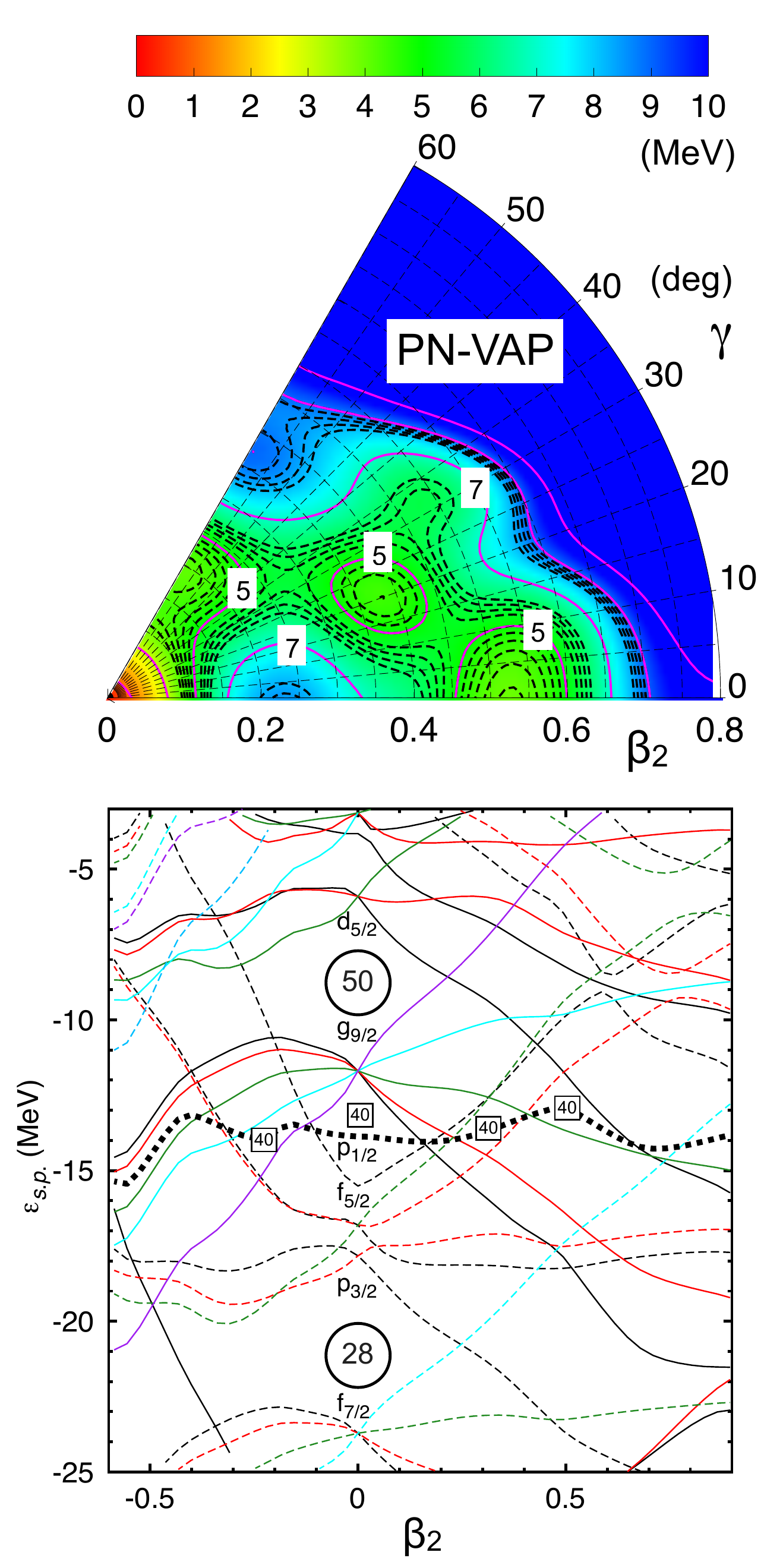}
\caption{(a) Particle number projected (PN-VAP) potential energy 
surface -normalized to their absolute minima respectively- calculated 
for $^{80}$Zr using the Gogny D1S interaction. Contour lines are 
separated by 0.2 MeV (dashed) and 1 MeV (continuous). (b) Single 
particle energies for neutrons as a function of the quadrupole 
deformation $\beta_{2}$. The Fermi level is represented by a thick 
dotted line. Figure adapted from Ref.~\cite{Rodriguez2011a}.}
\label{80Zr_PNVAP}
\end{center}
\end{figure}
%%%%%%%%%%%%%%%%%%%%%%%%%%%%%%%%%%%%%%  
%%%%%%%%%%%%%%%%%%%%%%%%%%%%%%%%%%%%%%%
\begin{figure*}[t]
\begin{center}
\includegraphics[width=\textwidth]{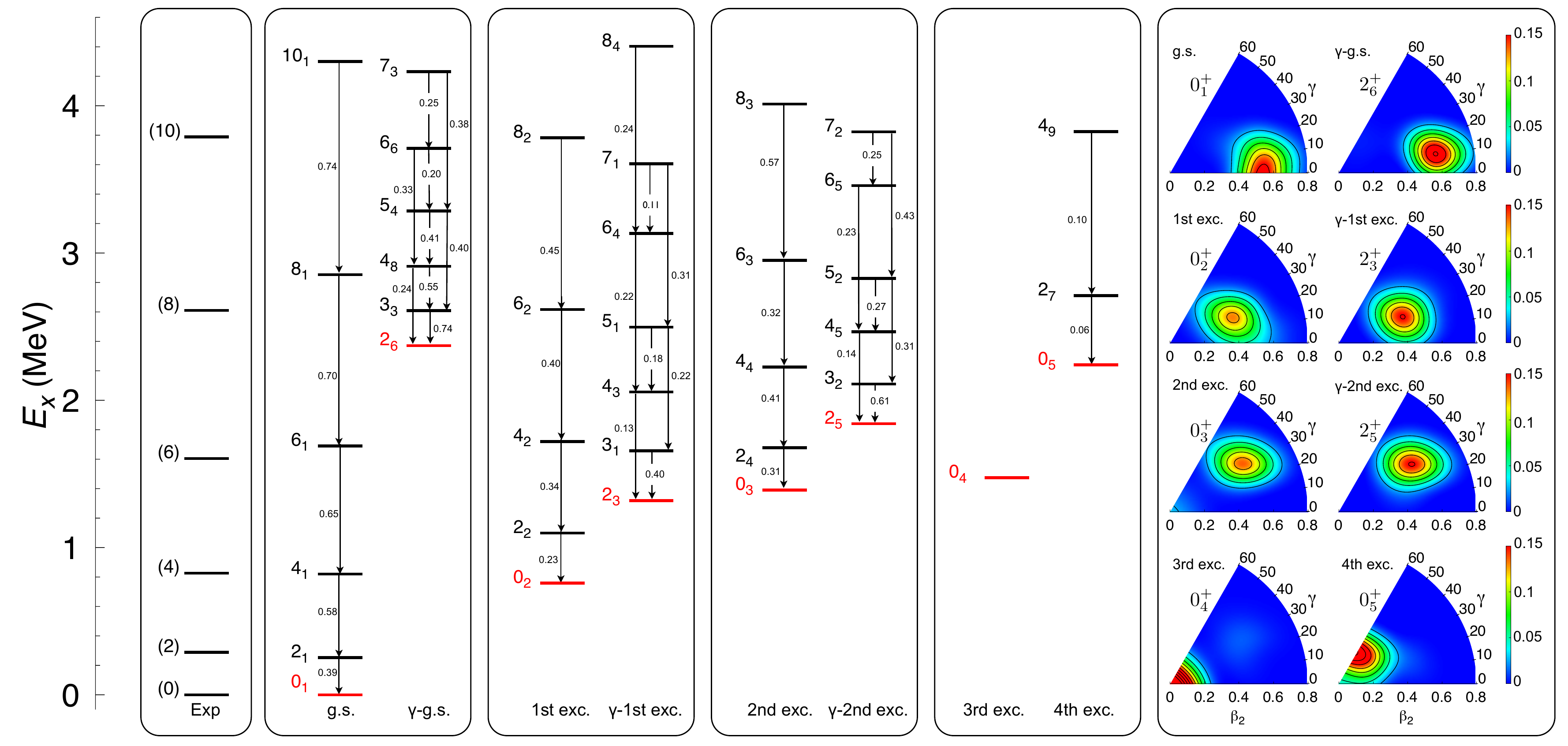}
\caption{Left panel: Experimental and theoretical (with Gogny D1S) 
spectra for $^{80}$Zr. $B(E2)$ values are given in $e^{2}b^{2}$. Right 
panel: Collective wave functions for the band heads of the bands. 
Figure taken from Ref.~\cite{Rodriguez2011a}.}
\label{80Zr_GCM}
\end{center}
\end{figure*}
%%%%%%%%%%%%%%%%%%%%%%%%%%%%%%%%%%%%%% 
To illustrate the steps described above, we analyze the nucleus 
$^{80}$Zr following Ref.~\cite{Rodriguez2011a}. In 
Fig.~\ref{80Zr_PNVAP}(a) the PES calculated with the PN-VAP method 
along the triaxial $(\beta_{2},\gamma)$ plane is plotted. Here, we 
observe several minima (spherical, axial prolate, axial oblate, 
triaxial) that could indicate the appearance of shape coexistence 
and/or shape mixing in this isotope. The origin of these minima in 
terms of the underlying shell structure is shown in 
Fig.~\ref{80Zr_PNVAP}(b) where the single-particle energies (s.p.e.) 
computed for this nucleus are plotted along the axial quadrupole 
deformation. Hence, the Fermi level crosses some gaps in these s.p.e. 
at the position of the minima obtained in the PES. These gaps are 
formed by the $pf$ and $g_{9/2}$ spherical shells (around the spherical 
point $\beta_{2}=0$) and by the evolution of these orbits whenever the 
quadrupole deformation is increased. 

The final results after performing the particle number and angular 
momentum projection, and the quadrupole shape mixing (axial and 
non-axial), are shown in Fig.~\ref{80Zr_GCM}. Several bands are found 
in the theoretical spectrum. The character of these bands can be 
primarily characterized by their level spacing and electromagnetic 
properties ($B(E2)$ and spectroscopic quadrupole moments). For example, 
the ground state, first excited and second excited bands are rotational 
bands with side bands with a $\gamma$-band character associated to 
them. Experimentally, only some ground state band levels have been 
measured and they also indicate a rotational behavior. 

As mentioned above, SCCM methods provide a very useful theoretical tool 
to identify the character of each state, namely, the collective wave 
functions (c.w.f.'s). These are represented in the right panel of 
Fig.~\ref{80Zr_GCM} only for the band-head states since the rest of the 
states belonging to the same band show similar c.w.f.'s as their 
corresponding band-heads. We observe that the most relevant shapes in 
the ground state band are located at an axial prolate deformation, the 
second and third bands are triaxial deformed, and the $0^{+}_{4}$ and 
$0^{+}_{5}$ are spherical and slightly axial oblate deformed 
respectively. Hence, these states are related to the minima found in 
the PN-VAP PES and the reordering of the energy is mainly due to the 
correlations obtained by the angular momentum restoration of the 
system. Moreover, the c.w.f.'s do not show mixing between these minima 
and, therefore, this nucleus is an example of shape coexistence but not 
of shape mixing.

Shapes of both individual nuclei and systematics along isotopic chains 
have been thoroughly studied with SCCM methods with Gogny EDFs. First 
applications were done without particle number projection and assuming 
only the axial quadrupole deformation as the collective coordinate. 
Nevertheless, a good quantitative agreement was found in the 
description of normal deformed and superdeformed bands in 
$^{32}$S~\cite{Rodriguez-Guzman2000}, the shape evolution in $N=20$ and 
$N=28$ isotones and magnesium 
isotopes~\cite{Rodriguez-Guzman2000a,Rodriguez-Guzman2000b,Rodriguez-Guzman2002,Rodriguez-Guzman2002a}, 
and the shape coexistence in neutron deficient lead 
isotopes~\cite{Rodriguez-Guzman2004}. In these works, intrinsic HFB 
instead of PN-VAP states were used. As mentioned above, the use of 
PN-VAP and simultaneous particle number and angular momentum 
restoration with Gogny EDFs was then implemented in 
Ref.~\cite{Rodriguez2007}, assuming again axial symmetry.  
The shape evolution of cadmium isotopes in the whole $N=50-82$ shell 
was well reproduced with these calculations, including the anomalous 
behavior of the $2^{+}$ excited state in the nucleus 
$^{128}$Cd~\cite{Rodriguez2008a}. Moreover, the transition from 
spherical (U(5) symmetry) to axial prolate (SU(3) symmetry) shapes in 
neodymium and samarium isotopes, and its interpretation as a quantum 
phase transition (with X(5) as the critical symmetry), was also 
analyzed with axial SCCM calculations~\cite{Rodriguez2008}.

\begin{figure}
\begin{center}
\includegraphics[width=0.5\textwidth]{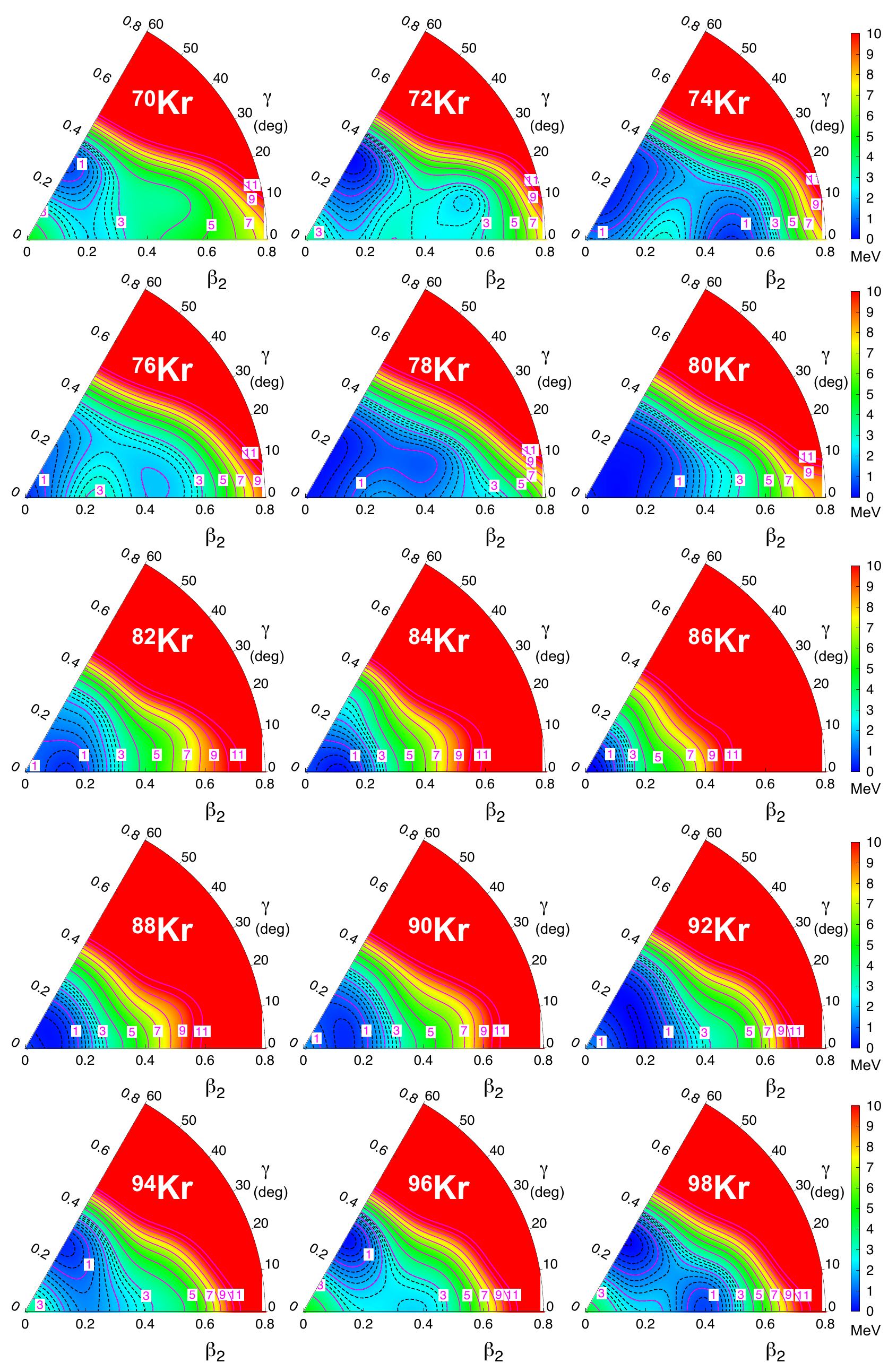}
\caption{Potential energy surfaces in the triaxial plane calculated 
with PN-VAP method and the Gogny D1S interaction for Kr isotopes. 
Figure adapted from Ref.~\cite{Rodriguez2014}.}
\label{Kr_PES}
\end{center}
\end{figure}

A breakthrough in the range of applicability of SCCM methods with Gogny 
interactions to study nuclear shapes was the inclusion of the 
quadrupole triaxial deformation as a collective 
coordinate~\cite{Rodriguez2010}.  At the same time, the access to high 
performance computing facilities has allowed the calculations of 
systematics along isotopic/isotonic chains within this formalism. A 
paradigmatic example is the study of the structure of krypton isotopes 
from neutron deficient to neutron rich nuclei~\cite{Rodriguez2014}. 
Potential energy surfaces computed with Gogny D1S interaction in the 
$(\beta_{2},\gamma)$ plane reveal a rather involved shape evolution as 
it is shown in Fig.~\ref{Kr_PES}. The semi-magic nucleus $^{86}$Kr is 
spherical -as expected- and their closest neighbors are slightly 
prolate deformed ($^{82-84,88-90}$Kr). Adding more neutrons above 
$N=50$ results in the appearance of an oblate minimum ($^{94-96}$Kr) 
and a potential shape coexistence (oblate-prolate) in $^{98}$Kr.  Clear 
candidates for shape coexistence are also obtained in the neutron 
deficient part where several nuclei show two distinct minima in their 
PES ($^{72-78}$Kr).   

\begin{figure}
\begin{center}
\includegraphics[width=0.5\textwidth]{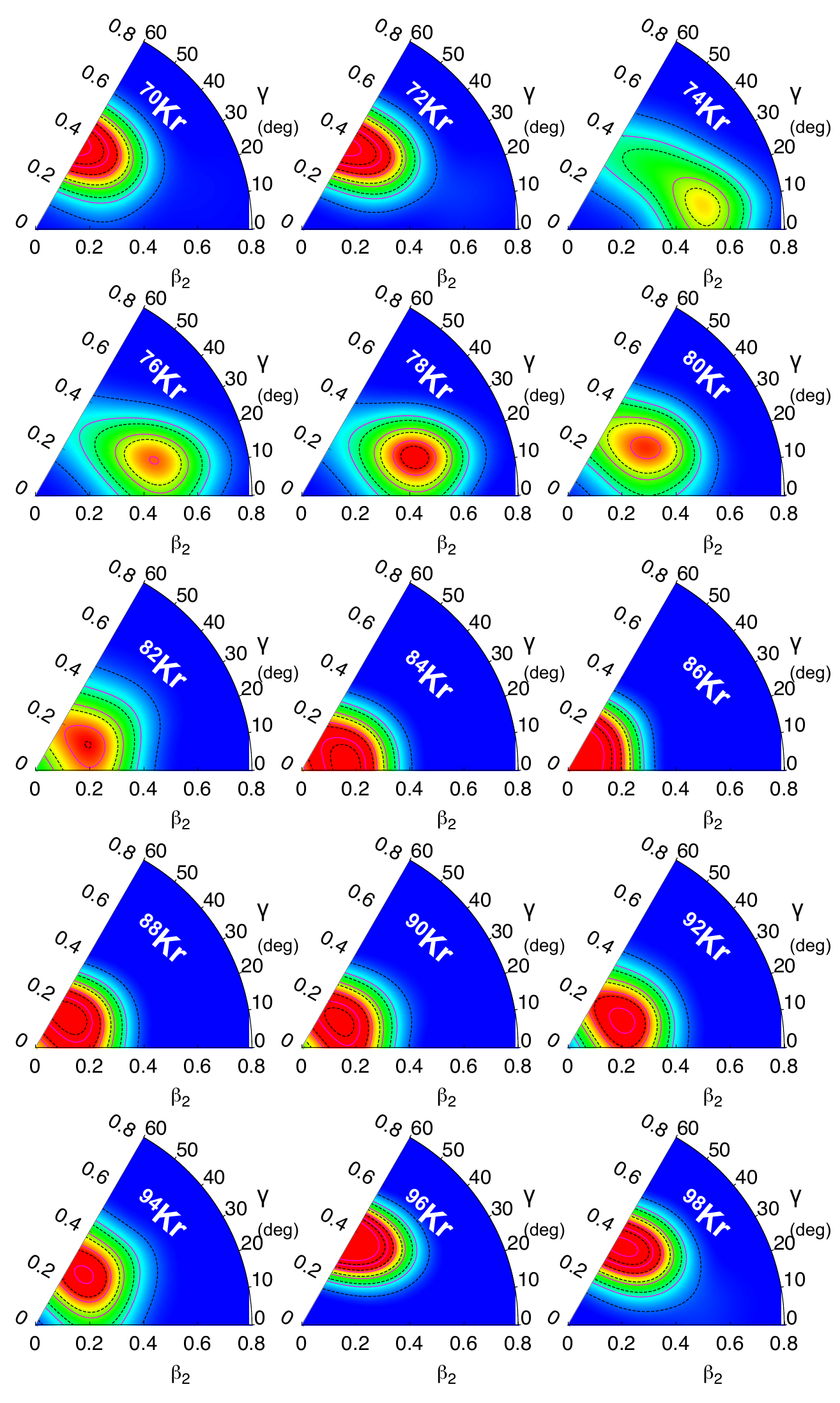}
\caption{Collective wave functions for the ground states calculated 
with the SCCM method and the Gogny D1S interaction for Kr isotopes. 
Figure adapted from Ref.~\cite{Rodriguez2014}.}
\label{Kr_cwf}
\end{center}
\end{figure}

The shape evolution obtained after applying the full SCCM method can be 
seen in Fig.~\ref{Kr_cwf} where the ground state collective wave 
functions are represented. A clear transition from oblate to 
triaxial-prolate states is observed from  $^{70-72}$Kr to $^{76-78}$Kr 
with the nucleus $^{74}$Kr being the transitional isotope. In this 
case, the ground state c.w.f. is an example of shape mixing along the 
$\gamma$ degree of freedom.  Around the spherical $N=50$ isotope, 
nuclei are less deformed as it could inferred from the PES 
($^{82-84}$Kr and $^{88-90}$Kr). Then, the deformation increases again 
towards oblate deformed nuclei ($^{94-98}$Kr).

\begin{figure}
\begin{center}
\includegraphics[width=0.5\textwidth]{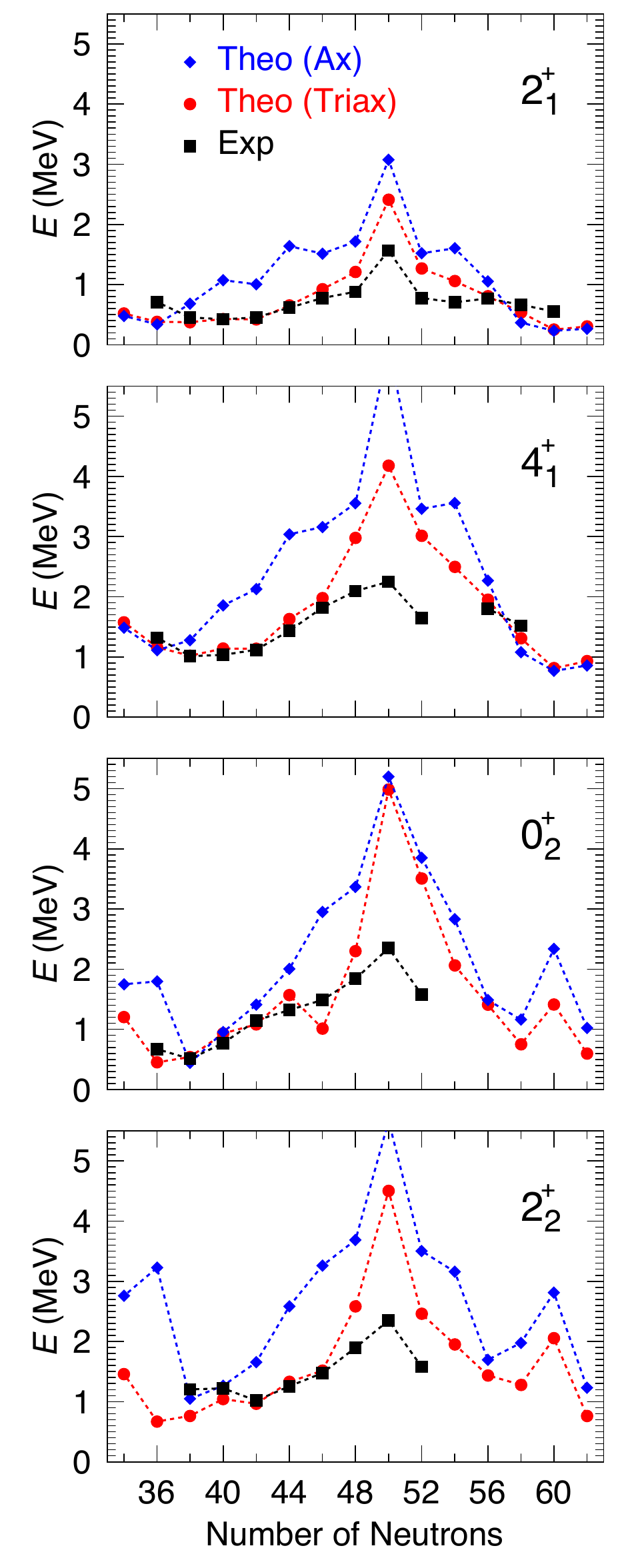}
\caption{Excitation energies along the krypton isotopic chain 
$^{70-98}$Kr for (a) $2^{+}_{1}$, (b) $4^{+}_{1}$, (c) $0^{+}_{2}$, 
and, (d) $2^{+}_{2}$ states. Black boxes, blue diamonds, and red 
bullets represent the experimental values (taken from 
Ref.~\cite{DataBase}), and the results of SCCM axial and SCCM triaxial 
calculations, respectively. Figure taken from 
Ref.~\cite{Rodriguez2014}.}
\label{Kr_EXC}
\end{center}
\end{figure}

Apart from this analysis, the results obtained with the SCCM can be 
compared to the experimental data as it is plotted in 
Fig.~\ref{Kr_EXC}. The agreement is rather good except for the region 
around $N=50$ where explicit quasiparticle excitations will play a 
relevant role to account for these excited states. Furthermore, the 
improvement achieved by the inclusion of the triaxial degree of freedom 
with respect to purely axial SCCM calculations is also represented in 
Fig.~\ref{Kr_EXC}. 
%%%%%%%%%%%%%%%%%%%%%%%%%%%%%%%%%%%%%%%
%%%%%%%%%%%%%%%%%%%%%%%%%%%%%%%%%%%%%% 
%%%%%%%%%%%%%%%%%%%%%%%%%%%%%%%%%%%%%%%
%%%%%%%%%%%%%%%%%%%%%%%%%%%%%%%%%%%%%% 
%%%%%%%%%%%%%%%%%%%%%%%%%%%%%%%%%%%%%%%
%%%%%%%%%%%%%%%%%%%%%%%%%%%%%%%%%%%%%% 

Hence, this kind of SCCM calculations with Gogny EDFs including 
particle number and angular momentum projection and triaxial quadrupole 
shape mixing have been extensively used recently in collaboration with 
experimental groups. For example, the role of triaxiality in the shape 
evolution, shape mixing and/or shape coexistence has been studied in 
$^{42}$Ca~\cite{Hadynska2016,Hadynska2018}, neutron rich 
Zn~\cite{Illana2014}, Ge~\cite{Sieja2013,Lettmann2017}, 
Se~\cite{Chen2017}, Kr~\cite{Flavigny2017}, Mo~\cite{Ralet2017}, 
Os~\cite{John2014} and Pt~\cite{John2017} isotopes, comparing the most 
recent experimental data with theoretical predictions provided not only 
by Gogny SCCM methods but also by large scale shell model calculations, 
5DCH approaches, and/or other SCCM methods with Skyrme functionals. 
%%%%%%%%%%%%%%%%%%%%%%%%%%%%%%%%%%%%%%%
\begin{figure}
\begin{center}
\includegraphics[width=0.8\textwidth]{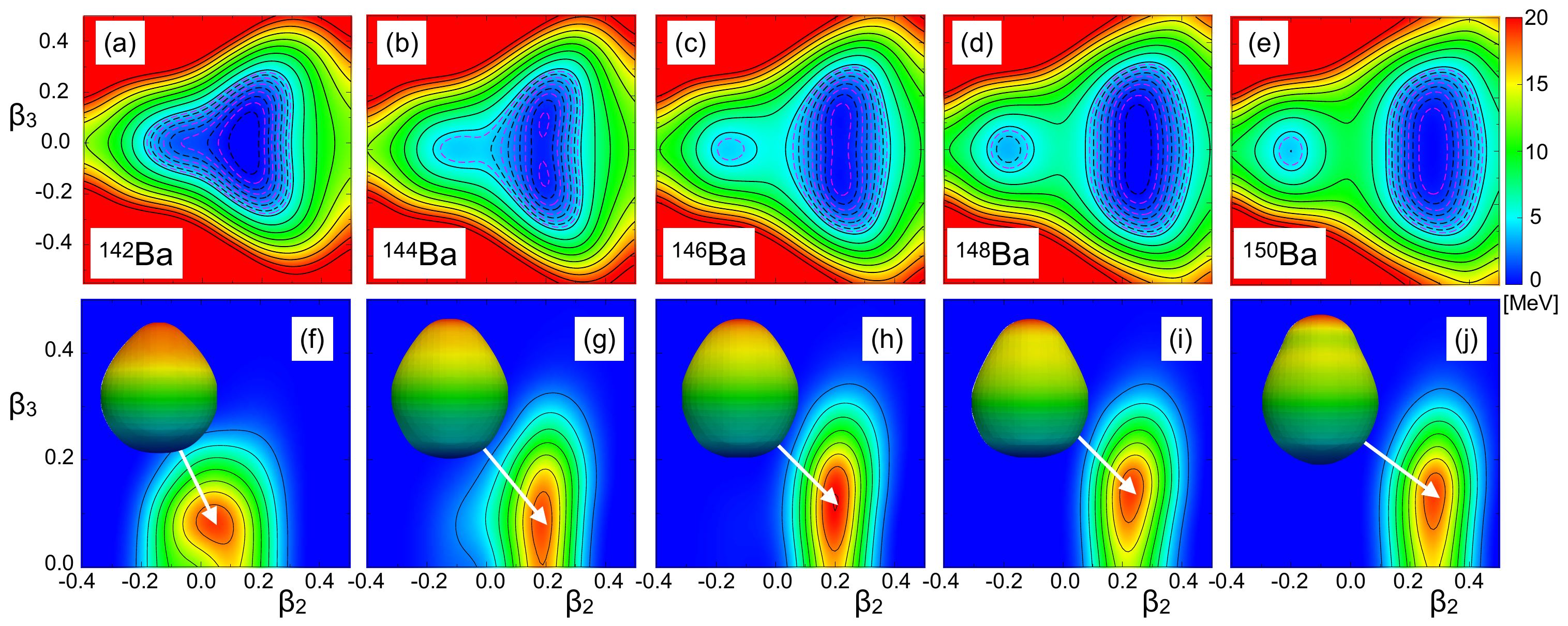}
\caption{(a)-(e) Potential energy surfaces in the 
$(\beta_{2},\beta_{3})$ plane for the Ba isotopic chain for $N=86-94$. 
(f)-(j) Calculated collective wave functions of the ground states 
($0^{+}_{1}$) are represented for $\beta_{3}>0$. To better visualize 
the shape of the isotopes, the surface that joins the points with 
constant spatial density is also shown for each isotope. These 
densities are computed with the HFB wave functions that correspond to 
the maximum of each collective wave function (see the arrows). Figure 
taken from Ref.~\cite{Lica2018}.}
\label{Ba_chain1}
\end{center}
\end{figure}
%%%%%%%%%%%%%%%%%%%%%%%%%%%%%%%%%%%%%% 
%%%%%%%%%%%%%%%%%%%%%%%%%%%%%%%%%%%%%%%
\begin{figure}
\begin{center}
\includegraphics[width=0.8\textwidth]{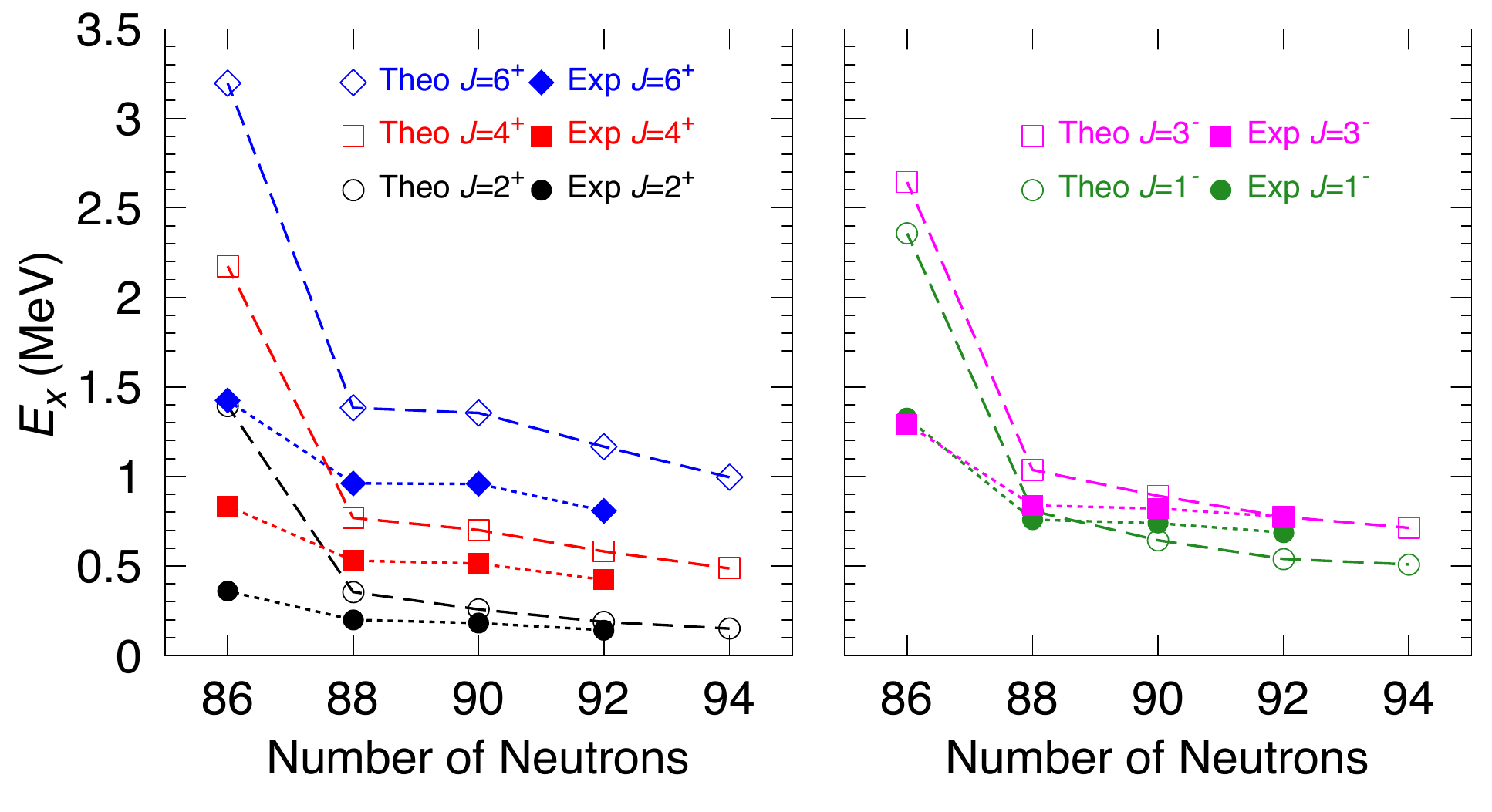}
\caption{Experimental (filled symbols) and theoretical (open symbols) 
energy spectra in the Ba isotopic chain. Positive-parity states are 
grouped in panel (a) while the negative-parity ones are grouped in 
panel (b). Figure taken from Ref.~\cite{Lica2018}.}
\label{Ba_chain2}
\end{center}
\end{figure}
%%%%%%%%%%%%%%%%%%%%%%%%%%%%%%%%%%%%%% 

On the other hand, SCCM calculations including particle number, parity 
and angular momentum projection and both axial quadrupole and octupole 
shape mixing have been also applied to study the interplay between 
quadrupole and octupole degrees of freedom, in particular, in neutron 
rich barium isotopes~\cite{Bernard2016,Bucher2017,Lica2018}. The 
mean-field PES and the ground state collective wave functions in the 
$(\beta_{2},\beta_{3})$ plane are represented for $^{142-150}$Ba nuclei 
in Fig.~\ref{Ba_chain1}. Here, we see the increase of quadrupole 
deformation whenever the number of neutrons departs from $N=82$. More 
interestingly, we observe the appearance of a non-negligible octupole 
deformation in this region that remains even after carrying out the 
shape mixing. The comparison between the theoretical predictions and 
the experimental data is shown in Fig.~\ref{Ba_chain1} both for 
positive and negative parity states. The qualitative agreement with the 
experiment is rather good although the transition between less 
quadrupole deformed ($^{142}$Ba) to well-deformed states 
($^{144-150}$Ba) is sharper in the theoretical results.
%%%%%%%%%%%%%%%%%%%%
\subsubsection{Shell closures}
SCCM methods are also suited to analyze the appearance/degradation of 
magic numbers in exotic nuclei. This property is intimately linked to 
the previous section since closed-shell (open-shell) nuclei are 
spherical (deformed). From the experimental point of view, large values 
of the excitation energies, $E(2^{+}_{1})$,  and small values of the 
reduced transition probabilities, $B(E2)$, indicate the presence of 
shell closures in even-even nuclei. The opposite is considered as 
fingerprints of a collective behavior typical from open-shell systems. 
SCCM methods have access to show the most relevant intrinsic shapes to 
build a given nuclear state (i.e., the collective wave function), and, 
on the other hand, the excitation energies and electromagnetic 
properties computed in the laboratory system. The latter can be 
directly compared to the experimental data. 

First applications to the study of shell closures within this framework 
were also the ones referred above with angular momentum projection and 
axial quadrupole deformation mixing but without particle number 
projection~\cite{Rodriguez-Guzman2000a,Rodriguez-Guzman2000b,Rodriguez-Guzman2002,Rodriguez-Guzman2002a}. 
These calculations showed that $N=20$ and $N=28$ are not good magic 
numbers in $^{32}$Mg, and $^{44}$S, $^{42}$Si and $^{40}$Mg, 
respectively, since small excitation energies, large $B(E2)$ values and 
deformed ground state c.w.f.'s were predicted. These results were in 
qualitative agreement with the available experimental data.

%%%%%%%%%%%%%%%%%%%%%%%%%%%%%%%%%%%%%%%
\begin{figure}
\begin{center}
\includegraphics[width=0.5\textwidth]{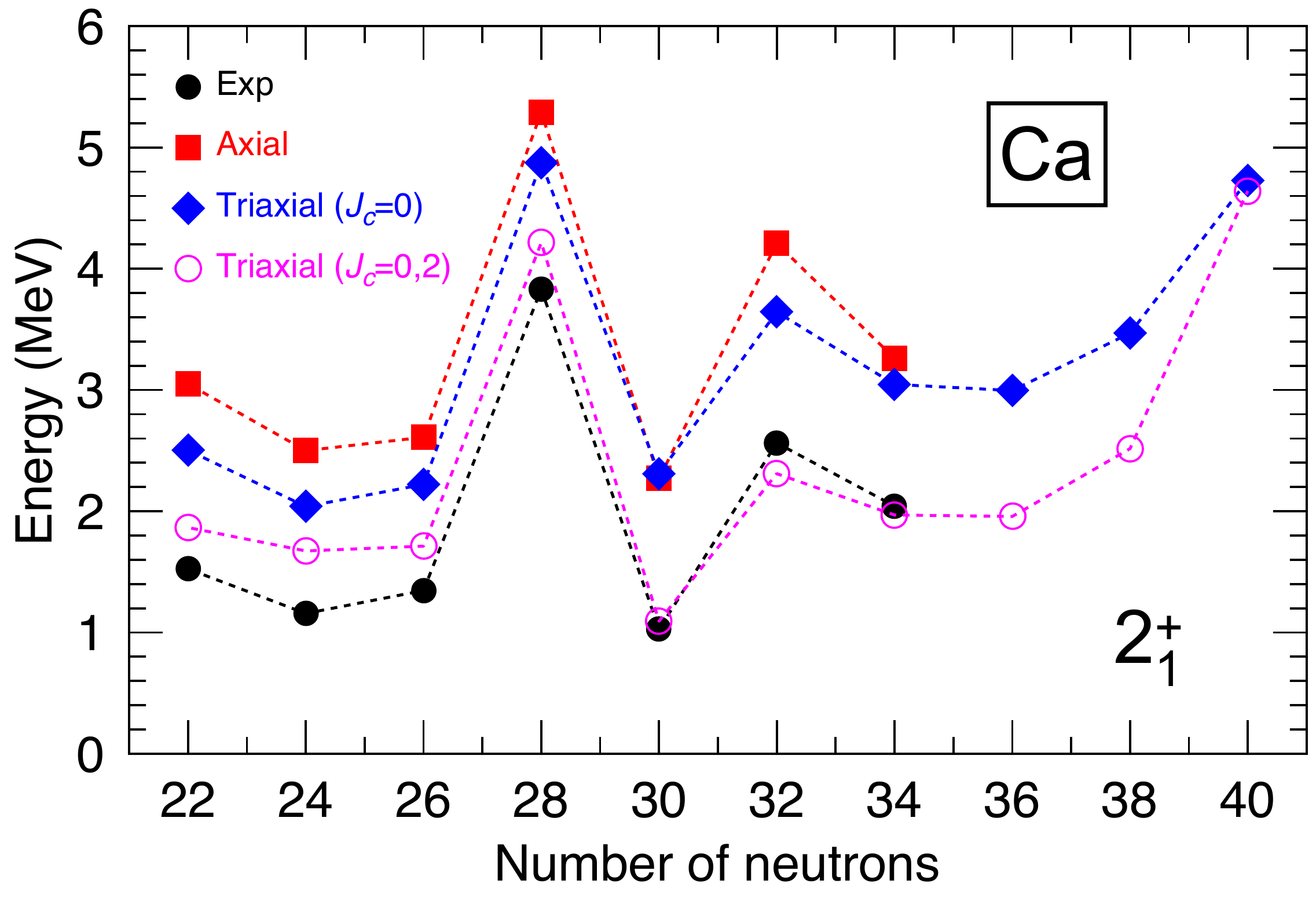}
\caption{Experimental (black bullets) and theoretical $2^{+}_{1}$ 
excitation energies for calcium isotopes. Calculations are performed 
with SCCM methods including particle number and angular momentum 
projection, and axial quadrupole shapes (red squares), axial and 
triaxial quadrupole shapes (blue diamonds) and axial and triaxial with 
and without cranking states (magenta circles). Gogny D1S is used.}
\label{Ca_chain}
\end{center}
\end{figure}
%%%%%%%%%%%%%%%%%%%%%%%%%%%%%%%%%%%%%% 
The potential new shell closures at $N=32$ and $N=34$ in neutron rich 
calcium isotopes were studied with axial SCCM including particle number 
projection~\cite{Rodriguez2007}. Again, a qualitative agreement with 
the experimental data was obtained for the $E(2^{+}_{1})$ and reduced 
transition probabilities in the Ca, Ti and Cr chains around this number 
of neutrons. These calculations forecast, apart from the well-known 
$N=28$ shell closure at $^{48}$Ca, a $N=32$ robust shell closure in 
$^{52}$Ca and a decrease of the $E(2^{+}_{1})$ excitation energy in 
$N=34$ (see Fig.~\ref{Ca_chain}). The latter prediction in the nucleus 
$^{54}$Ca was confirmed experimentally later on~\cite{Steppenbeck2013}. 
Furthermore, Fig.~\ref{Ca_chain} shows again the compression of the 
theoretical spectrum whenever triaxial and cranking states are included 
within the SCCM method (see also Fig.~\ref{Mg_cranking}). In the most 
sophisticated case, the theoretical results are in a good quantitative 
agreement with the experimental values and anticipate a $N=40$ 
(spherical harmonic oscillator) shell closure in $^{60}$Ca. In fact, 
the degradation of spherical harmonic oscillator shell closures has 
been also analyzed with SCCM methods with the Gogny interaction, e.g., 
in Zirconium isotopes ($Z=40$)~\cite{Rodriguez2011a,Paul2017} and in 
neutron rich $N=40$ isotones ($^{66}$Fe, $^{64}$Cr and, to a lesser 
extent, $^{62}$Ti)~\cite{Rodriguez2016a}. 

%%%%%%%%%%%%%%%%%%%%%%%%%%%%%%%%%%%%%%%
\begin{figure}
\begin{center}
\includegraphics[width=0.5\textwidth]{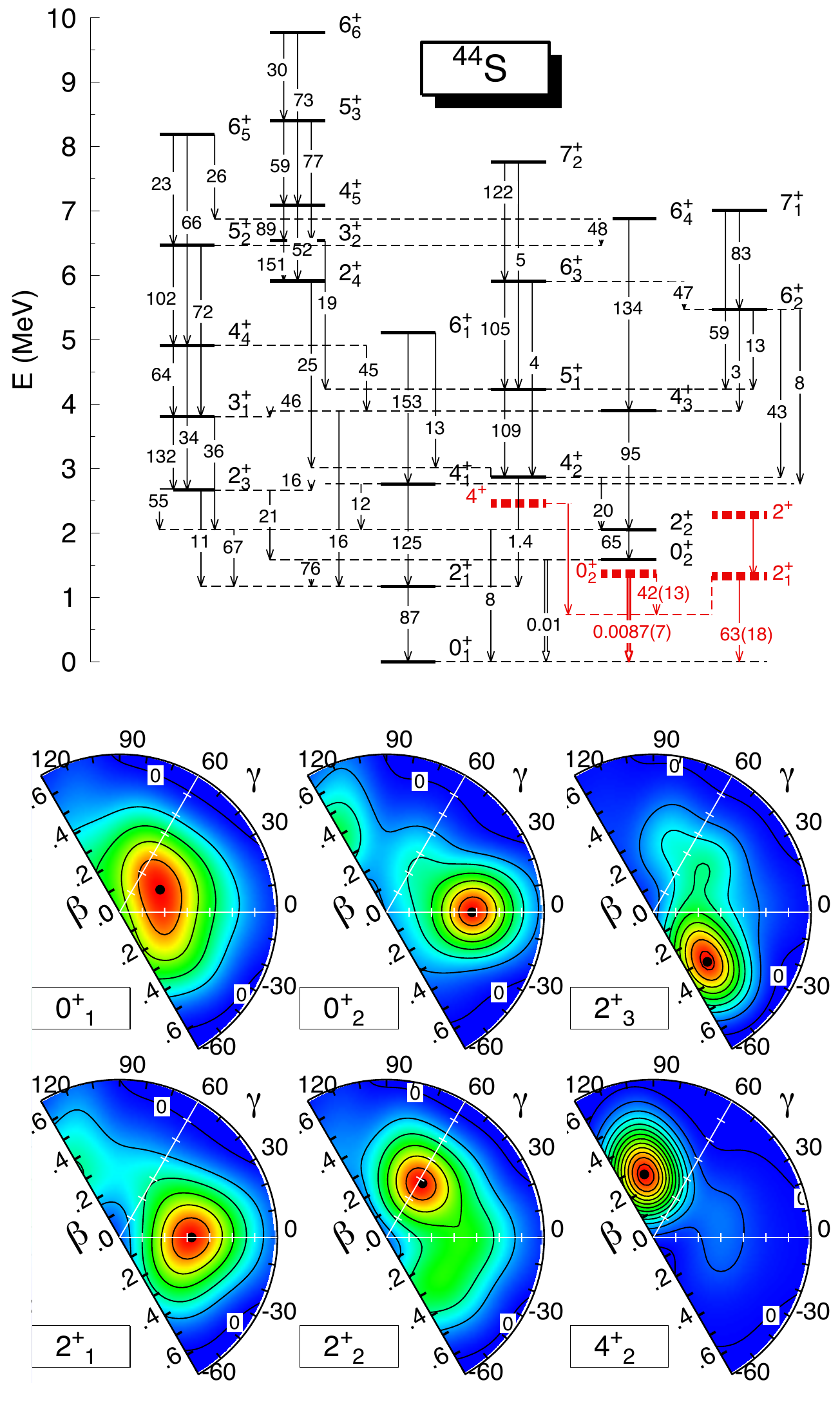}
\caption{Top panel:  Excitation energies, $B(E2)$ (in e$^{2}$fm$^{4}$) 
and $\rho^{2}(E0)$ for $^{44}$S calculated with a SCCM method with 
Gogny D1S that includes static and cranking intrinsic states in the 
$(\beta_{2},\gamma)$ plane. Experimental data are shown as thick dashed 
lines. Bottom panel: Collective wave functions for the selected states 
shown in the legend. Figure adapted from Ref.~\cite{Egido2016}.}
\label{44S_cranking}
\end{center}
\end{figure}
%%%%%%%%%%%%%%%%%%%%%%%%%%%%%%%%%%%%%% 
Very recently, SCCM calculations including three sextants in the 
$(\beta_{2},\gamma)$ plane and states with different cranking frequency 
have been performed to study the structure of $^{44}$S~\cite{Egido2016} 
and $^{42}$Si~\cite{Egido2016a}. In both cases deformed configurations 
have been found in the ground state, proving the degradation of $N=28$ 
magic number in these neutron rich nuclei. In addition, the inclusion 
of cranking states in the three sextants allows, on the one hand, an 
outstanding agreement with the experimental data, and, on the other 
hand, the description of isomeric states related to aligned states 
produced by rotations near the cranking 
axis~\cite{Egido2016,Egido2016a}. This is represented in 
Fig.~\ref{44S_cranking} as an example. In the top panel, the 
theoretical spectrum shows a ground state band and a first excited band 
with $\Delta J=2$ and built on top of $0^{+}_{1}$ and $0^{+}_{2}$ 
states respectively. Additionally, three $\Delta J=1$ bands are 
obtained on top of $2^{+}_{3}$, $2^{+}_{4}$ and $4^{+}_{2}$ states. The 
band-head of the latter is very close in energy to the $4^{+}_{1}$ 
state but it has a very different character. This result is in a very 
good agreement with the most recent experimental data plotted in red. 
In the bottom panel of Fig.~\ref{44S_cranking} the collective wave 
functions of the relevant states are shown to give a more detailed 
physical insight. Here, the ground state is found to be deformed and 
have a significant shape mixing along the triaxial degree of freedom. 
In addition, the $2^{+}_{1}$ and $0^{+}_{2}$ states are axial prolate 
deformed. On the whole, this is a fingerprint of the degradation of 
$N=28$ shell closure in this nucleus. A similar result was already 
obtained with Gogny SCCM triaxial calculations without 
cranking~\cite{Rodriguez2011}. However, those time-reversal symmetry 
conserving calculations could not reproduce the low-lying $4^{+}_{2}$ 
state almost degenerated with the $4^{+}_{1}$ state since it 
corresponds to a less collective state peaked at a deformation close to 
the intrinsic (cranking) rotation axis. Thus, this state is considered 
as an aligned state with $K=4$ that is only obtained thanks to the 
inclusion of the cranking degree of freedom.
%%%%%%%%%%%%%%%%%%%%%%%%%%
\subsubsection{Lepton number violating processes}
%%%%%%%%%%%%%%%%%%%%%%%%%%
Symmetry conserving configuration mixing methods with Gogny EDFs can be 
also used to evaluate nuclear matrix elements of lepton number 
violating processes such as neutrinoless double beta decay 
($0\nu\beta\beta$)~\cite{Rodriguez2010a,Rodriguez2011b,Rodriguez2013,Vaquero2013,Menendez2014}. 
This decay is the most promising process to disentangle the Majorana 
nature of the neutrino, its effective mass and the mass 
hierarchy~\cite{Avignone2008,Engel2017}. Here, an even-even nucleus 
decays into its even-even neighbor with two protons (neutrons) more 
(less). This process is energetically possible and is not hindered by 
single-beta decay in only very few cases across the nuclear chart. 
$2\nu\beta\beta$ decays, where two neutrinos are also emitted in the 
process, have been already observed with very long half-lives 
($~10^{19-21}$ years), but the neutrinoless channel, the relevant one 
to study those properties of the neutrinos, has not been experimentally 
detected yet. In the most plausible scenario in which a light Majorana 
neutrino is exchanged in the $0\nu\beta\beta$ decay, the inverse of the 
half-life is~\cite{Avignone2008}:
\begin{equation}
\left[T_{1/2}^{0\nu}(0_{i}^{+}\rightarrow0_{f}^{+})\right]^{-1}=G_{0\nu}|M^{0\nu}|^{2}\left(\frac{\langle
    m_{\beta\beta}\rangle}{m_{e}}\right)^{2} 
\label{half_life}
\end{equation}
where $G_{0\nu}$ is a kinematic phase space factor, $m_{e}$ is the 
electron mass and $\langle m_{\beta\beta}\rangle$ is the effective 
Majorana neutrino mass. The nuclear physics part of this process is 
encoded in the so-called nuclear matrix element (NME), $M^{0\nu}$. 
Several nuclear structure methods have been used so far to calculate 
these NMEs (see Ref.~\cite{Engel2017} and references therein), namely, 
large scale shell model (SM), quasiparticle random phase approximation 
(QRPA), interacting boson model (IBM), projected 
Hartree-Fock-Bogoliubov (PHFB) and energy density functional methods 
(EDF). 

In most of the cases, these NMEs are computed as the matrix elements 
between the $0^{+}$ ground states of the mother and granddaughter 
nuclei connected by the corresponding transition operator. This is the 
so-called closure approximation since it by-passes the virtual 
transitions to the intermediate states in the odd-odd nucleus. 
Therefore, SCCM methods with Gogny EDF, including particle number and 
angular momentum restoration, and axial quadrupole shape mixing, can be 
used to compute the ground states of the two neighboring nuclei 
involved in the process, and then, the corresponding transition matrix 
elements, as it is regularly done in electromagnetic decays. 

The transition operator of this electroweak process is obtained 
assuming additionally a non-relativistic approach. Thus, the NMEs is 
given by the sum of Fermi (F), Gamow-Teller (GT) and tensor (T) 
terms~\cite{Avignone2008}, although the latter are normally neglected 
in Gogny EDF applications:
\begin{equation}
M^{0\nu}=-\left(\frac{g_{V}}{g_{A}}\right)^{2}M^{0\nu}_{F} +M^{0\nu}_{GT}
\label{MMM}
\end{equation}
with $g_{V}=1$ and $g_{A}=1.25$ being the vector and axial coupling constants, and:
\begin{equation}
M^{0\nu}_{F/GT}=\langle0^{+}_{f}|\hat{M}^{0\nu}_{F/GT}|0^{+}_{i}\rangle\label{NME_EV}
\end{equation} 
with:
\begin{eqnarray}
\label{eq:1}
\hat{M}^{0\nu}_{F}& = &\left(\frac{g_{A}}{g_{V}}\right)^{2}\sum_{i<j} \hat{V}_{F}(r_{ij})\hat{\tau}^{(i)}_{-}\hat{\tau}^{(j)}_{-}, \\
\hat{M}^{0\nu}_{GT} & = & \sum_{i<j} \hat{V}_{GT}(r_{ij})(\hat{\bm{\sigma}}^{(i)} \cdot\hat{\bm{\sigma}}^{(j)})\hat{\tau}^{(i)}_{-}\hat{\tau}^{(j)}_{-}     
\end{eqnarray}
In these expressions, $\hat{\tau}_{-}$ is the isospin ladder operator 
that changes neutrons into protons and $\hat{\bm{\sigma}}$ are the 
Pauli matrices acting on the spin part of the wave functions. The 
so-called neutrino potentials $\hat{V}_{F/GT}$ depend on the relative 
distance between two nucleons (see Refs.~\cite{Avignone2008,Engel2017} 
and references therein for more details). 

NMEs within the SCCM framework are computed in the following manner. 
Initial ($i$) and final ($f$) states correspond to the ground states 
given by Eq.~\ref{GCM_ansatz} with $(J=0,\sigma=1)$, or, in their 
respective natural basis (Eq.~\ref{GCM_ansatz2}):
\begin{equation}
|0^{+}_{i/f}\rangle=|\Psi^{0+}_{1,i/f}\rangle=\sum_{\Lambda_{i/f}}g^{0+}_{1,i/f}(\Lambda_{i/f})|\Lambda^{0+}_{i/f}\rangle
\label{GCM_ansatz3} 
\end{equation}

Then, each component of the NME is calculated using an expression 
similar to Eq.~\ref{GCM_Expec_Val}. Using the natural basis that 
corresponds to each nucleus, the NME is evaluated 
as~\cite{Rodriguez2010a}:
\begin{equation}
M^{0\nu}_{F/GT}=\sum_{\Lambda_{f}\Lambda_{i}}\sum_{\mathbf{q}_{f}\mathbf{q}_{i}}
(g^{0+}_{1}(\Lambda_{f}))^{*}\left(\frac{u^{0+}_{\Lambda_{f}}(\mathbf{q}_{f})}{\sqrt{n^{0+}_{\Lambda_{f}}}}\right)^{*}\bar{M}^{0\nu}_{F/GT}(\mathbf{q}_{f},\mathbf{q}_{i})\left(\frac{u^{0+}_{\Lambda_{i}}(\mathbf{q}_{i})}{\sqrt{n^{0+}_{\Lambda_{i}}}}\right)g^{0+}_{1}(\Lambda_{i})
\label{NME_final}
\end{equation}
Here, the quantity 
$\bar{M}^{0\nu}_{F/GT}(\mathbf{q}_{f},\mathbf{q}_{i})=\langle\Phi^{0+}_{f}(\mathbf{q}_{f})|\hat{M}^{0\nu}_{F/GT}|\Phi^{0+}_{i}(\mathbf{q}_{i})\rangle$, 
where $|\Phi^{0+}_{i/f}(\mathbf{q}_{i/f})\rangle$ are projected states 
defined in Eq.~\ref{PPNAMP_state}, gives the dependence of the NMEs 
with the collective coordinates of the mother and daughter whenever is 
properly normalized:
\begin{equation}
M^{0\nu}_{F/GT}(\mathbf{q}_{f},\mathbf{q}_{i})=\frac{\bar{M}^{0\nu}_{F/GT}(\mathbf{q}_{f},\mathbf{q}_{i})}{\sqrt{\langle\Phi^{0+}_{i}(\mathbf{q}_{i})|\Phi^{0+}_{i}(\mathbf{q}_{i})\rangle\langle\Phi^{0+}_{f}(\mathbf{q}_{f})|\Phi^{0+}_{f}(\mathbf{q}_{f})\rangle}} 
\label{NME_qfqi}
\end{equation}
Therefore, the value of the NME can be understood as the convolution of 
the collective wave functions of initial and final states with the 
intensity of the NMEs as a function of the collective coordinates given 
in the previous expression. 

%%%%%%%%%%%%%%%%%%%%%%%%%%%%%%%%%%%%%%%
\begin{figure}
\begin{center}
\includegraphics[width=0.5\textwidth]{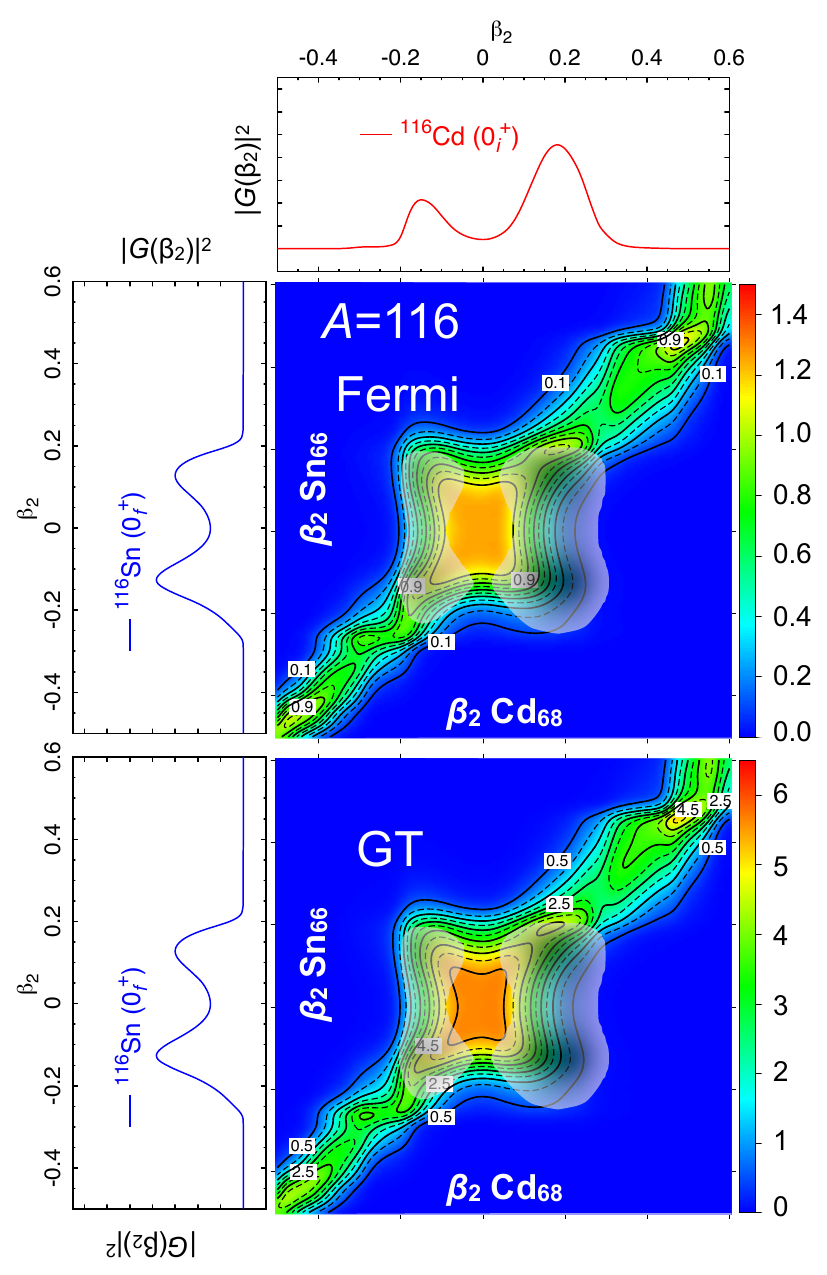}
\caption{Contour plots: Strength of the Fermi and Gamow-Teller parts of 
$0\nu\beta\beta$ nuclear matrix elements as a function of the axial 
quadrupole deformation of the initial $^{116}$Cd and final $^{116}$Sn 
nuclei. Ground state collective wave functions for $^{116}$Cd (top) and 
$^{116}$Sn (left) nuclei. The shaded area shows the product of the two 
collective wave functions. Gogny D1S is used.}
\label{116Cd_0nu_def}
\end{center}
\end{figure}
%%%%%%%%%%%%%%%%%%%%%%%%%%%%%%%%%%%%%%  
SCCM methods with Gogny EDFs are very well suited to study nuclear 
structure effects in the $0\nu\beta\beta$ NMEs. All of these 
calculations assume particle number and angular momentum restoration, 
and axial quadrupole shape 
mixing~\cite{Rodriguez2010,Rodriguez2011b,Rodriguez2013,Beller2013,Menendez2014}. 
In Ref.~\cite{Vaquero2013} the fluctuations in the number of particles 
are also added as a collective coordinate. Thus, the role of 
deformation, pairing correlations and shell effects in the NMEs have 
been studied both in the actual candidates to detect $0\nu\beta\beta$ 
decays~\cite{Rodriguez2010,Rodriguez2011b,Vaquero2013} as well as in 
virtual decays in isotopic chains~\cite{Rodriguez2013,Menendez2014}. 
The latter studies are important to identify the origin of the 
discrepancies found in the NMEs computed with different many-body 
methods, in particular, between SCCM and SM calculations.  

In general, SCCM calculations have shown that the strength of the NMEs 
is very much reduced when the deformations of the initial and final 
states are different. In fact, the decay is more probable between 
states with intrinsic spherical shapes. This is illustrated in 
Fig.~\ref{116Cd_0nu_def} where the strength of the Fermi and GT parts 
in the decay of the nucleus $^{116}$Cd are plotted as a function of the 
axial quadrupole deformation of the mother ($^{116}$Cd) and of the 
granddaughter ($^{116}$Sn) (Eq.~\ref{NME_qfqi}). The largest values 
correspond to the diagonal part of the figure, and, more specifically, 
to the area around the spherical shape. However, the final value of the 
NME considers the regions of these plots explored by the ground state 
collective wave functions of the initial and final nuclei. These are 
plotted on the top and on the left of Fig.~\ref{116Cd_0nu_def} and the 
shaded areas correspond to the product of the two collective wave 
functions. 

%%%%%%%%%%%%%%%%%%%%%%%%%%%%%%%%%%%%%%%
\begin{figure}
\begin{center}
\includegraphics[width=0.5\textwidth]{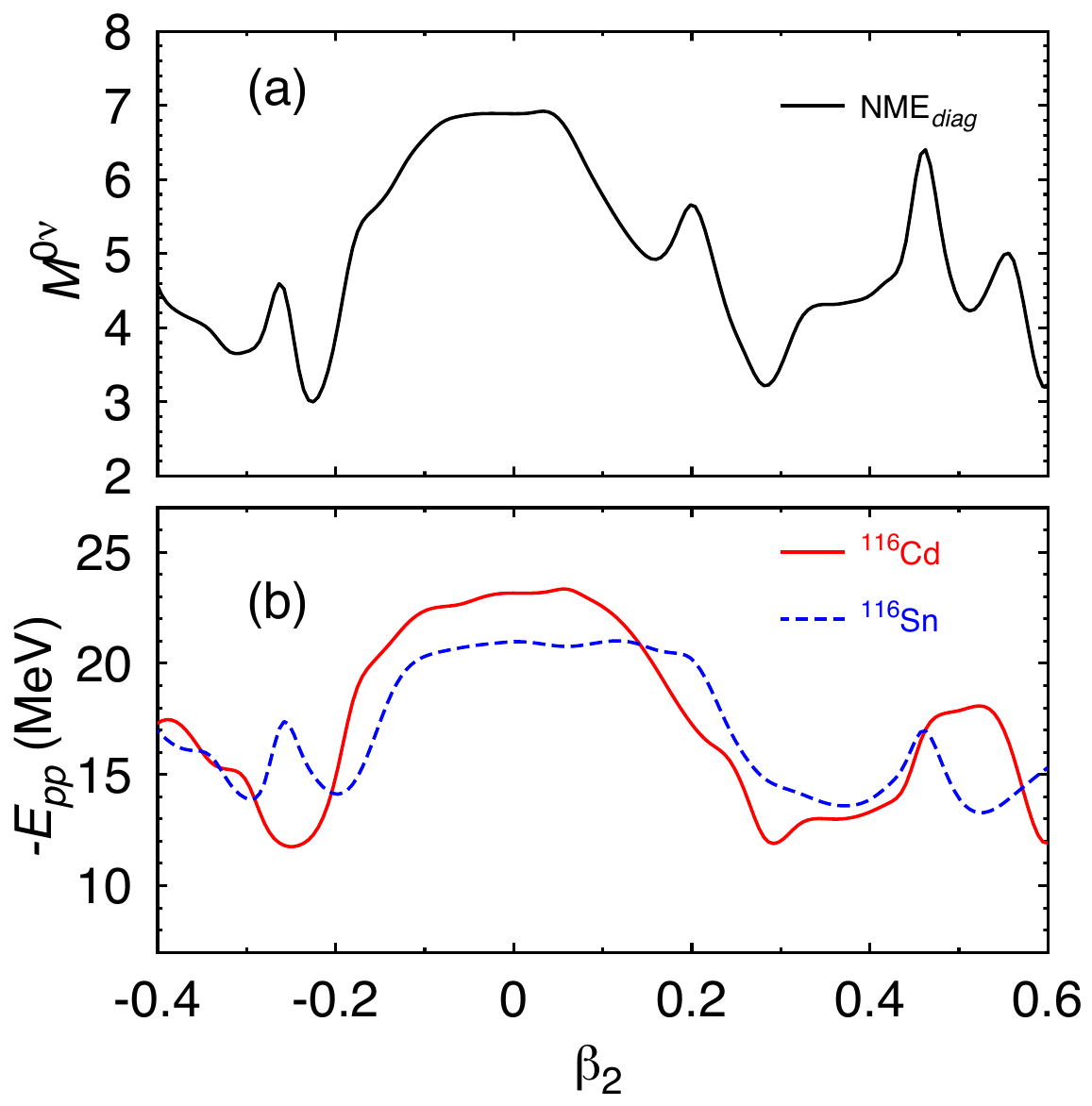}
\caption{(a) Diagonal part of the $0\nu\beta\beta$ nuclear matrix elements, and, (b) pairing energies for $^{116}$Cd and $^{116}$Sn nuclei, as a function of the axial quadrupole deformation. Gogny D1S is used.}
\label{116Cd_0nu_pair}
\end{center}
\end{figure}
%%%%%%%%%%%%%%%%%%%%%%%%%%%%%%%%%%%%%%  
The other degree of freedom playing a major role in the 
$0\nu\beta\beta$ NMEs is the nuclear pairing. In 
Fig.~\ref{116Cd_0nu_pair} the pairing energies of the initial and final 
states as a function of the quadrupole deformation are represented as 
well as the diagonal part of the NMEs for the decay of $^{116}$Cd. 
Here, a clear correlation between these quantities are observed and the 
largest values of the NMEs are obtained at deformations where larger 
pairing energies are found. This effect was explicitly shown in 
Ref.~\cite{Vaquero2013}, i.e., the increase of the NMEs with increasing 
the pairing content (proton-proton and neutron-neutron pairing) of the 
initial and final states. However, the actual values of the NMEs could 
be reduced by including two degrees of freedom that destroy 
like-particle pairing correlations, namely, higher seniority components 
(see, e.g., Ref.~\cite{Menendez2014}) and proton-neutron pairing (see, 
e.g., Ref.~\cite{Hinohara2014}). This kind of states have not been 
included so far in SCCM Gogny EDF calculations.

Finally, the same formalism can be applied to the evaluation of NMEs of 
resonant neutrinoless double electron capture ($0\nu 
ee$)~\cite{Bernabeu1983}. In this case, the SCCM calculations of the 
NMEs in the possible few candidates to detect this lepton number 
violating process predict much longer half-lives (of the order of 
$10^{31}$ years) than the $0\nu\beta\beta$ 
ones~\cite{Smorra2012,Rodriguez2012}.
%%%%%%%%%%%%%%%%%%%%%%%%%%
\subsubsection{Occupation numbers}
%%%%%%%%%%%%%%%%%%%%%%%%
Nuclear shell model (SM) and EDF methods are the most widely used 
frameworks to describe microscopically the nuclear structure. Thus, 
searching for connections between these formalisms are very useful due 
to their complementarity. SM works in the laboratory frame, 
diagonalizing effective nuclear interactions defined in valence spaces 
-normally assuming a core- and using Slater determinants -built upon a 
spherical mean-field- as the basis of the many-body Hilbert 
space~\cite{Caurier2005}. On the other hand, EDF methods are defined in 
an intrinsic frame, i.e., tend to break most of the symmetries of the 
interaction at the mean-field level. Furthermore, these are no-core 
calculations and the number of major harmonic oscillator shells 
included is larger than in SM approaches. These aspects make SM and EDF 
states difficult to connect. Nevertheless, single-particle energies in 
the deformed basis with Nilsson-like plots are routinely found in EDF 
calculations to understand qualitatively the orbits that play a role 
for a given nucleus. In some cases, relevant deformed mean-field states 
have been studied in terms of the particle-hole structure in a 
spherical basis~\cite{Caurier1995,Egido2004}.

A more quantitative way of analyzing the underlying shell structure of 
individual nuclear states within the SCCM framework has been recently 
proposed in Ref.~\cite{Rodriguez2016a}. It is based on computing the 
number of particles contained in each self-consistent spherical 
mean-field orbit when the states are defined with the SCCM ansatz 
(Eq.~\ref{GCM_ansatz}). These occupation numbers are not 
observables~\cite{Duguet2012,Duguet2015} and are model dependent. 
However, they can provide a qualitative comparison of the internal 
structure of the SCCM states with SM states. On the other hand, they 
can also serve as a guidance for defining SM valence spaces because 
they reveal the importance of each spherical orbit in the description 
of, for example, rotational bands.

To calculate the occupation numbers of spherical orbits one needs to 
define first those single-particle levels. One reasonable choice is the 
canonical basis of the spherically-symmetric self-consistent mean-field 
solution of the nucleus under study. Thus, the operator associated to 
the number of particles occupying a given spherical orbit, $\alpha$, 
defined by the quantum numbers $(n_{\alpha}l_{\alpha}j_{\alpha})$ is:
\begin{equation}
\hat{n}_{\alpha}=\sum_{m_{j_{\alpha}}}a^{\dagger}_{n_{\alpha}l_{\alpha}j_{\alpha}m_{j_{\alpha}}}a_{n_{\alpha}l_{\alpha}j_{\alpha}m_{j_{\alpha}}}
\label{occ_num_op}
\end{equation}
These creation and annihilation single-particle operators 
$(a^{\dagger}_{\alpha},a_{\alpha})$ are obtained from the 
diagonalization of the density-matrix, $\rho^{sph}_{ab}$~\cite{Ring1980}:
\begin{equation}
\rho^{sph}_{ab}=\langle\phi^{sph}|c^{\dagger}_{b}c_{a}|\phi^{sph}\rangle
\label{density_matrix_c}
\end{equation}
where $|\phi^{sph}\rangle$ is the spherical quasiparticle vacuum and 
$(c^{\dagger}_{a},c_{a})$ are creation and annihilation single-particle 
operators that correspond to the arbitrary working basis used to define 
the HFB transformation~\cite{Ring1980}. As it is done in previous 
sections, the occupation numbers are now calculated as the expectation 
values of the operator defined in Eq.~\ref{occ_num_op} between SCCM 
states, i.e., substituting $\hat{O}=\hat{n}_{\alpha}$ in 
Eq.~\ref{GCM_Expec_Val}. Moreover, the dependence of the occupation 
numbers along the collective coordinates can be computed as:
\begin{equation}
n^{J_{\sigma}\pi}_{\alpha}(\mathbf{q})=\frac{\langle\Phi^{J_{\sigma}\pi} (\mathbf{q})|\hat{n}_{\alpha}|\Phi^{J_{\sigma}\pi}(\mathbf{q})\rangle}{\langle\Phi^{J_{\sigma}\pi} (\mathbf{q})|\Phi^{J_{\sigma}\pi}(\mathbf{q})\rangle}
\label{occ_num_q}
\end{equation}

%%%%%%%%%%%%%%%%%%%%%%%%%%%%%%%%%%%%%%%
\begin{figure}
\begin{center}
\includegraphics[width=\textwidth]{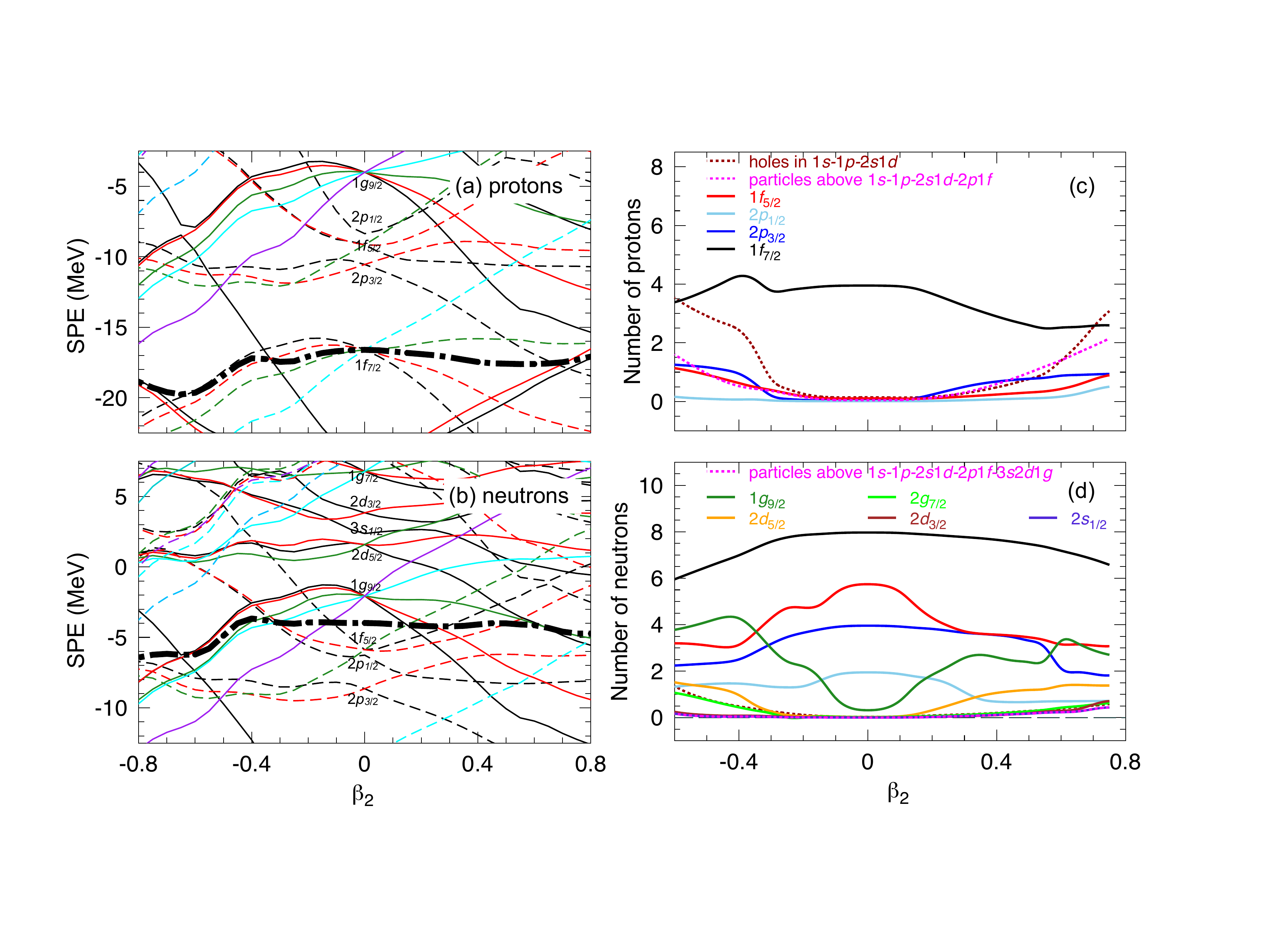}
\caption{(a)-(b) Single-particle energies and (c)-(d) occupation 
numbers of spherical orbits as a function of the axial quadrupole 
deformation for protons and neutrons calculated for $^{64}$Cr with the 
Gogny D1S interaction. Figure adapted from Ref.~\cite{Rodriguez2016a}}
\label{Nilsson_64Cr}
\end{center}
\end{figure}
%%%%%%%%%%%%%%%%%%%%%%%%%%%%%%%%%%%%%%   
Thus, in the particular case where the collective coordinate is 
$\mathbf{q}=\beta_{2}$, Eq.~\ref{occ_num_q} allows a quantitative 
evaluation of the number of particles occupying the different orbits 
along the quadrupole moment, improving the qualitative description that 
is usually given with the analysis of Nilsson-like plots. This is shown 
in Fig.~\ref{Nilsson_64Cr} for the nucleus $^{64}$Cr ($Z=24$, $N=40$) 
taken as an example of the performance of the method. In the left 
panel, the HFB single-particle energies and the Fermi level are 
represented for protons and neutrons. We observe that, around the 
spherical point, protons (neutrons) are expected to occupy up to the 
$f_{7/2}$ ($pf$-shell) orbits. Once the quadrupole deformation 
increases, some neutron levels coming from the $g_{9/2}$ and the 
$pf$-shell orbits cross the Fermi level, as well as one proton level 
coming from the $sd$-shell. The quantitative counterpart of this 
analysis is shown in the right panel. Here, we observe that the neutron 
$g_{9/2}$ orbit is clearly being filled in and the $f_{5/2}$ orbit is 
being emptied when the deformation departures from the spherical point. 

%%%%%%%%%%%%%%%%%%%%%%%%%%%%%%%%%%%%%%%
\begin{figure}
\begin{center}
\includegraphics[width=0.5\textwidth]{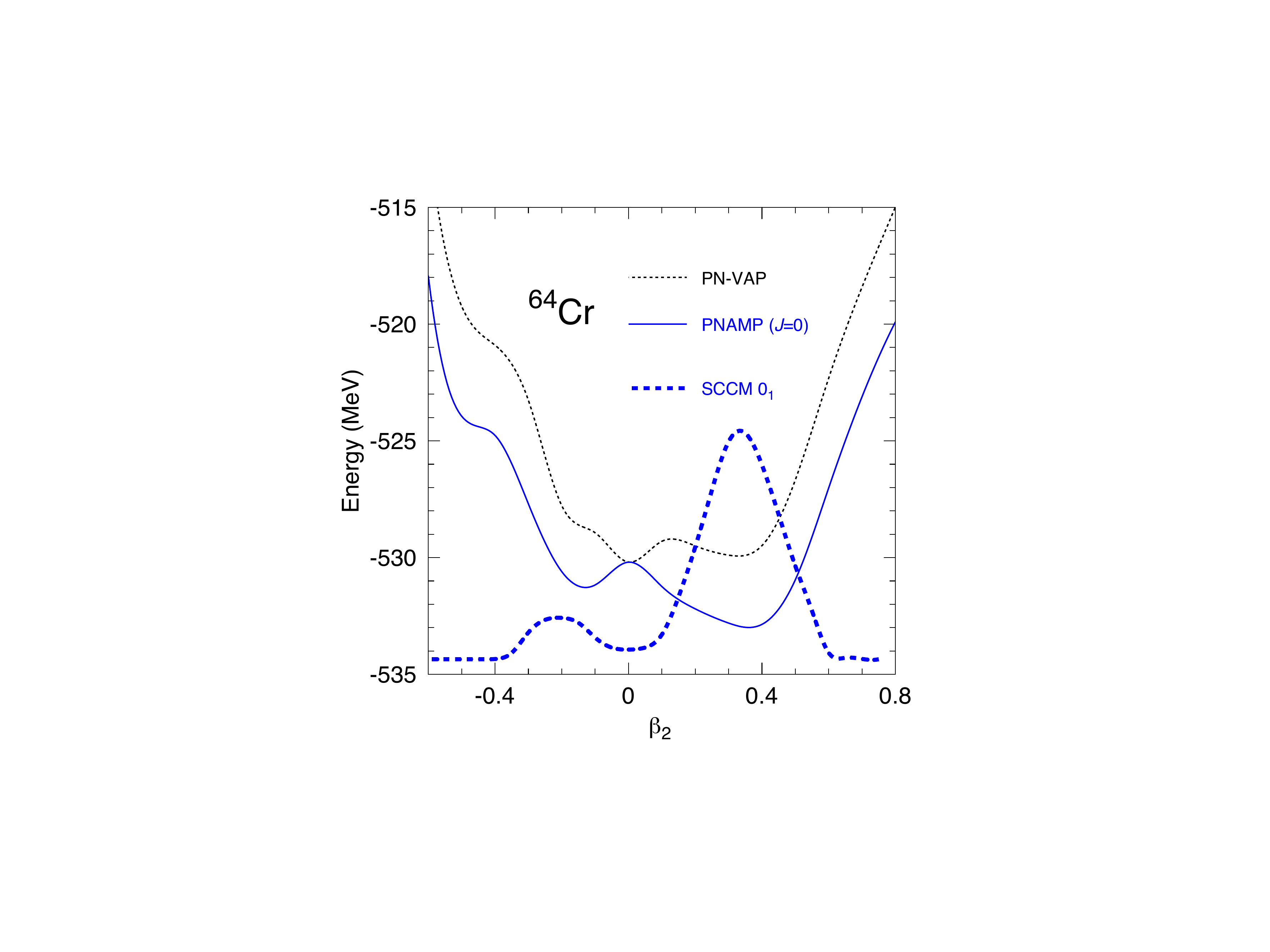}
\caption{Potential energy surfaces as a function of the axial 
quadrupole deformation calculated for the nucleus $^{64}$Cr with PN-VAP 
(dashed), and particle-number and angular momentum, $J=0$, projection 
(continuous) approaches. With the thick-dashed line, the ground state 
collective wave-function is shown. Gogny D1S is used. Figure adapted 
from Ref.~\cite{Rodriguez2016a}.}
\label{64Cr_PES_CWF}
\end{center}
\end{figure}
\begin{figure}
\begin{center}
\includegraphics[width=\textwidth]{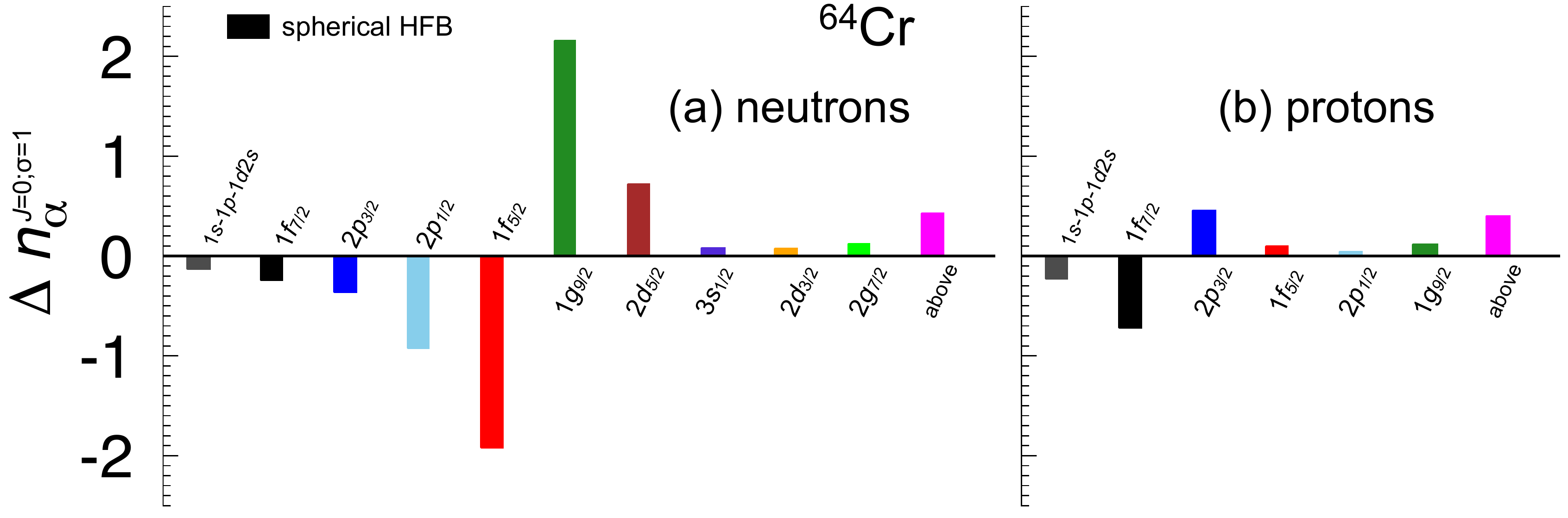}
\caption{
Difference between the occupation numbers of spherical orbits for the 
ground state of $^{64}$Cr calculated with the SCCM method and the 
occupation numbers of those orbits in the spherical HFB configuration. 
Figure adapted from Ref.~\cite{Rodriguez2016a}.}
\label{64Cr_occ}
\end{center}
\end{figure}

%%%%%%%%%%%%%%%%%%%%%%%%%%%%%%%%%%%%%%  
The final values for the occupation numbers are given by the evaluation 
of Eq.~\ref{GCM_Expec_Val} that includes the effect of the shape 
mixing. In Fig.~\ref{64Cr_PES_CWF} the particle number variation after 
projection (PN-VAP), the particle number and angular momentum ($J=0$) 
projection potential energy surfaces, and the ground state collective 
wave function are plotted. We see that, even though the $N=40$ 
spherical harmonic oscillator shell closure produces an absolute 
minimum in the PN-VAP PES, the angular momentum restoration and shape 
mixing produce that the deformed prolate configuration becomes the 
ground state. The occupation numbers computed in the ground state are 
represented in Fig.~\ref{64Cr_occ}. To better visualize the relevant 
orbits needed to describe this deformed state, the difference between 
the occupation numbers provided by the SCCM method and the occupancies 
given by the HFB solution at the spherical point are plotted. Thus:
\begin{equation}
\Delta n^{J;\sigma}_{\alpha}=\langle\Phi^{J;\sigma}|\hat{n}_{\alpha}|\Phi^{J;\sigma}\rangle-\langle\phi^{sph}|\hat{n}_{\alpha}|\phi^{sph}\rangle
\end{equation}
Positive (negative) values of $\Delta n^{J;\sigma}_{\alpha}$ mean 
particles (holes) in a given level $\alpha$ with respect to the filling 
in the spherical HFB configuration. 

%%%%%%%%%%%%%%%%%%%%%%%%%%%%%%%%%%%%%%%
%%%%%%%%%%%%%%%%%%%%%%%%%%%%%%%%%%%%%%
We observe that orbits below $1f_{7/2}$ are almost fully occupied and 
orbits above the neutron $sdg$ (proton $1g_{9/2}$) orbits have small 
(although non-negligible) occupancies. The neutron $pf$ ($1g_{9/2}$ and 
$2d_{5/2}$) orbits are depopulated (populated). Most of the 
depopulation of the neutron $pf$ shell comes from the $1f_{5/2}$ and 
$2p_{1/2}$ levels and, to a lesser extent, the $2p_{3/2}$ and 
$1f_{7/2}$ orbits. It is also interesting to see that $3s_{1/2}$, 
$2d_{3/2}$ and $2g_{7/2}$ are not very much populated in the ground 
state. Furthermore, the $1f_{7/2}$ orbit does not contain anymore the 
four valence protons as in the spherical case but it accommodates 
roughly one proton less, while the $2p_{3/2}$ and $2f_{5/2}$ starts to 
be slightly occupied. This analysis is in a rather good agreement with 
SM results and points out to the role of the $g_{9/2}$ neutron orbit in 
the onset of deformation in neutron rich $N=40$ isotopes (see 
Ref.~\cite{Rodriguez2016a} and references therein). 

In summary, this method is a perfect tool to interpret the results of 
SCCM methods not only in terms of collective variables (with the 
collective wave functions) but also to study the underlying shell 
structure. Additionally, it allows a closer comparison with SM results, 
and, indirectly, with nucleon transfer/removal reaction experiments as 
it has been done in the case of $^{76}$Ge and 
$^{76}$Se~\cite{Rodriguez2017}.

%% file: Summary.tex
\section{Summary and future prospects}

The success of the Gogny force in describing many different facets of 
nuclear structure along the years is a clear indication that it still 
represents a rock solid alternative for a microscopic description of 
the physics of the atomic nucleus in all regions of the nuclide chart 
with special emphasis on the interpretation of experimental findings.

For the years to come, one might think of several improvements to be 
made to the Gogny force. Some of them are minor, like including a 
finite range spin--orbit term or increasing the number of Gaussians. 
Some others require more work, like finding an operator based finite 
range replacement to the density dependent term, exploring the time-odd 
sector of the interaction or finding observables to fit a tensor term. 
The fitting protocol can also be improved as to take into account 
observables sensitive to proton-neutron pairing. Also, inclusion of 
beyond mean field effects in the fitting protocol should be a priority 
in the years to come. The exchange of ideas with the \textit{ab initio} 
community could also seed some light on new forms of effective energy 
density functionals. Work in many of these considerations is already 
underway and hopefully we will witness some advances soon.

Concerning the improvement of the quantum mechanic many body methods to 
be used along with the Gogny force there are again minor developments 
that mostly involve the writing of new computer codes breaking all 
possible symmetries. In this category we could think of a triaxial QRPA 
code with the ability to compute also collective inertias exactly. New 
developments expanding beyond the use of a HO basis  will be helpful to 
handle exotic shapes like to ones found in fission, the ones required 
to describe super-heavy elements or the accuracy needed for nuclear 
astrophysics. Also a proper handling of the continuum would be highly 
welcome to describe near drip-line physics and reactions from a more 
fundamental perspective. 

Among the major improvements, a sound implementation of configuration 
interaction methods based on deformed intrinsic states which are 
subsequently projected to have the proper quantum numbers would be the 
main development. It will allow to treat consistently the physics of 
all kind of nuclei irrespective of their mass number and the collective 
character of the excitation under study. The challenge in this case is 
how to implement those methods for density dependent effective 
interactions. This will allow, among many other applications, to deal 
with odd mass systems at the same level of sophistication and 
applicability as the even-even case. Another major breakthrough would 
be the use of the Gogny force to analyze the time evolution of nuclear 
reactions using time dependent formalisms like TDHFB or TDGCM.

%% file: MatElem.tex
\section{Matrix elements of the Gogny force in the harmonic oscillator basis: basic ideas}
\label{App:A}

The evaluation of matrix elements in mean field calculations with phenomenological effective
interactions is of great relevance given the huge amount of them required, specially for
heavy nuclei calculations. Therefore, it is highly desirable to find procedures to
compute them in the quickest and most accurate possible way. In general, the matrix
elements of two body operators are given as double three dimensional integrals (sixfold
integrals) and different strategies have been used to reduce the complexity of their
evaluation. A popular one in nuclear structure is to use contact interactions
(as in the generic Skyrme case) to carry out analytically three of the integrals
in order to reduce the sixfold integrals to threefold ones. Another strategy is
to use interactions that factorize as products of the integration variables. This is
the approach taken with the Gogny force, where the finite range central potential
is given in terms of a Gaussian form factor
$$
e^{-(\vec{r}_{1}-\vec{r}_{2})^{2}/\mu^{2}} = \prod_{i=1}^{3}e^{-(x_{i\,1}-x_{i\,2})^{2}/\mu^{2}}
$$
By using basis states with the same factorization property it is possible to
reduce the sixfold integration to the product of twofold integrals reducing
thereby the amount of integrals to compute. The choice of a Gaussian form
factor is advantageous for two additional reasons: First, it is intimately connected
with the harmonic oscillator basis, as the Hermite polynomials are orthogonal
with respect to a Gaussian weight and their generating functions are also
Gaussians. As a consequence, the evaluation of the matrix elements in this
basis can be carried out analytically. Second, it is possible to expand many smooth functions as a 
linear combination of tempered Gaussians (i.e. Gaussians with different
widths) as it is common practice with the Coulomb and Yukawa potentials \cite{Quentin1972,Gir83}
expanding enormously the range of applicability of the method. 

An additional advantage of the harmonic oscillator basis is that, due to a nice
property, it is possible to reduce the number of matrix elements to be computed
in one dimension from $O(N^{2})$ to $O(N)$ where $N$ is the number of HO states
\cite{TALMAN1970273,Gog75,Parrish2013}.

For all the reasons above, calculations with the Gogny force are carried out
in a harmonic oscillator basis. Depending on the underlying self-consistent symmetries of
the calculation, spherical, axially symmetric or triaxial harmonic oscillator
basis are used. The three representations are connected by means of unitary
transformations \cite{TALMAN1970273,CHASMAN1967401} and it is straightforward to go
from one to another. In the following, we will illustrate the calculation
of the matrix elements of a Gaussian form factor in the triaxial case, that
is, assuming that the HO wave functions are factorized as the product of
three one-dimensional HO states along each spatial direction $x$, $y$ and $z$
\begin{equation}
	\varphi_{\alpha}(\vec{r}) = \varphi_{a_{x}}(x)\varphi_{a_{y}}(y)\varphi_{a_{z}}(z)
\end{equation}
The other two cases corresponding to spherical \cite{Gog75} ($r$, $\theta$ and $\varphi$ as
spatial coordinates )and axial \cite{You09} ($r_\perp$, $\varphi$ and $z$)
symmetries will not be considered here. The derivation presented here
follows that of Ref \cite{Gir83} but taking into account the caveats discussed in
\cite{Egido1997} concerning the accuracy in the numerical evaluation of 
alternating sums.

The goal is to compute the matrix element of the
two-body Gaussian interaction of the central part
\begin{equation}
I^{BB}_{abcd} = \int d\vec{r}_1 d\vec{r}_2 \varphi_a^*(\vec{r}_1)\varphi_b^*(\vec{r}_2)
e^{-\frac{(\vec{r}_1-\vec{r}_2)^2}{\mu^2}}
\varphi_c(\vec{r}_1)\varphi_d(\vec{r}_2)
\end{equation}
where $\varphi_\alpha(\vec{r})$ factorizes as $\varphi_{a_x} (x)\varphi_{a_y} (y)\varphi_{a_z} (z)$. 
The one-dimensional harmonic oscillator wave functions are given by the 
traditional expression in terms of the Hermite polynomials 
\begin{equation}
\varphi_{n}(x;b)=\frac{1}{\sqrt{\sqrt{\pi}2^{n}n!b}}e^{-\frac{x^{2}}{2b^{2}}}H_{n}(x/b)
\end{equation}
The generic matrix element of the Gaussian is then written as the product 
of three 1-D matrix elements
\begin{equation}
I_{abcd}^{BB}=\prod_{i=1}^3I_{a_ib_ic_id_i}^{(i)}
\end{equation}
where $I_{a_ib_ic_id_i}^{(i)}$ is the two-dimensional integral
\begin{equation}
I_{a_ib_ic_id_i}^{(i)}=\int dx_1dx_2\varphi _{{a_i}}(x_1)\varphi
_{{b_i}}(x_2)e^{-\frac{(x_1-x_2)^2}{\mu ^2}}\varphi _{{c_i}}(x_1)\varphi
_{{d_i}}(x_2)
\end{equation}
The strategy to compute this integral is to use the following property 
of the harmonic oscillator wave functions \cite{TALMAN1970273,Gog75,Parrish2013}
\begin{equation}
\varphi_{n_a}(x)\varphi_{n_c}(x)=\frac 1{b^{1/2}\pi ^{1/4}}\sum_{n_\mu
}\tau (n_an_cn_\mu )e^{-\frac{x^2}{2b^2}}\varphi _{n_\mu }(x)
\end{equation}
where the sum on $n_{\mu}$ contains a finite number of terms 
($|n_{a}-n_{c}|\leq n_{\mu}\leq n_{a}+n_{c}$)
and the geometric coefficient $\tau (n_an_cn_\mu )$ is given by
\begin{equation}
\tau(n_{a},n_{b},n_{\mu})=\frac{\left(n_{a!}n_{b}!n_{\mu}!\right)^{1/2}}{(s-n_{\mu})!(s-n_{a})!(s-n_{b})!}\Delta_{n_{a}+n_{b},n_{\mu}}
\end{equation}
with $s=\frac{1}{2}(n_{a}+n_{b}+n_{\mu})$ and 
\begin{equation}
\Delta_{n,m}=\frac{1}{2}\left(1+(-)^{n+m}\right)	
\end{equation}
Preservation of the parity symmetry (expressed by the $\Delta$ symbol) guarantees 
that in all the non-zero coefficients
the combinations in the denominator are integers and therefore the factorial
makes sense.
The result can be easily understood by noticing that, apart from the Gaussian
factors, the 1D HO oscillator wave functions are polynomials and the product
of two polynomials of degrees $n_1$ and $n_{2}$ can unambiguously be expressed 
as a linear combination of polynomials of degree $n_{1}+n_{2}$ at most.
Using the above result the previous integral is expressed as
\begin{equation}
I_{abcd}^{(i)}=\frac 1{b\sqrt{\pi }}\sum_{n_\mu n_\nu }\tau (n_an_cn_\mu
) J (n_bn_dn_\mu )
\end{equation}
where
\begin{equation}
J(n_{b},n_{d},n_{\mu})=\frac{1}{b^{1/2}\pi^{1/4}}
\int dx_{1}\varphi_{n_{\mu}}(x_{1})e^{-\frac{x_{1}^{2}}{2b^{2}}}
\int dx_{2}e^{-\frac{(x_{1}-x_{2})^{2}}{\mu^{2}}}\varphi_{n_{b}}(x_{2})
\varphi_{n_{d}}(x_{2})	
\end{equation}
To compute this set of coefficients there are essentially two alternatives:
(a) to perform a transformation to the center of mass $(x_{1}+x_{2})/2$ and
relative coordinate $x_{1}-x_{2}$ using Moshinsky coefficients or (b) to use
the Parseval theorem of Fourier transforms to write the integral in momentum space as
\begin{equation}
	\int d{k}F^{*}_{n_{\mu}}({k})\widetilde{G}_{n_{b}n_{d}}({k})
\end{equation}
where $F_{n_{\mu}}(k)$ is the Fourier transform of 
$\varphi_{n_{\mu}}(x_{1})e^{-\frac{x_{1}^{2}}{2b^{2}}}$
and $\widetilde{G}_{n_{b},n_{d}}(k)$ is the Fourier transform of a convolution. Using
these ideas we arrive to the final expression
\begin{equation}
	J(n_{b},n_{d},n_{\mu})=\frac{\sqrt{2\pi}}{\sqrt{b\sqrt{\pi}}}\int dkF_{n_{\mu}}^{*}(k)E(k)G_{n_{b}n_{d}}(k)
\end{equation}
where the corresponding quantities in momentum space are given by
\begin{equation}
F_{n_{\mu}}(k)=\mathcal{F}\left(\varphi_{n_{\mu}}(x_{1})e^{-\frac{x_{1}^{2}}{2b^{2}}}\right)={}
\frac{b}{\sqrt{\sqrt{\pi}2^{n_{\mu}}n_{\mu}!b}}\frac{(-i)^{n_{\mu}}}{\sqrt{2}}\left(kb\right)^{n_{\mu}}e^{-(kb)^{2}/4}	
\end{equation}
for the Fourier transform of the 1-D HO wave function times a Gaussian,
\begin{equation}
	E(k)=\mathcal{F}\left(e^{-x^{2}/\mu^{2}}\right)=\frac{\mu}{\sqrt{2}}e^{-(k\mu)^{2}/4}
\end{equation}
for the Fourier transform of a Gaussian, and
\begin{eqnarray}
G_{n_{b}n_{d}}(k) & = & \mathcal{F}\left(\varphi_{n_{b}}(x)\varphi_{n_{d}}(x)\right)={}
\frac{n_{b}!n_{d}!(-i)^{n_{b}+n_{d}}}{\sqrt{\pi2^{n_{b}+n_{d}+1}n_{b}!n_{d}!}}e^{-(kb)^{2}/4} \\ \nonumber
& \times & \sum_{p}\frac{(-2)^{p}}{p!(n_{b}-p)!(n_{d}-p)!}(kb)^{n_{b}+n_{d}-2p}
\end{eqnarray}
for the Fourier transform of the product of two 1-D harmonic oscillator
wave functions. The momentum integral is straightforward and it imposes the selection
rule $n_{\mu}+n_{b}+n_{d}=\textrm{even}=2s$. The final result is \cite{Egido1997}
\begin{equation}\label{eq:J}
J(n_{b},n_{d},n_{\mu})  =  \frac{\mu}{b\sqrt{2\pi}}\frac{(-1)^{s+n_{\mu}}}{G^{s+1/2}}
\left(\frac{n_{b!}n_{d}!}{n_{\mu}!}\right)^{1/2}\sum_{p}\frac{\Gamma(s-p+1/2)}{p!(n_{b}-p)!(n_{d}-p)!}\left(-G\right)^{p}
\end{equation}
with the coefficient $G=1+\frac{\mu^{2}}{2b^{2}}$.
The  quantity $ J(n_{b},n_{d},n_{\mu})$ can be written using
hypergeometric series as
\begin{eqnarray}
J(n_{b},n_{d},n_{\mu}) & = & \frac{\mu}{b\sqrt{2\pi}}\frac{(-1)^{s+n_{\mu}}}{G^{s+1/2}}
\left(\frac{1}{n_{b!}n_{d}!n_{\mu}!}\right)^{1/2}\Gamma(s+1/2)\\ \nonumber
& \times &  _{2}F_{1}(-n_{b},-n_{d},1/2-s;G)
\end{eqnarray}
The advantage of this result is the possibility to use the vast literature 
around hyper-geometric functions to find efficient and accurate ways to 
evaluate the coefficients. In the case $\mu=0$ it is even possible to 
find an explicit analytical expression for the above matrix element. In others,
use of any  of the 24 Gauss recursion relations is useful  to simplify
or make more accurate the evaluation of the $J$ coefficients.

The previous results can be used also to compute the matrix elements of
a contact interaction $\delta(\vec{r}_{1}-\vec{r}_{2})$ given the 
relation between the delta function and a Gaussian of zero range
\begin{equation}
\delta(x_{1}-x_{2})=\lim_{\mu\rightarrow0}\frac{1}{\sqrt{\pi}}\frac{e^{-(x_{1}-x_{2})^{2}/\mu^{2}}}{\mu}
\end{equation}

It is also possible to use the above results to compute the matrix elements
of the Coulomb interaction by using its expansion as a linear
combination of tempered Gaussians \cite{Quentin1972,Gir83}
\begin{equation}
\frac 1{|\overline{r}_1-\overline{r}_2|}=\frac 2{\sqrt{\pi }}\int_0^\infty
\frac{d\mu }{\mu ^2}e^{-|\overline{r}_1-\overline{r}_2|^2/\mu ^2}.
\end{equation}
Using this expansion it is possible to express the matrix element of 
the Coulomb potential $I_{abcd}^C$ in terms of the integral of 
$I_{abcd}^\mathrm{Gauss}(\mu )$ with respect to $\mu${}
\begin{equation}
I_{abcd}^C=\frac 2{\sqrt{\pi }}\int_0^\infty \frac{d\mu }{\mu ^2}%
I_{abcd}^\mathrm{Gauss}(\mu )
\end{equation}
The only dependence on $\mu $ of $I^\mathrm{Gauss}$ is  in the coefficients $J$
of Eq. (\ref{eq:J}). The integral on $\mu$ cannot be evaluated analytically,
but an appropriate change of variables and the use of the Gauss-Legendre
numerical integration allows for an accurate evaluation of the required
integrals \cite{Gir83}.

Using the same ideas of expanding in a linear combination of Gaussian
functions it is also possible to compute the matrix elements of the Yukawa 
potential in this way \cite{Quentin1972,Dobaczewski2005}

The evaluation of the matrix elements of the density dependent part of 
the interaction or the Slater approximation to the Coulomb exchange potential
have necessarily to be carried out numerically. For this purpose, the
Gauss-Hermite (or the appropriate one in the axially symmetric or spherical
case) method is very handy at a moderate computational cost.

%% file: Codes.tex
% -------------------------------------------------------------------------------------------------------
%                                                                                          Computer codes
% -------------------------------------------------------------------------------------------------------

\section{Computer codes}
\label{App:B}

To our knowledge, computer codes to carry out mean field and beyond 
mean field calculations with the Gogny force have been developed by 
several different groups: the Bruy\`eres Le Ch\^{a}tel (BlC) group in France, 
the Madrid (Mad) group in Spain, several groups in  Japan (see 
\cite{Hashimoto2013,Tagami2016} for two different implementations) and 
finally the collaborative effort around the HFBtho \cite{Per17} and 
HFBodd \cite{Schunck17} codes involving many scientists from different 
institutions. Of all the codes mentioned, only the HFBtho and HFBodd 
ones have been published \cite{Per17,Schunck17} in a journal or made publicly 
available. Therefore, after briefly discussing the published codes, we 
will be focusing in this section on the codes produced by the present 
authors which are available upon request to the authors.

The HFBtho is a computer code primarily developed to implement Skyrme HFB calculations
in an axially symmetric transformed harmonic oscillator basis. The peculiarity 
of the transformed HO it  that it allows for
more flexibility in describing the asymptotic properties of the single
particle wave functions. In the most recent version \cite{Per17}, the Gogny force has
been incorporated as an option, but in the more traditional harmonic oscillator basis.

The HFBodd is a computer code that was born as an implementation
of the HFB equation with the zero range Skyrme interaction. The recent versions of
the code are flexible enough as to break all possible symmetries including
simplex, reflection symmetry and time reversal. More recent versions also
allow the calculation of Hamiltonian overlaps and projected energies for
a variety of broken symmetries. The HFB amplitudes are expressed in a triaxial
HO basis. In the last iteration of the code \cite{Schunck17}, the Gogny force has been 
incorporated as an option into the code.

The two codes above are written in fortran90 and include lots of computational
tricks to improve performance like OpenMP or MPI for production in a 
parallel environment. However, in a single computer their performance is
far from satisfactory in contrast with the computer codes developed by
the present authors.

Along the years the authors have developed several computer codes in 
FORTRAN 77 to implement HFB and beyond calculations with the Gogny force.
Let us first mention the family of codes preserving axial symmetry (but
breaking reflection symmetry if required): 

\begin{itemize}
	\item {\it HFBax} is a HFB solver implementing all different terms of the
	 Gogny force, including Coulomb exchange and Coulomb antipairing. It
	 is based on an expansion on an axially symmetric harmonic oscillator
	 basis with a single center. Thanks to it implementation of the
	 evaluation of matrix elements using hypergeometric series (see 
	 Appendix \ref{App:A}) it can 
	 go quite far in basis size without encountering any numerical 
	 instability (see \cite{Egido1997}). It can even handle fission \cite{War02,Baran2015}
	 or even cluster emission \cite{War11} up to the scission point and 
	 beyond (two well separated fragments) without any problem. The method of choice to solve the HFB
	 equation is a second order gradient method \cite{Robledo2011} that reduces
	 substantially the iteration count. A procedure for an unattended 
	 minimization of the oscillator length parameters is also available. 
	 \item {\it Atb} is similar to HFBax but with the pecularity that time reversal invariance
	 can be broken. This opens up the possibility to solve the HFB equations 
	 with full blocking and the self-consistent treatment of  
	 odd-A nuclei \cite{Rob12,Dobaczewski2015} or even the study of high-K isomers \cite{Rob15}
	 in super-heavy nuclei. One of the key elements of this code is the
	 implementation of an orthogonality constraint to prevent the iterative
	 solution of the HFB equation to always converge to the lowest lying
	 solution for a given $K$ quantum numbers. 
	 \item {\it HFBaxT} represents the extension to finite temperature 
	 \cite{Egi00,Mar03,Mar09} of the HFBax code. It can also handle
	 fission at finite temperature \cite{Mar09} as well as the calculation
	 of level densities in the finite temperature framework.
     \item {\it PNPax} computes the particle number projected energy of a HFB wave function
     obtained with HFBax. It also allows for a search, in the spirit of the RVAP
     with the particle number fluctuation as variational parameter, of
     the lowest energy PNP solution. 
     \item {\it GCMax} is a computer code to compute the Hamiltonian overlap
     between any two general HFB wave functions obtained from HFBax. This
     computed code together with an auxilliary program (HW) can be used
     to solve the Hill-Wheeler equation of the GCM procedure for axially
     symmetric wave functions. Applications to GCM calculations with the
     octupole degree of freedom as generating coordinate \cite{Robledo2011}
     or the quadrupole and octupole degree of freedom \cite{Rodriguez-Guzman2012,Robledo2013} are
     common applications of this computer code.
\end{itemize}

A triaxial code breaking time reversal symmetry, but preserving simplex
is also available. It is dubbed {\it HFBTri } and it can handle 
both  even and odd number parity type of wave functions. Together with
the possibility to break time reversal, the handling of odd number parity
states opens up the opportunity to deal with odd mass nuclei using the
HFB blocking procedure. The constraint in orthogonality is fully implemented
as well as several other constraints on two body operators like $\langle \Delta N^{2}\rangle${}
or $\langle \Delta J_{z}^{2}\rangle$.  The method of choice to solve the HFB
equation is a second order gradient method \cite{Robledo2011} that reduces
substantially the iteration count. It shares with the {\it HFBax} the same protocols to
compute the matrix elements and therefore rather large basis sizes up to
20 shells can be used.

%% file: GWT.tex
% -------------------------------------------------------------------------------------------------------
%                                                                                Generalized Wick Theorem
% -------------------------------------------------------------------------------------------------------

\section{Overlaps of operators between HFB states: the Generalized Wick Theorem}
\label{App:C}

In the implementation of symmetry restoration and/or configuration 
mixing the evaluation of overlaps of operators between arbitrary HFB 
states is required. This is so because the symmetry operators are 
exponentials of one body operators (which are part of the Lie algebra 
generating the HFB states) and therefore its action on arbitrary HFB 
states leads again to HFB states.

The evaluation of the overlaps is greatly simplified by using the Generalized Wick Theorem
(GWT). The theorem, derived in many different ways \cite{Oni66,Balian.69,Hara95,Ber12}, 
states that generic overlaps of the form
\begin{equation}
\frac{\langle \Phi | \hat O | \Phi ' \rangle}{\langle \Phi | \Phi ' \rangle}
\end{equation}
can always be written as the sum of all possible two-quasiparticle contractions 
\begin{eqnarray}\label{eq:cont}
\frac{\langle \Phi | \beta_\mu \beta_\nu | \Phi ' \rangle}{\langle \Phi | \Phi ' \rangle} & = & C_{\mu \nu}\\ 
\frac{\langle \Phi | \beta_\mu \beta_\nu^\dagger | \Phi ' \rangle}{\langle \Phi | \Phi ' \rangle} & = & \delta_{\mu \nu} \\
\frac{\langle \Phi | \beta_\mu^\dagger \beta_\nu^\dagger | \Phi ' \rangle}{\langle \Phi | \Phi ' \rangle} & = & 0 
\end{eqnarray}
In the above expressions $|\Phi\rangle$ and $|\Phi '\rangle$ are arbitrary, 
non orthogonal, HFB wave functions
with amplitudes $U$, $V$ and $U'$, $V'$ respectively and $\hat{O}$ is a 
general multi-body operator.
Typically, the operator
$\hat{O}$ is a linear combination of  products of creation and annihilation 
quasiparticle operators belonging to a given
HFB state. To simplify the results, we will assume that those quasiparticle operators are
the ones associated to $|\Phi\rangle$. As an example, considering $\hat O$ to be a
one-body operator we obtain 
 \begin{eqnarray}
\frac{\langle \Phi | \hat O | \Phi ' \rangle}{\langle \Phi | \Phi ' \rangle} & = &  \langle \Phi | \hat O | \Phi \rangle
+ \frac{1}{2} \sum_{mu nu} O^{20}_{\mu \mu }\frac{\langle \Phi | \beta_{\mu} \beta_{nu} | \Phi ' \rangle}{\langle \Phi | \Phi ' \rangle} \\ \nonumber{}
& = &  \langle \Phi | \hat O | \Phi \rangle
+ \frac{1}{2} \sum_{mu nu} O^{20}_{\mu \mu }C_{\mu \nu}
\end{eqnarray}

The only non-trivial contraction required in the previous overlap is 
given in terms of the skew-symmetry matrix $C_{\mu \nu}$ which is the 
product of two matrices
\begin{equation}
C_{\mu \nu} = A^{-1}B 
\end{equation}
where $A=U^\dagger U' + V^\dagger V'$ and $B=U^\dagger V' + V^\dagger U'$. Using the GWT we can
write, for instance, the non-trivial matrix element entering the evaluation of the
Hamiltonian overlap as 
\begin{equation}
\frac{\langle \Phi | \beta_\mu \beta_\nu \beta_\sigma \beta_\rho | \Phi ' \rangle }
{\langle \Phi | \Phi ' \rangle } =
C_{\mu \nu} C_{\sigma \rho} - C_{\mu \sigma} C_{\nu \rho} + C_{\mu \rho} C_{\nu \sigma}.
\end{equation}
The overlap between the two HFB wave functions is given by the expression
\begin{equation}
\langle \Phi | \Phi ' \rangle = \pm \sqrt{\det A}.
\end{equation}
The presence of the square root implies an indeterminacy in the sign that 
requires further consideration. The importance
of the undetermined sign comes from the fact that the overlap is part of the integrand of
the integrals characteristic of the symmetry restoration or configuration mixing methods. A
wrong assignment of the sign can substantially change the value of the integral.
The sign problem has been addressed in the past using different strategies
\begin{itemize}
\item Continuity: The wave function $|\Phi '\rangle$ belongs to a dense 
set that includes $|\Phi\rangle$. By fixing the initial sign by  
$\langle \Phi | \Phi  \rangle =1$, the one of $ \langle \Phi | \Phi ' 
\rangle $ for a $|\Phi '\rangle$ close to $|\Phi \rangle$ is fixed by 
continuity. The method is complemented by the knowledge of the 
derivative of the overlap with respect to some parameter $\alpha$ 
labeling the set of the $|\Phi '\rangle$. However, the method requires 
an adaptive mesh to work in arbitrary situations. 
\item Time reversal 
invariant systems: in those cases where there is Kramer's degeneracy 
the Bogoliubov amplitudes have a block structure that also renders the 
matrix $A$ as a block matrix with equal entries $\bar{A}$ in the 
diagonal. As a consequence the $\det A$ is given by a square $\det A = 
\det^2 \bar{A}$ and therefore $\langle \Phi | \Phi ' \rangle = \det 
\bar{A}$. 
\item Neergard and W\"ust method: Factoring out the matrices 
$U^\dagger$ and $U'$ off the matrix $A$ we can express the overlap in 
terms of the determinant $\det (1+Z^\dagger Z')$ where the skew 
symmetric matrices are simply given by $Z=VU^{-1}$. As shown in Ref. 
\cite{Neergard1983} the matrix $Z^\dagger Z' $ has doubly degenerate 
eigenvalues and again $\det (1+Z^\dagger Z')$ can be written as a 
square. 
\item Pfaffian method: Using techniques of fermion coherent 
states it is possible to compute the overlap of two arbitrary HFB 
states without any ambiguity in the sign \cite{Robledo2009}. The final 
results is expressed in terms of the Pfaffian of a skew-symmetric 
matrix. The Pfaffian of a skew-symmetrix matrix is defined in a similar 
fashion to the determinant of an arbitrary matrix and shares with the 
determinant many properties. An interesting property of the Pfaffian is 
the one connecting the Pfaffian with the square of a determinant. The 
numeric evaluation of the Pfaffian has a computational cost very 
similar to the one of the determinant and therefore it represents a very useful 
alternative to any of the methods mentioned above 
\cite{Gonzalez-Ballestero2010}.
\end{itemize}

There are a couple of difficulties in the application of the GWT or 
overlap formulas in practical implementations. One has to do with the common 
situation where the occupancies $v_{k}^{2}$ of the HFB wave functions involved are 
close to  one. In this case  some of the expressions involving the inverse of the 
$U$ amplitudes are not well defined. This problem has been 
solved in the literature \cite{Bonche1990,Val01,Rob11a}. Also 
useful formulas to deal with unoccupied (and therefore irrelevant) 
states have been given. The other difficulty is connected with the 
situation where the overlap of the two HFB wave functions is zero and 
therefore the contractions of the GWT as given in Eq (\ref{eq:cont}) are 
ill defined (in fact, divergent). This might be thought as an expectional
situation but it is indeed common in PNP or in other similar contexts 
involving cranking wave functions \cite{Anguiano2001,Oi05}. The 
consequences of this failure of the GWT have been discussed in many 
places \cite{Anguiano2001,Lacroix2009} in the context of EDFs and is 
still an unresolved aspect of the theory that has to be further 
clarified. If the overlap $\langle \Phi | \hat O | \Phi ' \rangle$ is 
still required, a way to compute it is going to a sort of canonical 
basis \cite{Taj92,Lacroix2009} or use the Pfaffian technique as 
discussed in \cite{Ber12}. In the latter reference, a way to compute 
efficiently overlaps of multiquasiparticle excitations of the different 
HFB vacuums without the combinatorial explosion characteristic of the 
GWT has been considered. It is again based on the Pfaffian ideas and 
could represent a breakthrough in applications of the projected shell 
model of Hara and Sun~\cite{Hara95}.

Another practical aspect of the GWT has to do with the common situation
when the two HFB wave functions are expanded on different single particle basis not
spanning the same subspace of the whole Hilbert space. In this case,
the formulas above have to modified to take into account this peculiarity. The
idea is to formally enlarge both basis as to expand the same subspace but setting
to zero the occupancy of the extended orbitals \cite{Bonche1990,Robledo1994,Ber12}.

In applications involving statistical ensembles, where traces with a 
density matrix operator are used instead of mean values, it is still possible 
to used the ideas of the GWT in those cases where a given operator is 
multiplied by the exponential of an one body operator (a symmetry 
transformation or a Bogoliubov transformation operator). The corresponding 
expression is similar to the one of the GWT (linear combination of 
contractions) but the contractions are given by a different expression 
\cite{Gau60,Per07}.

%% file: Abbreviations.tex
% -------------------------------------------------------------------------------------------------------
%                                                                                          Abbreviations
% -------------------------------------------------------------------------------------------------------

\section{Abbreviations and acronyms}
\label{App:D}

In the following table a list of abbreviations and acronyms used in the paper is given

\begin{tabular}{l|l}
5DCH     & 5D Collective Hamiltonian                          \\
AMP      & Angular momentum projection                        \\
AM-PAV   & Angular momentum PAV                               \\
AM-VAP   & Angular momentum VAP                               \\
BCS      & Bardeen Cooper Schrieffer                          \\
BHF      & Bruckner HF                                        \\
CHFB     & Cranked Hartree Fock Bogoliubov                    \\
COM      & Center of Mass                                     \\
CSE      & Collective Schr\"odinger equation                  \\
EDF      & Energy density functional                          \\
EFA      & Equal filling approximation                        \\
EOS      & Equation of state                                  \\
FTHFB    & Finite temperature HFB                             \\
GCM      & Generator coordinate method                        \\
GOA      & Gaussian overlap approximation                     \\
HFB      & Hartree Fock Bogoliubov                            \\
HF       & Hartree Fock                                       \\
HO       & Harmonic Oscillator                                \\
HWG      & Hill-Wheeler-Griffins                              \\
IBM      & Interacting boson model                            \\
LN       & Lipkin Nogami                                      \\
PAV      & Projection After Variation                         \\
PES      & Potential energy surface                           \\
PLN      & Projected LN                                       \\
PNAMP    & Particle number AMP                                \\
PN-PAV   & Particle number PAV                                \\
PNP      & Particle Number Projection                         \\
PN-VAP   & Particle number VAP                                \\
PPNAMP   & Parity PNAMP                                       \\
QRPA     & Quasiparticle RPA                                  \\
RMF      & Relativistic Mean Field                            \\
RPA      & Random Phase Approximation                         \\
R-PN-VAP & Restricted PN-VAP                                  \\
RVAP     & Restricted VAP                                     \\
SCCM     & Symmetry conserving configuration mixing           \\
SCK2     & Self-consistent Kamlah 2                           \\
TRSC     & Time reversal symmetry conserving                  \\
TR       & Time reversal                                      \\
VAP      & Variation After Projection                         
\end{tabular}

The parametrizations of the Gogny force discussed in Sec. \ref{Sec:Gogny_force}
are also listed here for a compilation: 

\begin{itemize}
\item Traditional D1 family: D1, D1', D1S, D1N, D1M, D1P,  DS280, D1M$^{*}$
\item New D2 family: D2
\item Including tensor term: D1ST, D1MT, D1ST2a, D1ST2b, D1MT2c, GT2
\item Similar to Gogny: SEI
\end{itemize}